\documentclass[12pt,epsf,amssymb]{article}

\usepackage{graphicx}
\usepackage{amsfonts}
\usepackage{amsmath}
\usepackage[dvipsnames]{xcolor}
\usepackage{amssymb}
\usepackage{amsthm}
\usepackage[mathscr]{eucal} 
\usepackage{url}
\usepackage[bookmarks, colorlinks]{hyperref}
\hypersetup{
	linkcolor=blue,
	citecolor=blue,
	filecolor=black,
	urlcolor=blue}

\def\N{\mathbb{N}}
\def\Z{\mathbb{Z}}
\def\Q{\mathbb{Q}}
\def\R{\mathbb{R}}
\def\C{\mathbb{C}}
\def\P{\mathbb{P}}

\begin{document}

\baselineskip 0.6cm
\newcommand{\nequiv}{\mbox{\ooalign{\hfil/\hfil\crcr$\equiv$}}}
\newcommand{\nsupset}{\mbox{\ooalign{\hfil/\hfil\crcr$\supset$}}}
\newcommand{\nni}{\mbox{\ooalign{\hfil/\hfil\crcr$\ni$}}}
\newcommand{\nin}{\mbox{\ooalign{\hfil/\hfil\crcr$\in$}}}

\newcommand{\vev}[1]{ \left\langle {#1} \right\rangle }
\newcommand{\bra}[1]{ \langle {#1} | }
\newcommand{\ket}[1]{ | {#1} \rangle }
\newcommand{\Dsl}{\mbox{\ooalign{\hfil/\hfil\crcr$D$}}}
\newcommand{\Slash}[1]{{\ooalign{\hfil/\hfil\crcr$#1$}}}
\newcommand{\EV}{ {\rm eV} }
\newcommand{\KEV}{ {\rm keV} }
\newcommand{\MEV}{ {\rm MeV} }
\newcommand{\GEV}{ {\rm GeV} }
\newcommand{\TEV}{ {\rm TeV} }

\def\diag{\mathop{\rm diag}\nolimits}
\def\tr{\mathop{\rm tr}}

\def\Spin{\mathop{\rm Spin}}
\def\SO{\mathop{\rm SO}}
\def\O{\mathop{\rm O}}
\def\SU{\mathop{\rm SU}}
\def\U{\mathop{\rm U}}
\def\Sp{\mathop{\rm Sp}}
\def\SL{\mathop{\rm SL}}

\def\change#1#2{{\color{blue}#1}{\color{red} [#2]}\color{black}\hbox{}}

\theoremstyle{definition}
\newtheorem{thm}{Theorem}[section]
\newtheorem{defn}[thm]{Definition}
\newtheorem{exmpl}[thm]{Example}
\newtheorem{props}[thm]{Proposition}
\newtheorem{lemma}[thm]{Lemma}
\newtheorem{rmk}[thm]{Remark}
\newtheorem{notn}[thm]{Notation}
\newtheorem{qstn}[thm]{Question}
\newtheorem{excs}[thm]{Exercise}
\newtheorem{cor}[thm]{Corollary}
\newtheorem{conj}[thm]{Conjecture}
\newtheorem{anythng}[thm]{}

 \begin{titlepage}
  
 
  \vskip 1cm
  \begin{center}
  {\bf {\large Notes on Characterizations of 2d Rational SCFTs: Algebraicity, Mirror Symmetry and Complex Multiplication}}
  \vskip 1.2cm
 
Abhiram Kidambi,$^{1,2}$ Masaki Okada$^3$ and Taizan Watari$^3$
 
 \vskip 0.4cm
  {\it $^1$ Max Planck Institute for Mathematics in the Sciences, \\ Inselstrasse 22, 04103 Leipzig Germany

 $^2$ Erwin Schr\"odinger Institute, Boltzmanngasse 9A, 1090 Vienna, Austria

    $^3$
    Kavli Institute for the Physics and Mathematics of the Universe, 
    University of Tokyo, Kashiwa-no-ha 5-1-5, 277-8583, Japan
   }

   
 \abstract{
These notes combine results from two papers by the present authors viz.,
Part I (arXiv:2205.10299) and Part II (arXiv:2212.13028) into one streamlined
version for better readability, along with a review on theory of complex
multiplication for non-singular complex projective varieties and complex tori
that is aimed at string theorists.
We think that it is worth posting this edition as a separate entry in
arXiv for those reasons, although this edition contains no essential
progress beyond Part I and Part II. \\[0.2cm]

S.\ Gukov and C.\ Vafa proposed 
 a characterization of rational $N=(1,1)$ superconformal field theories (SCFTs) on 1+1 dimensions with Ricci-flat K\"{a}hler target spaces in terms of the Hodge structure of the target space, extending 
 an earlier observation by G.\ Moore.
We refined this idea and obtained a conjectural statement on necessary and sufficient conditions for 
such SCFTs to be rational, which we indeed prove to be true in the case the target space is $T^4$.
In the refined statement, the algebraicity of the geometric data of the target space turns out to be essential, and the Strominger--Yau--Zaslow fibration in the mirror correspondence also plays a vital role.
 } 
 \end{center}
 \end{titlepage}

 \tableofcontents
 

\section{Introduction}

This manuscripts combines the materials and results in
 \cite{Kidambi:2022wvh} and \cite{Okada:2022jnq} into a single, streamlined vesion together with an extensive pedagogical section. Portions of discussions in \cite{Kidambi:2022wvh} 
have been dropped, while more review materials have been added
(Expansion of \S \ref{sssec:Moore-GV-idea} and inclusion of appendix \ref{sec:guide}).

\subsection{String Theoretic Introduction}
\label{ssec:Phys-Intro}

\subsubsection{Motivation and Background}
\label{sssec:Intro-background}

$N=(1,1)$ supersymmetric non-linear sigma models in $1+1$ dimensions (2$d$)
have a non-trivial moduli space when the target space $M$ is a Ricci--flat 
K\"{a}hler manifold. These theories, which are superconformal field 
theories (SCFTs), define a compactification of Type II string theory. 
These SCFTs are \textit{rational} CFTs\footnote{
There are many ways to define a rational CFT. A simple way of explaining 
what rational CFTs are to string theorists is that the left-moving and the 
right-moving superchiral algebras both have a finite number of irreducible 
representations (see \cite{Moore:1989vd} and references therein). 
}
 only at special points in the moduli space. The question that this note addresses is: \emph{At what points in the moduli 
space of $M$ are those SCFTs rational?}

A complete analysis has been done for the case of $M=T^n$ in the bosonic string and superstring case in \cite{Wendland:2000ye}, and in the Heterotic 
string \cite{Moore:1998pn,Moore:1998zu}.
It seems almost impossible to go beyond 
toroidal compactifications. G. Moore pointed out, however, that 
$M=T^2$-target bosonic and superstring are rational SCFTs if and only if 
both the elliptic curve $M$ (as a complex manifold) and its mirror $W$ 
have complex multiplications (reviewed in
section \ref{sssec:Moore-GV-idea}), and 
the imaginary quadratic fields of $M$ and $W$ are isomorphic to each other.  
In Gepner constructions for K3 and Calabi--Yau compactifications, which 
are also rational SCFTs, the non-trivial rational Hodge structure of the 
cohomology groups are also characterized by {\it number fields} 
(finite dimensional algebraic extension fields over the field of rational 
numbers $\Q$). 
Based on these observations, S. Gukov and C. Vafa 
conjectured the following \cite{Gukov:2002nw}:
\begin{conj}
\label{conj:GV-original} \cite[\S7]{Gukov:2002nw}
{\it Consider the $N=(1,1)$ supersymmetric non-linear sigma model 
for a Ricci-flat K\"{a}hler manifold $(M;G,B;I)$, where 
$M$ is a differentiable manifold of real $2n$-dimensions, 
$G$ a Riemannian metric on $M$, $B$ a closed 2-form on $M$ (called  \underline{the $B$-field}) 
and $I$ a complex structure so that $(M;G;I)$ is K\"{a}hler. 
This SCFT is rational if and only if (i) both the complex manifold 
$(M;I)$ and its mirror $(W;I^\circ)$ are of CM-type, and (ii)
the endomorphism fields of $(M;I)$ and $(W;I^\circ)$ are isomorphic. }
\end{conj}
\noindent
Being {\it CM-type} is a property of complex structure of a manifold 
(or of Hodge structure on a cohomology group) that generalizes the 
notion of complex multiplication of $M=T^2$. We believe that 
the \autoref{sec:guide} of this manuscript (combined with 
the appendix of of \cite{Okada:2023udq}) 
will be useful for string theorists who are not familiar with 
the theory of complex multiplication beyond the case of elliptic curves covered 
in \cite{Moore:1998pn}, \cite{Moore:1998zu}. 

Giving (and establishing) such a characterization is a well-defined 
question in mathematical physics, which may be of interest in its own right.  
Apart from the idea in Conj. \ref{conj:GV-original}, our knowledge of rational 
SCFTs has been mostly construction based. 
Not much is known beyond the Gepner constructions, and  
lattice vertex operator algebras and their orbifolds. 
If the criteria for rationality of SCFTs are proven to be something close to 
the one in Conj. \ref{conj:GV-original}, that means that there are more 
rational SCFTs than those obtained by these constructional approaches. 
Those rational SCFTs will include the ones in the small-volume limit 
region in the moduli space; rational SCFTs will be an ideal (and a rare) 
tool to study how string theory captures geometry at the short-distance 
(high-energy) region in a situation where classical Einstein gravity 
is not a good approximation.\footnote{
Open problem: Understand how close such rational points are to each other in the moduli space of SCFTs.
}

Further progress can be expected in a couple of other directions, 
when the observation in Conj. \ref{conj:GV-original} is understood 
better, with more systematically constructed examples of rational SCFTs 
with geometric interpretations. For example, on the side of arithmetic 
geometry, it is known that complex analytic CM-type manifolds $M$ are 
known to admit arithmetic models \cite{Schnider1937,shimura2016abelian,piatetski1973arithmetic,rizov2005complex}, 
at least when $M$ is either an abelian variety or a K3 surface, and
some of the $L$-functions defined for these arithmetic models are expected to have modular transformation properties.
It will be an interesting subject of research to explore relations  
between such modular objects in arithmetic geometry and $g=1$ 
chiral correlation functions of the corresponding rational SCFTs in string 
theory. References \cite{Kondo:2018mha,Kondo:2019jpi} studied that for the case 
where the target space is a CM elliptic curve. To undertake a similar study 
for abelian varieties and K3 surfaces, a better understanding of 
the relation between CM manifolds and rational SCFTs is a necessary starting point. 

In particle phenomenology, too, the work of Ref. \cite{Gukov:2002nw} is of 
significance.  
In Type IIB Calabi--Yau orientifold compactifications, the gravitino 
mass and the cosmological constant are not generically 
much smaller than the Planck scale of the effective theory on $3+1-$dimensions
due to non-zero fluxes \cite{Denef:2008wq}. When the complex structure of 
the Calabi--Yau threefold has period integrals characterized by 
number fields (such as in Refs. \cite{Gukov:2002nw, DeWolfe:2004ns, Aspinwall:2005ad}), 
then the gravitino mass can be much smaller than the Planck scale 
for a much larger fraction of flux configurations \cite{Moore:2004fg, 
DeWolfe:2004ns, Kanno:2017nub}. 
If the observation by Gukov--Vafa is true, then we may attribute particle phenomenologies 
such as electroweak gaugino dark matter, gauge coupling unification and 
small gravitino mass, to large chiral algebra on the worldsheet theory, 
and not to a larger symmetry of the spacetime field theory. 

\subsubsection{Prior-, Present-, and Future Work}
\label{ssec:Intro-What-We-Do}

The characterization conditions in Conj. \ref{conj:GV-original} have been 
tested 
only for $M=T^2$ and a few Gepner constructions. An obvious 
direction of research to test if those conditions are satisfied for other 
known constructions; such a study has not been done systematically 
even for $M=T^4$ so far (apart from one article \cite{Chen:2005gm}, on which 
we say more shortly). Whereas Ref. \cite{Wendland:2000ye} already 
worked out in the case of $M=T^m$ when the bosonic CFT (and SCFT) 
of $(M;G,B)_{M=T^m}$ is rational, the conditions on $G$ and $B$ are 
stated there in a language that makes sense only for tori, not in 
a language as in Conj. \ref{conj:GV-original} that makes sense for a broader 
class of target-space manifolds. One may also be interested in 
whether all the data $(M;G,B;I)$ satisfying the conditions in 
Conj. \ref{conj:GV-original} yield a rational SCFT. One may further be interested 
in constructing such rational SCFTs explicitly. This would broaden the 
list of rational SCFTs tremendously. 

In being realistic and formal, however, and one can
easily find that some concepts in Conj. \ref{conj:GV-original} need to be clarified 
and the conditions there to be stated more precisely, in order for the 
Conjecture to be testable even in the case $M = T^4$ or a Calabi--Yau threefold. Refinement of these ideas formally will allow us come up with a functioning version of the conjecture which we can test, prove or disprove.    

Reference \cite{Chen:2005gm} is the only published record\footnote{
Reference \cite{Okada:2023udq} is also an effort to make progress along 
this line. It addresses a math problem 
(Question \ref{qstn:weak-vs-srong} in p. \pageref{pg:Qstn-strng-vs-weak}) 
related to Conj. \ref{conj:GV-original}.
} %
the present authors are aware of that has pushed the frontier 
forward along the line of the idea in Conj. \ref{conj:GV-original} after 
Ref. \cite{Gukov:2002nw}. The work \cite{Chen:2005gm}
by Meng Chen in fact contains an example of $(M;G,B)|_{M=T^4}$
that seems to have the properties stated in Conj. \ref{conj:GV-original}
but the corresponding CFT/SCFT is \emph{not} rational. The present 
authors do not regard the example as an indication that the idea 
in Conj. \ref{conj:GV-original} is entirely wrong, but that the conditions 
stated there are not enough. We will explain in more details 
in \autoref{sec:metric} in this article.  

In this article, we begin by pointing out the aspect in which Conj. \ref{conj:GV-original} needs to be refined 
for it to be a functioning conjectural statement applicable to 
a general Ricci-flat K\"{a}hler manifold (section \ref{sssec:need2refine}). 
\label{pg:no-written-record}
Although those observations are not particularly difficult, 
discussions pertaining to these observations have (as far as the authors are aware) not been recorded. It is therefore a task worth undertaking. 

We will use the case $M=T^4$, where it is known which sets 
of data $(M;G,B)|_{M=T^4}$ are for rational SCFT \cite{Wendland:2000ye}, 
to refine the idea Conj. \ref{conj:GV-original} both in concepts as well as in 
the precise choice of conditions. The \emph{refined conjecture} which will arrive at is the following: 
\begin{conj}
(for self-mirror $M$): 
\label{conj:forT4}
{\it Let $M$ be a real $2n$-dimensional manifold which admits a Ricci-flat 
Riemannian metric $G$ such that there exists a complex structure $I'$ 
on $M$ with which $(M;G;I')$ is K\"{a}hler. Let $B$ be a closed 2-form on $M$. 
Suppose, further, that the family of such $(M;G,B)$ is self-mirror 
in that $h^{p,q}(M) = h^{n-p,q}(M)$ for $0 \leq p, q\leq n$. 

The non-linear sigma model $N=(1,1)$ SCFT associated with the data 
$(M; G, B)$ is a rational SCFT if and only if the following conditions 
are satisfied: }
\begin{enumerate}
\item {\it There exists a polarizable complex structure $I$ so that 
$(M; G; I)$ is K\"{a}hler, $B^{(2,0)}=0$, and the following four conditions 
are satisfied; }
\item {\it The rational Hodge structure on $H^*(M;\Q)$ is of strong CM-type; }
\item {\it The complexified K\"{a}hler parameter $(B+i\omega)$, where 
 $\omega(-,-) := 2^{-1}G(I-,-)$ is the K\"{a}hler form, is in the 
algebraic part  $(H^2(M;\Q) \cap H^{1,1}(M;\R))\otimes \tau^r_{(n0)}(K^r)$, 
where $K^r$ is the endomorphism CM field of the level-$n$ 
rational Hodge structure $[H^n(M;\Q)]_{\ell=n}$ and $\tau^r_{(n0)}$ its 
embedding associated with the Hodge $(n,0)$ component; }
\item {\it There exists a geometric SYZ-mirror of the $N=(1,1)$ SCFT;
  there may be more than one; the data of such a mirror is denoted by 
  $(W;G^\circ, B^\circ; I^\circ)$; }
\item (Strong) {\it For \underline{any one} of the geometric SYZ-mirror SCFTs, there is 
a Hodge isomorphism $\phi^*: H^*(M;\Q) \rightarrow H^*(W;\Q)$ such that 
$\phi^*$ is the identity map on the vector subspaces 
$\pi_M^*: H^*(B;\Q) \hookrightarrow H^*(M;\Q)$ and 
$\pi_W^*: H^*(B;\Q) \hookrightarrow H^*(W;\Q)$; here, 
$\pi_M: M \rightarrow B$ and $\pi_W: W \rightarrow B$ are the 
SYZ $T^n$-fibrations over a common base manifold $B$ of real dimension $n$. 
In other words, }
\begin{align}
  \phi^* \circ \pi_M^* = \pi^*_W  \qquad \qquad \left( {\rm i.e.,~}
  \phi^*|_{\pi^*(H^*(B;\Q))} = {\rm id}_{\pi^*(H^*(B;\Q))} \right). 
  \label{eq:cond-8}
\end{align}
\setcounter{enumi}{4}
\item (Weak) 
{\it There \underline{exists} a geometric SYZ-mirror SCFT for which 
there is a Hodge isomorphism $\phi^*: H^*(M;\Q) \rightarrow H^*(W;\Q)$ 
such that $\phi^*$ is the identity map on the vector subspaces 
$\pi_M^*: H^*(B;\Q) \hookrightarrow H^*(M;\Q)$ and 
$\pi_W^*: H^*(B;\Q) \hookrightarrow H^*(W;\Q)$. }
\end{enumerate}
\end{conj}
\noindent
In these notes, we will prove Conj. \ref{conj:forT4} when $M=T^4$. This version Conj. \ref{conj:forT4} makes 
sense only for families of $M$ that are self-mirror.
So, as a wild speculation, a couple of trial versions of the 
characterization conditions are presented also for cases for 
$M$ that are not self-mirror (Conj. \ref{conj:gen} in 
section \ref{sec:towards}). 

The characterization condition 2 for a rational SCFT in  
Conj. \ref{conj:forT4} was already in Conj. \ref{conj:GV-original}; 
this article only adds a proof that it is a part of necessary and sufficient 
conditions in the case $M=T^4$. Condition 5 (weak/strong) in Conj. \ref{conj:forT4} 
is stronger than condition (ii) in Conj. \ref{conj:GV-original}; 
without the extra property (\ref{eq:cond-8}), we find that there are sets of data 
$(M;G,B;I)|_{M=T^4}$ satisfying all the other conditions in Conj. \ref{conj:forT4}, 
whereas the corresponding CFT/SCFTs are not rational (see section \ref{ssec:one-more} 
(section \ref{ssec:illustrate} in particular)). 
This observation also proves that the 
SYZ mirror fibration (not just mirror symmetry) plays a vital role 
in characterizing the geometric data for SCFTs that are rational. 
Conditions 1 and 3 both state that the geometric data $(M;G,B;I)$ 
for a rational SCFT have to be not just complex analytic Riemannian 
manifolds but have an implementation as algebraic varieties, and have 
K\"{a}hler metrics that are also of algebraic nature (should be in the 
cone generated by divisors); the results of   
Meng Chen \cite[Thm. 2.5 \& Prop. 3.10]{Chen:2005gm} 
play a vital role in finding that condition 1 in Conj. \ref{conj:forT4} is essential 
(see discussion in section \ref{ssec:pol}), and the present authors read 
that the apparent counter example of \cite[\S4]{Chen:2005gm} to 
Conj. \ref{conj:GV-original} as the necessity for adding condition 3 in 
Conj. \ref{conj:forT4}. Overall, we observe that all of complex 
multiplications, algebraicity of the target manifold and the SYZ-mirror
fibration morphisms are a vital part of the conditions for the 
corresponding SCFTs to be rational, at least for the case $M=T^4$, 
and possibly for a broader class of target manifolds if the version 
Conj. \ref{conj:forT4} (or Conj. \ref{conj:gen}) is true.  

It may seem that a case study for
$M=T^4$ is nothing more than a straightfoward generalization of the
case $M=T^2$. In fact, it is not. Conceptual issues such as polarizability
of complex structure (condition 1) and the difference between the
algebraic part $H^2(M;\Q) \cap H^{1,1}(M;\R)$ and the whole $H^2(M;\Q)$
(condition 3) are non-trivial when $M=T^4$ and K3, while they are not
when $M=T^2$. The case study with $T^4$ is practically the best way to figure
out the right choice of statements when it comes to those conceptual issues.

Our discovery using $T^4$ that condition (\ref{eq:cond-8})---best
stated in terms of the Strominger--Yau--Zaslow torus fibration---is
a vital part of the suffcient condition lends support for the idea
that mirror symmetry might play an important role in the characterization
of rational CFTs (Conj. \ref{conj:GV-original}). If the characterization
conditions were to involve only the rational Hodge structure of
the cohomology group $H^*(W;\Q)$ of the mirror manifold, the same
conditions may well be translated into those on $H^*(M;\Q)$ with
one more set of rational Hodge structure without referring to
a mirror geometry (cf. \cite{Kidambi:2022wvh} and references therein).
We now know that necessary and sufficient conditions should involve
both the rational Hodge structure on $H^*(W;\Q)$ and the compatibility
condition (\ref{eq:cond-8}), at least when $M=T^4$.
It is therefore much more economical to state the conditions by referring
to mirror symmetry. 

The goal of these notes is to find a version of statements
(such as Conj. \ref{conj:forT4} and Conj. \ref{conj:gen}) that is
testable and is not ``obviously" false for a broader class of Ricci-flat
K\"{a}hler manifolds $M$, for which purpose we used a case study with $T^4$.
Although section \ref{sec:towards} of this article runs a few checks
on formal perspectives, providing positive and concrete evidence for
Conj. \ref{conj:forT4} and \ref{conj:gen} in string theory (or even
at the level of mathematical rigour)
with $M$ such as a K3 surface and Calabi--Yau manifolds
is beyond the scope of these notes. The word ``Conjecture''
is synonymous to ``A Testable Version of the Statements''
in these notes. 

\subsection{Mathematical Introduction and Summary}
\label{ssec:Math-Intro}

Superstring theory does a couple of things. One among them is 
to set up a dictionary between the three totally different 
setups: physical systems in a spacetime, modular invariant 
superconformal field theories on Riemann surfaces, and geometric data of a compactification manifold. 
The latter two can be treated formally in a purely mathematical framework. 
So string theory provides a non-trivial map between the 
moduli spaces of SCFTs, and geometric data. 
Even without the physical motivation described in 
section \ref{sssec:Intro-background}, these maps between the two aforementioned objects can be a question of mathematical interest. 

To be more specific, let $M$ be a manifold of certain topology that 
admits a Riemannian metric $G$ and complex structure $I$ so that $(M;G;I)$ 
is K\"{a}hler. For such a manifold $M$, we have a moduli space of 
$(M;G;B)$ of Ricci-flat Riemannian metric $G$ and a closed 2-form $B$ on $M$. 
There is also a moduli space of the SCFTs given by the $N=(1,1)$ non-linear 
sigma model for the target space $(M;G;B)$, and there is a 
map between the two moduli spaces. The question to be addressed 
in this article is to identify in the moduli space of geometric data 
the points that are mapped to the moduli space of SCFTs where the SCFTs 
are rational.

Virtually nothing has been known on this question, except for the complete 
answer in the case of $M=T^m$ (see \cite{Moore:1998pn,Moore:1998zu,Wendland:2000ye}), 
some scattered examples of other rational SCFTs 
and an idea Conj. \ref{conj:GV-original} for a possible answer to the question 
intended for manifolds $M$ of more general topological types. 
From the point of view of mathematical formality there is, however, much room to polish the idea 
in Conj. \ref{conj:GV-original}, for the reasons reviewed in 
section \ref{sssec:need2refine}.  
These notes primarily report research in string theory, where the 
authors find most value in identifying the way one should polish 
and refine the idea Conj. \ref{conj:GV-original} into a form (or forms) of 
mathematical conjectures that are functional and testable. 
For that reason, Conj. \ref{conj:forT4} (and Conj. \ref{conj:gen})
is the main string theoretic result. One will be able 
to come up with various ideas on further research in string theory 
made possible by those Conjectures. 
It is not hard to imagine, at the same time, that readers with a mathematical background would 
be interested in whether any solid result has been obtained here. 
From that perspective, the following is the main result:
\begin{thm}
\label{thm:forT4}
{\it There is a proof for Conj. \ref{conj:forT4} for the case $M=T^4$, 
where $n=2$.
Conditions 1--5 (strong) on $(M;G,B;I)|_{M=T^4}$ are necessary when 
the corresponding SCFT is rational, and the set of conditions 1--5 (weak) 
is sufficient for the corresponding SCFT to be rational. }
\end{thm}
\noindent 
Note that the $N=(1,1)$ SCFT for $(T^m;G,B)$ is rational 
if and only if the bosonic CFT for $(T^m;G,B)$ is rational. 

\subsection{How to Read These Notes}
\label{ssec:read-guide}

We begin in section \ref{sssec:need2refine} with identifying 
which aspects of the original idea Conj. \ref{conj:GV-original}
need to be refined (cf p. \pageref{pg:no-written-record}). 
After some review materials in sections \ref{sssec:Moore-GV-idea}, 
\ref{ssec:torus-RCFT}, \ref{ssec:CM-abel-surface} and the 
appendix \ref{sec:guide}, we will prove that conditions 1--5 
are {\it necessary} for the $N=(1,1)$ SCFT of $(T^4;G,B)$ to be rational  
in sections \ref{sec:choose-cpx-str}--\ref{sec:SYZ-mirror} (except 
section \ref{ssec:1st-step-4-converse}). The proof that the set of 
conditions 1--5 is also {\it sufficient} is completed within 
sections \ref{ssec:1st-step-4-converse}--\ref{ssec:one-more}.
Some of necessary technical calculations are found in the 
appendix \ref{sec:details}. 
 
In section \ref{sec:towards}, we present a few variations of how 
Conj. \ref{conj:forT4} may be generalized to a case the target-space 
manifold $M$ is not necessarily self-mirror; the arguments there 
are a little more speculative. Section \ref{ssec:distrib} collects 
immediate consequences that we can extract after the study in 
sections \ref{sec:choose-cpx-str}--\ref{sec:towards}, assuming that 
Conj. \ref{conj:forT4} is true. 

String theorists who just want to dive into the main content of these notes are encouraged to have a look at Conj. \ref{conj:GV-original}, Conj. \ref{conj:forT4}, 
Thm. \ref{thm:forT4} and Conj. \ref{conj:gen}.
Section \ref{sssec:need2refine}, Rmk. \ref{rmk:cntr-ex-MC} and 
section \ref{ssec:illustrate} are also recommended to a string theorist. 
The lead-in paragraphs of sections \ref{sec:choose-cpx-str} and \ref{sec:metric}, 
and also section \ref{ssec:brief-comm} are easier to read, and will explain 
what is being done in other technical parts of this article. 
Although discussions in 
sections \ref{sssec:5classes-B+iomega}--\ref{ssec:one-more} and 
the appendix \ref{sec:details} are long, bulk of the materials there 
is elementary computations in T-duality of torus compactifications, 
Hodge theory and Galois theory; almost all the ideas are captured 
in Lemma \ref{lemma:mirrorhHdgH2L2=vHdgOnAlg} and the illustrative example 
in section \ref{ssec:illustrate}.

For string theorists who intend to read this manuscript seriously, it is important to have some level of familiarity with Hodge theory and the theory 
of complex multiplication. Appendix \ref{sec:guide} will be a useful 
guide for this purpose. 

\section{Preliminaries}
\label{sec:prelim}

\subsection{The Gukov--Vafa Conjecture}
\label{ssec:GV-conj}

\subsubsection{Observations}
\label{sssec:Moore-GV-idea}

{\bf Elliptic Curves:} 
To get started, we review the observation of \cite{Moore:1998pn,Moore:1998zu}
that established a possible connection between rational CFTs and 
complex multiplication. The following pedagogical review does not 
add anything new to the contents of \cite{Moore:1998pn,Moore:1998zu}; 
we simply recall relevant details here. 

Let ($M;G,B)|_{M=T^2}$ be a set of data of a $T^2$-compactification 
(a bosonic CFT and also an $N=(1,1)$ SCFT). In this case of $M=T^2$, 
one can always find a complex structure $I$ uniquely so that 
$(M;G;I)$ is K\"{a}hler; $\omega(-,-) = 2^{-1}G(I-,-)$ is the K\"{a}hler 
form. The combination $\rho: = \int_{M} (B+i\omega)/[(2\pi)^2\alpha']$ 
is called the complexified K\"{a}hler parameter. 
The complex manifold $(M;I)$ has a presentation 
$\C/(\Z \oplus \tau \Z)$ for some complex parameter $\tau$ in the complex upper 
half plane $\in {\cal H}$. 
So, one may extract a pair of complex parameters $\tau$ and $\rho$ 
(their ${\rm SL}_2\Z$ orbits in fact) from $(M;G,B)$ for 
a $T^2$ compactification. The mirror CFT (and the mirror geometry) 
of this set-up is known to be the one where $\tau$ and $\rho$ are 
exchanged; the mirror complex manifold $\C/(\Z \oplus  \rho \Z)$ is denoted 
by $(W;I^\circ)$ where $W \simeq T^2$ and $I^\circ$ the complex structure 
on $W$. 

For the CFT (and SCFT) of $(M;G,B)$ to be rational, it is necessary 
and sufficient that there is a rank-2 subgroup in the space of 
\begin{align}
 H_1(M;\Z) \oplus H^1(M;\Z) = \left\{ 
    w^1 \alpha + w^2 \beta + n_1 \hat{\alpha} + n_2 \hat{\beta} \; | 
  \; w^1, w^2, n_1,n_2 \in \Z \right\}
\end{align}
of winding and Kaluza--Klein charges where the complexified right-moving 
momentum 
\begin{align}
  p_R^\C \propto \left( n_2 - n_1 \tau + w^1 \rho + w^2 \tau \rho \right)
\end{align}
vanishes. This mathematical condition is equivalent to the condition 
that (*1) $\rho$ is in the ${\rm GL}_2(\Z)$ orbit of $\tau$ and that 
(*2) $\tau \in {\cal H}$ is subject to a quadratic equation  
\begin{align}
  a_\tau \tau^2 + b_\tau \tau + c_\tau = 0 
 \label{eq:quadratic-eq-EC-tau}
\end{align}
for some set of mutually prime integers $(a_\tau, b_\tau, c_\tau)$. 

The combination of the two conditions (*1) and (*2) can also be 
reorganized into a form that is democratic between $\tau$ and $\rho$. 
Both of the field $\Q(\tau) \subset \C$ and $\Q(\rho) \subset \C$,
the minimum subfield of $\C$ obtained by adjoining $\tau$ and $\rho$ 
respectively to the field $\Q$ of rational numbers, should be quadratic 
and totally imaginary extensions (**1):
\begin{align}
 \Q(\tau) & \; \cong \Q[x]/(a_\tau x^2+b_\tau x+c_\tau) \cong_{\rm vect.space} 
   \left\{ p + q x \; | \; p,q \in \Q \right\} , \\
 \Q(\rho) & \; \cong \Q[y]/(a_\rho y^2 + b_\rho y + c_\rho) \cong_{\rm vect.space} 
   \left\{ p + q y \; | \; p,q \in \Q \right\}  
\end{align}
for some mutually prime integers $(a_\rho, b_\rho, c_\rho)$, and the two fields should be isomorphic 
(**2), i.e., $\Q(\tau) \cong \Q(\rho)$.\footnote{\cite[Appendix A]{Kanno:2017nub} provides a quick guide 
to the relevant concepts of field theory}

Condition (**1) on $\Q(\tau)$ is in fact equivalent
 \cite{Moore:1998pn,Moore:1998zu} to the presence 
of complex multiplication in the complex manifold 
$(M;I) \cong \C/(\Z \oplus \tau \Z)$; condition (**1) on $\rho$ is 
for $(W;I^\circ)$.  Here is a very first step of introduction to the 
theory of complex multiplication, using the complex manifold 
$(M;I) = \C/(\Z \oplus \tau \Z)$ as an example. Consider the set of 
holomorphic and addition-law preserving maps of $(M;I)$ to itself, 
where the addition law descends from the addition law of vectors in the complex plane 
$\C=\R^2$ before the quotient. These set of maps ${\rm End}(M;I)$ forms a ring. 
Whatever the value of the parameter $\tau \in {\cal H}$ is, this ring 
always contain a ring $\Z\vev{(1\times)} = \{ (n \times) \; | \; n \in \Z \}$ 
of multiplying an integer $n$ to the complex analytic coordinate; 
the period rank-2 lattice $\Z \oplus \tau \Z$ is mapped to itself
(not necessarily surjectively) so the map $T^2 \rightarrow T^2$ is well-defined.
When the complex parameter $\tau$ is subject to the 
condition (\ref{eq:quadratic-eq-EC-tau}), however, there is one more 
generator in the ring ${\rm End}(M;I)$, $(a_\tau \tau \times)$ that multiplies 
the complex number $a_\tau \tau$ on the complex analytic coordinate of 
$\C/(\Z \oplus \tau \Z)$; the period lattice $(\Z \oplus \tau \Z)$ is mapped 
to $a_\tau \tau \Z \oplus a_\tau \tau^2 \Z 
= \tau (a_\tau \Z) \oplus (-b_\tau \tau-c_\tau) \Z \subset \Z \oplus \tau \Z$, 
so a map $T^2 \rightarrow T^2$ is well-defined. It is the definition 
of having complex multiplications for an elliptic curve $(M;I)$ that 
the ring ${\rm End}(M;I)$ is of rank-2 rather than rank-1; 
see Def. \ref{def:CM-geom-4CT}, Thm. \ref{thm:ellC-criteria-CMorNot} 
and Ex. \ref{ex:CM-EllC}. Generalization of this notion to 
higher-dimensional complex tori is reviewed in the 
appendix \ref{ssec:Guide-CM4CT-PartI}. 

The ring ${\rm End}(M;I)$ was introduced above by using the holomorphic 
and addition-law preserving maps from $(M;I) = \C/(\Z \oplus \tau \Z)$ 
to itself, but the same ring can be introduced as that of 
Hodge-structure preserving endomorphisms of $H^1(M;\Z)$; the geometric maps
of $\C/(\Z \oplus \tau \Z)$ have their action on $H^1(M;\Z)$ 
by the pull-backs. The complex multiplication $(a_\tau \tau \times) \in 
{\rm End}(M;I)$, for example, acts on the homology cycles through 
\begin{align}
  (a_\tau \tau \times) : (1, \tau) \longmapsto 
 (a_\tau \tau, \; a_\tau \tau^2) = (1, \; \tau) \left( \begin{array}{cc}
   0 & -c_\tau \\ a_\tau & -b_\tau \end{array} \right)
\end{align}
with integer coefficients, and on the integral cohomology basis by pull-back 
through 
\begin{align}
 (a_\tau \tau \times)^*: \; ( \hat{\alpha}, \hat{\beta} ) \longmapsto (\hat{\alpha}, \hat{\beta})
   \left( \begin{array}{cc} 0 & a_\tau \\ -c_\tau & -b_\tau \end{array} \right),
\end{align}
so (1,0)-forms and (0,1)-forms in $H^1(M;\C)$ are eigenvectors,  
\begin{align}
\hat{\alpha} + \tau \hat{\beta}  \longmapsto 
   a_\tau \tau (\hat{\alpha} + \tau \hat{\beta} ), \\
\hat{\alpha} + \bar{\tau} \hat{\beta} \longmapsto 
  a_\tau \bar{\tau} (\hat{\alpha} + \bar{\tau} \hat{\beta}),  
\end{align}
and the eigenvalues $a_\tau \tau$ and $a_\tau \bar{\tau}$ of 
the action $(a_\tau \tau \times)^*$ are Galois 
conjugates of one another. For any geometric map $(p + q a_\tau \tau)\times$
in ${\rm End}(M;I)$, with $p, q \in \Z$, both the (1,0)-form 
$(\hat{\alpha} + \tau\hat{\beta})$ and the (0,1)-form 
$(\hat{\alpha} + \bar{\tau}\hat{\beta})$ are simultaneous eigenvectors, 
with their eigenvalues $(p+q a_\tau \tau)$ and $(p+q a_\tau \bar{\tau})$ 
Galois conjugate to each other. The same argument can be repeated 
for the mirror $(W; I^\circ)$.  
Discussion in this paragraph is just 
an elementary translation of condition (**1) without introducing  
anything new, but is meant to be a very pedagogical introduction 
to the theory of complex multiplication. 
Thm. \ref{thm:smCM4CT-via-EndAlgCoh}, 
Rmk. \ref{rmk:announce-CMdef-use-EndAlg-onCoh}, 
Lemma \ref{lemma:very-useful}, 
Rmk. \ref{rmk:CM-Intuitv-summary4AV} and 
Prop. \ref{props:CM-Intuitv-summary4CT}
deal with the materials along this line\footnote{
Although there is a string-theorist-friendly quick review of 
theory of complex multiplication in the appendix B of \cite{Kanno:2017nub}, 
large fraction of the materials there are covered and superseded by the 
combination of the appendix \ref{sec:guide} of this article and 
the appendix A of \cite{Okada:2023udq} now. The appendix A 
of \cite{Kanno:2017nub} provides a quick summary on rudiments 
on number fields, which some of readers with non-math background may still 
find useful.  
} %
 for complex tori with 
sufficiently many complex multiplications of arbitrary dimensions; 
we will come back in section \ref{ssec:CM-abel-surface} and elaborate 
a little more 
on Lemma \ref{lemma:very-useful} and Rmk. \ref{rmk:CM-Intuitv-summary4AV}.

Both conditions (**1) and (**2) deal with the field of fractions 
$\Q(\tau)$ of the ring ${\rm End}(M;I)$ and the field of fractions 
$\Q(\rho)$ of the ring ${\rm End}(W;I^\circ)$. The fields $\Q(\tau)$ and 
$\Q(\rho)$ are in fact the algebra of Hodge-structure preserving 
$\Q$-linear endomorphisms of $H^1(M;\Q)$ and $H^1(W;\Q)$, respectively, 
instead of $H^1(M;\Z)$ and $H^1(W;\Z)$. Such algebras are denoted by 
${\rm End}(H^1(M;\Q))^{\rm Hdg}$ and ${\rm End}(H^1(W;\Q))^{\rm Hdg}$, 
respectively, and are called the {\it endomorphism algebra}, which are 
determined for the information how Hodge decomposition over $\C$ is introduced 
relatively to the reference frame of the vector spaces over $\Q$ such as 
$H^1(M;\Q)$ and $H^1(W;\Q)$ ({\it rational Hodge structure}).\footnote{
Appendix \ref{ssec:rHstr-Pol-alg} develops the concepts of rational Hodge 
structure, endomorphism algebras and related things and their math 
for cases more general than elliptic curves. We will need some of them 
already in Discussion \ref{statmnt:subtlty-justL=n?} 
and the rest of them heavily 
in section \ref{ssec:CM-abel-surface} onwards. 
} %
Condition (**2) in the present context 
of $M \cong W \cong T^2$ states that the rational Hodge structure 
on $H^1(M;\Q)$ and $H^1(W;\Q)$---both with complex multiplications---are 
isomorphic. 
We therefore arrive at the following statement: the SCFT for $(M;G,B)$ is rational if and only if the rational Hodge structures on $H^1(M;\Q)$ and $H^1(W;\Q)$ are both with complex multiplications and are isomorphic.\\[0.5cm]
{\bf The Fermat Quintic Gepner Construction, and the Likes:}
Let us consider another example: the $\Z_5$-orbifold 
of the tensor product of five copies of the 2$d$ $N=(2,2)$ 
minimal model with the central charge $c_L=c_R=3k/(k+2)$ and $k=3$, 
denoted by $3^{\otimes 5}/\Z_5$. 
This SCFT is rational, and is interpreted as a 2$d$ 
non-linear sigma model whose target space is a quintic Calabi--Yau threefold 
with a very special complex structure and a very special complexified 
K\"{a}hler parameter. The complexified K\"{a}hler parameter 
is chosen at the small volume limit within the complex 1-dimensional 
moduli space, and the complex structure parameter is chosen 
at the Fermat point\footnote{
The 101-dimensional moduli space of complex structure corresponds to 
choosing an arbitrary homogeneous function $F(\Phi_{i=1,\cdots,5})$ 
of degree-5 on $\P^4$ that defines a threefold $M$ through $M = 
\{ [\Phi_i] \in \P^4 \; | \; F(\Phi)=0 \} \subset \P^4$. 
The Fermat point in the moduli space corresponds to the choice 
$F = \sum_{i=1}^5 (\Phi_i)^5$.
} %
  of the complex 101-dimensional moduli space.  
So, both the 4-dimensional vector space $H^3(W;\Q)$ and the 204-dimensional 
vector space $H^3(M;\Q)$ are given very specific Hodge decomposition 
(rational Hodge structure).

The notion of complex multiplication for $T^2$ has been generalized 
to complex manifolds of higher dimension; it is conventional to use the 
word {\it CM-type} for projective varieties other than elliptic curves
instead of complex multiplications. There are variations 
in how the notion is generalized, in fact, as we will review 
in Discussion \ref{statmnt:subtlty-justL=n?} below; see also 
Ref. \cite{Okada:2023udq} and references therein.  
If we set those issues aside for the moment, then it is known that the rational Hodge 
structure on the 4-dimensional 
$H^3(W;\Q)$ is of CM-type in the sense of Def. \ref{def:CM-rat-Hdg-str-byEnd}, 
with the endomorphism field being the cyclotomic field $\Q(\zeta_5)$ of 
a primitive fifth root of unity $\zeta_5$;
the cohomology group $H^3(M;\Q)$ also contains a rational Hodge substructure 
that is $4$-dimensional over $\Q$, level-3, and is of CM-type (Def. \ref{def:CM-rat-Hdg-str-byEnd}); 
the CM-field is $\Q(\zeta_5)$ on this substructure. So, the endomorphism fields 
of both sides are isomorphic to each other, just like in (**2) in the case 
of $M=T^2$. It was this observation that led Gukov and Vafa to propose 
the idea Conj. \ref{conj:GV-original} with an intention to cover a broad 
class of Ricci-flat K\"{a}hler manifolds. 

\subsubsection{Necessary Refinements}
\label{sssec:need2refine}

Although Conj. \ref{conj:GV-original} seems elegant, 
a closer inspection on its formal aspects reveals that 
there are a few points that need clarification in order to have a 
conjecture that we can test. The observations/issues raised in  
Discussions \ref{statmnt:subtlty-choose-cpx-str}, 
\ref{statmnt:subtlty-justL=n?} 
 and \ref{statmnt:subtlty-mirror?} below are not particularly difficult ones, 
but they should be written down as we start off. The study 
in section \ref{sec:choose-cpx-str} and onwards is not just for testing/refuting 
Conj. \ref{conj:GV-original}, but also for resolving conceptual issues 
in Discussions \ref{statmnt:subtlty-choose-cpx-str}, \ref{statmnt:subtlty-justL=n?}
and \ref{statmnt:subtlty-mirror?}. 

\begin{anythng}
\label{statmnt:subtlty-choose-cpx-str}
{\bf Continuous deformation of complex structure:}
Since Conjecture \ref{conj:GV-original} tries to characterize 
rational SCFTs by using a Hodge structure, there is no way of interpreting 
the conditions there without choosing a complex structure. 
Now, consider a case where the target space $(M; G)$ is either a torus $T^{2n}$ 
of real $2n$ dimensions with $n\geq 2$, or a hyper-K\"{a}hler manifold. 
On one hand, for such a smooth manifold $M$ and a Riemannian metric $G$ on it, 
there is a continuous freedom in choosing a complex structure $I$ with 
which the metric $G$ is compatible.  
On the other hand, a 2$d$ non-linear sigma model is specified by only the 
data $(M; G)$, without referring to a complex structure on $M$. 
Whether the SCFT is rational or not should therefore be a property 
of $(M; G)$, not of the data $(M; G; I)$. 

One option is to give up on applying the idea along the line of 
Conj. \ref{conj:GV-original} to such Riemannian manifolds 
$(M;G)$ where the choice of complex structure $I$ is not unique. 
Those geometries are just as valid as target spaces of string theory
as Calabi--Yau $n$-folds with $n>2$, however. If the observations/ideas in
Conj. \ref{conj:GV-original} are not mere coincidence but reflection
of deeper connection between SCFTs and arithmetic geometry/number theory,
the observation will {\it somehow} be extended to cover broader class of
qualified target spaces. 
If we wish to find a version of conjecture that is applicable to  
the class of manifolds we are referring to here, then we should 
add a few more conditions\footnote{
It is impossible to read condition (i) in Conj. \ref{conj:GV-original}
for all $I$ forming a continuous family. Even when $(M;G;I)$ is of CM-type, 
the K\"{a}hler manifold $(M;G;I')$ with a complex structure $I'$ 
infinitesimally deformed from $I$ does not have sufficiently many 
complex multiplications.  
} %
on the choice of $I$ that is implicit in Conj. \ref{conj:GV-original}; 
our proposal\footnote{
In the previous preprint version of this manuscript \cite{Kidambi:2022wvh},
an alternative choice of conditions on $I$ is also discussed. See \S5.2.3 
of \cite{Kidambi:2022wvh}.
} %
is to include condition 1 in Conj. \ref{conj:forT4}, based on 
discussions in sections \ref{ssec:pol} and \ref{ssec:B-transc}. 
\end{anythng}

\begin{anythng}
\label{statmnt:subtlty-justL=n?}
{\bf Variations in the definition of CM-type:}
(the authors assume that the readers are already familiar with 
materials reviewed in the appendices \ref{ssec:rHstr-Pol-alg} and 
\ref{ssec:def-CM-rHstr})
Let $(M;G;I)$ be a Ricci-flat K\"{a}hler manifold of complex dimension $n$; the 
cohomology group $H^n(M;\Q)$ is endowed with a rational Hodge structure 
by the complex structure of $(M;I)$. The rational Hodge structure on $H^n(M;\Q)$
is not necessarily simple, but has a decomposition into simple Hodge 
substructures
\begin{align}
  H^n(M;\Q) \cong \oplus_{a\in A} [H^n(M;\Q)]_a ~,
\end{align}
if $(M;I)$ is projective (like all the Calabi--Yau $n$-folds with $n>2$ are);
see Prop. \ref{props:rHstr-indcmp-dcmp}.  
Let $D_a := {\rm End}([H^n(M;\Q)]_a)^{\rm Hdg}$ be the algebra of Hodge 
endomorphisms of a simple component $[H^n(M;\Q)]_a$; it is always a 
division algebra.
The endomorphism algebra of $(H^n(M;\Q),I)$ is of the form
(see Lemmas \ref{lemma:End-alg-is-CM-wPol-ala-Milne},
 \ref{lemma:Wedderburn}, 
Rmk. \ref{rmk:EndAlg-str-notby-Wedbb-butby-HdgIsoClssDcmp} and 
Prop. \ref{props:rHstr-indcmp-dcmp}) 
\begin{align}
  {\rm End}(H^n(M;\Q))^{\rm Hdg} \cong \oplus_{\alpha \in {\cal A}} M_{n_\alpha}(D_\alpha); 
\end{align}
simple Hodge substructures labeled by $a \in A$ are grouped 
into classes labeled by $\alpha \in {\cal A}$ by Hodge isomorphisms, 
${\cal A} = A/\sim$; the endomorphism algebras $D_a$ for $a \in A$ that 
belong to $\alpha \in {\cal A}$ are all common, and are denoted by $D_\alpha$;
$n_\alpha$ is the number of simple components ($a$'s) in a Hodge isomorphism
class $\alpha$. 

Take the Fermat quintic Calabi--Yau threefold $M$ as an example.   
The 204-dimensional vector space $H^3(M;\Q)$ has a decomposition 
into simple rational Hodge substructures \cite[\S3]{shioda1982geometry}, 
\begin{align}
 [H^3(M;\Q)]_{\ell=3} \oplus
     \left( \oplus_{a=1}^{50} [H^3(M;\Q)]_{\ell=1, a} \right), 
     \label{eq:Fermat-5-wgt3-coh-dcmp}
\end{align}
and each one of the components is of 4-dimensional over $\Q$, 
supporting a simple rational Hodge substructure. Just one of them, one 
with the subscript $\ell =3$, contains the 1-dimensional Hodge (3,0) component, 
while all the other 50 substructures only contain a 2-dimensional Hodge (2,1) 
component and a 2-dimensional Hodge (1,2) component. Those 50 simple 
substructures are Hodge isomorphic to each other, and share 
the same endomorphism field $D_\alpha \cong \Q(\zeta_5)$. So, the 
endomorphism algebra as a whole has the form of 
$\Q(\zeta_5) \oplus M_{50}(\Q(\zeta_5))$. 

The notion of complex multiplications of elliptic curves has been 
generalized to the notion of CM-type of rational Hodge structure 
(see Def. \ref{def:CM-rat-Hdg-str-byEnd}). In the example of the Gepner 
construction $3^{\otimes 5}/\Z_5$ referred to in 
section \ref{sssec:Moore-GV-idea}, all the 51 rational Hodge 
substructures of $H^3(M;\Q)$ are of CM-type, and the rational Hodge 
structure of $H^3(W;\Q)$ is also of CM-type. 

\vspace{5mm}

There is a variation in how we introduce the notion of CM-type 
to complex projective non-singular varieties $(M;I)$ of higher dimensions, 
instead of Def. \ref{def:CM-rat-Hdg-str-byEnd} on 
rational Hodge (sub)structures of $H^*(M;\Q)$. 
\begin{defn}
\label{def:strongCM}
A complex projective non-singular variety $M$ is said to be 
of {\it strong CM-type} when all the rational Hodge substructures of 
$H^*(M;\Q)$ are of CM-type in the sense of 
Def. \ref{def:CM-rat-Hdg-str-byEnd}.

A complex $n$-dimensional projective non-singular variety $M$ 
with a choice of a polarization $D_P$ is said to be of 
{\it CM-type(mid-dim-primCoh)} when the primitive component of 
the middle dimensional cohomology group $[H^n(M;\Q)]^{(0)}$
(see footnote \ref{fn:prim-dcmp}) is of CM-type 
in the sense of Def. \ref{def:CM-rat-Hdg-str-byEnd}. 
\end{defn}
\noindent
For $M$ to be of CM-type(mid-dim-primCoh), the Hodge structure on 
$H^k(M;\Q)$ with $0< k  <n$ do not have to be of CM-type; the two 
conditions may be different when $M$ is an abelian variety 
of $n \geq 3$ dimensions,\footnote{
In the case $M$ is an abelian surface, the two conditions (and even the 
weak-CM condition below) are known to be equivalent \cite{Okada:2023udq}. 
} %
a hyper-K\"{a}hler manifold of real 8-dimensions and higher, or 
a Calabi--Yau fourfold with $h^{2,1}(M) >0$ (see \cite[(90)]{Braun:2014xka}), 
for example. Even a weaker version is available when the variety $(M;I)$ has 
a trivial canonical bundle (as in the case $(M;I)$ has a Ricci-flat metric).
\begin{defn}
\label{def:weakCM-level}
Let $(M;I)$ be a complex $n$-dimensional projective non-singular variety with 
a trivial canonical bundle; $h^{n,0}(M)=1$ then. The weight-$n$ 
rational Hodge structure on $H^n(M;\Q)$ has just one simple rational Hodge 
substructure on the vector subspace, denoted by $[H^n(M;\Q)]_{\ell = n}$, 
whose $\otimes \C$
contains the Hodge $(n,0)$ and $(0,n)$ components. 
This substructure is called the {\it level-$n$ component}.
The variety $M$ 
is said to be of {\it weak CM-type} when the level-$n$ rational Hodge 
substructure on $[H^n(M;\Q)]_{\ell = n}$ is of CM-type (in the sense of Def. 
\ref{def:CM-rat-Hdg-str-byEnd}). 
\end{defn}
\noindent 
The Hodge structure on the level-$n$ component is of special importance 
in the perspective of string theory, because this substructure always 
exists and can be extracted uniquely from $(H^*(M;\Q),I)$,   
and also because this particular Hodge substructure carries the information 
of the complexified K\"{a}hler parameter of the mirror $(W;I^\circ)$. 

The variations in the notion of CM-type of a geometry are in fact 
absent if ``ye'' is the right answer to the following 
\label{pg:Qstn-strng-vs-weak}
\begin{qstn}
\label{qstn:weak-vs-srong}
whether a projective non-singular variety $(M;I)$ with a trivial canonical 
bundle is of strong CM-type when it is of weak CM-type.
\end{qstn}
\noindent
This remains to be an open question in mathematics; one can find an 
update on this math question in \cite{Okada:2023udq}, which has proved for certain cases---including 
abelian surfaces---that the weak CM-type condition on $M$ implies 
the strong CM-type property (trivial for K3 surfaces). 
The Fermat quintic Calabi--Yau threefold is also regarded as another 
non-trivial example. To the best of knowledge of the authors, 
there is no known example of a weak CM-type that is not of strong CM-type.\footnote{
We thank Profs. Goto and Yui for letting us updated in the state of affairs 
in this field.
} %

In the absence of a known example that sits in between different definitions, 
we cannot determine at this moment 
whether we should read condition (i) in Conj. \ref{conj:GV-original} 
as strong CM-type, weak CM-type, or any other variations in between.\footnote{
\label{fn:BV}
Calabi--Yau threefolds of the form of Borcea--Voisin 
orbifolds \cite{borcea1998calabi,voisin1993miroirs} will be a good 
testing ground in resolving this issue in math; see \cite{MR3228297}, for example. 
To work on this class of cases in string theory, however, we should work on K3 surfaces first. 
} %
We have chosen to state the conditions in Conj. \ref{conj:forT4} 
with the strong CM-type property, but that can also be the weak CM-type
property, because the test study in this article uses $M \cong T^4$ 
(abelian surfaces) where the weak CM property implies the strong CM property
(and vice versa) \cite{Okada:2023udq}. In another trial version of the 
Conjecture, 
Conj. \ref{conj:gen}, we will choose to employ the weak CM-type property 
and seek for theoretically possible condition statements. At least, 
we aim to raise the level of alertness to the subtlety involved 
in the definition of CM-type property of a variety. If Question \ref{qstn:weak-vs-srong}
is resolved affirmatively, this subtlety on the conditions in 
Conjs. \ref{conj:forT4} and \ref{conj:gen} is also gone, however. 

Condition (ii) in Conj. \ref{conj:GV-original} refers 
to the fields of CM-type Hodge structures of $M$ and the mirror 
manifold $W$. An easy way to make sense of this condition is to think 
of them as the endomorphism fields of the level-$n$ rational Hodge 
substructures $[H^n(M;\Q)]_{\ell=n}$ and $[H^n(W;\Q)]_{\ell = n}$. 
Condition (ii) is almost equivalent to existence of a Hodge isomorphism 
between the two level-$n$ components (the latter is a little stronger; 
see a review in \ref{statmnt:CM-pair}).  

If one should require other simple components to be of CM-type, 
as discussed before, then one may also have to refine the 
condition (ii); whether it is read as a Hodge isomorphism of 
the level-$n$ components on both sides, or as Hodge isomorphisms of 
some pairs of simple components of $H^*(M;\Q)$ and $H^*(W;\Q)$. 
In general, $\dim_\Q[H^n(M;\Q)]$ is not necessarily equal to 
$\dim_\Q[H^n(W;\Q)]$, so there is no natural choice of pairs of 
simple components besides the pair of the level-$n$ components. 
\end{anythng}

\begin{anythng}
\label{statmnt:subtlty-mirror?} {\bf The mirror:} 
The statement of Conjecture \ref{conj:GV-original} is written 
by referring to a mirror manifold $W$. It is not always guaranteed, however, 
that an $N=(2,2)$ SCFT with a non-linear sigma model interpretation 
$(M;G,B;I)$ has 
a mirror-equivalent $N=(2,2)$ SCFT that can be interpreted as a non-linear 
sigma model of a mirror geometry $(W;G^\circ, B^\circ;I^\circ)$.
Even when there is, it is not guaranteed that there is a unique choice of the 
mirror geometry data. 

It is an interesting question whether there is always such a mirror 
manifold when the $(M;G,B;I)$-target $N=(2,2)$ SCFT is rational. 
Given the absence of guarantee of existence of a mirror, or 
absence of uniqueness of mirrors, it is an option to write down 
the conditional statements of Conj. \ref{conj:GV-original} without 
referring to the mirror geometry; that is possible by formally introducing 
yet another rational Hodge structure on $H^*(M;\Q)$ by using the complexified 
K\"{a}hler parameter instead of the complex structure of $(M;I)$; the 
present authors indeed did that in the preprint 
version \cite{Kidambi:2022wvh}.  

Further investigation in the preprint version \cite{Okada:2022jnq} has made us 
change our minds, however. In the study using $M\cong T^4$ in this article, 
we will find that we have to impose a condition---the latter half of 
condition 5 in Thm. \ref{thm:forT4}---to have a sufficient 
condition for the SCFTs to be rational, and the condition is best 
stated by using the language of Strominger--Yau--Zaslow (SYZ) torus fibration 
of mirror correspondence. This observation based on the case $M\cong T^4$
encouraged us to formulate the refined version Conj. \ref{conj:forT4}
as it is, especially in conditions 4 and 5. 
\end{anythng}

\subsection{Rational CFTs with Torus Target}
\label{ssec:torus-RCFT}

Since rational CFTs in torus compactifications have been completely classified, 
we may use the established results to refine and test 
Conjecture \ref{conj:GV-original}.  
In this section \ref{ssec:torus-RCFT}, we quote results from 
Ref. \cite{Wendland:2000ye} relevant to our analysis.  

\begin{props} 
\cite[Lemma 4.5.1]{Wendland:2000ye}
\label{props:Wendland-451}
{\it Let $T^{m} = \R^m/\Z^{\oplus m}$ be a real $m$-dimensional torus
with a smooth structure, and $X^I$s with $I=1,\cdots, m$ a 
set of coordinates of $\R^m$ with periodicity $\Delta X$ 
(i.e., $X^I \sim X^I + \Delta X$). Let $G = G_{IJ} dX^I \otimes dX^J$
be a constant Riemannian metric on $T^m$ (i.e., $G_{IJ} \in \R$ are 
independent of the coordinates $X^K$'s), and $B = 2^{-1} B_{IJ} dX^I \wedge dX^J$
a 2-form on $T^m$ where $B_{IJ}$ are independent of the coordinates.  
 
The bosonic CFT for the data $(T^m; G, B)$ is rational if and only if }\footnote{
The author of \cite{Wendland:2000ye} adopts the convention 
$\Delta X = 2 \pi R$, 
$R=\sqrt{\alpha'}$ and $\alpha' =2$. We will use the convention 
$\Delta X = 2\pi \sqrt{\alpha'}$ throughout this article. The metric 
and $B$-field satisfying (\ref{eq:cond-GnB-rational}) are therefore 
said to be {\it rational}. 
} %
\begin{align}
 \left( \frac{\Delta X}{(2\pi) \sqrt{\alpha'}} \right)^2 G_{IJ} \in \Q, \qquad
 \left( \frac{\Delta X}{(2\pi) \sqrt{\alpha'}} \right)^2 B_{IJ} \in \Q. 
  \label{eq:cond-GnB-rational}
\end{align}
{\it The condition for the 
$N=(1,1)$ SCFT associated with the data $(T^m;G,B)$ to be rational 
is also the same as above. } 
\end{props}
 
We are interested in the cases where $m=2n$, when there is a possibility of 
introducing a complex structure on the target space $T^m$. 
Wendland has further derived this
\begin{cor}\cite[Thm. 4.5.5]{Wendland:2000ye}
\label{cor:Wend-ExEx}
{\it Let $(T^{2n};G, B)$ be a set of data for which the (S)CFT is rational. 
Then there exists a surjective homomorphism 
$\varphi: T^{2n} \cong \R^{2n}/\Z^{\oplus 2n} \longrightarrow
 \prod_{a=1}^n \C/(\Z \oplus \tau_a \Z)$ with respect to the abelian group 
law on $\R^{2n}$ and $\C^n$ that has the following properties. 
Each one of $\C/(\Z \oplus \tau_a \Z)$ is 
a CM elliptic curve (i.e., $[\Q(\tau_a):\Q] = 2$), 
and there is a metric on $\C/(\Z \oplus \tau_a \Z)$, given by 
$ds^2 = g_a (du^a \otimes d\bar{u}^{\bar{a}} + {\rm h.c.})$ with\footnote{
$u^a$ with $a=1, \cdots, n$ are the complex coordinates of the $a$-th
elliptic curve $\C/(\Z\oplus \tau_a \Z)$ with the periodicity 
$u^a \sim u^a + 1 \sim u^a + \tau_a$. 
} %
$g_a \in \Q$ so that the pull-back of the metric $ds^2$ by $\varphi$ 
agrees with the metric $G$ on $T^{2n}$. }
\end{cor}

This result was a part of supporting evidence for Conj. \ref{conj:GV-original} 
in the following sense. 
Firstly, there is already an implicit choice of complex structure $I_0$ on 
$\prod_a \C/(\Z \oplus \tau_a \Z)$, with which the metric $ds^2$ is compatible. 
The metric $G$ on $T^m$ is compatible with the complex structure $I= \varphi^*(I_0)$. 
The complex torus $(T^{2n};I)$ is of CM-type since 
$\prod_a \C/(\Z \oplus \tau_a \Z)$ is of CM-type. The metric $g_a \in \Q$ should be 
split into $g_a = {\rm Im}(\rho_a)/{\rm Im}(\tau_a)$ so that 
${\rm Im}(\rho_a)$ parametrizes the volume of $\C/(\Z \oplus \tau_a \Z)$. 
It then also follows that $\Q(i {\rm Im}(\rho_a)) \cong \Q(\tau_a)$.

It remains to be an 
open question whether the class of complex structures of the form 
$I = \varphi^*(I_0)$ are all of those (or a part of those) where 
a Gukov--Vafa-like statement holds true (this issue was raised 
in Discussion \ref{statmnt:subtlty-choose-cpx-str}).
We will discuss those issues, starting in section \ref{sec:choose-cpx-str}. 

\subsection{Coarse Classification of CM-type Abelian Surfaces}
\label{ssec:CM-abel-surface}

In our attempt to prove the refined version of the conjecture 
(Conj. \ref{conj:forT4}) for $M\cong T^4$, which includes the 
condition 2 on the CM-type Hodge structure, we need to deal with 
complex abelian surfaces of CM-type.
As a preparation for the study in the following sections, therefore, 
we quote a known result on classification of CM-type abelian surfaces. 
Abelian varieties of $n$-dimensions constitute 
a small special subclass of complex tori of $n$-dimensions.
The following quoted results only cover abelian surfaces, not 
complex torus $(T^4;I)$; although we will add in the appendix \ref{sec:guide} 
a little more information on complex tori other than abelian varieties, 
eventually we do not need them in the study after section \ref{ssec:pol}. 

For the purpose of this article, we do not need to work on each one of 
infinitely many isomorphism classes of abelian surfaces of CM-type, 
it is enough to deal with their modulo-isogeny classes.  
Here, as we have already explained in section \ref{sssec:Moore-GV-idea}, 
Def. \ref{def:isogeny} and Rmk. \ref{rmk:CM-vs-isogeny}, 
one complex torus $(M;I)$ has sufficiently many complex 
multiplications if and only if (iff) another complex torus $(M';I')$ isogenous 
to $(M;I)$ does. Equivalently, the rational Hodge structure on $H^1(M;\Q)$ 
by $I$ has sufficiently many complex multiplications iff the rational 
Hodge structure $H^1(M';\Q)$ by $I'$ has and there is a Hodge isomorphism 
between the two rational Hodge structures (Def. \ref{def:rHdgStr}).  
In fact, we even find that we can write down just one proof for infinitely many 
isogeny-classes so long as their endomorphism algebra share the same 
Galois-theoretical structure. So, the result quoted below\footnote{
The primary resource of this classification 
is \cite[\S5 + pp.64--65 Ex.8.4.(2)]{shimura2016abelian}; 
one may also look at \cite{birkenhakeabelian}. 
The combination of Prop. \ref{props:rHstr-indcmp-dcmp}, 
Rmk. \ref{rmk:cmplt-reducibl-abel-var} and 
Rmk. \ref{rmk:CM-Intuitv-summary4AV} reviewed in the appendix of 
this article also has enough information. 
} %
is the classification of CM-type abelian surfaces 
by the Galois-theoretical structure of their endomorphism algebras.\footnote{
Interested readers might find in \cite[\S4.1]{Okada:2023udq}
and references therein a theory for similar classification of CM-type 
abelian varieties of higher dimensions. The study in this article 
does not need them, however.  
} %

\begin{anythng} {\bf (Classification)}
\label{statmnt:classify-abel-surf}
Any complex abelian varieties $(M;I)$ is isogenous to the product of 
simple abelian varieties (Prop. \ref{props:rHstr-indcmp-dcmp} 
and Rmk. \ref{rmk:cmplt-reducibl-abel-var}). When it comes to 
an abelian surface, it is either simple on its own, or isogenous 
to the product of two elliptic curves. 

Think of the case an abelian surface $(M;I)$ is simple, first. 
When such $(M;I)$ is of CM-type, then the algebra ${\rm End}(H^1(M;\Q))^{\rm Hdg}$ 
is a CM-field $K$ that is a degree-4 extension over $\Q$. 
It is known that there are two 
possibilities \cite[pp.64--65 Ex.8.4.(2)]{shimura2016abelian}
\begin{itemize}
\item [(B)] The extension $K/\Q$ is Galois. In this case,\footnote{
\label{fn:cycl-5}
The cyclotomic field $K=\Q(\zeta_5)$, where $\zeta_5 := e^{2\pi i/5}$,
is an example. To see this, note first that
$\Q(\zeta_5) \cong \Q[x]/(x^4+x^3+x^2+x+1)$. It contains a totally real
subfield $K_0:= \Q(\zeta_5+\zeta_5^{-1})\cong \Q[y]/(y^2-5/4)$ (note that
$(x+1/x)^2 + (x+1/x)-1=0$, equivalently $(x+1/x+1/2)^2-5/4=0$); 
the field $K=\Q(\zeta_5)$ is generated by adjoining $\xi:= x-x^{-1}$ to $K_0$;
$\xi^2 = (x-1/x)^2 = -(x+1/x)-3 = -y-5/2$. So, $p=-5/2$, $q=-1$ and $d=5/4$.
In this example, $d' = 5 \in d (\Q^\times)^2$ indeed. 
} %
the field $K$ is isomorphic to 
\begin{align}
\Q[x,y] \; /& \; (y^2-d, \; x^2-p-qy), \label{eq:endK-str-caseBC} \\
  & \; d \in \N_{>1} \backslash (\N_{>1})^2, \quad p,q\in \Q, \quad p<0, \; q \neq 0, 
 \quad  d' := p^2-q^2d > 0 \nonumber 
\end{align}
satisfying $d' \in d (\Q^\times)^2$. The totally real subfield is 
$\Q[y]/(y^2-d) =: K_0$.
The Galois group is ${\rm Gal}(K/\Q) \cong \Z_4$. 
\item [(C)] The extension $K/\Q$ is not Galois. 
The field $K$ in this case is also isomorphic to (\ref{eq:endK-str-caseBC})
with $d' \nin d (\Q^\times)^2$ instead. 
The totally real subfield is $\Q[y]/(y^2-d) =: K_0$. 
The Galois group of the normal closure $K^{\rm nc}$ of $K$ 
is of the form ${\rm Gal}(K^{\rm nc}/\Q) \cong \Z_4 
 \rtimes \Z_2$.
\end{itemize}
In most of the situations that we encounter in this article, 
it suffices to use the common algebraic structure (\ref{eq:endK-str-caseBC})
of the endomorphism fields in the cases (B) and (C), 
so the two cases can often be treated simultaneously. 

When an abelian surface $(M;I)$ is isogenous to the product 
of two elliptic curves $E_1 \times E_2$, instead, and is of CM-type, 
there are two other possibilities in the Galois theoretical property 
of the endomorphism algebra; the case (A) is when $E_1$ and $E_2$ are 
isogenous to each other (and are both of CM-type), and the case 
(A') when $E_1$ and $E_2$ are not isogenous to each other (but both are of 
CM-type). In the case (A), therefore, $(M;I)$ is isogenous to $E \times E$ 
where $E$ is a CM elliptic curve. It is known that 
\begin{itemize}
\item [(A)] ${\rm End}(H^1(M;\Q))^{\rm Hdg} \cong M_2(K^{(2)})$ where 
$K^{(2)}={\rm End}(H^1(E;\Q))^{\rm Hdg}$ has the structure of  
an imaginary quadratic field $K^{(2)} \cong \Q[x]/(x^2-p)$ for some $p \in \Q_{<0}$, 
\item [(A')] ${\rm End}(H^1(M;\Q))^{\rm Hdg} \cong K^{(2)}_1 \oplus K^{(2)}_2$
where $K^{(2)}_i = {\rm End}(H^1(E_i;\Q))^{\rm Hdg}$ is an imaginary quadratic 
field $K^{(2)}_i \cong \Q[x]/(x^2-p_i)$ for some $p_i \in \Q_{<0}$ for each 
$i=1,2$. In the case (A'), $p_2/p_1 \nin (\Q^\times)^2$. 
\end{itemize}
An isogeny between $(M;I)$ and $E_1 \times E_2$ is chosen, fixed and used implicitly 
in the cases (A) and (A'), already here, and also in the rest of this article. 
For both of case (A) and (A'), the imaginary quadratic fields $K^{(2)}/\Q$
are Galois, with ${\rm Gal}(K^{(2)}/\Q) \cong \Z_2$ generated by the 
complex conjugation. The totally real subfield $K_0$ is $\Q$.  
\end{anythng}

\begin{anythng}
{\bf (Weight-1 rational Hodge structure and embeddings)} 
The weight-1 rational Hodge structure of complex tori with sufficiently 
many complex multiplications (incl. abelian varieties of CM-type) is 
tightly constrained so much that all the abelian surfaces that belong 
to each one of the four cases can be studied simultaneous in this article. 
The tight constraint on the weight-1 Hodge structure on $H^1(M;\Q)$ 
is already reviewed in Lemma \ref{lemma:very-useful} along with 
an elementary proof; here, we illustrate how to use in practice, 
while preparing notations to be used in the following sections. 

The CM-field $K$ of degree-4 acts on the four-dimensional vector 
space $H^1(M;\Q)$ faithfully in the case (B, C), and the CM-field 
(imaginary quadratic field) $K^{(2)}$ [resp. $K^{(2)}_i$] of degree-2
does on the two-dimensional vector space $H^1(E;\Q)$ [resp. $H^1(E_i;\Q)$]
in the case (A) [resp. case (A')]. As explained in
Lemma \ref{lemma:very-useful} 
and Rmk. \ref{rmk:Phi-p-q}, the action of those fields 
can be diagonalized over $\C$ simultaneously, and there is one-to-one 
correspondence between the set of the eigenvectors/spaces and the set of 
embeddings\footnote{
To take the CM field $K=\Q[x]/(x^4+x^3+x^2+x+1)$ in footnote \ref{fn:cycl-5}
as an example, the set of embbedings of $K$ is
$\{ \rho_a \; | \; a \in 1,2,3,4\}$, where $\rho_a: x \mapsto e^{2\pi i a/5}$. 
In the component $[H^3(M;\Q)]_{\ell=3}$ of the Fermat quintic Calabi--Yau
threefold $M$ (see (\ref{eq:Fermat-5-wgt3-coh-dcmp})), for example,
the Hodge (3,0) component has a generator $\sum_{I=1}^4 e_I e^{2\pi i I /5}
= \sum_{I=1}^4 e_I \rho_{a=1}(x^I)$, and the Hodge (2,1) component
a generator $\sum_{I=1}^4 e_I \rho_{a=2}(x^I)$, for some basis $\{ e_I \}$
of the vector space $[H^3(M;\Q)]_{\ell=3}$ over $\Q$; this is an example of
the structure (\ref{eq:egVec-from-very-useful-L}).
} %
of the fields into $\C$; each of the eigenspaces is in 
just one Hodge (p,q) component; each of the Hodge (p,q) components---only 
the Hodge (1,0) and (0,1) components in the case of $H^1(M;\Q)$---has a basis 
that consists of a part of those eigenvectors. This dictionary, while 
elementary, will be exploited countless times in this article.  
If a reader does not find that this explanation is enough, we 
recommend to take time to read Lemma \ref{lemma:very-useful} and 
Rmk. \ref{rmk:Phi-p-q}. 

We have stated in Lemma \ref{lemma:very-useful} that the simultaneous 
eigenstates of the action of a field $F$ on a vector space $V_\Q = {\rm Span}_\Q\{ e_I\}$, 
with $[F:\Q] = \dim_\Q V_\Q$, is of the form 
\begin{align}
   v_{\rho_a} := \sum_{I=1}^{[F:\Q]} e_I \rho_a(\eta_I ), \qquad 
      \rho_a \in {\rm Hom}_{\rm field}(F,\overline{\Q})
  \label{eq:egVec-from-very-useful-L}
\end{align}
for some basis $\{ \eta_I \}$ of $F$ over $\Q$; when we change 
a $\Q$-basis $\{ e_I \}$ of $V_\Q$, the basis $\{ \eta_I \}$ also changes 
accordingly so that the eigenspace $\C v_{\rho_a}$ remains to be the same.
Conversely, we may choose a basis $\{ \eta_I \}$ of $F/\Q$ in any way 
we like, and then a $\Q$-basis $\{ e_I \}$ of $V_\Q$ changes accordingly. 
It will be convenient to set $\{ \eta_I \} = \{ 1,y,x,xy \}$ in the 
cases (B, C) and to set $\{ 1,x_i \}$ in the cases (A, A') in our analysis. 
The corresponding basis $\{ e_I \}$ of $V_\Q$ is denoted by 
$\{ \hat{\alpha}^1, \hat{\alpha}^2, \hat{\beta}_1, \hat{\beta}_2\} \subset 
H^1(M;\Q)$ in the cases (B, C), and by 
$\{ \hat{\alpha}^i, \hat{\beta}_i \} \subset H^1(E_i;\Q)$ 
in the cases (A) and (A'). Although there are still infinitely 
many CM-type abelian surfaces (see also section \ref{ssec:distrib} and 
Rmk. \ref{rmk:hierarchical-class-CM-EC}) 
 after specifying the endomorphism algebra ${\rm End}(H^1(M;\Q))^{\rm Hdg}$, 
the uniform treatment by the basis 
$\{\hat{\alpha}^1,\hat{\alpha}^2,\hat{\beta}_1,\hat{\beta}_2\}$ of $H^1(M;\Q)$
and the convenient basis $\{ \eta_I \}$ of $K/\Q$ is possible for 
each of the cases (B,C), (A) and (A'); the difference among those 
infinitely many CM-type abelian surfaces is hidden in how each of the elements of the basis $\{ \hat{\alpha}^1,\hat{\alpha}^2,\hat{\beta}_1,\hat{\beta}_2\}$ 
is placed relatively to the integral subspace $H^1(M;\Z) \subset H^1(M;\Q)$.  
\end{anythng}

\begin{anythng}
\label{statmnt:CM-pair}
{\bf (CM pair)} When there are two abelian varieties 
$(M; I)_1$ and $(M;I)_2$ of dimension $n$ that are both CM-type and 
are isogenous to each other (equivalently, the weight-1 rational 
Hodge structure of $(H^1(M_1;\Q), I_1)$ and $(H^1(M_2;\Q), I_2)$ are
Hodge isomorphic), the endomorphism algebra ${\rm End}(H^1(M_1;\Q))^{\rm Hdg}$
and ${\rm End}(H^1(M_2;\Q))^{\rm Hdg}$ are isomorphic. The converse is not true, 
so the endomorphism algebra ${\rm End}(H^1(M;\Q))^{\rm Hdg}$ alone is not 
enough in specifying an isogeny class of abelian varieties. 

Let $(M;I)$ be an $n$-dimensional complex abelian variety of CM-type 
where there is a totally imaginary field $F$ of degree $2n$ in 
${\rm End}(H^1(M;\Q))^{\rm Hdg}$ acting on $H^1(M;\Q)$. Then the $2n$ 
embeddings of $F$ into $\C$, ${\rm Hom}_{\rm field}(F,\overline{\Q})$, 
is grouped into those---$\Phi^{(1,0)}$---that correspond to the 
Hodge (1,0) component and $\Phi^{(0,1)}$ to the Hodge (0,1) component
(cf Rmk. \ref{rmk:Phi-p-q}). 
The pair $(F, \Phi^{(1,0)})$, called a {\it CM pair}, contains enough 
information to specify an isogeny class of CM-type abelian 
varieties.\footnote{
This notion can be generalized to a rational Hodge structure $(V_\Q,\phi)$
that are not necessarily polarizable, not necessarily of weight-1. 
It is a pair $(F, \amalg_{p,q} \Phi^{(p,q)})$ of a number field 
$F \subset {\rm End}(V_\Q, \phi)$ with $[F:\Q] = \dim_\Q V_\Q$ 
and a decomposition of embeddings ${\rm Hom}_{\rm field}(F,\overline{\Q})$ 
such that the complex conjugation of $\Phi^{(p,q)}$ is $\Phi^{(q,p)}$. 
} %

The Galois theoretical classification of CM-type abelian varieties 
of $n$-dimensions should therefore deal with CM pairs, not CM fields. 
In the case $n=2$ (and also $n=1$), 
however, the classification of CM pairs is no finer than that of the 
endomorphism algebras that ended up with (B), (C), (A) and (A').
Let us explain why in the following, while preparing 
notations to be used in the analysis in this article.

Let us begin with an imaginary quadratic field $K^{(2)} \cong \Q[x]/(x^2-p)$, 
where $p \in \Q_{<0}$. The endomorphism field of a CM elliptic curve is 
always of this form, and this field is also relevant to the abelian surfaces 
in the cases (A) and (A'). There are two embeddings of the field 
$\Q[x]/(x^2-p)$:
\begin{align}
  \tau_\pm : x \longmapsto \pm \sqrt{p} = \pm i \sqrt{-p} \in \C; 
\end{align}
here, and throughout this article, we adopt a convention that 
$\sqrt{d} \in \R$ for $d \in \R_{>0}$ is the square root in the 
real positive axis, and $\sqrt{p} \in \C$ for $p \in \R_{<0}$ the 
square root in the upper complex half plane.\footnote{
In this convention, $\sqrt{p_1}\sqrt{p_2} = - \sqrt{p_1p_2}$ for 
$p_1, p_2 \in \R_{<0}$.
}  %
The two embeddings of $K^{(2)}$, one in $\Phi^{(1,0)}$ 
and the other in $\Phi^{(0,1)}$, are regarded as the two embeddings 
$\{ \tau_{\pm} \}$ of the field $\Q[x]/(x^2-p)$, but which one of 
$\tau_+$ and $\tau_-$ corresponds to $\Phi^{(1,0)}$ depends on 
the choice of an isomorphism $K^{(2)} \cong \Q[x]/(x^2-p)$. 
It is always possible to let the embedding $\Phi^{(1,0)}$ of $K^{(2)}$ 
correspond to $\tau_+$ by modifying the isomorphism by 
${\rm Aut}(K^{(2)} / \Q) \cong \Z_2$. It is for this reason that the 
classification for $n=1$ and the cases (A) and (A') for $n=2$ is fine enough.  
 
Let us now have a look at the degree-4 field in (\ref{eq:endK-str-caseBC})
that is isomorphic to the endomorphism field of simple CM-type abelian 
surfaces (i.e., cases (B, C)). The four embeddings 
of the field $\Q[x,y]/(y^2-d, x^2-p-qy)$ are denoted by $\tau_{\pm \pm}$, where 
\begin{align}
 \tau_{\pm *}: y \mapsto \pm \sqrt{d}, \qquad 
 \tau_{\pm +}:& \;  x \mapsto \sqrt{p \pm q\sqrt{d}}
                         = i \sqrt{-p\mp q\sqrt{d}}, \\
 \tau_{\pm -}: & \; x \mapsto - \sqrt{p \pm q \sqrt{d}}
                         = -i\sqrt{-p\mp q\sqrt{d}}.
  \label{eq:introduce-sh-notation-below}
\end{align}
We introduce a short-hand notation $\sqrt{+} := \sqrt{p+q\sqrt{d}}$ and 
$\sqrt{-} := \sqrt{p-q\sqrt{d}}$ for the pure imaginary complex numbers 
in the upper half plane and use it for the sake of compactness of notation 
in this article. For one fixed isomorphism ${\rm End}(H^1(M;\Q))^{\rm Hdg}=
K \cong \Q[x,y]/(y^2-d, x^2-p-qy)$, the two embeddings in $\Phi^{(1,0)}$ 
of ${\rm End}(H^1(M;\Q))^{\rm Hdg}=K$ may correspond to any one of the 
four pairs that are not the complex conjugation pairs 
$\{ \tau_{++}, \tau_{+-} \}$ or $\{ \tau_{-+}, \tau_{--}\}$. By 
replacing the totally real generator $y$ and totally imaginary generator 
$x$ of the field $\Q[x,y]/(y^2-d, x^2-p-qy)$ appropriately (so the 
rational numbers $d$, $p$ and $q$ will change), however, it is always 
possible to have $\Phi^{(1,0)} = \{ \tau_{++}, \tau_{-+} \}$ 
in the new choice of the generators. It is for this reason that 
the classification ending up with (B) and (C) is enough. 
In the analysis in this article, we understand that this change 
in the presentation is always made, and work only with the CM pairs 
$(K, \{ \tau_{++}, \tau_{-+} \})$ in the cases (B, C).  

With those conventions, we may choose 
$\{dz^1, dz^2\} := \{ v_{a=++}, v_{a=-+} \}$ as a basis of $H^{1,0}(M;\C)$
in the case (B, C):
\begin{align}
  (dz^1, dz^2) = (\hat{\alpha}^1, \hat{\alpha}^2, \hat{\beta}_1, \hat{\beta}_2) \left( \begin{array}{cc} 1 & 1 \\ \sqrt{d} & - \sqrt{d} \\
   \sqrt{p+q\sqrt{d}} & \sqrt{p-q\sqrt{d}} \\
   \sqrt{p+q\sqrt{d}}\sqrt{d} & - \sqrt{p-q\sqrt{d}}\sqrt{d}
  \end{array} \right) =:
 (\hat{\alpha}^i, \hat{\beta}_i ) \left( \begin{array}{c} Z^T \\ \alpha^T
    \end{array} \right);  
  \label{eq:hol-rat-basis-trnsf-BC}
\end{align}
$Z$ is real-valued and $\alpha$ pure-imaginary valued; both are $2\times 2$ 
matrices. In the case (A) and (A'), we may choose 
$\{ dz^1, dz^2 \} := \{ \hat{\alpha}^1 + \sqrt{p_1}\hat{\beta}_1, \;
\hat{\alpha}^2 + \sqrt{p_2} \hat{\beta}_2 \}$ as a basis of $H^{1,0}(M;\C)$; 
in the case (A), $\sqrt{p_1}=\sqrt{p_2}=\sqrt{p}$. In the matrix form, 
\begin{align}
  (dz^1, dz^2) = (\hat{\alpha}^i, \hat{\beta}_i) \left( \begin{array}{c} Z^T \\ \alpha^T \end{array} \right), \qquad 
   Z :=\diag(1,1), \quad \alpha := \diag(\sqrt{p_1}, \sqrt{p_2}) . 
  \label{eq:hol-rat-basis-trnsf-A}
\end{align}
\end{anythng}

\begin{anythng}
\label{statmnt:rflx-field-levelN}
{\bf (The level-$n$ Hodge structure and the reflex field)}
Let $(M;I)$ be a abelian variety of complex $n$ dimensions that 
is of CM-type under the definition Def. \ref{def:CM-geom-4CT}, and 
equivalently Rmk. \ref{rmk:geomCM=strCM-CTAV}.
This definition based on the rational Hodge structure 
of weight-1 implies that $(M;I)$ is of strong CM-type, and consequently, 
of weak CM-type.
The level-$n$ rational Hodge structure, supported 
on the vector subspace $[H^n(M;\Q)]_{\ell = n} \subset H^n(M;\Q)$, is always 
simple; the endomorphism field of this level-$n$ substructure is known to 
be calculable systematically from the weight-1 rational Hodge structure 
$(H^1(M;\Q), I)$ of the abelian variety $(M;I)$, as we review in the following 
(cf \cite[Prop. 1.9.2]{MR257031}).  

Let ${\cal K} = \oplus_i F_i \subset {\rm End}(H^1(M;\Q))^{\rm Hdg}$ be 
an algebra in the form of the direct sum of totally imaginary fields 
with the property $\sum_i [F_i:\Q] = 2n$. Let $\Phi^{(1,0)}_{F_i}$ be 
the $[F_i:\Q]/2$ embeddings of $F_i$ that correspond to the Hodge (1,0) 
components. Then the {\it reflex field} for 
the pair $(\oplus_i F_I, \{ \Phi^{(1,0)}_{F_i} \})$ is defined by 
(e.g., \cite[\S8]{shimura2016abelian} and \cite[Def. 1.17]{milnecm}) 
the field generated by 
\begin{align}
 K^r := \Q \left( 
  \cup_i \left\{ \sum {}_{\tau \in \Phi^{(1,0)}_{F_i}} \tau(x_i) \; | 
    \; x_i \in F_i \right\} \right) \subset \C.
\end{align}
Although there is ambiguity in the choice of the algebra 
${\cal K} = \oplus_i F_i \subset {\rm End}(H^1(M;\Q))^{\rm Hdg}$ 
in a case more than one simple Hodge substructures 
are mutually Hodge-isomorphic, the reflex field $K^r$
in this definition is known not to depend on the 
ambiguity \cite[\S8]{shimura2016abelian}. 
 
The number field $K^r$ here is introduced as a subfield of $\C$. 
On the other hand, the endomorphism field of the CM-type simple 
rational Hodge structure on $[H^n(M;\Q)]_{\ell = n}$ has 
$\dim_\Q [H^n(M;\Q)]_{\ell=n}$ embeddings into $\C$. Because 
$h^{n,0}(M)=1$, there is just one embedding of this endomorphism field, 
$\Phi^{(n,0)} = \{\tau^r_{(n0)} \}$; the image of the field 
${\rm End}([H^n(M;\Q)]_{\ell=n})^{\rm Hdg}$ by this embedding $\tau^r_{(n0)}$ 
is the subfield $K^r \subset \C$ (\cite[Prop. 1.9.2]{MR257031}). 
So, we will often identify $K^r$ with the endomorphism field 
of the level-$n$ component with this embedding $\tau^r_{(n0)}$ in this article. 
\end{anythng}

Here is a little more detail on the reflex field (the endomorphism 
field of the level-$n$ component), which we use in the analysis 
in this article. Just a straightforward computation is enough to 
verify the following statements. 

In the case (A), the reflex field is isomorphic to 
$K^r \cong \Q[\xi]/(\xi^2 - p)$ when the endomorphism algebra 
of the weight-1 rational Hodge structure is $M_2(\Q(\sqrt{p}))$. The level-2 component 
(a reminder: $\dim_\C M=n=2$ now) must be of 2-dimensions over $\Q$ 
because $[K^r:\Q] =2$.  

In the case (A'), $K^r \cong \Q[\xi_1, \xi_2]/(\xi_1^2-p_1, \; \xi_2^2-p_2)$.
It is also possible to adopt $\xi := \xi_1$ and $\eta := \xi_1\xi_2$ 
as generators, when $K^r \cong \Q[\xi,\eta]/(\xi^2-p_1, \eta^2-p_1p_2)$. 
The four embeddings of $K^r$ may be denoted by $\tau^r_{\epsilon',\epsilon^r}$, 
where $\tau^r_{ \pm *}:\eta \mapsto \pm \sqrt{p_1p_2}$ and 
$\tau^r_{*\pm}: \xi \mapsto \pm \sqrt{p_1}$.  

In the case (B, C), 
\begin{align}
 K^r \cong \Q[\eta , \xi]/(\eta^2-d', \xi^2-2p+2\eta).
\end{align}
Let us introduce notations $\tau^r_{\pm \pm}$
for the four embeddings of the reflex field 
$K^r \hookrightarrow \overline{\Q}$. 
\begin{align}
 \tau^r_{\pm *}: \eta \mapsto \pm \sqrt{d'}, \qquad 
 \tau^r_{\pm +}: & \; \xi \mapsto \sqrt{p+q\sqrt{d}} \pm \sqrt{p-q\sqrt{d}}, \\
 \tau^r_{\pm -}: & \; \xi \mapsto
    - \left( \sqrt{p+q\sqrt{d}} \pm \sqrt{p-q\sqrt{d}} \right).
\end{align}
We will use the fact that 
\begin{align}
 \tau^r_{\pm +} (2q/\xi) = \left( \sqrt{p+q\sqrt{d}} \mp \sqrt{p-q\sqrt{d}}\right)/\sqrt{d}. 
\end{align}
%
%

\begin{defn}
For a general compact K\"{a}hler manifold $(M;I)$ of dimension $n$, 
\begin{align}
  {\cal H}^2(M_I) := H^{1,1}(M_I;\R) \cap H^2(M;\Q)
\end{align}
is said to be the {\it algebraic part} of $H^2(M;\Q)$. When $n=2$, 
$H^2(M;\Q)$ is not just a vector space, but is endowed with a bilinear 
form $H^2(M;\Q) \times H^2(M;\Q) \longrightarrow H^4(M;\Q) \cong \Q$. 
The orthogonal complement 
\begin{align}
  T_M \otimes \Q := \left[ {\cal H}^2(M_I)^\perp \subset H^2(M;\Q) \right]
\end{align}
with respect to the bilinear form is said to be the {\it transcendental part}.
It is known that bilinear form restricted to the algebraic part 
${\cal H}^2(M_I)$ is non-degenerate when the K\"{a}hler manifold $(M;I)$ 
is polarizable (i.e., an algebraic variety, such as an abelian variety 
(cf. Rmk. \ref{rmk:pol-exists-in-algV})).
So, there is an orthogonal decomposition of the vector space 
$H^2(M;\Q) \cong {\cal H}^2(M_I) \oplus (T_M\otimes \Q)$. This is a decomposition 
of rational Hodge structure; the substructure on ${\cal H}^2(M_I)$ is of 
level-0 and that on $T_M\otimes \Q$ of level-2 (cf Ex. \ref{ex:rHdgSubStr}). 

For a 2-form $\psi \in H^2(M;\R)$, its decomposition into 
$({\cal H}^2(M_I)\otimes \R) \oplus (T_M \otimes \R)$ is denoted by 
$\psi^{\rm alg} + \psi^{\rm transc}$, and are called the {\it algebraic 
and transcendental parts/components}. 
\end{defn}

\begin{lemma}
\label{lemma:gen-Sm} [well known in math literatures 
(e.g., \cite{shimura2016abelian,milnecm,Chen:2005gm})]
Let $(M;I)$ be a CM-type abelian variety of $n$ dimensions, 
with a CM field $F \subset {\rm End}(H^1(M;\Q))^{\rm Hdg}$ such that 
$[F:\Q] = 2n$. Then the algebraic part ${\cal H}^2(M_I) \subset H^2(M;\Q)$ 
contains an $n$ dimensional subspace ${\cal H}^2(M_I)_{\rm gen}$ specified below 
($h^{1,1}(M) = n^2$, so that is possible).   The proof also 
introduces a basis on ${\cal H}^2(M_I)_{\rm gen}$ and also explains 
how to construct a polarization within ${\cal H}^2(M_I)_{\rm gen} \subset 
{\cal H}^2(M_I)$. 
\end{lemma}
\proof Let $F_0$ be the totally real subfield of $F$, 
and $\Phi^{(1,0)} = \{ \tau_{a=1,\cdots, n}\}$ be the embeddings 
of $F$ corresponding to $H^{1,0}(M;\C)$. 
There must be a basis $\{e_{I=1,\cdots, 2n} \}$ of $H^1(M;\Q)$
and a basis $\{ \eta_{I=1,\cdots, 2n} \}$ of $F/\Q$ so that 
$\{dz^a := e_I \tau_a(\eta_I) \; | \; a=1,\cdots, n\}$ are the $n$ 
independent holomorphic 1-forms (cf Lemma \ref{lemma:very-useful}). 

Now, let $\xi_* \in F$ be a generator of $F/F_0$ (i.e., $F=F_0(\xi_*)$)
 so that $\xi_*^2 \in F_0$ (there is not such structure in a 
general totally imaginary field, so there is not an analogue for 
complex tori with sufficiently many complex multiplications). 
Then for any element $\xi \in \xi_*F_0^\times$, 
\begin{align}
 {\cal Q}^{(\xi)} & \; := 
        \sum_{a=1}^n 2\tau_a(\xi) dz^a \wedge d\bar{z}^{\bar{a}}, \\
   &\; = e_I \wedge e_J \sum_{a=1}^n \left( \tau_a(\xi \eta_I \bar{\eta}_J)
         - \tau_a(\xi \bar{\eta}_I  \eta_J) \right),  \nonumber \\
   & \; = e_I \wedge e_J \sum_{a=1}^n \left( \tau_a(\xi \eta_I \bar{\eta}_J)
       + \tau_a(\bar{\xi}\bar{\eta}_I \eta_J ) \right)
    = e_I \wedge e_J {\rm Tr}_{F/\Q}[\xi \eta_I \bar{\eta}_J] \in {\cal H}^2(M)
\end{align}
(for any field extension $E/F$, ${\rm Tr}_{E/F}[x] \in F$ for $x \in E$; 
see \cite[A.1.15]{Kanno:2017nub} or any introductory textbook on field 
theory).
Linearly independent choices of $\xi$ from $\xi_* F_0$ generate 
an $n$-dimensional subspace of ${\cal H}^2(M_I)$, which is denoted 
by ${\cal H}^2(M_I)_{\rm gen}$. 

For the (1,1) form ${\cal Q}^{(\xi)}$ to be a polarization 
(see Def. \ref{def:abelianVar-vs-coxTrs}--Rmk. \ref{rmk:refer2GriffHarr} 
and Rmk. \ref{rmk:pol-exists-in-algV}), 
first, choose $\{e_I\}$ to be an integral basis of $H^1(M;\Z)$, 
and restrict $\xi$ such that ${\rm Tr}_{F/\Q}[\xi \eta_I \bar{\eta}_J]
 \in \Z$ for all the pairs $(I, J)$; the basis $\{ \eta_I \}$ should be 
those that correspond to the integral basis $\{ e_I \}$. 
Second, impose inequalities on $\xi \in \xi_* K_0$ so that it is positive 
definite.  \qed 

\begin{lemma}
 \label{lemma:gen-Sm-nonSimpl}
Let $M_I = (M;I)|_{M=T^{2n}}$ be an abelian variety of CM-type. Then 
\[
 \dim_\Q {\cal H}^2(M_I) \geq \dim_\C M_I.
\] 
\end{lemma}
\proof We can split the vector space $H^1(M;\Q)$ into its 
components $\oplus_{a \in A} [H^1(M;\Q)]_a$ supporting simple Hodge 
substructures; let $K_a$ be the CM field ${\rm End}([H^1(M;\Q)]_a)^{\rm Hdg}$.
Thus, it is enough to prove the statement for a simple abelian variety, and 
that was done in Lemma \ref{lemma:gen-Sm}.    \qed 

Lemmas \ref{lemma:gen-Sm}
and \ref{lemma:gen-Sm-nonSimpl} above imply that 
\begin{align}
    {\cal H}^2(M)\otimes \C & \; \supset {\cal H}^2(M)_{\rm gen} \otimes \C
   = {\rm Span}_\C \left\{ dz^a \wedge d\bar{z}^{\bar{a}} \right\},  
\end{align}
which can be accompanied by 
\begin{align}
 T_M \otimes \C & \; \subset T_M^{\rm gen}\otimes \C = 
   {\rm Span}_\C \left\{ (dz^a\wedge dz^b)_{a<b}, \; 
       (d\bar{z}^{\bar{a}}\wedge d\bar{z}^{\bar{b}})_{a<b}, 
    \; (dz^a\wedge d\bar{z}^{\bar{b}})_{a\neq b} \right\}, 
  \label{eq:formula-T-gen} 
\end{align}
to have $H^2(M;\C) \cong ({\cal H}^2(M_I)_{\rm gen}\otimes \C) \oplus (T_M^{\rm gen}\otimes \C)$.
For a CM abelian surface $(M;I)$, where $n=2$, 
$(T_M^{\rm gen}\otimes \C)$ is orthogonal to ${\cal H}^2(M_I)_{\rm gen}\otimes \C$
with respect to the bilinear form $H^2(M)\times H^2(M)\rightarrow H^4(M)\cong \C$. 

\begin{anythng}
\label{statmnt:Tm}
{\bf $n=2$, Cases (B, C, A')}: In the cases (B, C, A')
of CM abelian surfaces $(M;I)$, we already know that 
$T_M\otimes \Q = [H^2(M;\Q)]_{\ell = 2}$ is of dimension $[K^r:\Q]=4$. 
This means that $T_M\otimes \C = T_M^{\rm gen}\otimes \C$, which 
also means that ${\cal H}^2(M_I)_{\rm gen}\otimes \C = {\cal H}^2(M_I)\otimes \C$. 

In the case (B, C), 
\begin{align}
 dz^1 \wedge d\bar{z}^{\bar{1}} & \; = -2\sqrt{p+q\sqrt{d}} \left\{ 
    (\hat{\alpha}^1\hat{\beta}_1) + d(\hat{\alpha}^2\hat{\beta}_2)
     + \sqrt{d}(\hat{\alpha}^1\hat{\beta}_2 + \hat{\alpha}^2\hat{\beta}_1)
    \right\},  \label{eq:dz1-dzbar1-caseBC} \\
 dz^2 \wedge d\bar{z}^{\bar{2}} & \; = -2\sqrt{p-q\sqrt{d}} \left\{ 
    (\hat{\alpha}^1\hat{\beta}_1) + d(\hat{\alpha}^2\hat{\beta}_2)
     - \sqrt{d}(\hat{\alpha}^1\hat{\beta}_2 + \hat{\alpha}^2\hat{\beta}_1)
    \right\},  \label{eq:dz2-dzbar2-caseBC}
\end{align}
so
\begin{align}
  {\cal H}^2(M_I) &\; = {\rm Span}_\Q \left\{ 
     (\hat{\alpha}^1\hat{\beta}_1) + d(\hat{\alpha}^2\hat{\beta}_2), \; 
     (\hat{\alpha}^1\hat{\beta}_2 + \hat{\alpha}^2\hat{\beta}_1) \right\},
     \label{eq:Sm-Qbasis-caseBC} \\
  T_M \otimes \Q & \;  = {\rm Span}_\Q \left\{ 
     (\hat{\alpha}^1\hat{\alpha}^2), \; (\hat{\beta}_1\hat{\beta}_2), \;
     (\hat{\alpha}^1\hat{\beta}_2-\hat{\alpha}^2\hat{\beta}_1), \; 
     (\hat{\alpha}^1\hat{\beta}_1 - d \hat{\alpha}^2\hat{\beta}_2) \right\}. 
  \label{eq:Tm-Qbasis-caseBC}
\end{align}
We will use the following notation later:
\begin{align}
 e_1 := \hat{\alpha}^1\hat{\beta}_1 +d\hat{\alpha}^2\hat{\beta}_2, 
  \quad e_2 := \hat{\alpha}^1\hat{\beta}_2 + \hat{\alpha}^2\hat{\beta}_1. 
    \label{eq:def-ratBasis-H2-caseBC}
\end{align}

The following generator of the Hodge $(2,0)$ component is in $T_M\otimes \C$ indeed:
\begin{align}
-\frac{dz^1 \wedge dz^2}{2\sqrt{d}}  & \; = \left[ 
    \hat{\alpha}^1\hat{\alpha}^2
 + (\sqrt{+}-\sqrt{-})/(2\sqrt{d}) \; [\hat{\alpha}^1\hat{\beta}_1-d\hat{\alpha}^2\hat{\beta}_2]  \right. \nonumber \\
 & \qquad \qquad \quad \left. 
 + (\sqrt{+}+\sqrt{-})/2 \; [
  \hat{\alpha}^1\hat{\beta}_2- \hat{\alpha}^2\hat{\beta}_1]
 +\sqrt{d'}\; \hat{\beta}_1\hat{\beta}_2 \right].
  \label{eq:dz1-dz2-caseBC}
\end{align}
Here, we used the notation introduced 
below (\ref{eq:introduce-sh-notation-below}). 
This generator has the form 
of (\ref{eq:egVec-from-very-useful-L}), where 
$\{ \eta_I \} = \{ 1, \; \eta, \; \xi/2, \; q/\xi \}$ is the 
basis of $K^r/\Q$ when the basis $\{e_I \}$ of $T_M \otimes \Q$ 
is the one in (\ref{eq:Tm-Qbasis-caseBC}), and the embedding $\tau^r_{++}$
corresponds to the 1-dimensional ($h^{2,0}=1$) eigenspace for the Hodge 
(2,0) component, i.e., $\tau^r_{(20)} = \tau^r_{++}$.   

In the case (A'), 
\begin{align}
  dz^1 \wedge d\bar{z}^{\bar{1}} = -2\sqrt{p_1}\hat{\alpha}^1 \hat{\beta}_1, 
    \qquad 
  dz^2 \wedge d\bar{z}^{\bar{2}} = -2\sqrt{p_2}\hat{\alpha}^2 \hat{\beta}_2, 
\end{align}
so $dz^1 \wedge dz^2 = (\hat{\alpha}^1\hat{\alpha}^2)
 - \sqrt{p_1p_2} (\hat{\beta}_1\hat{\beta}_2)
 + \sqrt{p_2}(\hat{\alpha}^1\hat{\beta}_2)
 + \sqrt{p_1}(\hat{\beta}_1\hat{\alpha}^2)$ is in $T_M \otimes \C$ below:
\begin{align}
 {\cal H}^2(M_I) & \; = {\rm Span}_\Q \left\{ 
     (\hat{\alpha}^1\hat{\beta}_1), \; (\hat{\alpha}^2\hat{\beta}_2)
          \right\},
       \label{eq:Sm-Qbasis-caseAprm} \\
  T_M \otimes \Q & \;  = {\rm Span}_\Q \left\{ 
     (\hat{\alpha}^1\hat{\alpha}^2), \; (\hat{\beta}_1\hat{\beta}_2), \;
     (\hat{\alpha}^1\hat{\beta}_2), \; (\hat{\alpha}^2\hat{\beta}_1) 
        \right\}. 
   \label{eq:Tm-Qbasis-caseAprm}    
\end{align}
The generator $dz^1 \wedge dz^2$ of the Hodge (2,0) component in 
$T_M \otimes \Q$ has the form of (\ref{eq:egVec-from-very-useful-L}), 
with the embedding $\tau^r_{++}$ for the Hodge (2,0) component (i.e., 
$\tau^r_{(20)} = \tau^r_{++}$); 
$\{ 1, -\eta, -\eta/\xi, \xi\}$ is the basis of $K^r/\Q$ that 
corresponds to the basis (\ref{eq:Tm-Qbasis-caseAprm}) of $T_M \otimes \Q$. 

{\bf In the case (A)},\footnote{
\label{fn:SM}
The case (A), where $(M;I)$ is isogenous to a product of two copies 
of a CM elliptic curve, is known \cite{shioda1974singular} 
to be the case of rank-2 $T_M$. 
} %
\begin{align}
  T_M\otimes \Q & \; = {\rm Span}_\Q \left\{
     (\hat{\alpha}^1\hat{\alpha}^2+ p \hat{\beta}_1\hat{\beta}_2), \; 
     (\hat{\alpha}^1\hat{\beta}_2 + \hat{\beta}_1\hat{\alpha}^2) \right\}, 
   \label{eq:Tm-Qbasis-caseA}
\end{align}
generated by the real and imaginary part of $dz^1 \wedge dz^2 = (\hat{\alpha}^1+\sqrt{p}\hat{\beta}_1)(\hat{\alpha}^2+\sqrt{p}\hat{\beta}_2)$.  
\begin{align}
{\cal H}^2(M_I) & \; = {\rm Span}_\Q \left\{ (\hat{\alpha}^1\hat{\beta}_1), \; 
    (\hat{\alpha}^2\hat{\beta}_2), \; 
    (\hat{\alpha}^1\hat{\alpha}^2-p\hat{\beta}_1\hat{\beta}_2), \;
    (\hat{\alpha}^1\hat{\beta}_2 - \hat{\beta}_1\hat{\alpha}^2) \right\}, 
    \label{eq:Sm-Qbasis-caseA}
\end{align}
generated by $dz^1\wedge d\bar{z}^{\bar{1}}/(-2\sqrt{p})$ and  
$dz^2\wedge d\bar{z}^{\bar{2}}/(-2\sqrt{p})$ in ${\cal H}^2(M)_{\rm gen}$, along 
with the real and imaginary part of $dz^1 \wedge d\bar{z}^{\bar{2}}$. 
\qed 
\end{anythng}

\section{Choice of Complex Structure (Conditions 1 \& 2)}
\label{sec:choose-cpx-str}

As we have remarked in Discussion \ref{statmnt:subtlty-choose-cpx-str}, 
there is no way not to choose a complex structure on $T^{2n}$
when we wish to establish a Gukov--Vafa-like characterization 
of rational $T^{2n}$-target (S)CFTs. On one hand, it is desirable to 
find a characterization statement that works well for a broader class 
of complex structures. On the other hand, if there is a choice of 
complex structure $I$ that is well-motivated in string theory, 
there is a chance that we have a sharper/clearer characterization statement
for such a smaller class of complex structure $I$. 

Mathematicians tend favor of algebraic varieties over 
general complex analytic manifolds, of abelian varieties over 
general complex tori (e.g. \cite{Chen:2005gm}). 
String theorists, however, do not have any apriori reason to restrict 
attention only to the objects in the category of algebraic varieties; 
it is rather desirable that no set of geometric data $(M;G,B)$ for an 
SCFT is overlooked because of limited range of attention.
As reviewed in the appendix \ref{ssec:rHstr-Pol-alg}, 
a complex manifold $(M;I)$ is regarded as an algebraic variety if and only 
if $I$ is polarizable. So, this issue is whether it is righteous to 
restrict our attention to the choice of complex structure 
that is polarizable. 

In the case $M \cong T^{2n}$, there is always a choice 
of complex structure $I$ that is polarizable 
(isogenous to the product of $n$ CM elliptic curves), 
if a set of data $(M;G;B)$ that corresponds to a rational CFT
(Cor. \ref{cor:Wend-ExEx}). So we may argue that no set of data 
$(M;G,B)$ for a rational SCFT is lost immediately by demanding 
that there exists a polarizable complex structure $I$ (with which 
$(M;I)$ is of strong CM-type), at least in the case study for $M=T^{2n}$.  
That is a passive support for demanding that $I$ is polarizable. 
In section \ref{ssec:pol}, however, we will argue that there is 
also a positive support for restricting our attention only to 
polarizable complex structures, even from the perspective of string theorists. 

There is one more choice of complex structure $I$ for a set of data 
$(M;G,B)$ that is felt better motivated by string theorists. Instead of 
demanding that $I$ is polarizable, we may reduce arbitrariness in the choice 
of $I$ by demanding that the Hodge (2,0) component of the two-form $B$ 
vanishes with respect to $I$; when the two-form $B$ is generic, this demand 
uniquely determines the choice of $I$, and the arbitrariness in Discussion 
\ref{statmnt:subtlty-choose-cpx-str} is gone. 
That is also a choice that is often made when string theorists deal with 
a case $M$ is a K3 surface, or a hyper-K\"{a}hler manifold. Those two 
restrictions on $I$ are independent, however, and are not compatible in 
general. We will see in section \ref{ssec:B-transc}, however, that the two 
requirements are always compatible when $M=T^4$, and the SCFT 
for $(M;G,B)$ is rational. 
 
That was how the present authors were led to set condition 1 in 
Conj. \ref{conj:forT4} as it is. Existence of such a complex structure 
is therefore non-trivial for a generic data $(M;G,B)$ of an SCFT; 
a proof of existence (condition 1), done in this section for $M=T^4$, 
already provides a non-trivial characterization for geometric data $(M;G,B)|_{M=T^4}$ 
that correspond to rational SCFTs. 

\subsection{Polarization}
 \label{ssec:pol}

\begin{anythng}
{\bf (Complex tori with sufficiently many complex multiplications 
vs CM-type abelian varieties)} Definitions of {\it sufficiently many 
complex multiplications} of a complex torus and the {\it CM-type} property 
of an abelian variety are quoted and reviewed in 
the appendix \ref{ssec:Guide-CM4CT-PartI}, 
Thm. \ref{thm:smCM4CT-via-EndAlgCoh} and 
Rmk. \ref{rmk:announce-CMdef-use-EndAlg-onCoh}, 
Def. \ref{def:abelianVar-vs-coxTrs}--Rmk. \ref{rmk:refer2GriffHarr} and  
Def. \ref{def:CM-rat-Hdg-str-byEnd}.  The key intuitive properties of 
those complex tori and abelian varieties, which also carry the same information 
as the definitions, are stated in Prop. \ref{props:CM-Intuitv-summary4CT} 
and Rmk. \ref{rmk:CM-Intuitv-summary4AV}, respectively. 
  
To string theorists, CM-type abelian varieties must 
look very much like ordinary complex tori with sufficiently many complex 
multiplications. Whether or not they have an implementation as algebraic varieties 
(than complex analytic manifolds) would not usually matter in physics. 
Although one can introduce a notion of {\it polarization} and argue whether 
a geometry/Hodge structure has it or not, possible existence 
of a polarization\footnote{
In math literature, discussions are developed sometimes for a triple 
$(M; I; {\cal Q})$ of a complex torus $(M;I)$ that admits polarizations
along with a choice ${\cal Q}$ of a polarization (alternatively, along with 
an embedding of $(M;I)$ into a projective space), not just demanding 
existence of ${\cal Q}$'s. The automorphism group of $(M;I;{\cal Q})$
is finite in the case of abelian varieties \cite[Cor. 5.1.9]{birkenhakeabelian}, 
for example, and the moduli space of $(M;I;{\cal Q})$ has a nicer math property 
than that of $(M;I)$. In this article as a research in string theory, however, only 
the possible existence of polarizations matters, not a choice of it. 
} %
does not immediately give any physics intuition for something non-trivial to exist. In fact, that is not right. 
To see this, let us consider the following 
result by Meng Chen \cite{Chen:2005gm}.
\end{anythng}

\begin{props}\cite[Thm. 2.5]{Chen:2005gm}
\label{props:MChen-thm2.5}
{\it Let $(M;I)|_{M=T^{2n}}$ be an abelian variety, i.e., a complex 
torus that admits a polarization. If there exists a constant 
metric $G$ compatible with $I$ that is rational in the sense 
of (\ref{eq:cond-GnB-rational}), then 
the polarizable rational Hodge structure on $H^1(M;\Q)$ is of CM-type. }
\qed
\end{props}

For a set of data $(T^{2n}; G, B)$ for which the (S)CFT is rational, 
there is always a complex structure $I$ with which $G$ is compatible 
and which admits a polarization (see Cor. \ref{cor:Wend-ExEx} and 
the discussion that follows). So, Prop. \ref{props:MChen-thm2.5} above 
is not an empty statement for {\it any} $T^{2n}$-target rational (S)CFTs. 
This is already a proof that condition 2 in Conj. \ref{conj:forT4} 
(in conjunction with condition 1) is a necessary condition for the 
rationalness of the SCFT.  

The statement of Prop. \ref{props:MChen-thm2.5} does not say what happens 
if we choose a non-polarizable complex structure $I$ for a set of data 
$(M;G,B)|_{M=T^{2n}}$ of a rational SCFT. By reading between the lines 
of the proof of \cite[Thm. 2.5 + Prop. 2.4]{Chen:2005gm}, however, 
we can find an answer to this question, as follows.  

The following two conditions on a complex torus $(M;I)|_{M=T^{2n}}$ 
of $n$ dimensions are known to be equivalent:\footnote{
The proof of Props. 17.3.4 and 17.3.5 of \cite{birkenhakeabelian} does not 
assume that the Hodge structure in question admits a polarization.

We do not write down a definition of a Hodge group in this article,
but is found in many literatures. It is enough in reading the main text 
of this article just to know that there is a widely accepted definition 
for this jargon. If a reader wants to find a review on a Hodge group 
or a Mumford--Tate group written by string theorists for string theorists, 
the appendix A.3 of \cite{Okada:2023udq} will be an option.
} %
\begin{itemize}
\label{pg:CM-def-1n2}
 \item [(i)] The algebra ${\rm End}(H^1(M;\Q))^{\rm Hdg}$ over $\Q$ 
    contains a commutative semi-simple subalgebra of dimension $2n$.
 \item [(ii)] The Hodge group ${\rm Hdg}(M_I)$ of the Hodge structure is 
   commutative. 
\end{itemize}
Condition (i) is adopted as the definition for a complex torus to have 
sufficiently many complex multiplications in 
Def. \ref{def:CM-rat-Hdg-str-byEnd} and Rmk. \ref{rmk:geomCM=strCM-CTAV} 
in this article. 
Prop. 2.4 of \cite{Chen:2005gm} proves, with an abelian variety $(M;I)$ 
in mind, that the properties (i) and (ii) above are equivalent to the property 
\begin{itemize}
\item [(iii)] the set of real points ${\rm Hdg}(M_I)(\R)$ of the Hodge group is compact.
\end{itemize}
Thm. 2.5 of \cite{Chen:2005gm} proves the compactness (iii) of 
${\rm Hdg}(M_I)(\R)$ when there exists a rational metric $G$ that is 
compatible with $I$, and hence the CM property (i) and (ii). 
In proving the equivalence between the properties (iii) and (i, ii), however, 
Ref. \cite{Chen:2005gm} uses the fact that ${\rm Hdg}(M_I)(\R)^{{\rm Ad}(h(i))}$ 
is compact; to prove the compactness of this group, Thm. 1.3.16 
of \cite{rohde2009cyclic}\footnote{
We refer to the LNM version, not to its arXiv versions. 
} %
uses the positive definiteness of a polarization of the rational Hodge 
structure.\footnote{
For a stringy reader who wants a more down-to-earth presentation on how 
polarizability of a rational Hodge structure makes a difference, an example 
in the appendix \ref{ssec:Hdg-grp-ex} might serve as an antidote to the 
chain of abstract logic here. 
} %
To conclude, the equivalence between the properties (iii) 
and (i, ii) breaks down when the rational Hodge structure $(H^1(M;\Q),I)$
does not necessarily have a polarization. 

In our context, even when there is a constant rational metric $G$ 
of $(M;G;I)|_{M=T^{2n}}$, 
we cannot derive the property (i), the presence of sufficiently many 
complex multiplications (endomorphisms), if $I$ is not polarized. 
We could pay attention to complex tori $(M;I)|_{M=T^{2n}}$ satisfying the 
property (i) above, but it is not obvious whether 
there exists\footnote{
Although Thm. 2.5 of \cite{Chen:2005gm} constructs rational metrics 
satisfying (\ref{eq:cond-GnB-rational}) by (\ref{eq:basis-KahlerF-forRatG}), this 
construction exploits properties of CM fields that are not available to a general 
totally imaginary field.
} %
a constant rational metric compatible with the complex structure $I$. 
For this reason, we pay attention only to complex structures $I$ that 
admit polarization in the rest of this article. 

\begin{anythng}
\label{statmnt:enumrt-polarizable-I-forT4nG}
For a given $M=T^4$ and a not-necessarily-rational constant metric $G$ on it,
there are infinitely many choices of complex structure $I$ that is 
compatible with $G$ and is polarizable. We can see this as follows. 
Recall, first, that the metric $G$ determines the real 3-dimensional 
vector subspace $\Pi_G$ of $H^2(T^4;\R)$ that consists of 2-forms that 
are self-dual under the Hodge-* operation with respect to the metric $G$. 
Choice of a complex structure $I$ compatible with $G$ is to specify 
one semi-line $\R_{>0} \omega$ for $\omega = 2^{-1} G(I-,-)$ within the subspace 
$\Pi_G$; so, the choice of $I$ comes with a variety 
$S^2$ \cite[\S2]{Aspinwall:1996mn}; 
the two directions in $\Pi_G$ orthogonal to $\omega$ with respect to the 
wedge product supports the holomorphic (2,0) form $\Omega_M$ on $T^4$. 
Recall also that any 2-form $\psi \in H^2(T^4;\R)$ can be decomposed into the 
self-dual component and the anti-self-dual component under the Hodge-* 
operation, $\psi = \psi_{\parallel} + \psi_{\perp}$.  

Now, choose any $\psi \in H^2(T^4;\Q)$ with $\int \psi \wedge \psi >0$; 
then $\psi_\parallel \neq 0$ because $\int \psi \wedge \psi >0$. 
One may then choose a complex structure so that 
$\omega \in \R_{>0} \psi_\parallel$ or $\omega \in - \R_{>0} \psi_\parallel$;
with this choice, $\psi$ is in ${\cal H}^2(T^4_I)$; the condition 
$\int \psi \wedge \psi >0$ implies that the Hermitian $2\times 2$ matrix 
$\{ (\psi)_{a\bar{b}} \}_{a,b=1,2}$ in 
$\psi = i\psi_{a\bar{b}}dz^a \wedge d\bar{z}^{\bar{b}}$ has a positive determinant, 
meaning that the product of the two real-valued eigenvalues is positive.    
So, either $\psi$ or $-\psi$ is a polarization ($D_P$ in 
Rmk. \ref{rmk:pol-exists-in-algV}). This procedure exhausts all the 
possible polarizable and compatible $I$ for a given $(T^4; G)$. 
Prop. \ref{props:MChen-thm2.5} by Meng Chen implies that 
$(T^4;I)$ is of CM-type for all those $I$'s. 

For a given $M=T^{2n}$ and a constant {\it rational} metric $G$ on it, 
there is yet another way to realize that there are countably infinitely 
many complex structures $I$ compatible with $G$ and is polarizable; 
one can find countably infinitely many choices of isogenies $\varphi$ 
in Cor. \ref{cor:Wend-ExEx} (proven in \cite{Wendland:2000ye}), 
and hence infinitely many such complex structures $\varphi^*(I_0)$; 
there can be more of those outside of those of the form of $\varphi^*(I_0)$. 
\end{anythng}

\subsection{Transcendental Part of the B-field}
\label{ssec:B-transc}

Restricting attention only to complex structures $I$ that are polarizable 
may certainly fit for the taste of algebraic geometers, and 
is also necessary in setting up characterization of rational SCFTs 
in terms of complex multiplications along the line of Conj. \ref{conj:forT4}. 
There is an alternative choice of complex structure $I$ that has been 
often employed in studies of string compactification on K3 surface 
(where $h^{2,0}(M_I)$ is non-zero, just like in the case $M=T^{2n}$);
that is to set $I$ so that the 2-form $B$ of the SCFT data 
$(M;G,B)$ is purely of Hodge (1,1) component, $B^{(2,0)}=0$.  
The two preferred choices of a complex structure are actually compatible
as we see in 
\begin{props}
\label{props:I-polNalgB}
This is for the case $n=2$. 
{\it Let $(T^{2n};G, B)|_{2n=4}$ be a set of data for which the (S)CFT is 
rational. 
Then there exists a polarizable complex structure $I$ on $T^{4}$ with which 
$G$ is compatible, and the $B$-field only has the Hodge $(1,1)$ component 
with respect to that $I$. In particular, the $B$-field is in the algebraic part 
${\cal H}^2(T^4_I)$.}
\end{props}
\proof   
When $B_\parallel=0$, automatically there is no Hodge (2,0) or (0,2) component 
in $B=B_\perp$, regardless of which direction in $\Pi_G$ is chosen (and of how 
a compatible complex structure $I$ is chosen). We just have to choose any $I$ in $S^2$
such that a polarization exists (such an $I$ exists; we have already seen 
that at the end of section \ref{ssec:torus-RCFT}). 

When $B_\parallel \neq 0$, there is virtually no free choice for $I$
after requiring that the Hodge (2,0) component is absent; we have to choose 
$\omega \in \R B_\parallel$. Choosing $\omega \in \R_{<0} B_\parallel$ instead of 
$\omega \in \R_{>0} B_\parallel$ is nothing more than declaring holomorphic 
coordinates on $T^4$ as anti-holomorphic coordinates instead. So, we 
fix $\omega = 2^{-1}G(I-,-)$ by the condition $\omega \in \R_{>0} B_\parallel$, 
and prove that there is a polarization under $I$. 

To this end, note that 
$\int \Omega_M \wedge B_\parallel = 0$ and $\int \Omega_M \wedge B_\perp=0$, 
which is equivalent to 
\begin{align}
\int_{T^4} \Omega_M \wedge B = 0, \qquad \int_{T^4} \Omega_M \wedge (*B)=0. 
  \label{eq:temp-B-1}
\end{align}
So, both $B$ and $*B$ are in $H^{1,1}(T^4_I;\R)$. 
We already know that $B$ is also in $H^2(T^4;\Q)$ 
when $(T^4;G,B)$ is for a rational (S)CFT (see (\ref{eq:cond-GnB-rational})).

If $\R (*B) = \R B$, then either $*B=B$ or $*B=-B$. 
As we are in the case $B_\parallel \neq 0$ now, $*B=B$ is the only option, 
and $B_\parallel = B = * B$. In that situation, either $B$ or $-B$ 
is a polarization (repeat the argument in 
Discussion \ref{statmnt:enumrt-polarizable-I-forT4nG}).
This proves that either $B$ or $-B$ is positive definite, besides being 
rational.  

If $*B$ and $B$ are linearly independent in $H^2(T^4;\R)$, 
then ${\rm Span}_\R\{ B, *B\} \subset H^{1,1}(T^4;\R)$ is a 
2-dimensional subspace, with signature (1, 1). 
Now, we claim that $\R(*B) \cap H^2(T^4;\Q)$ is not $\{0\}$. 
To see this, it is enough to note that 
\begin{align}
 (*B)_{IJ} = \sqrt{{\rm det}(G)} \; \epsilon_{IJMN}\; G^{MK}G^{NL}B_{KL}\frac{1}{2}, 
  \label{eq:temp-B-2}
\end{align}
where $\epsilon_{IJKL}$ is the $\{ \pm 1\}$-valued totally anti-symmetric 
tensor of rank-4; $\R(*B)$ contains such 2-forms 
as $\sqrt{{\rm det}(G)}^{\pm 1} (*B)$, which are rational, as promised. 
This means that ${\cal H}^2(T^4_I) = H^2(T^4;\Q) \cap H^{1,1}(T^4_I;\R)$ 
is at least of 2-dimensions over $\Q$ of signature (1, 1). Moreover, 
within the 2-dimensional ${\cal H}^2(T^4_I)\otimes \R$, there is a line 
$\R \omega$ along the K\"{a}hler form, and there is a rational point 
of ${\cal H}^2(T^4_I)$ arbitrarily close to the line $\R \omega$ in 
${\cal H}^2(T^4_I)\otimes \R$. Such a rational point is a polarization.  

The last statement in Prop. \ref{props:I-polNalgB} follows from 
Lemma \ref{lemma:no20-noTransc} below. We review it below for the benefit 
of the reader not familiar with it.  \qed

Such a complex structure in Prop. \ref{props:I-polNalgB} 
is almost unique when $B_\parallel \neq 0$, 
and there will be infinitely many when $B_\parallel =0$ (all of those 
described in Discussion \ref{statmnt:enumrt-polarizable-I-forT4nG}
are qualified).

\begin{lemma}
\label{lemma:no20-noTransc}
Let $T_M \otimes \Q$ be the transcendental part of a K\"{a}hler 
surface $(M;I)$ that has a polarization in $H^2(M;\Q)$. 
When $\psi \in T_M \otimes \Q$ is decomposed into 
$\psi^{(2,0)} + \psi^{(0,2)} + \psi^{(1,1)}$ and $\psi^{(2,0)}=0$, 
then $\psi = 0$.
\end{lemma}
\proof\footnote{
This lemma is well established in Hodge theory.
} %
The input $\psi^{(2,0)}=0$ implies $\psi^{(0,2)}=0$, 
because $\psi \in T_M\otimes \Q$ is real. This means that 
$\psi = \psi^{(1,1)}$ is in ${\cal H}^2(M_I)$. 

Since we have assumed that $(M;I)$ admits a polarization, 
$M_I = (M;I)$ is algebraic, so the intersection form on ${\cal H}^2(M_I)$ is 
non-degenerate (Hodge index theorem); 
$H^2(M;\Q) \cong {\cal H}^2(M_I) \oplus (T_M\otimes \Q)$ then. 
So, $\psi = \psi^{(1,1)}$ is both in ${\cal H}^2(M_I)$ and 
$T_M \otimes \Q$, which is possible only if $\psi = 0$. 
\qed 

\section{K\"{a}hler Form in the Algebraic Cone (Condition 3)}
\label{sec:metric}

Meng Chen found an example of a mirror pair of CM-type 
abelian surfaces $(M;I)$ and $(W;I^\circ)$ that are mutually 
isogenous. This set of data satisfies both conditions (i) 
and (ii) in Conj. \ref{conj:GV-original}. One can also translate 
the information in $(M;I)$ and $(W;I^\circ)$ into a set of data 
$(M;G,B;I)$ for which an SCFT is specified. Meng Chen pointed out 
that the SCFT in an example is {\it not} rational
\cite[Prop. 4.1 and Cor. 5.11]{Chen:2005gm}. 
The present authors take this observation/example as an indication that 
there are more necessary conditions besides (i) and (ii). 

In order to avoid various issues being mixed up, we postpone discussing 
mirror SCFTs to the next section, and deal with the complexified 
K\"{a}hler form $(B+i\omega)$ on $(M;I)$ instead of the complex 
structure of $(W;I^\circ)$ in this section \ref{sec:metric}. 
We will extract one necessary condition on $(B+i\omega) \in H^2(M;\C)$ 
for the SCFT of $(M;G,B; I)$ to be rational, as presented 
in Thm. \ref{thm:Kahler-is-alg}. We have already seen in 
Prop. \ref{props:I-polNalgB} that $B \in {\cal H}^2(T^4_I)$ when 
we choose a complex structure $I$ that is polarizable and compatible 
with $G$ such that $B^{(2,0)}=0$; Thm. \ref{thm:Kahler-is-alg} says, 
among other things, that the K\"{a}hler form for a {\it rational} SCFT 
is also in the subspace ${\cal H}^2(T^4_I)\otimes \R \subset H^{1,1}(T^4;\R)$
although that for a general SCFT can be in $H^{1,1}(T^4;\R)$.  
In the counter example to Conj. \ref{conj:GV-original} found by 
Meng Chen, one can verify by computation that its K\"{a}hler form is 
in $H^{1,1}(M;\R)$ but is outside of ${\cal H}^2(M)\otimes \R$, so 
the data $(M;G,B)$ of the example do not satisfy the necessary 
condition in Thm. \ref{thm:Kahler-is-alg}.  
This is now added as condition 3 in Conj. \ref{conj:forT4}. 

Note also that condition 3 in Conj. \ref{conj:forT4} is phrased 
by referring only to the K\"{a}hler form in $H^2(M;\Q) \otimes \overline{\Q}$, 
an object available in any complex K\"{a}hler manifold. That is to be 
contrasted with condition (\ref{eq:cond-GnB-rational}), which 
relies on periodic coordinates of $T^{m} \cong \R^m/\Z^{\oplus m}$ and 
the the globally constant values of the component fields of the metric $G$
and the 2-form $B$. So, although Thm. \ref{thm:forT4} for the cases $M=T^4$
only reorganizes the information contained in the original condition 
(\ref{eq:cond-GnB-rational}), the study being done here is motivated 
primarily to come up with a version (Conj. \ref{conj:forT4} or 
Conj. \ref{conj:gen}) that may work for a broader class of target 
space geometry. 

\begin{anythng}
The first one of the conditions in (\ref{eq:cond-GnB-rational})---one 
for the {\it metric}---involves an integral basis of $H^1(T^{2n};\Z)$. 
This condition is still in the same form---the components 
are rational numbers when we use a rational basis of $H^1(T^{2n};\Q)$. 
Let us translate this condition into that for the {\it K\"{a}hler form}.

The metric $G$ is Hermitian under the complex structure $I$, that is, 
\begin{align}
 G = h_{a\bar{b}} dz^a\otimes d\bar{z}^{\bar{b}}
    + h_{\bar{a}b}d\bar{z}^{\bar{a}}\otimes dz^b
\end{align}
for some constant Hermitian $n\times n$ matrix $h = (h_{a\bar{b}})$. 
Here, we use a basis $\{ dz^a, d\bar{z}^{\bar{a}} \}$ of $H^1(T^{2n}_I;\C)$ 
that are eigenvectors of the CM-field action.  
Using the linear relations such as (\ref{eq:hol-rat-basis-trnsf-BC},
 \ref{eq:hol-rat-basis-trnsf-A}), the rationality of the components $G_{IJ}$
in a rational basis is translated to the rationality of all the components 
of the matrix 
\begin{align}
   \left( \begin{array}{cc} Z^T & \overline{Z}^T \\
       \alpha^T & \overline{\alpha}^T \end{array} \right)
   \left( \begin{array}{cc} & h \\ h^T & \end{array} \right)
   \left( \begin{array}{cc} Z & \alpha \\ \overline{Z} & \overline{\alpha} 
   \end{array} \right) = \left( \begin{array}{cc} 
     Z^T h \overline{Z} + \overline{Z}^T h^T Z & 
      Z^T h \overline{\alpha} + \overline{Z}^T h^T \alpha \\
     \overline{\alpha}^T h^T Z + \alpha^T h \overline{Z}  &
      \alpha^T h \overline{\alpha} + \overline{\alpha}^T h^T \alpha
    \end{array} \right).
\end{align}
That is, 
\begin{align}
  Z^T h \overline{Z} + \overline{Z}^T h^T Z  & \; \in M_n(\Q)^{\rm sym} , 
     \label{eq:cond-met-ratnl-n2-1} \\
  \alpha^T h \overline{\alpha} + \overline{\alpha}^T h^T \alpha & \;
   \in M_n(\Q)^{\rm sym}, 
     \label{eq:cond-met-ratnl-n2-2}  \\
  Z^T h \overline{\alpha} + \overline{Z}h^T \alpha & \; \in M_n(\Q).
    \label{eq:cond-met-ratnl-n2-3}
\end{align}
We wish to solve those conditions in terms of the matrix $h = (h_{a\bar{b}})$; 
this is done by working separately for each of the cases (B, C), (A') and (A).
So, the analysis leading to Thm. \ref{thm:Kahler-is-alg} 
is only for $T^{2n}|_{2n=4}$. 
We will use the following parametrization of the $2\times 2$ matrix $h$: 
\begin{align}
  h = \left( \begin{array}{cc} h_1 & c_1-ic'_2 \\
            c_1+ic'_2 & h_2 \end{array} \right),
      \qquad h_{1,2}, \; c_1, \; c'_2 \in \R.
\end{align}
\end{anythng}

{\bf Case (B, C): } Condition (\ref{eq:cond-met-ratnl-n2-1}) implies 
that 
\begin{align}
h_1+h_2 \in \Q, \qquad h_1 - h_2 \in \sqrt{d}\Q, \qquad 
c_1 \in \Q, \qquad {}^\forall c'_2 \in \R,  
\end{align}
and condition (\ref{eq:cond-met-ratnl-n2-2}) on top of this implies 
that 
\begin{align}
 c_1=0. 
\end{align}
On the other hand, condition (\ref{eq:cond-met-ratnl-n2-3}) 
is equivalent to $c'_2=0$ with arbitrary $h_{1,2}$ and $c_1$. 
So, we have\footnote{
For the metric to be positive definite, $a>0$ and $a^2-b^2d>0$. 
} %
\begin{align}
  h = \diag(a+b\sqrt{d}, a-b\sqrt{d}), \qquad {}^\exists a, \; b\in \Q;
\end{align}
the corresponding K\"{a}hler form $\omega(-,-) = 2^{-1}G(I-,-)$ is 
\begin{align}
 \omega = i (a+b\sqrt{d}) dz^1\wedge d\bar{z}^{\bar{1}}
   + i (a-b\sqrt{d}) dz^2\wedge d\bar{z}^{\bar{2}}.
   \label{eq:Kahler-form-BC-4RCFT}
\end{align}

{\bf Cases (A') and (A):} 
Condition (\ref{eq:cond-met-ratnl-n2-1}) is translated 
to $h_{1,2}, c_1 \in \Q$, and condition (\ref{eq:cond-met-ratnl-n2-2}) 
on top of this imposes $c_1 \in \sqrt{p_1p_2} \Q$.
So we should have $c_1 =0$ in the case (A'), while 
$c_1 \in \sqrt{p_1p_2}\Q$ is equivalent to $c_1\in \Q$ in the case (A). 
On the other hand, condition (\ref{eq:cond-met-ratnl-n2-3}) 
implies $c'_2 \in \sqrt{-p_1}\Q \cap \sqrt{-p_2}\Q$; so we should have 
$c'_2=0$ in the case (A'), while we just have $c'_2 \in \sqrt{-p}\Q$.  
To summarize, we should have 
\begin{align}
 (A'): & \qquad  h = \diag( a_1, a_2), \qquad a_{1,2} \in \Q, 
    \label{eq:parametr-Kahler-caseApr} \\
 (A): & \qquad h = \left( \begin{array}{cc} h_1 & c_1 -c_2 \sqrt{p} \\
      c_1+c_2 \sqrt{p} & h_2 \end{array} \right), \qquad 
       h_{1,2}, \; c_1, \; c_2 \in \Q, 
    \label{eq:parametr-Kahler-caseA}
\end{align}
and the corresponding K\"{a}hler forms are 
\begin{align}
  \omega & \; = i a_1 dz^1\wedge d\bar{z}^{\bar{1}}
     + i a_2 dz^2\wedge d\bar{z}^{\bar{2}},  
    \label{eq:Kahler-form-Apr-4RCFT} \\
  \omega & \; = i (dz^1, dz^2) \wedge \left( \begin{array}{cc} 
      h_1 & c_1-c_2\sqrt{p} \\ c_1+c_2 \sqrt{p} & h_2 \end{array} \right)
    \left( \begin{array}{c} d\bar{z}^{\bar{1}} \\ 
        d\bar{z}^{\bar{2}} \end{array} \right), 
   \label{eq:Kahler-form-A-4RCFT}
\end{align}
respectively. 
Having done this analysis, we are ready for this
\begin{thm}
\label{thm:Kahler-is-alg}
This is for $n=2$. {\it Let $(T^{2n}; G, B)|_{2n=4}$ be a set of data for 
which the (S)CFT is rational. For a polarizable complex structure $I$ on 
$T^{4}$ with which $G$ is compatible, the K\"{a}hler form 
$\omega = 2^{-1}G(I-,-)$ is always in the algebraic part of the 2-forms, 
${\cal H}^2(T^4_I)\otimes \R $.

Moreover, the combination $i\omega$ is in ${\cal H}^2(T^4_I)\otimes
  \tau^r_{(20)}(K^r)$, where $K^r$ is the reflex field 
  in Discussion \ref{statmnt:rflx-field-levelN}.
and $\tau^r_{(20)}$ its embedding for the Hodge (2, 0) component 
in $T_{T^4_I}\otimes \C$. }
\end{thm}
\proof It is just necessary to write down the K\"{a}hler forms 
in (\ref{eq:Kahler-form-BC-4RCFT}, \ref{eq:Kahler-form-Apr-4RCFT}, 
\ref{eq:Kahler-form-A-4RCFT}) in the rational basis in 
Discussion \ref{statmnt:Tm}.  In the case (B, C), 
\begin{align}
 i \omega & \; = 2 \tau^r_{++}(a \xi_r + bqd/\xi_r) e_1 
   + 2\tau^r_{++}(bd\xi_r + aqd/\xi_r) e_2. 
  \label{eq:Kahler-form-BC-4RCFT-2}   
\end{align}
The basis $\{ e_1, e_2\}$ of ${\cal H}^2(T^4_I)\otimes \Q$ has been 
introduced in (\ref{eq:def-ratBasis-H2-caseBC}). In the case (A'), 
\begin{align}
 i \omega & \; = 2a_1 \sqrt{p_1}(\hat{\alpha}^1\hat{\beta}_1)
    + 2a_2 \sqrt{p_2}(\hat{\alpha}^2\hat{\beta}_2),
   \label{eq:Kahler-form-Aprm-4RCFT-2}
\end{align}
while 
\begin{align}
 i\omega & \; = \sqrt{p} \left[
   2h_1 (\hat{\alpha}^1\hat{\beta}_1) + 2h_2 (\hat{\alpha}^2\hat{\beta}_2)
   + 2c_1(\hat{\alpha}^1\hat{\beta}_2 - \hat{\beta}_1\hat{\alpha}^2)
   + 2c_2(\hat{\alpha}^1\hat{\alpha}^2-p\hat{\beta}_1\hat{\beta}_2)\right]
  \label{eq:Kahler-form-A-4RCFT-2}
\end{align}
in the case (A).  \qed

\begin{rmk}
\label{rmk:cntr-ex-MC}
(stated already at the beginning of this section \ref{sec:metric})
Meng Chen presents in \cite[\S4]{Chen:2005gm} an example of a pair 
of CM-type abelian surfaces $(T^4;I)$ and $(T^4;I^\circ)$ that 
are geometric SYZ-mirror of each other, and are also isogenous to 
each other, and yet the SCFT corresponding to $(T^4;G,B;I)$ is {\it not} 
rational.  This example satisfies both conditions (i) and (ii) 
in Conj. \ref{conj:GV-original}. It also satisfies most of the conditions 1--5
in Conj. \ref{conj:forT4}/Thm. \ref{thm:forT4}. In the example 
of \cite[\S4]{Chen:2005gm}, however, one can verify by computations 
that the K\"{a}hler form is in $H^{1,1}(T^4_I;\R)$, but is not within the 
algebraic part ${\cal H}^2(T^4_I) \otimes \R$; condition 3 of Thm. \ref{thm:forT4}
is not satisfied. So, this is a counter example of 
Conj. \ref{conj:GV-original}, but it is not for Conj. \ref{conj:forT4}/Thm. \ref{thm:forT4}.
\end{rmk}

\section{Geometric SYZ-mirror and Hodge Isomorphisms}
\label{sec:SYZ-mirror}

\subsection{Implications of Meng Chen's Paper (on Condition 5)}
\label{ssec:quote-MC}

Besides Prop. \ref{props:MChen-thm2.5}, 
here we quote two more results from \cite{Chen:2005gm}. 
\begin{props}\cite[Prop. 3.10]{Chen:2005gm}
\label{props:MChen-prp3.10}
{\it Let $(T^{2n};G,B;I)$ and $(T^{2n};G^\circ, B^\circ; I^\circ)$ be a 
mirror pair. When $G$ and $B$ satisfy the 
condition (\ref{eq:cond-GnB-rational}), then the complex torus 
$(T^{2n};I)$ and $(T^{2n};I^{\circ})$ are isogenous to each other. }  \qed
\end{props}
\noindent Here, 
\begin{defn}
\label{def:geometric-SYZ-mirror}
Consider an $N=(1,1)$ SCFT that is further endowed with one U(1) current 
$J_L$ in the left-moving sector and one more $J_R$ in the right-moving sector 
so that the $N=1$ superconformal algebra is extended to an $N=2$ superconformal 
algebra in both the left-moving sector and the right-moving sector;
we call such a collection of data {\it an $N=(1,1)$ SCFT with an $N=(2,2)$ 
superconformal structure}. 
A pair of such $N=(1,1)$ SCFTs with an $N=(2,2)$ superconformal structure 
(by $(J_L, J_R)$) are said to be a {\it mirror pair} when there is an 
isomorphism between those two $N=(1,1)$ SCFTs such that $J_L$ and $J_R$ of 
one of the $N=(1,1)$ SCFTs is mapped to $J_L$ and $-J_R$ of the other. 

When an $N=(1,1)$ SCFT with an $N=(2,2)$ superconformal structure is given by 
a set of geometric data $(M;G,B;I)$ (the currents $J_L, J_R$ are given 
by using $I$), we say that this $N=(1,1)$ SCFT with an $N=(2,2)$ 
superconformal structure {\it has a geometric SYZ-mirror}, if there is 
another $N=(1,1)$ SCFT with an $N=(2,2)$ superconformal structure 
given by a geometric data $(W;G^\circ, B^\circ;I^\circ)$ such that 
the two $N=(1,1)$ SCFTs with an $N=(2,2)$ superconformal structure
forms a mirror pair. 
\end{defn}
Note that there always exists a mirror pair for an $N=(1,1)$ SCFT 
with an $N=(2,2)$ superconformal structure, because we may always 
reset $J_R$ by $-J_R$. Existence of a geometric SYZ-mirror, on the 
other hand, is non-trivial. 
It is a belief widely accepted in string theory 
community that a geometric SYZ-mirror pair  
has $T^n$-fibrations $\pi_M: M \rightarrow B$
and $\pi_W: W \rightarrow B$ over a common real 
$n$-dimensional manifold $B$ (separately from the 
closed 2-form in $(M;G,B;I)$; apologies for duplicate notations) so that the mirror correspondence is regarded as the T-duality along 
the torus fiber. 

In the proof of \cite[Prop. 3.10]{Chen:2005gm}, the complex structures 
$I$ and $I^\circ$ are not assumed to be polarizable, or to have 
the property that $B^{(2,0)}=0$ or $(B^\circ)^{(2,0)}=0$. It assumes, 
on the other hand, a situation where there exists a set of data 
$(T^{2n};G^\circ, B^\circ;I^\circ)$ forming a geometric SYZ-mirror pair 
with the SCFT for the set of data $(T^{2n};G,B;I)$.

Here is another result to quote:
\begin{props}(\cite[Thm. 3.11]{Chen:2005gm})
{\it Let $(T^{2n};G,B;I)$ and $(T^{2n};G^\circ, B^\circ;I^\circ)$ be sets of data 
so that their $N=(1,1)$ SCFTs with an $N=(2,2)$ superconformal structure 
forms a mirror pair. Suppose further that $I$ is polarizable, 
and that $G$ and $B$ satisfy condition (\ref{eq:cond-GnB-rational}).
Then $(T^{2n};I^\circ)$ is also of CM-type. }
\end{props}
\proof (as in \cite{Chen:2005gm}) It follows from the conditions 
given here that the pair of complex tori $(T^{2n};I)$ and $(T^{2n};I^\circ)$ 
are isogenous to each other (Prop. \ref{props:MChen-prp3.10}). This means that the rational Hodge 
structure on $H^1(T^{2n};\Q)$ by $I$ and $I^\circ$ are isomorphic to each other. 
Now, recalling that the rational Hodge structure on $H^1(T^{2n};\Q)$ by $I$ is of CM-type 
(Prop. \ref{props:MChen-thm2.5}), 
we can conclude that the rational Hodge structure $(H^1(T^{2n};\Q),I^\circ)$ 
is also of CM-type.  \qed

Given this situation, existence of 
a Hodge isomorphism (without imposing eq. (\ref{eq:cond-8})) in condition 5 
of Thm. \ref{thm:forT4} can also be derived for a set of data 
$(T^4;G,B;I)$ when its SCFT is rational, 
once we confirm that there always exists a geometric SYZ-mirror pair.  

\subsection{Brief Comments on Condition 4}
\label{ssec:brief-comm}

Let $(T^{2n};G,B;I)$ be a set of data for an $N=(1,1)$ SCFT with an 
$N=(2,2)$ superconformal structure. We may take T-duality 
along $n$ directions $\Gamma_f \cong \Z^{\oplus n} \subset H_1(T^{2n};\Z)$ while 
keeping $n$ other directions $\Gamma_b \cong \Z^{\oplus n} \subset H_1(T^{2n};\Z)$ 
intact, when $\Gamma_f \oplus \Gamma_b \cong H_1(T^{2n};\Z)$. 
This T-duality yields a geometric SYZ-mirror of 
the $N=(1,1)$ SCFT with the $N=(2,2)$ superconformal structure for 
$(T^{2n};G,B;I)$ if and only if \cite[Prop. 8]{VanEnckevort:2003qc}
\begin{align}
 \omega|_{\Gamma_{f}\otimes \R} = 0, \qquad \qquad B|_{\Gamma_{f}\otimes \R}=0. 
  \label{eq:cond-Enckevort}
\end{align}
For a generic choice of moduli $(G, B)$ and along with a choice of $I$, 
it is not possible to find such $\Gamma_f$.
Condition 4 in Conj. \ref{conj:forT4} / Thm. \ref{thm:forT4} is 
therefore non-trivial. 

Note that condition (\ref{eq:cond-Enckevort}) on the T-dual 
directions $\Gamma_f$ depends on the complexified K\"{a}hler parameter 
$B+i\omega$. In proving the existence of a geometric SYZ-mirror for a set 
of data $(T^4;G,B;I)$ of a rational SCFT (the property in condition 4), 
we may exploit the property that 
$(B+i\omega) \in {\cal H}^2(T^4_I) \otimes \tau^r_{(20)}(K^r)$, as we have seen 
in Prop. \ref{props:I-polNalgB} and Thm. \ref{thm:Kahler-is-alg}, 
especially the parametrization of $i\omega$ given explicitly in 
the proof of Thm. \ref{thm:Kahler-is-alg}. 
In proving that condition 5 (strong) is also a necessary condition 
for the SCFT to be rational, we also need to be able to list up 
all the geometric SYZ-mirrors. 
We will do this in section \ref{ssec:list-SYZ-mirror}. 

When one wants to prove that the set of conditions 1--5 on $(T^{2n};G,B;I)$ 
is sufficient for its SCFT to be rational, on the other hand, 
we can still use condition 3 on $(B+i\omega)$,  
but we should not use the parametrization of $i\omega$ 
in Thm. \ref{thm:Kahler-is-alg} (i.e., (\ref{eq:Kahler-form-BC-4RCFT-2}, 
\ref{eq:Kahler-form-Aprm-4RCFT-2}, \ref{eq:Kahler-form-A-4RCFT-2})). 
%
%
For this purpose, it is necessary to make the most out of condition 5, 
and we need to list the geometric SYZ-mirrors along the way. 

We therefore pause for a moment in section \ref{ssec:1st-step-4-converse}, and 
take the first step of analysis in the converse 
direction, to prove that the set of conditions 1--5 is sufficient.  
By exploiting conditions 1--4 and a part of 5, we will arrive 
at a parametrization of $i\omega$ given in (B,C)--(\ref{eq:BC-B+iomega};$\pm$), 
(A')--(\ref{eq:Aprime-B+iomega-RCFT}), (A')--(\ref{eq:Aprime-B+iomega-nonRCFT}) and 
(A)--(\ref{eq:A-B+iomega}) in section \ref{ssec:1st-step-4-converse}; 
those five classes of $i\omega$ is a little broader than those in the three 
classes of $i\omega$ in (B,C)--(\ref{eq:Kahler-form-BC-4RCFT-2}), 
(A')--(\ref{eq:Kahler-form-Aprm-4RCFT-2}) and (A)--(\ref{eq:Kahler-form-A-4RCFT-2}), 
hinting that the full content of condition 5 is essential 
part of a sufficient condition in Conj. \ref{conj:forT4}.  
In section \ref{ssec:list-SYZ-mirror}, 
we will list up all the geometric SYZ-mirrors for this broader 
class of $(B+i\omega)$; the result in section \ref{ssec:list-SYZ-mirror} 
is used in section \ref{ssec:one-more} for proving both 
the necessity and sufficiency of the conditions.

\subsection{A First Step in the Converse Direction}
\label{ssec:1st-step-4-converse}

This section \ref{ssec:1st-step-4-converse} is for the proof in 
the converse direction, so it is not assumed that a set of data 
$(T^4;G,B;I)$ is for a rational SCFT. Instead, we impose conditions 
1--3; we also assume that at least there is 
one geometric SYZ mirror (condition 4) whose set of data is denoted by 
$(T^4;G^\circ,B^\circ;I^\circ)$; we also use a property 
that there is an isogeny $\phi^*: H^*(T^4;\Q) \rightarrow H^*(T^4_\circ;\Q)$, 
which is a part of condition 5; we will use $T^4_\circ$ for $T^4$ as the mirror 
geometry. The compatibility condition (\ref{eq:cond-8}) can be used only after 
we find the list of SYZ torus fibrations in section \ref{ssec:list-SYZ-mirror}. 

\subsubsection{Decomposition of the Complexified K\"{a}hler Form}
 \label{sssec:vert-coh-mapped2-alg-subsp}

\begin{anythng} ({\bf Decomposition of $\mho$}) 
Let us first extract various properties of the complexified K\"{a}hler form
$(B+i\omega)$ from the existence of a geometric SYZ-mirror. 
Think of a T-duality of $T^{2n}$ along $n$-directions 
$\Gamma_f \subset H_1(T^{2n};\Z)$ that fix $n$ other directions 
$\Gamma_b \subset H_1(T^{2n};\Z)$. D-branes in the original SCFT 
for $(T^{2n};G,B)$ have their corresponding D-branes that are physically 
equivalent in the T-dual SCFT for $(T^{2n}_\circ;G^\circ,B^\circ)$. 
This correspondence sets an isomorphism between $H_*(T^{2n};\Z)$ and 
$H_*(T^{2n}_\circ;\Z)$, and hence also an isomorphism 
$g^*: H^*(T^{2n}_\circ;\Z) \rightarrow H^*(T^{2n};\Z)$. This isomorphism 
has the property 
\begin{align*}
g^*( H^{2n}(T^{2n}_\circ;\Z) ) & \; = \wedge^n (\Gamma_b^\vee), \qquad 
 g^*(H^0(T^{2n}_\circ;\Z)) = \wedge^n(\Gamma_f^\vee), \\
 g^*(H^{2n-1}(T^{2n}_\circ;\Z)) & \; = \wedge^{n-1}(\Gamma_b^\vee)  \oplus 
   (\wedge^n \Gamma_b^\vee)\otimes \Gamma_f^\vee, \\
 g^*(H^{2n-2}(T^{2n}_\circ;\Z)) & \; = \wedge^{n-2}(\Gamma_b^\vee) \oplus 
  (\wedge^{n-1}\Gamma_b^\vee)\otimes \Gamma_f^\vee \oplus 
  (\wedge^n \Gamma_b^\vee) \otimes (\wedge^2 \Gamma_f^\vee), 
\end{align*}
where $\Gamma_f^\vee$ and $\Gamma_b^\vee$ in $H^1(T^{2n};\Z)$ 
are the 1-cocycles that are non-trivial on $\Gamma_f$ and $\Gamma_b$, 
respectively, and trivial on $\Gamma_b$ and $\Gamma_f$, respectively. 

Now, think of the case the T-duality is a geometric SYZ-mirror. 
The Hodge structure on $H^{2n-1}(T^{2n}_\circ;\Z)$ 
by $I^\circ$ is such that 
\begin{align}
g^*(H^{n,n-1}(T^{2n}_\circ;\C)) & \; = {\rm Span}_\C \left\{
      \mho e \; | \; e \in \wedge^{n-1}\Gamma_b^\vee  \right\}, \\
g^*(H^{n-1,n}(T^{2n}_\circ;\C)) & \; = {\rm Span}_\C \left\{
      \overline{\mho} e \; | \; e \in \wedge^{n-1}\Gamma_b^\vee \right\}, 
\end{align}
where $\mho := e^{(B+i\omega)/2}$. The description of the Hodge structure 
on $H^{2n-m}(T^{2n};\Z)$ for $m\geq 2$ is a little more involved; the 
condition (\ref{eq:cond-Enckevort}) only implies that 
\begin{align*}
 (B+i\omega) \in \left[ (\Gamma_f^\vee \otimes \Gamma_b^\vee) 
    \oplus (\wedge^2 \Gamma_b^\vee) \right]\otimes \C, 
\end{align*}
so (cf \cite{Golyshev:1998vzz}, \cite{Kapustin:2000aa}, \cite{Hitchin:2003cxu})  
\begin{align}
 \left\{ \mho e,  \; \overline{\mho}e \; | \; 
   e \in \wedge^{n-2}\Gamma_b^\vee \right\} \subset 
  g^*\left(H^{2n-2}(T^{2n}_\circ;\C) \oplus H^{2n}(T^{2n}_\circ;\C)\right). 
\end{align}

What is always true, in the case $n=2$, is that 
the Hodge (2,0) component and the (0,2) component of $g^*(H^{2}(T^4_\circ;\C))$
are generated by the projection images of $\mho$ and $\overline{\mho}$, 
respectively, i.e., by $\mho_2$ and $\overline{\mho}_2$ in 
\begin{align}
 \mho & \; = \mho_4 g^*(e_{4,\circ}) + \mho_2 \in g^*(H^4(T^{4}_\circ;\C)) 
   \oplus g^*(H^2(T^4_\circ;\C)), \\
 \overline{\mho} & \; = \overline{\mho}_4 g^*(e_{4,\circ}) + \overline{\mho}_2 
  \in g^*(H^4(T^{4}_\circ;\C)) \oplus g^*(H^2(T^4_\circ;\C)); 
\end{align}
$e_{4,\circ}$ is a generator of $H^4(T^4_\circ;\Z)$. 

Next, we combine the above property with condition 2 and the existence 
of a Hodge isomorphism $\phi^*$ in condition 5. Now, the rational Hodge structure 
on the level-2 component $[H^2(T^4;\Q)]_{\ell=2}$ is of CM-type (condition 2), 
and the existence of a Hodge isomorphism $\phi^*: [H^2(T^4;\Q)]_{\ell=2} 
\rightarrow [H^2(T^4_\circ;\Q)]_{\ell=2}$ (a part of condition 5) implies 
that the rational Hodge structure of the mirror, $g^*([H^2(T^4_\circ;\Q)]_{\ell=2})$
 by $I^\circ$ is also of CM-type. 
The rational Hodge structure of $g^*([H^2(T^4_\circ;\Q)]_{\ell=2})$ therefore 
has to be such that its Hodge (2, 0) component is generated by $\mho_2$, 
and also of CM-type. 

This then indicates that the generator $\mho_2$ has a form 
of (\ref{eq:egVec-from-very-useful-L}) or in Lemma \ref{lemma:very-useful}; 
it is not even necessary to rescale $\mho_2$ by multiplying some complex number,   
because $\mho_2$ and $\mho$ are identical 
in $(\wedge^{0} \Gamma_b^\vee)\otimes \C = H^0(T^4;\C)$ and 
$(\wedge^2 \Gamma_b^\vee \wedge^2 \Gamma_f^\vee)\otimes \C = H^4(T^4;\C)$
(though not in $(\wedge^2 \Gamma_b^\vee)\otimes \C$), 
which means in particular that $\mho = e^{(B+i\omega)/2}$ and $\mho_2$ have a rational 
coefficient with respect to a generator of $H^0(T^4;\Q)$.  Now, the following 
two facts follow. One is that $\mho_4 \in \tau^r_{(20)}(K^r)$, because 
both $(B+iJ)$ and the 2-form part of $\mho_2$ are in 
$H^2(T^4;\Q)\otimes \tau^r_{(20)}(K^r)$. The other fact is that the 
$(\mho_2)^\sigma$ for each $\sigma \in {\rm Gal}((K^r)^{\rm nc}/\Q)$---the 
Galois transformations on the coefficients of $\mho_2$ with respect to a 
rational basis of $g^*([H^2(T^4_\circ;\Q)]_{\ell=2})$---is in a definite 
Hodge component of the rational Hodge structure 
$(g^*(H^2(T^4_\circ;\Q)), I^\circ)$. We can choose the vectors of the form $(\mho_2)^\sigma$
to have a basis of each of the Hodge components (cf Lemma \ref{lemma:very-useful}). 
\end{anythng}

\begin{anythng}
\label{statmnt:introduce-TMv-byGal}
For a complex $n$-dimensional projective K\"{a}hler manifold $(M;I)$, 
a vector subspace 
\begin{align}
 A(M_I)\otimes \Q := \oplus_{m=0}^{n} \left( H^{2m}(M;\Q) \cap H^{m,m}(M_I;\R)
   \right) \subset H^{\rm even}(M;\Q)
\end{align}
is given a bilinear form 
\begin{align}
  (A(M_I)\otimes \Q) \times (A(M_I)\otimes \Q) \ni (\psi, \chi) \longmapsto 
   (-1)^{\frac{n(n-1)}{2}} \sum_{m=0}^n \int_M ((-1)^m \Pi_{2m} \psi) \wedge \chi,
 \label{eq:pairing-Mukai}
\end{align}
where $\Pi_{2m}$ is the projection to the component 
$H^{2m}(M;\Q) \cap H^{m,m}(M_I;\R)$. 

When $(M;I)$ is a complex torus $(T^{2n};I)$, and 
$(B+i\omega) \in {\cal H}^2(M_I) \otimes \tau^r_{(n0)}(K^r)$, 
a vector subspace $T_M^v\otimes \Q \subset 
A(M_I)\otimes \Q$ is specified canonically, as follows. 
Expand $\mho = e^{(B+i\omega)/2} \in (A(M_I)\otimes \Q)\otimes \tau^r_{(n0)}(K^r)$
in the form of $\sum_I e_I \tau^r_{(n0)}(\eta_I)$ for some basis $\{ \eta_I \}$
of $K^r/\Q$ and $\{ e_I \} \subset A(M_I)\otimes \Q$; then 
set $T_M^v\otimes \Q \subset A(M_I)\otimes \Q$ as ${\rm Span}_\Q \{ e_I \}$.    
  
Now, think of a set of data $(T^4;G,B;I)$ satisfying conditions 1--4 of 
Thm. \ref{thm:forT4} such that there exists a Hodge isomorphism 
$\phi^*: [H^2(T^4;\Q)]_{\ell=2} \rightarrow [H^2(T^4_\circ;\Q)]_{\ell=2}$. 
Then the homomorphism 
$T_M^v\otimes \Q \hookrightarrow H^{\rm even}(T^4;\Q) \twoheadrightarrow 
g^*([H^2(T^4_\circ;\Q)]_{\ell=2})$ of vector spaces, an isomorphism in fact, map 
\begin{align}
   (T_M^v\otimes \Q) \otimes \overline{\Q} \ni \mho^\sigma \longmapsto 
    (\mho_2)^\sigma \in g^*([H^2(T^4_\circ;\Q)]_{\ell=2}) \otimes \overline{\Q} 
\end{align}
for any $\sigma \in {\rm Gal}((K^r)^{\rm nc}/\Q)$. Each one of 
$(\mho_2)^\sigma$'s is in one definite Hodge component, say $(p,2-p)$ component
in the Hodge structure of $(g^*(H^2(T^4_\circ;\Q)),I^\circ)$, so we may also 
introduce a rational Hodge structure on the vector space $T_M^v\otimes \Q$
by declaring that $\mho^\sigma$ is in the Hodge $(p,2-p)$ component. 
The rational Hodge structure on $T_M^v\otimes \Q$ so defined has a 
Hodge isomorphism with that of $g^*([H^2(T^4_\circ;\Q)]_{\ell=2})$ by 
construction. 

The rational Hodge structure on $[H^2(T^4_\circ;\Q)]_{\ell=2} \subset 
H^2(T^4_\circ;\Q)$ has a pairing 
\[ 
(-,-): (\psi, \chi) 
\longmapsto \int_{T^4_\circ}\psi \wedge \chi \in \Q. 
\]
It is known that this 
pairing in the mirror description on $g^*([H^2(T^4_\circ;\Q)]_{\ell=2})$ agrees
with the pairing (\ref{eq:pairing-Mukai}) on 
$T_M^v \otimes \Q \subset A(M_I)\otimes \Q$; although general 
elements in $T_M^v \otimes \Q$ are not purely 
in $g^*(H^n(T^{2n}_\circ;\Q))$
they are still in $g^*(\oplus_{k\geq n} H^k(T^{2n}_\circ;\Q))$ because of  
the condition (\ref{eq:cond-Enckevort});
the discrepancy $\mho - \mho_2$, which is 
in $g^*(\oplus_{k>n}H^k(T^{2n}_\circ;\Q))|_{n=2}$, 
does not contribute in the paring above on $T^4_\circ$.
By combining all the discussion so far in this 
section \ref{ssec:1st-step-4-converse}, we have 
\end{anythng}
\begin{lemma}
\label{lemma:mirrorhHdgH2L2=vHdgOnAlg}
Let $(T^4;G,B;I)$ be a set of data satisfying conditions 1--4 in 
Thm. \ref{thm:forT4}, and suppose further that there is a Hodge isomorphism 
$\phi^*: (H^*(T^4;\Q), I) \rightarrow (H^*(T^4_\circ;\Q),I^\circ)$. 
Then the polarized rational Hodge structure on $[H^2(T^4_\circ;\Q)]_{\ell=2}$
has a Hodge isomorphism with the polarized rational Hodge structure 
on $T_M^v\otimes \Q$ constructed above. Since the pairing $(-,-)$ is a 
polarization on the $\ell=2$ component $[H^2(T^4_\circ;\Q)]_{\ell=2}$, 
so is the pairing (\ref{eq:pairing-Mukai}) on $T_M^v\otimes \Q$. 

In particular, the polarized rational Hodge structure on $T_M^v\otimes \Q$, 
which is determined by $\mho = e^{(B+i\omega)/2}$ and the Galois transformations 
on the coefficients, is of CM-type.  
The endomorphism field is that of $[H^2(T^4;\Q)]_{\ell=2}$ and its embedding 
associated with the Hodge $(2,0)$ component $\C \mho$ is that of 
the Hodge (2,0) component of $[H^2(T^4;\Q)]_{\ell=2}$.    \qed
\end{lemma}

\begin{rmk}
The construction in Discussion \ref{statmnt:introduce-TMv-byGal} for $M=T^4$
can be repeated for a complex projective K3 surface $(M;G,B;I)_{M=K3}$ with 
$(B+i\omega) \in {\cal H}^2(M_I)\otimes \tau^r_{(20)}(K^r)$. The resulting 
lattice $T_M^v := (T_M^v\otimes \Q) \cap H^{\rm even}(M;\Z)$ corresponds to 
$T(\varphi)|_{\varphi = \mho}$ in \cite[Def. 4.5]{MR2115675} and 
$L_{\varphi'}|_{\varphi' = \mho}$ in \cite[Def. 3.1]{kanazawa2024mirror}. 
The lattice $T_M^v$ does not necessarily contain $H^0(M;\Z) \oplus H^4(M;\Z) \subset A(M_I)\otimes \Q$;
the lattice $T_M := (T_M\otimes \Q) \cap H^2(M;\Z)$ does not necessarily contain a copy of 
a self-dual lattice of signature (1,1); their absence in $T_M^v$ and $T_M$
is not an issue at all in the 
mirror correspondence (cf \cite{kanazawa2024mirror}). 
\end{rmk}

\subsubsection{Constraints on the Complexified K\"{a}hler Form}
\label{sssec:5classes-B+iomega}

Lemma \ref{lemma:mirrorhHdgH2L2=vHdgOnAlg} is useful in narrowing down 
the possible choices of the complexified K\"{a}hler parameter $(B+i\omega)$
satisfying the conditions in Thm. \ref{thm:forT4}. Even though the derivation of 
Lemma \ref{lemma:mirrorhHdgH2L2=vHdgOnAlg} exploited {\it existence} of a geometric 
SYZ-mirror (Condition 4) as well as existence of a Hodge isomorphism 
$\phi^*: H^*(T^4;\Q) \longrightarrow H^*(T^4_\circ;\Q)$ with the mirror, 
the characterization on $(B+i \omega)$ extracted in Lemma \ref{lemma:mirrorhHdgH2L2=vHdgOnAlg} 
does not rely on the choice of a geometric SYZ-mirror (or a choice of the T-dual directions $\Gamma_f$). 

In the process of proving that the set of conditions 1--5 is sufficient 
for the SCFT of $(T^4;G,B;I)$ to be rational, of course we can use condition 2;
the complex torus $(T^4;I)$ is therefore in one of the four cases 
reviewed in section \ref{ssec:CM-abel-surface}. For each of the 
cases (B, C), (A') and (A), we may study the possible forms of $(B+i\omega)$ 
satisfying the properties in Lemma \ref{lemma:mirrorhHdgH2L2=vHdgOnAlg}. 
Details of the analysis is left to Lemmas \ref{lemma:GV-2-RCFT-caseBC-even} 
and \ref{lemma:GV-2-RCFT-caseAprm-even}; we quote the results here. 

{\bf Case (B, C):} 
when the CM-type abelian surface $(T^4;I)$ is in the case (B, C), 
we found in Lemma \ref{lemma:GV-2-RCFT-caseBC-even} that 
the properties in Lemma \ref{lemma:mirrorhHdgH2L2=vHdgOnAlg} allow 
the following form of $(B+i\omega)$:
\begin{align}
 (B+i\omega)/2 & \; = Z_1 e_1 + Z_2^\pm e_2,  
  \label{eq:BC-B+iomega}
\end{align}
where 
\begin{align*}
 Z_1 & \; := \tau^r_{++}\left( A+\frac{C'}{2}\xi +\frac{D'}{2d}\frac{2qd}{\xi}\right), \qquad 
 Z_2^{\pm} := \tau^r_{++} \left( \widetilde{A} \pm \frac{D'}{2}\xi
   \pm \frac{C'}{2}\frac{2qd}{\xi} \right), \\
 & \qquad \left( A, \widetilde{A}, C', D' \in \Q, \quad 
   (C', D') \neq (0,0) \right). \nonumber  
\end{align*}

In these two classes of $(B+i\omega)$, one with $Z_2^+$ and the other 
with $Z_2^-$ in (\ref{eq:BC-B+iomega}), the geometric data $B$
is the real part, $2(Ae_1+\widetilde{A}e_2)$, which is rational 
(as in (\ref{eq:cond-GnB-rational})). 
In the class of $(B+i\omega)$ with $Z_2^+$ in (\ref{eq:BC-B+iomega}), 
i.e., in (\ref{eq:BC-B+iomega};$+$), the imaginary part $i\omega$ is 
precisely of the form in (\ref{eq:Kahler-form-BC-4RCFT-2}), when we 
read $2a = C' \in \Q$ and $2b = D'/d \in \Q$. So, we see that the set of data 
$(T^4;G,B;I)$ is for a rational SCFT. On the other hand, in the class of 
$(B+i\omega)$ in (\ref{eq:BC-B+iomega};$-$), 
\begin{align}
 \frac{i}{2}\omega & \; = \tau^r_{++}\left( \frac{C'}{2}\xi+ \frac{D'q}{\xi}\right)
   e_1 - \tau^r_{++}\left( \frac{C'qd}{\xi}
                  +\frac{D'}{2} \xi \right) e_2,    \nonumber \\
  & \; = \frac{C'}{2} \left( e_1 (\sqrt{+}+\sqrt{-})
                  - e_2 (\sqrt{+}-\sqrt{-})\sqrt{d} \right)  \\
 & \qquad
     + \frac{D'}{2} \left( - e_2 (\sqrt{+}+\sqrt{-}) +e_1(\sqrt{+}-\sqrt{-})/\sqrt{d} \right).  \nonumber 
\end{align}
Rewriting this in terms of $dz^1\wedge d\bar{z}^{\bar{1}}$ and $dz^2\wedge d\bar{z}^{\bar{2}}$ according to (\ref{eq:dz1-dzbar1-caseBC}) and (\ref{eq:dz2-dzbar2-caseBC}),
\begin{align}
  \frac{i}{2}\omega  = \frac{p-q\sqrt{d}}{4\sqrt{d'}}\left(C'-\frac{D'}{\sqrt{d}}\right)dz^1\wedge d\bar{z}^{\bar{1}}+\frac{p+q\sqrt{d}}{4\sqrt{d'}}\left(C'+\frac{D'}{\sqrt{d}}\right)dz^2\wedge d\bar{z}^{\bar{2}}.
\end{align}
For this K\"{a}hler form to be fitted by the 
expression (\ref{eq:Kahler-form-BC-4RCFT-2}), we have
\begin{align}
  a = -\frac{1}{2\sqrt{d'}}(pC'+qD'),\qquad
  b = \frac{1}{2d\sqrt{d'}}(qdC'+pD').
\end{align}
The fitted parameters $a,b$ are not rational when $C', D' \in \Q$. 
The metric corresponding to this K\"{a}hler form does not satisfy 
the condition (\ref{eq:cond-GnB-rational}).  The resulting metric 
is positive definite for some region in $(C', D') \in \Q^2$, so 
the class of $(B+i\omega)$ in (\ref{eq:BC-B+iomega};$-$)
on $(T^4;I)$ of the case (B, C) include physically sensible $N=(1,1)$ SCFTs 
that are {\it not} rational. 

{\bf Case (A'):} when the CM-type abelian surface $(T^4;I)$ is in the case 
(A'), we found in Lemma \ref{lemma:GV-2-RCFT-caseAprm-even} that the properties 
in Lemma \ref{lemma:mirrorhHdgH2L2=vHdgOnAlg} allow 
the following form of $(B+i\omega)$:
\begin{align}
(B+i\omega)/2 = \left(A + C \sqrt{p_1} \right) \hat{\alpha}^1\hat{\beta}_1
    + \left(\widetilde{A} + \widetilde{C} \sqrt{p_2} \right)
       \hat{\alpha}^2\hat{\beta}_2,       \qquad 
    A, \widetilde{A} \in \Q, \;  C, \widetilde{C} \in \Q_{\neq 0},
  \label{eq:Aprime-B+iomega-RCFT}
\end{align}
or 
\begin{align}
(B+i\omega)/2 = \left(A + C \sqrt{p_2} \right) \hat{\alpha}^1\hat{\beta}_1
    + \left(\widetilde{A} + \widetilde{C} \sqrt{p_1} \right)
       \hat{\alpha}^2\hat{\beta}_2,       \qquad 
    A, \widetilde{A} \in \Q, \;  C, \widetilde{C} \in \Q_{\neq 0} .
   \label{eq:Aprime-B+iomega-nonRCFT}
\end{align}
In both classes of $(B+i\omega)$, the real part---$B$---is rational. 
The imaginary part $i\omega$ in the class (\ref{eq:Aprime-B+iomega-RCFT}) 
reproduces all the rational metric $G$ in (\ref{eq:parametr-Kahler-caseApr},
 \ref{eq:Kahler-form-Aprm-4RCFT-2}); 
$a_1 = C$ and $a_2 =\widetilde{C}$. In the second class
of solutions (\ref{eq:Aprime-B+iomega-nonRCFT}), the metric is not 
rational; if we are to fit $i\omega$ in (\ref{eq:Aprime-B+iomega-nonRCFT})
into the form of (\ref{eq:Kahler-form-Aprm-4RCFT-2}), then 
the dictionary is $a_1 = C\sqrt{p_2/p_1} \nin \Q$, 
and $a_2 = \widetilde{C}\sqrt{p_1/p_2} \nin \Q$. There is a region 
with a positive volume interpretation in the $(a_1,a_2)$ space; 
in such parameter space, the corresponding SCFT is {\it not} rational.  

{\bf Case (A):} when the CM-type abelian surface $(T^4;I)$ is in the case 
(A), the properties in Lemma \ref{lemma:mirrorhHdgH2L2=vHdgOnAlg} do
not yield an extra constraint on $(B+i\omega) \in 
{\cal H}^2(T^4_I) \otimes \tau^r_{(20)}(K^r)$, as we argue 
briefly below. In other words, 
\begin{align}
 B \in {\cal H}^2(T^4_I), \qquad 
 i \omega \in \sqrt{p} {\cal H}^2(T^4_I), \quad
       \int_{T^4} \omega\wedge \omega >0 .  
 \label{eq:A-B+iomega}
\end{align}
This class of $(B+i\omega)$ agrees with $B$ in (\ref{eq:cond-GnB-rational})
and $i\omega$ in (\ref{eq:Kahler-form-A-4RCFT-2}). 
So, we have already seen that the set of conditions in Thm. \ref{thm:forT4}
is sufficient, if $(T^4;I)$ is in the case (A), in guaranteeing 
that the SCFT for $(T^4;G,B;I)$ is rational. 

To see that all of $(B+i\omega)$ in (\ref{eq:A-B+iomega}) has the 
properties in Lemma \ref{lemma:mirrorhHdgH2L2=vHdgOnAlg}, note 
first that $\mho = e^{(B+i\omega)/2}$ and $\overline{\mho}$ in 
$A(T^4_I)\otimes \C$ satisfy $(\mho, \mho) =0$, 
$(\overline{\mho},\overline{\mho})=0$ in the pairing (\ref{eq:pairing-Mukai})
without imposing an extra constraint on $(B+i\omega)$. 
The underlying vector space $T_M^v\otimes \Q \subset A(T^4_I)\otimes \Q$
is generated by 
\[
{\rm Re}(\mho) = 1+B/2 + (B^2-\omega^2)/8, \qquad 
\frac{i{\rm Im}(\mho)}{\sqrt{p}} = (i\omega/2+iB\omega/4)/\sqrt{p}. 
\]
The rational Hodge structure on $T_M^v\otimes \Q$ has just 1-dimensional 
(2,0) component and 1-dimensional (0,2) component, nothing else, 
as the rational Hodge structure on $[H^2(T^4;\Q)]_{\ell=2}$ does. 
So, we already have all the properties expected for the Hodge decomposition 
on $T_M^v\otimes \Q$. \qed

\subsection{Listing Geometric SYZ Mirrors (Condition 4)}
\label{ssec:list-SYZ-mirror}

{\bf Recap:} 
To prove Thm. \ref{thm:forT4}, we both need to show that the 
properties 1--5 are necessary for the SCFT of $(T^4;G,B;I)$ to be 
rational (proof of necessity), and also to show that it is sufficient to 
impose the conditions there for the SCFT to be rational (the converse 
direction). In the proof of necessity, we still need to prove 
that a geometric SYZ-mirror exists (necessity of condition 4); 
although the existence of a Hodge isomorphism $\phi^*$ in condition 
5 then follows (see discussion in section \ref{ssec:quote-MC}), 
but still we need to prove the rest of the properties in condition 5.
To this end, we need to list up geometric SYZ-mirrors for the choice 
of $(B+i\omega)$ in (\ref{eq:BC-B+iomega};$+$) in the case (B, C), 
and also for $(B+i\omega)$ in (\ref{eq:Aprime-B+iomega-RCFT}) in the case (A'); 
we will not need such a full list for the case (A) in section \ref{ssec:one-more} as we will see.

In completing the proof in the converse direction, 
all that we still need to do is to show that the two classes of possibilities 
of $(B+i\omega)$, namely the ones in (B, C)--(\ref{eq:BC-B+iomega};$-$) and 
in (A')--(\ref{eq:Aprime-B+iomega-nonRCFT}), are eliminated by the conditions 
that we have not exploited in section \ref{ssec:1st-step-4-converse}.
What we have not exploited is the condition (\ref{eq:cond-8}) of compatibility 
of the Hodge isomorphisms $\phi^*$ with the SYZ torus fibration maps. 
So, we need to list up geometric SYZ-mirrors for $(B+i\omega)$ in (\ref{eq:BC-B+iomega};$-$) 
in the case (B, C), and for $(B+i\omega)$ in (\ref{eq:Aprime-B+iomega-nonRCFT}) in the case (A'), 
before examining the compatibility with the torus fibration maps. 

{\bf Preliminaries:}
In this section \ref{ssec:list-SYZ-mirror}, we present the list of all 
the geometric SYZ-mirrors for those four classes of $(B+i\omega)$.
Analysis is done in Lemmas \ref{lemma:list-SYZ-mirror-caseBC}, \ref{lemma:list-SYZ-mirror-caseAprm} and 
\ref{lemma:list-SYZ-mirror-caseA}; only the results are quoted here. 
A few elementary remarks are in order here, to set up a common language. 
First, it is enough to specify a 2-dimensional vector subspace 
$\Gamma_{f\Q} \subset H_1(T^4;\Q)$ to specify the T-dual directions; 
$\Gamma_f := \Gamma_{f\Q} \cap H_1(T^4;\Z)$. The condition (\ref{eq:cond-Enckevort}) 
is imposed on the vector subspace $\Gamma_{f\Q}$. 
Second, it is always possible to find $\Gamma_b \subset H_1(T^4;\Z)$ so that 
$\Gamma_{f} \oplus \Gamma_{b} \cong H_1(T^4;\Z)$. 
In the following discussion, we may use the notation 
$\Gamma_{b\Q} := \Gamma_b \otimes \Q \subset H_1(T^4;\Q)$.  
Thirdly, the rest of discussions in this article does not depend 
on a choice of the T-dual-fixed directions $\Gamma_b$, as we will see. 

For a mirror geometry $T^4_\circ$, we abuse notations and write 
\begin{align}
 H_1(T^4_\circ;\Z) \cong \Gamma_f^\vee \oplus \Gamma_b, \qquad 
 H^1(T^4_\circ;\Z) \cong \Gamma_{f} \oplus \Gamma_b^\vee; 
\end{align}
this is the map of the winding and Kaluza--Klein charges which 
identify closed string states on both sides of the T-duality.
The vector subspace $\Gamma_{b\Q}^\vee$ in $H^1(T^4;\Q)$ as well as 
in $H^1(T^4_\circ;\Q)$ can be seen as that of 1-cocycles of the base $B=T^2$
of the SYZ fibration, $\pi_M: T^4 \rightarrow B$ and 
$\pi_W: T^4_\circ \rightarrow B$, pulled back by $\pi_M$ and $\pi_W$. 

{\bf Case (B, C):} for any $(B+i\omega)$ in the form 
of (\ref{eq:BC-B+iomega};$+$) or of (\ref{eq:BC-B+iomega};$-$), 
the T-dualized directions $\Gamma_{f\Q} \subset H_1(T^4;\Q)$ of 
a geometric SYZ-mirror is of the form 
\begin{align}
\Gamma_{f\Q}={\rm Span}_\Q\{c,d\}, \quad
   & c:=c_1\alpha_1+c_2\beta^1+c_3\alpha_2+c_4\beta^2,
\nonumber\\
&d:=dc_3\alpha_1+dc_4\beta^1+c_1\alpha_2+c_2\beta^2,  \label{eq:BC-GammafQ}\\
&c_1,\ldots,c_4\in\Q, c_1^2+c_2^2+c_3^2+c_4^2\neq0.  \label{eq:BC-GammafQ-prm}
\end{align}
Computations leading to this result is found in 
Lemma \ref{lemma:list-SYZ-mirror-caseBC}.

The 2-dimensional subspace $\Gamma_{b\Q}^\vee$ of 1-cocycles 
of the base $B=T^2$ is where the pairing with the directions of 
the $T^2$-fiber $\Gamma_{f\Q} \subset H_1(T^4;\Q)$ vanishes. 
It is therefore generated by 
\begin{align}
\hat{e}&:=(c_1^2-dc_3^2)\hat{\beta}_1-(c_1c_2-dc_3c_4)\hat{\alpha}^1-d(c_1c_4-c_2c_3)\hat{\alpha}^2,
\label{eq:BC-c_1c_3nonzero-ehat}\\
\hat{f}&:=(c_1^2-dc_3^2)\hat{\beta}_2-(c_1c_4-c_2c_3)\hat{\alpha}^1-(c_1c_2-dc_3c_4)\hat{\alpha}^2,
\label{eq:BC-c_1c_3nonzero-fhat}
\end{align}
when $(c_1,c_3) \neq (0,0)$. There is a similar expression for 
$\hat{e}$ and $\hat{f}$ in the case $(c_2,c_4)\neq (0,0)$. 
Although we could write down an expression valid for 
a general $(c_1,c_2,c_3,c_4) \neq 0$, it is a mess. It is much more 
convenient to do computations in the rest of this article only 
in the case $(c_1,c_3) \neq (0,0)$, 
and confirm that the same line of argument also works in the case 
$(c_2,c_4) \neq (0,0)$, to cover all possible $(c_{1,2,3,4})$. 
Now, we may use 
$\{c,d,\hat{e},\hat{f}\}$ as a basis of $H^1(T^4_\circ;\Q)$. 

As for a basis of $H^1(T^4;\Q)$, we may use 
$\{ \hat{c}', \hat{d}', \hat{e}, \hat{f}\}$, where 
\begin{align}
\hat{c}' :=  c_1\hat{\alpha}^1 - dc_3\hat{\alpha}^2 , \qquad 
\hat{d}' :=  - c_3\hat{\alpha}^1 + c_1\hat{\alpha}^2 .
\label{eq:BC-c_1c_3nonzero-e'f'chat'dhat'}
\end{align}
Although $\hat{c}'$ and $\hat{d}'$ are not necessarily in the direction 
of $\Gamma_{f\Q}^\vee$, that does not matter anywhere in the rest of 
this article. 

{\bf Case (A'):} For any $(B+i\omega)$ in the form 
of (\ref{eq:Aprime-B+iomega-RCFT}) or of (\ref{eq:Aprime-B+iomega-nonRCFT}), 
the T-dualized directions $\Gamma_{f\Q} \subset H_1(T^4;\Q)$ of 
a geometric SYZ-mirror is of the form 
\begin{align}
\Gamma_{f\Q}={\rm Span}_\Q\{c,d\},\quad &c:=c_1\alpha_1+c_2\beta^1,\quad d:=c_3\alpha_2+c_4\beta^2,
   \label{eq:Aprime-GammafQ} \\
&c_1,\ldots,c_4\in\Q, (c_1,c_2),(c_3,c_4)\neq(0,0);
    \label{eq:Aprime-GammafQ-prm}
\end{align}
a little more details leading to this result are found in 
Lemma \ref{lemma:list-SYZ-mirror-caseAprm}. 

The 2-dimensional subspace $\Gamma_{b\Q}^\vee$ of the 1-cocycles on 
the base $B=T^2$ common to both is generated by 
\begin{align}
  \hat{e} := -c_2 \hat{\alpha}^1 + c_1 \hat{\beta}_1, \qquad 
  \hat{f} := -c_4 \hat{\alpha}^2 + c_3 \hat{\beta}_2; 
\label{eq:basis-1form-base-caseAprime}
\end{align}
they vanish on $\Gamma_{f\Q}$. Now we may use $\{c,d,\hat{e},\hat{f}\}$ 
as a basis of $H^1(T^4_\circ;\Q)$. 
As for a basis of $H^1(T^4;\Q)$ we use $\{ \hat{c}', \hat{d}',
 \hat{e}, \hat{f}\}$, where
\begin{align}
  \hat{c}' :=  c_1\hat{\alpha}^1 + c_2\hat{\beta}_1 , \qquad 
  \hat{d}' :=  c_3\hat{\alpha}^2 + c_4\hat{\beta}_2  ; 
  \label{eq:Aprime-e'f'chat'dhat'}
\end{align}
although these $\hat{c}'$ and $\hat{d}'$ are not necessarily chosen 
within $\Gamma_{f\Q}^\vee$, that is not an issue in the rest of this article. 

{\bf Case (A):} for any $(B+i\omega)$ in the form of (\ref{eq:A-B+iomega}), 
we have confirmed in Lemma \ref{lemma:list-SYZ-mirror-caseA} that  
there always exists a choice of $\Gamma_{f\Q} \subset H_1(T^4;\Q)$ 
for a geometric SYZ-mirror. Details are left to 
Lemma \ref{lemma:list-SYZ-mirror-caseA} in the appendix. 
In proving the compatibility property (\ref{eq:cond-8}) 
in condition 5 in this 
case (A)--(\ref{eq:A-B+iomega}), as we will do in 
section \ref{subsec:cond8-caseA}, an explicit list of choices 
of $\Gamma_{f\Q}$---one that corresponds to (\ref{eq:BC-GammafQ}, 
\ref{eq:Aprime-GammafQ})---is not necessary. 

{\bf Immediate consequences:} 
We have seen that there are indeed geometric SYZ-mirrors for $(B+i\omega)$ 
in any one of the five classes of $(B+i\omega)$. This means, on one hand, 
that the property in condition 4 of Thm. \ref{thm:forT4} is necessary; 
also the existence of the Hodge isomorphism $\phi^*$ in condition 5
of Thm. \ref{thm:forT4} follows because we can use 
Prop. \ref{props:MChen-prp3.10} now. 

In the converse direction of Thm. \ref{thm:forT4}, we had combined 
the properties in conditions 1--4 and the existence of $\phi^*$
in condition 5 to arrive at the characterization of $(B+i\omega)$
in Lemma \ref{lemma:mirrorhHdgH2L2=vHdgOnAlg}, and then pointed out 
that $(B+i\omega)$ has to be in one of the five classes quoted 
in section \ref{sssec:5classes-B+iomega}. 
Although the five classes of $(B+i\omega)$ has the property 
described in Lemma \ref{lemma:mirrorhHdgH2L2=vHdgOnAlg}, 
we did not check whether the existence of a geometric SYZ-mirror 
(condition 4) and a Hodge isomorphism $\phi^*$ (condition 5)
are guaranteed for any one of $(B+i\omega)$ in the five classes. 
Now, the results quoted above in this section \ref{ssec:list-SYZ-mirror}
guarantee that a geometric SYZ-mirror always exists. 

A Hodge isomorphism $\phi^*$ also 
exists for any one of $(B+i\omega)$ in the five classes. To see 
this,\footnote{
Instead of the argument in this paragraph, one may resort to 
explicit computations, as in Lemma \ref{lemma:mirror-H1-HdgStr-caseBC}. 
} %
 pick any geometric SYZ-mirror, and fix. The level-2 component 
$[H^2(T^4_\circ;\Q)]_{\ell=2}$ is of $[K^r:\Q]$ dimensions, with the 
signature (2,2) in the four classes in the case (A', B, C) and 
with the signature (2,0) in the case (A). So, the complex torus 
$(T^4_\circ;I^\circ)$ has non-trivial algebraic part 
${\cal H}^2(T^4_{I^\circ})$ with the signature (1,1) and (1,3), 
respectively; this means that the mirror complex torus $(T^4_\circ; I^\circ)$ 
has a polarization, and is an abelian surface. 
It is then possible to apply \cite[Thm. 1.4]{Okada:2023udq}
to see that $(T^4_\circ;I^\circ)$ is of CM-type; its proof 
for the case $[K^r:\Q] = 2, 4$ further implies that the weight-1 
rational Hodge structure is determined uniquely 
modulo Hodge isomorphisms. 
This is enough to guarantee that there is a Hodge isomorphism 
$\phi^*: H^1(T^4;\Q) \rightarrow H^1(T^4_\circ;\Q)$, 
and hence $\phi^*: H^*(T^4;\Q) \rightarrow H^*(T^4_\circ;Q)$. 

\subsection{Compatibility with the Torus Fibration (Condition 5)}
\label{ssec:one-more}

One of the crucial elements in the set of conditions 
in Conj. \ref{conj:forT4}/Thm. \ref{thm:forT4} that updates
Conj. \ref{conj:GV-original} is 
the compatibility (\ref{eq:cond-8}) of a Hodge isomorphism $\phi^*$ 
with the SYZ torus fibration morphisms 
$\pi_M: M \rightarrow B$ and $\pi_W:W\rightarrow B$. 
We have seen that the set of conditions 1--5 in Conj. \ref{conj:forT4}
except this additional characterizations on $\phi^*$ fails to be 
sufficient in guaranteeing that the SCFT of $(T^4;G,B;I)$ is rational. 
In the rest of this section \ref{ssec:one-more}, we will see 
that $\phi^*$ with the compatibility condition still exists 
for $(B+i\omega)$ in the three classes
(A')--(\ref{eq:Aprime-B+iomega-RCFT}), 
(B, C)--(\ref{eq:BC-B+iomega};$+$) and (A).
On the other hand, condition 5 including the additional 
characterization on $\phi^*$ eliminates the two classes 
of sets of data (A')--(\ref{eq:Aprime-B+iomega-nonRCFT}) and 
(B, C)--(\ref{eq:BC-B+iomega};$-$), where the SCFTs are not rational.
At the end of this section \ref{ssec:one-more}, therefore the proof 
of Thm. \ref{thm:forT4} is completed. 

Before writing down a systematic proof of this claim, however, 
let us illustrate why the additional characterization (\ref{eq:cond-8}) 
on $\phi^*$ makes a difference by using a simple pair of examples in the case (A'). 

\subsubsection{The Idea}
\label{ssec:illustrate}

Consider an abelian surface $M_I = T^4_I$ in the case (A') given 
by 
\begin{align}
 M_I = E_1 \times E_2, \qquad E_i := \C/(\Z \oplus \sqrt{p_i}\Z), \qquad 
   dz^i = \hat{\alpha}^i + \sqrt{p_i}\hat{\beta}_i \quad (i=1,2), 
   \label{eq:illustr-cpx-str}
\end{align}
where $H^1(E_i;\Z) \cong \Z \hat{\alpha}^i \oplus \Z \hat{\beta}_i$ for $i=1,2$. 
The integral basis of $H_1(E_i;\Z)$ dual to 
$\{ \hat{\alpha}^i,\hat{\beta}_i\}$ is denoted by $\{ \alpha_i, \beta^i \}$. 
In the presentation in section \ref{ssec:illustrate}, we work on  
two choices of a complexified K\"{a}hler parameter: 
\begin{align}
  (B+i\omega) = C \sqrt{p_1}\hat{\alpha}^1\hat{\beta}_1
      + \widetilde{C} \sqrt{p_2}\hat{\alpha}^2\hat{\beta}_2, 
   \qquad C, \widetilde{C} \in \Q_{>0}, 
    \label{eq:illustr-Kahler-1}
\end{align}
which belongs to the class of (A')--(\ref{eq:Aprime-B+iomega-RCFT}) (and 
the corresponding SCFT is rational), and the other 
\begin{align}
  (B+i\omega) = C \sqrt{p_2}\hat{\alpha}^1\hat{\beta}_1 
     +  \widetilde{C} \sqrt{p_1}\hat{\alpha}^2\hat{\beta}_2, 
    \qquad C, \widetilde{C} \in \Q_{>0}, 
     \label{eq:illustr-Kahler-2}
\end{align}
which belongs to the class of (A')--(\ref{eq:Aprime-B+iomega-nonRCFT})
(so the corresponding SCFT is not rational). 

The SYZ-mirror of the two $T^4$-target SCFTs are geometric,\footnote{
Make sure that $B|_{\Gamma_f} = 0$ and $\omega|_{\Gamma_f}=0$ 
in (\ref{eq:cond-Enckevort}). 
} %
 when we 
take T-duality along $\Gamma_f = {\rm Span}_\Z \{ \alpha_1, \alpha_2\}$ 
and fix $\Gamma_b = {\rm Span}_\Z\{ \beta^1, \beta^2\}$. Although there are 
infinitely many different choices of $\Gamma_f$ and $\Gamma_b$ for 
the two $T^4$-target SCFTs here, and although one has to deal with 
all those choices of $\Gamma_f$ and $\Gamma_b$ to discuss the 
the compatibility in condition 5(weak/strong), we will deal with just 
this one choice of $\Gamma_f$ and $\Gamma_b$ in this 
section \ref{ssec:illustrate} 
so that one can focus on the key point of the idea. 

With the first choice of $(B+i\omega)$, the mirror geometry 
$(T^4_\circ; G^\circ, B^\circ;I^\circ)$ is given by 
\begin{align}
 (T^4_\circ; I^\circ) = E_1^\circ \times E_2^\circ, \qquad 
  E_1^\circ = \C/(\Z \oplus C\sqrt{p_1}\Z), \quad 
  E_2^\circ = \C/(\Z \oplus \widetilde{C}\sqrt{p_2}\Z); 
\end{align}
with a notation $H_1(E_i^\circ;\Z) = \Z \alpha_i^\circ \oplus \Z\beta^i$
and $H^1(E_i^\circ;\Z) = \Z \hat{\alpha}^i_\circ \oplus \Z \hat{\beta}_i$, 
one may choose a set of generators of $H^{1,0}(E_i^\circ;\C)$ to be 
\begin{align}
  dz_\circ^1 = \hat{\alpha}_\circ^1 + C \sqrt{p_1} \hat{\beta}_1, 
   \qquad 
  dz_\circ^2 = \hat{\alpha}_\circ^2 + \widetilde{C} \sqrt{p_2} \hat{\beta}_2, 
\end{align}
and $(B^\circ+i\omega^\circ) =
 \sqrt{p_1}\hat{\alpha}^1_\circ \hat{\beta}_1
 + \sqrt{p_2}\hat{\alpha}^2_\circ \hat{\beta}_2$.  
There is a Hodge isomorphism $\phi^*: H^1(T^4;\Q)
 \longrightarrow H^1(T^4_\circ;\Q)$
\begin{align}
  \phi^*: \hat{\beta}_i \longmapsto \hat{\beta}_i, \quad 
   \hat{\alpha}^1 \longmapsto C^{-1} \hat{\alpha}^1_\circ, \quad 
   \hat{\alpha}^2 \longmapsto \widetilde{C}^{-1} \hat{\alpha}^2_\circ, \quad 
   dz^1 \mapsto C^{-1} dz_\circ^1, \quad 
  dz^2 \mapsto \widetilde{C}^{-1} dz_\circ^2. 
\end{align}
This map $\phi^*: H^1(T^4;\Q) \rightarrow H^1(T^4_\circ;\Q)$ is identity 
on the subspace $\Gamma_{b\Q}^\vee = 
{\rm Span}_\Q\{ \hat{\beta}_1,\hat{\beta}_2\}$
pulled back from the base of $\pi_M: (M=T^4)\rightarrow (T^2=B)$
and $\pi_W: (W=T^4_\circ) \rightarrow (T^2=B)$. 

With the second choice of $(B+i\omega)$, on the other hand, the mirror 
is given by 
\begin{align}
 (T^4_\circ;I^\circ)=E_1^\circ \times E_2^\circ, \quad 
 E_1^\circ = \C/(\Z \oplus C\sqrt{p_2}\Z), \quad 
 E_2^\circ = \C/(\Z \oplus \widetilde{C}\sqrt{p_1}\Z); 
\end{align}
with a notation $H_1(E_i^\circ;\Z) = \Z\alpha_i^\circ \oplus \Z\beta^i$
and $H^1(E_i^\circ;\Z) = \Z\hat{\alpha}^i_\circ \oplus \Z \hat{\beta}_i$, 
one may choose a set of generators of $H^{1,0}(E_i^\circ;\C)$ to be 
\begin{align}
 dz_\circ^1 = \hat{\alpha}^1_\circ + C \sqrt{p_2}\hat{\beta}_1, \qquad 
 dz_\circ^2 = \hat{\alpha}^2_\circ + \widetilde{C}\sqrt{p_1}\hat{\beta}_2,  
\end{align}
and $(B^\circ+i\omega^\circ) =
 \sqrt{p_1}\hat{\alpha}^1_\circ \hat{\beta}_1
 + \sqrt{p_2}\hat{\alpha}^2_\circ\hat{\beta}_2$. 
Any Hodge isomorphism $\phi^*: H^1(T^4;\Q) \rightarrow H^1(T^4_\circ;\Q)$
for this second choice of $(B+i\omega)$ has to be of the form 
\begin{align}
 (\hat{\alpha}^1,  \hat{\beta}_1) \longmapsto 
(\widetilde{C}^{-1} \hat{\alpha}^2_\circ, \hat{\beta}_2), \qquad 
 (\hat{\alpha}^2, \hat{\beta}_2) \longmapsto 
 (C^{-1} \hat{\alpha}^1_\circ, \hat{\beta}_1), 
\end{align}
preceded by an endomorphism (complex multiplication) in $\Q(\sqrt{p_1})\oplus \Q(\sqrt{p_2})$ 
on $H^1(T^4;\Q)$, and followed by an endomorphism on $H^1(T^4_\circ;\Q)$. 
That is precisely the existence of a rational Hodge isomorphism 
imposed as a part of condition 5 in Thm. \ref{thm:forT4}. 

In the case of the second choice of $(B+i\omega)$ in the class 
(A')--(\ref{eq:Aprime-B+iomega-nonRCFT}), however, every one of 
those rational Hodge isomorphisms $\phi^*$ maps $\hat{\beta}_1
 \in \Gamma_{b\Q}^\vee$ to ${\rm Span}_\Q\{\hat{\alpha}^2_\circ,
 \hat{\beta}_2\}$, 
and maps yet another 1-form of the base $\hat{\beta}_2 \in \Gamma_{b\Q}^\vee$ 
to ${\rm Span}_\Q\{ \hat{\alpha}^1_\circ, \hat{\beta}_1\}$. 
So there is no rational Hodge isomorphism $\phi^*$ that has the 
compatibility property (\ref{eq:cond-8}) in condition 5. 

We have confirmed that there exists a Hodge isomorphism $\phi^*: 
H^1(M;\Q)|_{M=T^4} \rightarrow H^1(W;\Q)|_{W=T^4_\circ}$ with the compatibility 
property (\ref{eq:cond-8}) on $H^1(B;\Q)|_{B=T^2} \cong \Gamma_{b\Q}^\vee$ 
for the first choice of $(B+i\omega)$. Condition 5 
of Conj. \ref{conj:forT4}/Thm. \ref{thm:forT4} demands, on the other hand, 
a Hodge isomorphism $\phi^*:H^*(T^4;\Q) \rightarrow H^*(T^4_\circ;\Q)$ 
with the property (\ref{eq:cond-8}) on the entire $H^*(B;\Q)_{B=T^2}$.  
In fact, a Hodge isomorphism on $\phi^*: H^1(T^4;\Q) \rightarrow 
H^1(T^4_\circ;\Q)$ can be extended by the wedge product (cup product) 
to a Hodge isomorphism on $H^*(T^4;\Q) \rightarrow H^*(T^4_\circ;\Q)$; 
the property that $\phi^*$ is identity on the whole $H^*(B;\Q)_{B=T^2}$ 
also follows automatically because this property is non-trivial 
only on $H^1(B;\Q)_{B=T^2}$. We will not repeat this argument in the 
rest of this section \ref{ssec:one-more}.

An elementary analysis so far shows that the set of data 
(\ref{eq:illustr-cpx-str}, \ref{eq:illustr-Kahler-1}) of a rational SCFT 
has $\phi^*$ in condition 5 (weak) including the compatibility
with the torus fibration; one may wonder if condition 5 (weak)---or 
even condition 5 (strong)---including the compatibility is a part 
of necessary conditions of all the sets of data in (B,C)--(\ref{eq:BC-B+iomega};$+$),
(A')--(\ref{eq:Aprime-B+iomega-RCFT}) and (A) of $T^4$-target rational SCFTs. 
We have also seen that the set of 
data (\ref{eq:illustr-cpx-str}, \ref{eq:illustr-Kahler-2})
of a non-rational SCFT does not have the compatibility in the 
condition 5(strong); one may therefore wonder if all the counter 
examples---$(B+i\omega)$ in (B, C)--(\ref{eq:BC-B+iomega};$-$) 
and in (A')--(\ref{eq:Aprime-B+iomega-nonRCFT})---may be eliminated 
by imposing the compatibility of $\phi^*$ with the torus 
fibration in condition 5 (strong), or even by imposing condition 5(weak). The rest of this section \ref{ssec:one-more} 
is a systematic proof of this expectation.

\subsubsection{The Case (A')}
\label{subsec:cond8-caseAprm}

In this section \ref{subsec:cond8-caseAprm}, we will work on the 
two classes of sets of data, (A')--(\ref{eq:Aprime-B+iomega-RCFT})
and (A')--(\ref{eq:Aprime-B+iomega-nonRCFT}). 
The essence of the following analysis is captured already in what we 
have done in section \ref{ssec:illustrate}. Here is one important 
difference, though. The analysis in section \ref{ssec:illustrate} was 
for only one choice of the T-dualized directions $\Gamma_f \subset 
H_1(T^4;\Z)$ and non-T-dualized directions $\Gamma_b \subset H_1(T^4;\Z)$. 
In order to prove that condition 5 (strong) is a part of necessary 
conditions for the SCFT of $(T^4;G,B;I)$ to be rational 
in case (A')--(\ref{eq:Aprime-B+iomega-RCFT}),   
we have to show that a rational Hodge isomorphism $\phi^*$ compatible 
with the torus fibration exists for each one of the possible 
choices of $\Gamma_f$ and $\Gamma_b$ (choices of geometric SYZ mirrors). 
In order to prove that the set of conditions in Thm. \ref{thm:forT4}
is sufficient (with condition 5 (weak) version), then 
we have to show that a rational Hodge isomorphism $\phi^*$ in the 
condition does not exist for each one of geometric SYZ mirrors 
of $(T^4;G,B;I)$ in the case (A')--(\ref{eq:Aprime-B+iomega-nonRCFT}). 
Possible choices of $\Gamma_f$ and $\Gamma_b$ have been studied 
and the results have been summarized in (\ref{eq:Aprime-GammafQ}) 
in section \ref{ssec:list-SYZ-mirror}
for the both cases of (A')--(\ref{eq:Aprime-B+iomega-RCFT})
and (A')--(\ref{eq:Aprime-B+iomega-nonRCFT}), so we can use them 
for the analysis below. 

First, we show that there is no Hodge isomorphism $\phi^*$
with the compatibility in condition 5 (weak) in case 
(A')--(\ref{eq:Aprime-B+iomega-nonRCFT}). This is a proof 
by contradiction. 

Assume that a set of data $(T^4;G,B;I)$ in the case 
(A')--(\ref{eq:Aprime-B+iomega-nonRCFT}) satisfies condition 5 (weak); 
let $(\Gamma_f,\Gamma_b)$ be the choice for a geometric SYZ-mirror 
and $\phi^\ast:H^1(T^4;\Q)\to H^1(T^4_\circ;\Q)$ be the 
Hodge isomorphism in the condition.
We will write down explicitly (i) what is implied by the condition 
$\phi^*|_{\Gamma_{b\Q}^\vee} ={\rm id}_{\Gamma_{b\Q}^\vee}$ and 
(ii) what is implied by $\phi^\ast$ being a Hodge isomorphism, and then 
show that there cannot be a $\phi^*$ satisfying both (i) and (ii). 

(i) The condition $\phi^\ast|_{\Gamma_{b\Q}^\vee}=\mathrm{id}_{\Gamma_{b\Q}^\vee}$ can be expressed in the form of 
\begin{align}
\phi^\ast
\begin{pmatrix}
\hat{c}' & \hat{d}' & \hat{e} & \hat{f}
\end{pmatrix}
=
\begin{pmatrix}
c & d & \hat{e} & \hat{f}
\end{pmatrix}
\left(\hspace{-5pt}\begin{array}{c|c}
r_{kl} & \hspace{-3pt}\begin{array}{c}0\\{\bf 1}\end{array}
\end{array}\hspace{-6pt}\right),
\label{eq:Aprime-cond8-asEq}
\end{align}
by using the basis $\{ \hat{c}', \hat{d}', \hat{e}, \hat{f} \}$ of $H^1(T^4;\Q)$ and the basis $\{c,d,\hat{e},\hat{f}\}$ of $H^1(T^4_\circ;\Q)$. 
Here, $(r_{kl})$ is a $4\times2$ $\Q$\hspace{1pt}-valued matrix.

Now, the condition (\ref{eq:Aprime-cond8-asEq}) implies that the 
holomorphic basis (cf Discussion \ref{statmnt:CM-pair} 
and (\ref{eq:basis-1form-base-caseAprime}, \ref{eq:Aprime-e'f'chat'dhat'}))
\begin{align}
\begin{pmatrix}
(c_1^2+c_2^2)dz^1 & (c_3^2+c_4^2)dz^2
\end{pmatrix}
& =
\begin{pmatrix}
\hat{c}' & \hat{d}' & \hat{e} & \hat{f}
\end{pmatrix}
\begin{pmatrix}
c_1+c_2\sqrt{p_1} & \\
& c_3+c_4\sqrt{p_2} \\
-c_2+c_1\sqrt{p_1} & \\
& -c_4+c_3\sqrt{p_2}
\end{pmatrix}
\end{align}
should be mapped by $\phi^\ast$ into
\begin{align}
\phi^\ast_\C
\begin{pmatrix}
(c_1^2+c_2^2)dz^1 & (c_3^2+c_4^2)dz^2
\end{pmatrix}
=
\begin{pmatrix}
c & d & \hat{e} & \hat{f}
\end{pmatrix}
\left(\hspace{-5pt}\begin{array}{c|c}
r_{kl} & \hspace{-3pt}\begin{array}{c}0\\{\bf 1}\end{array}
\end{array}\hspace{-6pt}\right)
\begin{pmatrix}
c_1+c_2\sqrt{p_1} & \\
& c_3+c_4\sqrt{p_2} \\
-c_2+c_1\sqrt{p_1} & \\
& -c_4+c_3\sqrt{p_2}
\end{pmatrix}.
\label{eq:Aprime-isogeny-on-Q-basis}
\end{align}

(ii) Next, we claim that $\phi^*$ being a Hodge isomorphism is equivalent to the presence of 
$\theta_1 \in \Q(\sqrt{p_1})^\times$ and $\theta_2 \in \Q(\sqrt{p_2})^\times$ such that 
\begin{align}
\phi^\ast_\C
\begin{pmatrix}
(c_1^2+c_2^2)dz^1 & (c_3^2+c_4^2)dz^2
\end{pmatrix}
& =
\begin{pmatrix}
dz^2_\circ & dz^1_\circ
\end{pmatrix}
\begin{pmatrix}
\theta_1 & \\
 & \theta_2
\end{pmatrix};
\label{eq:Aprime-isogeny-on-C-basis-preparation}
\end{align}
here, $dz^1_\circ$ and $dz^2_\circ$ are holomorphic 1-forms on 
$(T^4_\circ;I^\circ)$ where the Hodge endomorphism algebra 
${\cal K}\cong \Q(\sqrt{p_1}) \oplus \Q(\sqrt{p_2})$ acts through 
the factor $\Q(\sqrt{p_2})$ and $\Q(\sqrt{p_1})$, respectively. 
Those holomorphic 1-forms on the mirror 
must be in the form of
\begin{align}
\begin{pmatrix}
\lambda'_1dz^2_\circ & \lambda'_2 dz^1_\circ
\end{pmatrix}
&=
\begin{pmatrix}
d & \hat{f} & c & \hat{e}
\end{pmatrix}
\begin{pmatrix}
1 & \\
\rho'_1 & \\
 & 1\\
 & \rho'_2
\end{pmatrix}
\end{align}
for some $\lambda'_1, \rho'_1 \in \Q(\sqrt{p_1})^\times$ and 
$\lambda'_2, \rho'_2 \in \Q(\sqrt{p_2})^\times$.
Therefore, (\ref{eq:Aprime-isogeny-on-C-basis-preparation}) becomes
\begin{align}
\phi^\ast_\C
\begin{pmatrix}
(c_1^2+c_2^2)dz^1 & (c_3^2+c_4^2)dz^2
\end{pmatrix}
& =
\begin{pmatrix}
c & d & \hat{e} & \hat{f}
\end{pmatrix}
\begin{pmatrix}
 & 1\\
1 & \\
 & \rho'_2\\
\rho'_1  & 
\end{pmatrix}
\begin{pmatrix}
\theta'_1 & \\
 & \theta'_2
\end{pmatrix}, 
\label{eq:Aprime-isogeny-on-C-basis}
\end{align}
where $\theta'_1 := \theta_1/\lambda'_1 \in \Q(\sqrt{p_1})^\times$ and 
$\theta'_2:=\theta_2/\lambda'_2 \in \Q(\sqrt{p_2})^\times$.

\vspace{5mm}

Now that the properties (i) and (ii) have been paraphrased as (\ref{eq:Aprime-isogeny-on-Q-basis}) and (\ref{eq:Aprime-isogeny-on-C-basis}), respectively, let us see that there is no common solution $(r_{kl},\theta'_i)$ to (\ref{eq:Aprime-isogeny-on-Q-basis}) and (\ref{eq:Aprime-isogeny-on-C-basis}) indeed, or equivalently, no solution to 
\begin{equation}
\left(\hspace{-5pt}\begin{array}{c|c}
r_{kl} & \hspace{-3pt}\begin{array}{c}0\\{\bf 1}\end{array}
\end{array}\hspace{-6pt}\right)
\begin{pmatrix}
c_1+c_2\sqrt{p_1} & \\
& c_3+c_4\sqrt{p_2} \\
-c_2+c_1\sqrt{p_1} & \\
& -c_4+c_3\sqrt{p_2}
\end{pmatrix}
=
\begin{pmatrix}
 & 1\\
1 & \\
 & \rho'_2\\
\rho'_1 & 
\end{pmatrix}
\begin{pmatrix}
\theta'_1 & \\
 & \theta'_2
\end{pmatrix},
\label{eq:Aprime-isogeny-compare}
\end{equation}
for any $(c_\text{1--4},\rho'_{1,2})$. To see this, note that the both sides of (\ref{eq:Aprime-isogeny-compare}) 
are $4\times2$ matrices after multiplication, and look at the (3,1) entry: 
\begin{equation}
r_{31} (c_1+c_2\sqrt{p_1}) +(-c_2+c_1\sqrt{p_1})=0.
\end{equation}
Such $r_{31}\in\Q$ does not exist because of $(c_1,c_2)\neq(0,0)$ (see (\ref{eq:Aprime-GammafQ-prm})).
Similarly, we can also see that there is no way this equality holds for the (4,2) entry. 

\vspace{5mm}

The remaining task is to prove that $\phi^*$ with the compatibility in  condition 5 (strong) exists in the case 
(A')--(\ref{eq:Aprime-B+iomega-RCFT}).
That is to construct a Hodge isomorphism $\phi^\ast:H^1(T^4;\Q)\to H^1(T^4_\circ;\Q)$ satisfying $\phi^*|_{\Gamma_{b\Q}^\vee} ={\rm id}_{\Gamma_{b\Q}^\vee}$ for each one of geometric SYZ-mirrors for a set of data $(T^4;G,B;I)$ in the case of (A')--(\ref{eq:Aprime-B+iomega-RCFT}).

To do so, think of a geometric SYZ-mirror for $(\Gamma_f,\Gamma_b)$, and derive a condition for such a Hodge isomorphism $\phi^\ast$, first.
The derivation goes parallel to that for (\ref{eq:Aprime-isogeny-compare}), 
except (\ref{eq:Aprime-isogeny-on-C-basis-preparation})--(\ref{eq:Aprime-isogeny-on-C-basis}) need to be modified a little bit. Consequently we obtain
\begin{equation}
\left(\hspace{-5pt}\begin{array}{c|c}
r_{kl} & \hspace{-3pt}\begin{array}{c}0\\{\bf 1}\end{array}
\end{array}\hspace{-6pt}\right)
\begin{pmatrix}
c_1+c_2\sqrt{p_1} & \\
& c_3+c_4\sqrt{p_2} \\
-c_2+c_1\sqrt{p_1} & \\
& -c_4+c_3\sqrt{p_2}
\end{pmatrix}
=
\begin{pmatrix}
1 & \\
 & 1\\
 \rho'_1 & \\
 & \rho'_2
\end{pmatrix}
\begin{pmatrix}
\theta'_1 & \\
 & \theta'_2
\end{pmatrix},
\label{eq:Aprime-isogeny-compare-RCFT}
\end{equation}
where $r_{kl}\in\Q$ and $\theta'_i\in\Q(\sqrt{p_i})^\times$ are parameters for $\phi^\ast$, while $c_\text{1--4}\in\Q$ and $\rho'_i\in\Q(\sqrt{p_i})$ depend on $(\Gamma_f,\Gamma_b)$. 

Now, it is easy to see that a solution $(r_{kl},\theta'_i)$ to (\ref{eq:Aprime-isogeny-compare-RCFT}) exists.
The equation (\ref{eq:Aprime-isogeny-compare-RCFT}) at six out of the $4\times2$ entries imposes 
\begin{align}
&r_{12}=r_{21}=r_{32}=r_{41}=0,\\
&\theta'_1 = r_{11} (c_1+c_2\sqrt{p_1}),\quad \theta'_2 = r_{22} (c_3+c_4\sqrt{p_2}),
\end{align}
and the remaining variables $r_{11},r_{31},r_{22},r_{42}\in\Q$ 
are constrained by 
\begin{align}
-c_2+c_1\sqrt{p_1}&=\left( \rho'_1 r_{11}-r_{31}\right)(c_1+c_2\sqrt{p_1}),\\
-c_4+c_3\sqrt{p_2}&=\left( \rho'_2 r_{22}-r_{42}\right)(c_3+c_4\sqrt{p_2}), 
\end{align}
because of (\ref{eq:Aprime-isogeny-compare-RCFT}) at the 
two remaining entries. 
A solution $(r_{11},r_{31},r_{22},r_{42})$ is determined by elementary algebra in the imaginary quadratic fields $\Q(\sqrt{p_1})$ and $\Q(\sqrt{p_2})$.
Moreover, $r_{11}$ and $r_{22}$ turn out to be non-zero, which means 
that $\phi^\ast$ is invertible.
Since the above argument holds for arbitrary choice of $(\Gamma_f,\Gamma_b)$, 
this completes the proof of condition 5(strong) for 
case (A')--(\ref{eq:Aprime-B+iomega-RCFT}).

\subsubsection{Case (B, C)}
\label{subsec:cond8-caseBC}

In this section \ref{subsec:cond8-caseBC}, we will work for 
the two classes of $(B+i\omega)$, (B, C)--(\ref{eq:BC-B+iomega};$+$) 
and (B,C)--(\ref{eq:BC-B+iomega};$-$). 
Let us first show for $(B+i\omega)$ in the class (B,C)--(\ref{eq:BC-B+iomega};$-$)  
that a Hodge isomorphism $\phi^*$ in condition 5 cannot be 
chosen to be compatible with the SYZ torus fibration. 
We do so in a proof by contradiction.
The outline of the proof is the same as in section \ref{subsec:cond8-caseAprm}; 
it just takes a little more time to do so. 

Think of a set of data $(T^4;G,B;I)$ in the case 
(B,C)--(\ref{eq:BC-B+iomega};$-$). 
Let $(\Gamma_f,\Gamma_b)$ be the choice for a geometric SYZ-mirror, 
and $\phi^\ast:H^1(T^4;\Q)$ $\to H^1(T^4_\circ;\Q)$ 
the Hodge isomorphism satisfying the compatibility.  
We will write down explicitly (i) what is implied by the condition $\phi^*|_{\Gamma_{b\Q}^\vee} ={\rm id}_{\Gamma_{b\Q}^\vee}$ and (ii) what is implied by $\phi^\ast$ being a Hodge isomorphism, and 
then show that there cannot be $\phi^*$ satisfying both (i) and (ii). 

(i) The condition $\phi^\ast|_{\Gamma_{b\Q}^\vee}=\mathrm{id}_{\Gamma_{b\Q}^\vee}$ 
can be expressed in the form of 
\begin{align}
\phi^\ast
\begin{pmatrix}
\hat{c}' & \hat{d}' & \hat{e} & \hat{f}
\end{pmatrix}
=
\begin{pmatrix}
c & d & \hat{e} & \hat{f}
\end{pmatrix}
\left(\hspace{-5pt}\begin{array}{c|c}
r_{kl} & \hspace{-3pt}\begin{array}{c}0\\{\bf 1}\end{array}
\end{array}\hspace{-6pt}\right); 
\label{eq:BC-cond8-asEq}
\end{align}
here, $(r_{kl})$ is a $4\times2$ $\Q$\hspace{1pt}-valued matrix. 
We only deal with the cases with $(c_1,c_3)\neq(0,0)$ here; 
the following discussion 
can be repeated for the case of $(c_2,c_4)\neq(0,0)$ easily.

Now, the condition (\ref{eq:BC-cond8-asEq}) implies that the holomorphic basis 
in (\ref{eq:hol-rat-basis-trnsf-BC}) (cf 
also (\ref{eq:BC-c_1c_3nonzero-ehat}--\ref{eq:BC-c_1c_3nonzero-e'f'chat'dhat'}))
\begin{align}
\begin{pmatrix}
(c_1^2-dc_3^2)dz^1 & (c_1^2-dc_3^2)dz^2
\end{pmatrix}
& =
\begin{pmatrix}
\hat{c}' & \hat{d}' & \hat{e} & \hat{f}
\end{pmatrix}
\begin{pmatrix}
c_1 & c_3 & c_2 & c_4 \\
dc_3 & c_1 & dc_4 & c_2\\
 & & 1 & \\
 & & & 1
\end{pmatrix}
\begin{pmatrix}
1 & 1\\
\tau_{++}(y) & \tau_{-+}(y)\\
\tau_{++}(x) & \tau_{-+}(x)\\
\tau_{++}(xy) & \tau_{-+}(xy)
\end{pmatrix}
\label{eq:BC-isogeny-on-Q-basis-before}
\end{align}
should be mapped by $\phi^\ast$ into
\begin{align}
\phi^\ast_\C
\begin{pmatrix}
(c_1^2-dc_3^2)dz^1 & (c_1^2-dc_3^2)dz^2
\end{pmatrix}
=&
\begin{pmatrix}
c & d & \hat{e} & \hat{f}
\end{pmatrix}
\left(\hspace{-5pt}\begin{array}{c|c}
r_{kl} & \hspace{-3pt}\begin{array}{c}0\\{\bf 1}\end{array}
\end{array}\hspace{-6pt}\right)
\begin{pmatrix}
\tau_{++}(\Gamma) & \tau_{-+}(\Gamma) \\
\tau_{++}(\Gamma y) & \tau_{-+}(\Gamma y) \\
\tau_{++}(x) & \tau_{-+}(x) \\
\tau_{++}(xy) & \tau_{-+}(xy)
\end{pmatrix},
\label{eq:BC-isogeny-on-Q-basis}
\end{align}
where $\Gamma:=c_1+c_3y+c_2x+c_4xy$ is an element of $K={\rm End}(H^1(T^4;\Q))^{\rm Hdg}$ in the 
case (B, C).

(ii) Next, we will translate the condition that $\phi^\ast$ is a Hodge 
isomorphism. The images of the holomorphic basis elements 
\begin{align}
\phi^\ast_\C
\begin{pmatrix}
(c_1^2-dc_3^2)dz^1 & (c_1^2-dc_3^2)dz^2
\end{pmatrix}
=:
\begin{pmatrix}
dz^1_\circ & dz^2_\circ
\end{pmatrix}
%
%
\label{eq:BC-isom-diagonal}
%
\end{align}
generate the Hodge (1,0) component of $H^1(T^4_\circ;\C)$; 
when the endomorphism fields are identified through 
${\rm Ad}_{\phi^*}: K \cong {\rm End}(H^1(T^4;\Q))^{\rm Hdg} 
\ni x'\longmapsto \phi^* \circ x' \circ (\phi^*)^{-1} 
\in {\rm End}(H^1(T^4_\circ;\Q))^{\rm Hdg}$,
$dz^1_\circ$ and $dz^2_\circ$ are eigenvectors of the action of 
the field $K$,  
and the eigenvalues of $x' \in K$ are its images by the embeddings 
$\tau_{++}$ and $\tau_{-+}$, respectively. 
We note also that both $dz^1_\circ$ and $dz^2_\circ$ are normalized so that they 
return a rational value for some rational elements in $H_1(T^4_\circ;\Q)$.

On the other hand, it is also possible to work out the 
rational Hodge structure on the mirror cohomology group 
$H^1(T^4_\circ;\Q)$ by following the standard procedure for the mirror 
of complex tori (that are not necessarily of CM-type). 
The Hodge (1,0) subspace $H^{1,0}(T^4_\circ;\C)$ is generated by 
the two vectors
\begin{align}
  (dz_\circ^{1'}, dz_\circ^{2'}) = (c,d,\hat{e},\hat{f})
     \left( \begin{array}{cc} 
       \tau_{++}( \mp s_0 y) & \tau_{-+}(\mp s_0 y) \\
       \tau_{++}(-s_0) & \tau_{-+}(-s_0) \\
       \tau_{++}((c_1\pm c_3y)\Xi_{\pm}) & 
       \tau_{-+}((c_1\pm c_3y)\Xi_{\pm}) \\
       \tau_{++}((dc_3 \pm c_1y)\Xi_{\pm}) & 
       \tau_{-+}((dc_3 \pm c_1y)\Xi_{\pm}) \end{array} \right),
  \label{eq:BC-mirror-hol-basis-vs-rat-basis} 
\end{align}
where $s_0 := (c_1^2-dc_3^2) \in \Q^\times$. 
Here (and also until (\ref{eq:BC-isog-condition-final}) below), the signs shown 
below in $\pm$ and $\mp$ should be used for 
the case (B,C)--(\ref{eq:BC-B+iomega};$-$) under consideration now. 
The signs shown above 
in (\ref{eq:BC-mirror-hol-basis-vs-rat-basis}--\ref{eq:BC-isog-condition-final})
are for the case (B,C)--(\ref{eq:BC-B+iomega};$+$), which we will 
use in the latter half of this section \ref{subsec:cond8-caseBC}. 
Details of calculations leading to (\ref{eq:BC-mirror-hol-basis-vs-rat-basis})
is found in Lemma \ref{lemma:mirror-H1-HdgStr-caseBC}. 
The rational Hodge structure obtained in the computation manifestly 
has CM by the degree-4 CM field $K$, and the basis elements 
$\{dz^{1'}_\circ, dz^{2'}_\circ\}$ are the eigenstates of the action 
of $K$ (cf Lemma \ref{lemma:very-useful-2}).

The $K$-diagonal basis $\{ dz^1_\circ, dz^2_\circ\}$ 
in (\ref{eq:BC-isom-diagonal}) must be proportional to yet another  
$K$-diagonal basis $\{ dz^{1'}_\circ, dz^{2'}_\circ\}$ 
in (\ref{eq:BC-mirror-hol-basis-vs-rat-basis}); $dz^1_\circ$ 
[resp. $dz^2_\circ$] should be a complex multiple of $dz^{1'}_\circ$ 
[resp. $dz^{2'}_\circ$]. Furthermore, the proportionality constant 
should be found within $\tau_{++}(K)$ [resp. $\tau_{-+}(K)$], 
which follows from the fact that both $dz^1_\circ$ and $dz^{1'}_\circ$ return 
(not necessarily identical) rational numbers to (not necessarily identical) 
rational elements in $H_1(T^4_\circ;\Q)$. 
So, to conclude, the condition (ii) that $\phi^*$ is a Hodge isomorphism 
is translated to the existence of elements $\theta_+, \theta_- \in 
K^\times$ so that 
\begin{align}
&\phi^\ast_\C
\begin{pmatrix}
 s_0 dz^1 & s_0 dz^2
\end{pmatrix} \label{eq:BC-isogeny-on-C-basis} \\
&=
\begin{pmatrix}
c & d & \hat{e} & \hat{f}
\end{pmatrix}
\begin{pmatrix}
\tau_{++}(\mp s_0 y) & \tau_{-+}(\mp s_0 y) \\
-s_0 & -s_0 \\
\tau_{++}((c_1\pm c_3y)\Xi_\pm) & \tau_{-+}((c_1\pm c_3y)\Xi_\pm)\\
\tau_{++}((dc_3\pm c_1y)\Xi_\pm) & \tau_{-+}((dc_3\pm c_1y)\Xi_\pm)
\end{pmatrix}
\begin{pmatrix}
\tau_{++}(\theta_+) & \\
 & \tau_{-+}(\theta_-)
\end{pmatrix}.
  \nonumber
\end{align}

\vspace{5mm}

Now that the properties (i) and (ii) have been 
paraphrased as (\ref{eq:BC-isogeny-on-Q-basis}) 
and (\ref{eq:BC-isogeny-on-C-basis};$-$), respectively, let us see that 
there is no common solution $(r_{kl},\theta_+, \theta_-)$ 
to (\ref{eq:BC-isogeny-on-Q-basis}) and (\ref{eq:BC-isogeny-on-C-basis};$-$) 
for any $c_\text{1--4}$; this then implies that 
there is no choice $(\Gamma_f,\Gamma_b)$ for a geometric SYZ-mirror 
where condition 5 (weak) holds true in the case of 
(B,C)--(\ref{eq:BC-B+iomega};$-$).

Indeed, the compatibility condition on $(r_{k\ell}, \theta_+, \theta_-)$ is 
\begin{align}
&\left(\hspace{-5pt}\begin{array}{c|c}
r_{kl} & \hspace{-3pt}\begin{array}{c}0\\{\bf 1}\end{array}
\end{array}\hspace{-6pt}\right)
\begin{pmatrix}
\tau_{++}(\Gamma) & \tau_{-+}(\Gamma) \\
\tau_{++}(\Gamma y) & \tau_{-+}(\Gamma y) \\
\tau_{++}(x) & \tau_{-+}(x) \\
\tau_{++}(xy) & \tau_{-+}(xy)
\end{pmatrix}
\nonumber\\
&=
\begin{pmatrix}
\tau_{++}(\mp s_0 y) & \tau_{-+}(\mp s_0 y) \\
-s_0 & - s_0\\
\tau_{++}((c_1\pm c_3y)\Xi_\pm) & \tau_{-+}((c_1\pm c_3y)\Xi_\pm)\\
\tau_{++}((dc_3\pm c_1y)\Xi_\pm) & \tau_{-+}((dc_3\pm c_1y)\Xi_\pm)
\end{pmatrix}
\begin{pmatrix}
\tau_{++}(\theta_+) & \\
 & \tau_{-+}(\theta_-)
\end{pmatrix}
\label{eq:BC-isog-C-basis-vs-Q-basis}
\end{align}
when $(c_1,c_3) \neq 0$. The two elements $\theta_+, \theta_- \in 
K^\times$ have to be identical for the relations in the second row 
to hold. Now, the $4\times2$ relations among algebraic numbers 
in (\ref{eq:BC-isog-C-basis-vs-Q-basis}) are regarded as four relations 
among the elements in the CM field $K$.
To see that there is no solution $(r_{kl},\theta)$ to (\ref{eq:BC-isog-C-basis-vs-Q-basis};$-$), it suffices to focus on the lower two relations:
\begin{align}
&r_{31}\Gamma+r_{32}\Gamma y+x=(c_1\pm c_3y)\Xi_\pm\theta,
\label{eq:BC-isog-condition-1}\\
&r_{41}\Gamma+r_{42}\Gamma y+xy=(dc_3\pm c_1y)\Xi_\pm\theta.
\label{eq:BC-isog-condition-2}
\end{align}
Comparing $(\ref{eq:BC-isog-condition-1})\times(\pm y)$ and (\ref{eq:BC-isog-condition-2}), we can eliminate $\theta$ and obtain
\begin{align}
[\pm dr_{32}-r_{41}+(\pm r_{31}-r_{42})y]\Gamma=&\ axy, 
\label{eq:BC-isog-condition-final}\\
&\ a=\left\{\begin{array}{ll}
0 & \text{in the case (B,C)--(\ref{eq:BC-B+iomega};$+$)}, \\
2 & \text{in the case (B,C)--(\ref{eq:BC-B+iomega};$-$)}.
\end{array}\right. \nonumber
\end{align}
Recalling that $\Gamma=c_1+c_3y+c_2x+c_4xy$ and $(c_1,c_3)\neq(0,0)$, some easy algebra in the number field $K$ shows that this 
equation (\ref{eq:BC-isog-condition-final};$a=2$) for the case (B,C)--(\ref{eq:BC-B+iomega};$-$) does not have a solution $(r_{31},r_{32},r_{41},r_{42})$.

\vspace{5mm}

The remaining task is to prove that condition 5 (strong) is satisfied 
in the case (B, C)--(\ref{eq:BC-B+iomega};$+$).
That is to construct a Hodge isomorphism $\phi^\ast:H^1(T^4;\Q)\to H^1(T^4_\circ;\Q)$ satisfying $\phi^*|_{\Gamma_{b\Q}^\vee} ={\rm id}_{\Gamma_{b\Q}^\vee}$ for each one of geometric SYZ-mirrors for a set of data $(T^4;G,B;I)$ in the case 
(B, C)--(\ref{eq:BC-B+iomega};$+$).

To do so, we will find a solution $(r_{kl},\theta)$ to the equation (\ref{eq:BC-isog-C-basis-vs-Q-basis};$+$) for a given $c_\text{1--4}$.
Just as a reminder, $(r_{kl},\theta)$ are parameters for $\phi^\ast$, whereas 
$c_\text{1--4}$ are determined by a given $(\Gamma_f,\Gamma_b)$ for a geometric SYZ-mirror.

In fact, we can solve the equation (\ref{eq:BC-isog-C-basis-vs-Q-basis};$+$) as follows.
As already discussed above, the lower two rows of the equation (\ref{eq:BC-isog-C-basis-vs-Q-basis};$+$) lead to equations (\ref{eq:BC-isog-condition-1};$+$), (\ref{eq:BC-isog-condition-2};$+$), and (\ref{eq:BC-isog-condition-final}; $a=0$).
The last one (\ref{eq:BC-isog-condition-final}; $a=0$) is equivalent to
\begin{align}
r_{42}=r_{31},\quad r_{41}=dr_{32}.
\label{eq:BC-isog-solution-1}
\end{align}
In the same way, from the upper two rows of the equation (\ref{eq:BC-isog-C-basis-vs-Q-basis};$+$), we find
\begin{align}
  r_{11} = d r_{22},\quad  r_{12} = r_{21},
\label{eq:BC-isog-solution-2}
\end{align}
and
\begin{align}
&r_{21}\Gamma+r_{22}\Gamma y= - s_0 \theta.
\label{eq:BC-isog-condition-3}
\end{align}
Now, by eliminating $\theta$ from (\ref{eq:BC-isog-condition-1};$+$) and (\ref{eq:BC-isog-condition-3}), we have
\begin{align}
s_0 r_{31} + s_0 r_{32}y + r_{21}\Xi'_+ + r_{22}\Xi'_+y=-s_0 x\Gamma^{-1},
\label{eq:BC-isog-condition-4}
\end{align}
where $\Xi'_+:=(c_1+c_3y)\Xi_+$.
Since $1,y,\Xi'_+,\Xi'_+y\in K$ are linearly independent over $\Q$, 
this equation (\ref{eq:BC-isog-condition-4}) determines 
$(r_{31},r_{32},r_{21},r_{22})$ uniquely.

It is easy to confirm that the equation (\ref{eq:BC-isog-C-basis-vs-Q-basis};$+$) is satisfied whenever the parameters $(r_{kl},\theta)$ 
are subject to the relations:
\begin{align*}
&(\ref{eq:BC-isog-condition-4}): s_0 r_{31}+s_0 r_{32}y + r_{21}\Xi'_+ + r_{22}\Xi'_+y=-s_0 x\Gamma^{-1},\\
&(\ref{eq:BC-isog-solution-1}): r_{42}=r_{31},\quad r_{41}=dr_{32},\\
&(\ref{eq:BC-isog-solution-2}):  r_{11}= d r_{22},\quad r_{12} = r_{21},\\
&(\ref{eq:BC-isog-condition-3}): s_0 \theta= -(r_{21}\Gamma+r_{22}\Gamma y).
\end{align*}
So, indeed, there is a solution $(r_{k\ell}, \theta)$ common to both (i) and (ii). Furthermore, the Hodge morphism $\phi^\ast$ specified by this $(r_{kl},\theta)$ 
is an isomorphism; if $\theta$ were zero (and hence $\phi^\ast$ were not invertible), then $\Gamma$, $\Gamma y$, and $x$ would not be linear independent over $\Q$, 
according to (\ref{eq:BC-isog-condition-1}).
Since the above argument holds for arbitrary choice of a geometric SYZ-mirror, this completes the proof that condition 5 (strong)---including the compatibility---is a part of necessary conditions for the case 
(B,C)--(\ref{eq:BC-B+iomega};$+$).

\subsubsection{The Case (A)}  
\label{subsec:cond8-caseA}

%
We will show in section \ref{subsec:cond8-caseA} 
the existence of a Hodge isomorphism 
$\phi^\ast:H^1(T^4;\Q)\to H^1(T^4_\circ;\Q)$ satisfying 
$\phi^\ast|_{\Gamma_{b\Q}^\vee}={\rm id}_{\Gamma_{b\Q}^\vee}$ for any choice of 
$(\Gamma_f,\Gamma_b)$ for a geometric SYZ-mirror in the case (A). 

To get started, let us have a few words about rational and holomorphic bases of $H^1(T^4)$ and 
$H^1(T^4_\circ)$ to be used in the analysis. Let $\{c,d\}$ be a basis of $\Gamma_{f\Q}$, 
and $\{e,f\}$ that of $\Gamma_{b\Q}$;
the subspaces $\Gamma_{f\Q}^\vee$ and $\Gamma_{b\Q}^\vee$ of $H^1(T^4;\Q)$ are generated by 
$\{\hat{c},\hat{d}\}$ and $\{ \hat{e}, \hat{f}\}$, respectively; here, 
$\{\hat{c},\hat{d},\hat{e},\hat{f}\}$ is the dual basis of $\{ c,d,e,f\}$.  
The (1,0)-forms $dz^1$ and $dz^2$ in the case (A) entry in 
Discussion \ref{statmnt:classify-abel-surf} are related to the 
rational basis by 
\begin{align}
\begin{pmatrix}
dz^1 & dz^2
\end{pmatrix}
=
\begin{pmatrix}
\hat{c} & \hat{d} & \hat{e} & \hat{f}
\end{pmatrix}
\begin{pmatrix}
\lambda_{11} & \lambda_{12}\\
\vdots & \vdots\\
\lambda_{41} & \lambda_{42}
\end{pmatrix},
\label{eq:A-hol-vs-rat-basis}
\end{align}
with the coefficients $\lambda_{ij}$ in the imaginary quadratic field 
$\Q(\sqrt{p})$.

We may change the holomorphic basis and rational basis a little bit so that the 
analysis later in this section \ref{subsec:cond8-caseA} is easier. 
Note first that the upper $2 \times 2$ block of the coefficient matrix in (\ref{eq:A-hol-vs-rat-basis})
is invertible. This is because the holomorphic $n$-form on the $T^n$ fiber should have 
a non-zero period in the SYZ mirror correspondence. So, one may change the basis $\{ dz^1, dz^2\}$
by a ${\rm GL}_2(\Q(\sqrt{p}))$ transform so that 
the upper $2\times 2$ block in (\ref{eq:A-hol-vs-rat-basis})
is the identity matrix. Furthermore, it is possible to rearrange the rational basis of $\Gamma_{b\Q}^\vee$, 
denoted by $\{ \hat{e}', \hat{f}'\}$ now, so that the coefficient $\lambda_{31}$ becomes rational. To summarize, 
there are a rational basis $\{ \hat{c}, \hat{d}\}$ of $\Gamma_{f\Q}^\vee$, $\{ \hat{e}', \hat{f}'\}$ 
of $\Gamma_{b\Q}^\vee$, and a holomorphic basis $\{ d\tilde{z}^1, d\tilde{z}^2\}$ of $H^{1,0}(T^4;\C)$
so that 
\begin{align}
\begin{pmatrix}
d\tilde{z}^1 & d\tilde{z}^2
\end{pmatrix}
=
\begin{pmatrix}
\hat{c} & \hat{d} & \hat{e}' & \hat{f}'
\end{pmatrix}
\begin{pmatrix}
1 & 0\\
0 & 1\\
\widetilde{\lambda}_{31}' & \widetilde{\lambda}_{32}'\\
\widetilde{\lambda}_{41}' & \widetilde{\lambda}_{42}'
\end{pmatrix}, \quad \text{where} \quad \widetilde{\lambda}_{31}'\in\Q,
\label{eq:A-retake-rat}
\end{align}
and $\widetilde{\lambda}_{32}',\widetilde{\lambda}_{41}',\widetilde{\lambda}_{42}'\in\Q(\sqrt{p})$.

On the mirror side, we can regard  $\{c,d,\hat{e}',\hat{f}'\}$ as 
a rational basis of $H^1(T^4_\circ;\Q)$.  
%
Repeating the same argument as in the case of $T^4$, one can find a basis 
$\{ d\tilde{z}^1_\circ, d\tilde{z}^2_\circ \}$ of $H^{1,0}(T^4_\circ;\C)$
so that 
\begin{align}
\begin{pmatrix}
d\tilde{z}_\circ^1 & d\tilde{z}_\circ^2
\end{pmatrix}
=
\begin{pmatrix}
c & d & \hat{e}' & \hat{f}'
\end{pmatrix}
\begin{pmatrix}
1 & 0\\
0 & 1\\
\widetilde{\rho}'_{31} & \widetilde{\rho}'_{32}\\
\widetilde{\rho}'_{41} & \widetilde{\rho}'_{42}
\end{pmatrix}
\end{align}
for some coefficients $\widetilde{\rho}'_{ij}$ in $\Q(\sqrt{p})$.

Having set the stage, we now demonstrate that condition 5 (strong) is 
satisfied, including compatibility (\ref{eq:cond-8}). 
As we have done earlier, we will translate (i) condition (\ref{eq:cond-8}) 
and (ii) the Hodge-ness condition of the isomorphism $\phi^*: H^1(T^4;\Q) \rightarrow H^1(T^4_\circ;\Q)$, 
and then find a common solution.  

(i) The condition $\phi^\ast|_{\Gamma_{b\Q}^\vee}={\rm id}_{\Gamma_{b\Q}^\vee}$ is equivalent to the existence of a $4\times2$ $\Q$\hspace{1pt}-valued matrix $(r_{kl})$ such that
\begin{align}
\phi^\ast_\C
\begin{pmatrix}
d\tilde{z}^1 & d\tilde{z}^2
\end{pmatrix}
=&
\begin{pmatrix}
c & d & \hat{e}' & \hat{f}'
\end{pmatrix}
\left(\hspace{-5pt}\begin{array}{c|c}
r_{kl} & \hspace{-3pt}\begin{array}{c}0\\{\bf 1}\end{array}
\end{array}\hspace{-6pt}\right)
\begin{pmatrix}
1 & 0\\
0 & 1\\
\widetilde{\lambda}_{31}' & \widetilde{\lambda}_{32}'\\
\widetilde{\lambda}_{41}' & \widetilde{\lambda}_{42}'
\end{pmatrix}.
\label{eq:A-isogeny-on-Q-basis}
\end{align}
(ii) 
%
The condition that $\phi^\ast$ is a Hodge isomorphism is equivalent 
to the existence of an invertible matrix $(\theta_{ij})\in M_2(\Q(\sqrt{p}))$ 
such that
\begin{align}
\phi^\ast_\C
\begin{pmatrix}
d\tilde{z}^1 & d\tilde{z}^2
\end{pmatrix}
=&
\begin{pmatrix}
c & d & \hat{e}' & \hat{f}'
\end{pmatrix}
\begin{pmatrix}
1 & 0\\
0 & 1\\
\widetilde{\rho}'_{31} & \widetilde{\rho}'_{32}\\
\widetilde{\rho}'_{41} & \widetilde{\rho}'_{42}
\end{pmatrix}
\begin{pmatrix}
\theta_{11} & \theta_{12}\\
\theta_{21} & \theta_{22}
\end{pmatrix}.
\label{eq:A-isogeny-on-C-basis}
\end{align}

\vspace{5mm}

Now that the properties (i) and (ii) have been paraphrased as (\ref{eq:A-isogeny-on-Q-basis}) and (\ref{eq:A-isogeny-on-C-basis}), respectively, let us see that there exists a common solution $(r_{kl},\theta_{ij})$ to (\ref{eq:A-isogeny-on-Q-basis}) and (\ref{eq:A-isogeny-on-C-basis}) for a given $(\widetilde{\lambda}'_{ij},\widetilde{\rho}'_{ij})$.
Just as a reminder, $(r_{kl},\theta_{ij})$ are parameters of $\phi^\ast$, whereas $(\widetilde{\lambda}'_{ij},\widetilde{\rho}'_{ij})$ depend on a given $(\Gamma_f,\Gamma_b)$ for a geometric SYZ-mirror.

The compatibility condition of (\ref{eq:A-isogeny-on-Q-basis}) 
and (\ref{eq:A-isogeny-on-C-basis}) is 
\begin{align}
\left(\hspace{-5pt}\begin{array}{c|c}
r_{kl} & \hspace{-3pt}\begin{array}{c}0\\{\bf 1}\end{array}
\end{array}\hspace{-6pt}\right)
\begin{pmatrix}
1 & 0\\
0 & 1\\
\widetilde{\lambda}'_{31} & \widetilde{\lambda}'_{32}\\
\widetilde{\lambda}'_{41} & \widetilde{\lambda}'_{42}
\end{pmatrix}
=
\begin{pmatrix}
1 & 0\\
0 & 1\\
\widetilde{\rho}'_{31} & \widetilde{\rho}'_{32}\\
\widetilde{\rho}'_{41} & \widetilde{\rho}'_{42}
\end{pmatrix}
\begin{pmatrix}
\theta_{11} & \theta_{12}\\
\theta_{21} & \theta_{22}
\end{pmatrix}.
\label{eq:A-isog-compare}
\end{align}
This equation (\ref{eq:A-isog-compare}) is read as the following 
relations in the $4\times 2$ matrix entries:
\begin{align}
& \theta_{ij}=r_{ij} \qquad \qquad (i,j = 1,2),
\label{eq:A-solution-0}\\
& r_{31}+\widetilde{\lambda}'_{31}=\widetilde{\rho}'_{31}r_{11}+\widetilde{\rho}'_{32}r_{21},
\label{eq:A-solution-1}\\
& r_{41}+\widetilde{\lambda}'_{41}=\widetilde{\rho}'_{41}r_{11}+\widetilde{\rho}'_{42}r_{21},
\label{eq:A-solution-2}\\
& r_{32}+\widetilde{\lambda}'_{32}=\widetilde{\rho}'_{31}r_{12}+\widetilde{\rho}'_{32}r_{22},
\label{eq:A-solution-3}\\
& r_{42}+\widetilde{\lambda}'_{42}=\widetilde{\rho}'_{41}r_{12}+\widetilde{\rho}'_{42}r_{22}.
\label{eq:A-solution-4}
\end{align}

The parameters $\theta_{ij}$ are fixed relatively to $r_{k\ell} \in \Q$.
The parameters $(r_{11},r_{21},r_{31},r_{41})$ are determined from the 
equations (\ref{eq:A-solution-1}) and (\ref{eq:A-solution-2}) as follows.
From the imaginary parts of (\ref{eq:A-solution-1}) 
and (\ref{eq:A-solution-2}), we find
\begin{align}
\left\{\begin{array}{l}
\widetilde{\rho}'^{(2)}_{31}r_{11}+\widetilde{\rho}'^{(2)}_{32}r_{21}=\widetilde{\lambda}'^{(2)}_{31}(=0),\\
\widetilde{\rho}'^{(2)}_{41}r_{11}+\widetilde{\rho}'^{(2)}_{42}r_{21}=\widetilde{\lambda}'^{(2)}_{41},
\end{array}\right.
\end{align}
where we introduced a notation 
$\mu=\mu^{(1)}+\sqrt{p}\mu^{(2)}$ ($\mu^{(1)},\mu^{(2)}\in\Q$) for 
$\mu\in\Q(\sqrt{p})$.
This system of equations has a unique solution $(r_{11},r_{21})$, because $\begin{pmatrix}\widetilde{\rho}'^{(2)}_{31}&\widetilde{\rho}'^{(2)}_{32}\end{pmatrix}$ and $\begin{pmatrix}\widetilde{\rho}'^{(2)}_{41}&\widetilde{\rho}'^{(2)}_{42}\end{pmatrix}$ are linearly independent; if not, it contradicts the fact that the $4\times4$ matrix $\begin{pmatrix}{\bf 1}&{\bf 1}\\\widetilde{\rho}'_{ij}&\widetilde{\rho}'^{\mathrm{c.c.}}_{ij}\end{pmatrix}$ is a change-of-basis matrix from the rational one to the complex one of $H^1(T^4_\circ;\C)$, and hence invertible.
Now that we have obtained $(r_{11},r_{21})$, we can also determine the remaining $r_{31}$ and $r_{41}$ from (\ref{eq:A-solution-1}) and (\ref{eq:A-solution-2}), respectively. The parameters $(r_{12},r_{22},r_{32},r_{42})$ are also determined 
in the same way from (\ref{eq:A-solution-3}) and (\ref{eq:A-solution-4}).

This solution $(r_{kl},\theta_{ij})$ specifies a Hodge morphism 
$\phi^\ast:H^1(T^4;\Q)\to H^1(T^4_\circ;\Q)$ satisfying 
$\phi^\ast|_{\Gamma_{b\Q}^\vee}={\rm id}_{\Gamma_{b\Q}^\vee}$.
To make sure that this $\phi^\ast$ is an isomorphism, we have to show 
that $\begin{pmatrix}\theta_{11}&\theta_{12}\\\theta_{21}&\theta_{22}\end{pmatrix}$ or equivalently $\begin{pmatrix}r_{11}&r_{12}\\r_{21}&r_{22}\end{pmatrix}$ 
for this solution is invertible.
We will do this by contradiction as follows.

Assume that the $2\times2$ $\Q$\hspace{1pt}-valued matrix $\begin{pmatrix}r_{11}&r_{12}\\r_{21}&r_{22}\end{pmatrix}$ is not invertible.
This implies that the right-hand sides of the equations (\ref{eq:A-solution-1}) and (\ref{eq:A-solution-3}) coincide up to multiplication by rational constant, and therefore we find that $\widetilde{\lambda}'_{32}$ can be written as a $\Q$-linear combination of $r_{31}$, $r_{32}$ and $\widetilde{\lambda}'_{31}$.
Since $\widetilde{\lambda}'_{31}$ is a rational number as in (\ref{eq:A-retake-rat}), both $\widetilde{\lambda}'_{31}$ and $\widetilde{\lambda}'_{32}$ are real valued, in particular.
This contradicts the fact that the $4\times4$ matrix $\begin{pmatrix}{\bf 1}&{\bf 1}\\\widetilde{\lambda}'_{ij}&\widetilde{\lambda}'^{\mathrm{c.c.}}_{ij}\end{pmatrix}$ is invertible.
This ends the proof of $\phi^\ast$ being an isomorphism.

Since the above argument holds for arbitrary choice of a geometric SYZ-mirror, this completes the proof of condition 5 (strong)---including the compatibility (\ref{eq:cond-8})---for case (A).

\section{Towards Characterization of 2$d$ Rational SCFTs}
\label{sec:towards}

As is evident from what we wrote in Introduction, 
this section is still about 2d SCFTs that are interpreted 
as non-linear sigma models with the target spaces and 
Ricci-flat K\"{a}hler metrics.

Having confirmed towards the end of the previous section that 
the conditions listed in Conjecture \ref{conj:forT4} are necessary 
and sufficient for the $N=(1,1)$ SCFT for $(M;G,B;I)$ to be rational when $M=T^4$, Thm. \ref{thm:forT4} has been proven. 
The conditions in Conj. \ref{conj:forT4} still makes sense 
for a family of Ricci-flat K\"{a}hler manifolds that is self-mirror. 
It is therefore tempting to think whether the statement in Conj. \ref{conj:forT4}
may still be right as it is; there is no obvious evidence that it is wrong 
at this moment, at least to the present authors.
(Hence we describe it as a conjecture in Introduction.) 
The material in the remainder of this section \ref{sec:towards} is a speculative 
outlook on where and how Conj. \ref{conj:forT4} should be generalized. 

Properties in the conditions in Conj. \ref{conj:forT4} 
are still stated in the language of classical geometry, such as 
$(M; G, B; I)$. This is not particularly an issue for the families of manifolds 
with $M =T^{2n}$ and $M = {\rm K3}$; for a more general 
self-mirror family, however, treatment of data and conditions on them should be 
not in terms of classical geometry, but in terms of $N=(1,1)$ SCFT. 
A set of data $(M; G, B)$ corresponds to a point (one SCFT) in the moduli 
space of $N=(1,1)$ SCFTs of the target space $M$, and the homology 
groups $H_*(M;\Z)$ to D-brane charges in the $N=(1,1)$ SCFT. 
A choice of a complex structure $I$ must be encoded as the additional 
$N=(2,2)$ superconformal structure (Def. \ref{def:geometric-SYZ-mirror});
this choice determines spectral flow operators 
in the $N=(1,1)$ SCFT, and supersymmetry charges of the effective field 
theory after the compactification on ``$(M;G,B)$.''
The Hodge structures on $H^*(M;\Z)$ appearing in the conditions of 
Conjecture \ref{conj:forT4} 
should be phrased in terms of the central charges of the effective field 
theory supersymmetry algebra. 
The pull back $\pi^*_M: H^*(B;\Q) \longrightarrow H^*(M;\Q)$ of an SYZ 
torus fibration is captured by introducing a filtration in the space of 
D-brane charges. 
That will be an outline; details and gaps between the lines should still be 
filled here and there in writing down Conjecture \ref{conj:forT4} 
in the abstract language of 2$d$ SCFT.  

\vspace{8mm}

There is no chance, however, that the properties in Conj. \ref{conj:forT4}
hold true as they are, for all the families of Ricci-flat K\"{a}hler 
manifolds $M$ that are not self-mirror. A Hodge isomorphism $\phi^*$ 
in condition 5 would imply that $b_k(M) = b_k(W)$ for $k=0,1,\cdots, 2n$, 
but examples of rational SCFTs (e.g., Gepner constructions) are known 
where the Betti numbers of $M$ and $W$ are not the same.  
Relying on a random guess, we propose to modify the conditions as follows:
\vspace{3mm}
\begin{conj} (General): 
\label{conj:gen}
{\it Let $M$ be a real $2n$-dimensional manifold which admits a Ricci-flat 
K\"{a}hler metric $G$, and $B$ a closed 2-form on $M$. The non-linear sigma 
model $N=(1,1)$ SCFT associated with the data $(M; G, B)$ is 
conjectured to be a rational SCFT if and only if the following conditions 
are satisfied: }
\begin{enumerate}
\item {\it there exists\footnote{
When we think of $M$ with $h^{2,0}(M)=0$ (for example, when $M$ is a 
Calabi--Yau $n$-fold with $n>2$), the condition 1 is automatically satisfied. 
} %
 a polarizable complex structure $I$ so that $(M;G;I)$ 
is K\"{a}hler, $B^{(2,0)}=0$, and the following conditions are satisfied; }
\item {\it there exists a geometric SYZ-mirror of the $N=(1,1)$ SCFT with 
the $N=(2,2)$ superconformal structure for ($M;G,B;I)$; there may be more than 
one geometric SYZ-mirrors (geometric interpretations); the data of such a mirror 
is denoted by $(W;G^\circ,B^\circ;I^\circ)$; }
\item (strong) {\it for any one of the SYZ mirrors, one can find a 
decomposition of the rational Hodge structure 
$V_M^{(k)} \oplus \tilde{V}_M^{(k)} \cong H^k(M;\Q)$ and also a 
decomposition $V_W^{(k)} \oplus \tilde{V}_W^{(k)} \cong H^k(W;\Q)$ for each 
$0 \leq k \leq n$ that satisfy the following conditions (a--c); 
let $\Pi_M^{(k)}$ and $\Pi_W^{(k)}$ be the projection
 from $H^k(M;\Q)$ 
to $V_M^{(k)}$ and $H^k(W;\Q)$ to $V_W^{(k)}$, respectively;  }
\begin{itemize}
\item [(a)] {\it $\Pi_M^* \circ \pi_M^*: H^k(B;\Q) \rightarrow V_M^{(k)}$ and 
$\Pi_W^{(k)} \circ \pi_W^*: H^k(B;\Q) \rightarrow V_W^{(k)}$ are  
injective, }
\item [(b)] {\it the rational Hodge structure on $V_M^{(k)}$ is of CM-type, }
\item [(c)] {\it there exists a Hodge morphism $\phi^*: V_M^{(k)} \rightarrow V_W^{(k)}$
  such that $\phi^* \circ (\Pi_M^{(k)} \circ \pi_M^*) = \Pi_W^{(k)} \circ \pi_W^*$. }
\end{itemize}
\item {\it the complexified K\"{a}hler parameter $(B+i\omega)$, where 
$\omega(-,-) = 2^{-1}G(I-,-)$ is the K\"{a}hler form, is in 
$(H^2(M;\Q) \cap H^{1,1}(M;\R)) \otimes \tau^r_{(n0)}(K^r)$. Here, 
$K^r$ is the endomorphism field of the CM-type rational 
Hodge structure $([H^n(M;\Q)]_{\ell=n},I)$ and $\tau^r_{(n0)}$ the 
embedding of $K^r$ associated with the Hodge (n,0) component.  }
\end{enumerate}
\end{conj}
\vspace{3mm}

The conditions written above are only meant to be trial versions. 
 One may think of changing condition 3 from the one for {\it arbitrary} 
 geometric SYZ mirrors as above, to the one for {\it some} geometric 
SYZ mirrors, when the condition is referred to as condition 3 (weak). 
Besides this strong vs weak variation, there is still a wide variety 
in modifying the conditions; one might demand that the entire $H^k(M;\Q)$ 
is of CM-type instead of condition 3(b) in Conj.\ \ref{conj:gen}, 
or demand a Hodge morphism $\phi^*: H^k(M;\Q) \rightarrow H^k(W;\Q)$
such that $\phi^* \circ \pi_M^* = \pi_W^*$ on $H^k(B;\Q)$ instead 
of condition 3(c) in Conj.\ \ref{conj:gen}.  At this moment, 
there is not much evidence to pin down the right characterization conditions 
from study of SCFTs. 

One may still run a few tests on formal aspects of Conjecture \ref{conj:gen} 
to access its credibility. We do so in the rest of this 
section \ref{sec:towards}. 
Condition 3 in Conj.\ \ref{conj:gen} as it stands passes the two tests 
below; reference to a substructure $V_M^{(k)}$ is motivated in that context. 

Let us take a moment before jumping into the first test, to prepare 
notations and cultivate intuitions on what the substructure $V_M^{(k)}$
should be like. Suppose that 
\begin{align}
 H^k(M;\Q) \cong \oplus_{\alpha \in {\cal A}^k(M)} (\oplus_{a \in \alpha} V_a^{(k)})
    =: \oplus_{\alpha \in {\cal A}^k(M)} V_\alpha^{(k)}
    \label{eq:simple-Hdg-str-dcmp-gen-maintxt}
 \end{align}
is a decomposition into simple rational Hodge substructures
(cf. Prop. \ref{props:rHstr-indcmp-dcmp} and condition 1);
simple rational Hodge substructures labeled by $a$ are grouped 
together by Hodge isomorphisms among them (see (\ref{eq:Fermat-5-wgt3-coh-dcmp}) 
in Discussion \ref{statmnt:subtlty-justL=n?} for an example);
Hodge-isomorphism classes are 
labeled by $\alpha \in {\cal A}^k(M)$. First, whenever $V_a^{(k)}$ of one 
$a \in \alpha \in {\cal A}^k(M)$ is of CM-type, all the other $V_{a'}^{(k)}$ with 
$a' \in \alpha$ are of CM-type. So, whether a rational Hodge substructure 
of $H^k(M;\Q)$ is of CM-type or not can be asked for individual Hodge-isomorphism 
classes in ${\cal A}^k(M)$.  Secondly, one can see that the subspace 
$V_\alpha^{(k)} \subset H^k(M;\Q)$ for $\alpha \in {\cal A}^k(M)$ does not 
depend on a choice of a simple substructure decomposition. To see this, 
suppose that there is another simple substructure decomposition, 
$H^k(M;\Q) \cong \oplus_{\beta \in {\cal A}'} \oplus_{b \in \beta} U^{(k)}_b$; 
the set of Hodge-isomorphism classes ${\cal A}'$ should be the same as 
${\cal A}^k(M)$, and the Hodge isomorphism between 
$\oplus_{\alpha} \oplus_{a \in \alpha} V_a^{(k)}$ 
and $\oplus_{\beta} \oplus_{b \in \beta} U_b^{(k)}$ should be block-diagonal 
with respect to $\alpha, \beta \in {\cal A}^k(M)$. The subspace 
$\oplus_{b \in \alpha} U_b^{(k)} \subset H^k(M;\Q)$ is therefore identical to 
$V_\alpha^{(k)}$. As a candidate of a rational Hodge substructure 
$V_M^{(k)} \subset H^k(M;\Q)$ in condition 3 of Conj.\ \ref{conj:gen}, 
it is enough 
to think of the form $\oplus_{\alpha \in {\cal A}^k(M)_{\rm sub}} V_\alpha^{(k)}$, 
with the Hodge-isomorphism classes running over some subset ${\cal A}^k(M)_{\rm sub}$
of ${\cal A}^k(M)$. The complement $\tilde{V}_M^{(k)}$ is for 
${\cal A}^k(M) \backslash {\cal A}^k(M)_{\rm sub}$. Although one may choose 
a polarization to determine a decomposition (\ref{eq:simple-Hdg-str-dcmp-gen-maintxt}), 
the decomposition $V_M^{(k)} \oplus \tilde{V}_M^{(k)} \cong H^k(M;\Q)$ and the projection 
$\Pi_M^{(k)}$ are independent of the chosen polarization now.

There is a subset ${\cal A}^k_B(M) \subset {\cal A}^k(M)$ characterized 
as the set of the Hodge-isomorphism classes where the image of 
$\Pi^{(k)}_\alpha \circ \pi_M^*: H^k(B;\Q) \rightarrow V_\alpha^{(k)}$
is non-zero;
$\Pi^{(k)}_\alpha: H^k(M;\Q) \rightarrow V_\alpha^{(k)}$ is the projection. 
The same set of notations (such as ${\cal A}^k(W)$, ${\cal A}^k(W)_{\rm sub}$, ${\cal A}_B^k(W)$) 
is introduced for the SYZ torus fibration $\pi_W: W \rightarrow B$. 
For the purpose of condition 3(a) in Conj. \ref{conj:gen}, it is enough 
to choose ${\cal A}^k(M)_{\rm sub} \subset {\cal A}^k(M)$ as large as ${\cal A}^k_B(M)$. 
Condition 3(a) does not necessarily require that ${\cal A}^k(M)_{\rm sub}$ should 
contain all of ${\cal A}^k_B(M)$, because the images of $H^k(B;\Q)$ in $V_\alpha^{(k)}$'s 
with different $\alpha$'s may be correlated in general. 

Here, we begin with the first test. 
Conditions 3(b) and 3(c) do not treat $M$ and the mirror $W$ democratically, 
at least at first sight. If the Conj. \ref{conj:gen} is to provide 
a necessary and 
sufficient condition for an $N=(1,1)$ SCFT to be rational, then 
all of its mirror SCFTs must be rational. So, we have to make sure 
that condition 3 implies condition 3[$M\leftrightarrow W$]. 

Condition 3(b)[$M \leftrightarrow W$] follows from condition 3(a,b,c) 
in fact. 
To see this, think of the subset ${\cal A}^k(W)_{\rm sub~sub} \subset {\cal A}^k(W)_{\rm sub}$
where the image of $\phi^*$ in condition 3(c) is non-zero. Then we may replace 
$V_W^{(k)} = \oplus_{\alpha \in {\cal A}^k(W)_{\rm sub}} V_\alpha^{(k)}$ by 
$V_{W{\rm new}}^{(k)} = \oplus_{\alpha \in {\cal A}^k(W)_{\rm sub~sub}} V_\alpha^{(k)}$. 
This rational Hodge substructure of $H^k(W;\Q)$ still satisfies condition 3(a). 
Because they are in the non-zero image from CM-type simple rational Hodge substructures
(condition 3(b)), $V_{W{\rm new}}^{(k)}$ is also of CM-type. 

Condition 3(c)[$M\leftrightarrow W$] also follows from  
conditions 3(a,b,c). Think of the subset ${\cal A}^k(M)_{\rm sub~sub} \subset 
{\cal A}^k(M)_{\rm sub}$ where $\phi^*$ in condition 3(c) is non-zero. 
Then there is one-to-one correspondence between ${\cal A}^k(M)_{{\rm sub~sub}}$
and ${\cal A}^k(W)_{{\rm sub~sub}}$. We may replace $V_M^{(k)} = 
\oplus_{\alpha \in {\cal A}^k(M)_{\rm sub}} V_\alpha^{(k)}$ by 
$V_{M{\rm new}}^{(k)} = \oplus_{\alpha \in {\cal A}^k(M)_{\rm sub~sub}} V_\alpha^{(k)}$, 
and yet condition 3(a) is satisfied. Now, we construct a Hodge morphism 
$\psi^*: V_{W{\rm new}}^{(k)} \rightarrow V_{M{\rm new}}^{(k)}$ as follows, 
by focusing on each one-to-one correspondence pair 
$\alpha \in {\cal A}^k(M)_{{\rm sub~sub}}$
and $\beta \in {\cal A}^k(W)_{\rm sub~sub}$. 
The vector space $V^{(k)}_\beta \subset H^k(W;\Q)$ over $\Q$ can also be seen as 
a vector space over the CM field $K$ of the simple rational Hodge 
structures in $\alpha$ and $\beta$; choose any decomposition 
$V^{(k)}_\beta = {\rm Im}(\phi^*) \oplus [{\rm Im}(\phi^*)]^c$ 
as a vector space over $K$, and fix it. 
The vector space $V^{(k)}_\alpha$ also has a decomposition 
${\rm Ker}(\phi^*) \oplus [{\rm Ker}(\phi^*)]^c$ as a vector 
space over $K$; choose this decomposition in a way 
$[{\rm Ker}(\phi^*)]^c$ contains the image 
$\Pi^{(k)}_\alpha \circ \pi_M^* (H^k(B;\Q))$.
Then the Hodge morphism $\phi^*: [{\rm Ker}(\phi^*)]^c \longrightarrow {\rm Im}(\phi^*)$
is invertible; the inverse $\psi^*$ may be extended to 
$[{\rm Im}(\phi^*)]^c \subset V^{(k)}_\beta$ by zero, so we have 
a Hodge morphism $\psi^*: V^{(k)}_\beta \rightarrow V^{(k)}_\alpha$. 
This completes the construction of a Hodge morphism $\psi^*: V_{W{\rm new}}^{(k)} \rightarrow
V_{M{\rm new}}^{(k)}$. 
By construction, $\phi^*$ is an isomorphism between the injective images of 
$H^k(B;\Q)$ within $V_{M{\rm new}}^{(k)}$ and $V_{W{\rm new}}^{(k)}$, 
and $\psi^*$ gives the inverse of $\phi^*$ on those injective images. 
So, condition 3(c)[$M\leftrightarrow W$] follows indeed. 

The other test is to see if Conjecture \ref{conj:gen} (general) is consistent with 
Conjecture \ref{conj:forT4} (self-mirror); the latter is more reliable because it has been 
tested by the case $M=T^4$. When we read condition 3  of 
Conjecture \ref{conj:gen} for a self-mirror manifold $M$, they appear 
to be weaker than conditions 2 and 5 of Conj.\ \ref{conj:forT4}.  
This raises a concern that the conditions in Conj.\ \ref{conj:gen} might not be 
strong enough to be a sufficient condition for an SCFT to be rational. 

It is beyond the scope of this article to run this test on all 
the self-mirror families, but we can do it on the family $M=T^4$ here. 
To see that conditions 2 and 5 of Conj.\ \ref{conj:forT4} 
follow from condition 3 of Conj.\ \ref{conj:gen} for $M=T^4$, 
note first that the rational Hodge structure on 
$H^1(M;\Q)$ (not necessarily of CM-type) either is simple (I), or  
splits into two simple Hodge substructures (II). In the case (II), 
each of the two simple components of $H^1(M;\Q)$ contains 1-dimensional 
subspace of $\pi_M^*( H^1(B;\Q) )$; in any SYZ mirror, each complex 
coordinate contains both one SYZ fiber direction and one base direction.
This means that $V_M^{(k=1)}$ in condition 3 of Conj.\ \ref{conj:gen} 
should always be 
the entire $H^1(M;\Q)$, because of condition 3(a) in Conj.\ \ref{conj:gen}. 
Repeating the same argument, one finds that 
$V_W^{(k=1)} = H^1(W;\Q)$ ---(*). 
So, condition 3(b) of Conj.\ \ref{conj:gen} implies that the rational Hodge structure 
on the entire $H^1(M;\Q)$ is of CM-type, regardless of the case (I) or (II). 
The Hodge structures on $H^k(T^4;\Q)$ of all other $k$'s are also of CM-type then
(condition 2 of Conj. \ref{conj:forT4}). 
Secondly, the Hodge morphism $\phi^*: H^1(M;\Q) \rightarrow H^1(W;\Q)$
in condition 3(c) of Conj.\ \ref{conj:gen} cannot have a kernel when $M=T^4$; 
even in the case (II), 
the Hodge morphism $\phi^*$ should map the two simple substructures of $H^1(M;\Q)$
to the two simple substructures of $H^1(W;\Q)$ without a kernel, without a cokernel, 
to meet the requirement on $\phi^*$ in condition 3(c) (remember (*)). 
So, the Hodge morphism $\phi^*$ must be a Hodge isomorphism, as in the 
condition 5 of Conj.\ \ref{conj:forT4}. 

\vspace{5mm}

Here are three remarks. First, note that the first test motivated an idea of 
restricting the range (i.e., $V_M^{(k)} \subset H^k(M;\Q)$) 
in which the Hodge morphism $\phi^*$ is defined. 
One may wonder what happens if we set $V_M^{(k)}=H^k(M;\Q)$ and 
$V_W^{(k)}=H^k(W;\Q)$ in condition 3 of Conj. \ref{conj:gen}. We have not found 
how to prove existence of $\psi^*: H^k(W;\Q) \rightarrow H^k(M;\Q)$ 
for 3[$M \leftrightarrow W$] by using conditions 3(a,b,c), then; 
so we would not have an obvious $[M \leftrightarrow W]$ democracy 
if condition 3 does not allow freedom to choose $V_M^{(k)}$ in $H^k(M;\Q)$. 

Second, when $M$ is a Calabi--Yau $n$-fold and the base manifold $B$
of the SYZ torus fibration is $S^n$, condition 3(a) of
Conj. \ref{conj:gen} is non-trivial only in the middle dimensional
cohomology, $k=n$. Moreover, the generator of $H^k(B;\Q)|_{k=n}$---the
Poincare dual of the point class in $H_0(B;\Q)$---is pulled back
by $\pi^*_M$ and $\pi^*_W$ to the Poincare dual of the $T^n$ fiber
class $PD[{\rm fibr}]$ of $H^n(M;\Q)$ and $H^n(W;\Q)$, respectively.
Because the periods of the holomorphic ($n$,0)-form over the $T^n$ fiber
class is non-zero in an SYZ fibration, the projection of $PD[{\rm fibr}]$
to the level-$n$ components $[H^n(M;\Q)]_{\ell=n}$ and $[H^n(W;\Q)]_{\ell=n}$
are both non-zero. So, it is enough to choose $V_M^{(n)}=[H^n(M;\Q)]_{\ell=n}$
and $V_W^{(n)}=[H^n(W;\Q)]_{\ell=n}$, respectively, and condition 3(a)
is automatically satisfied. Condition 3(b) is equivalent to the
weak CM condition on $M$. 

Technically the conditions 3(a--c) as a whole are equivalent to
existence of the Hodge isomorphisms $\phi^*$ of the level-$n$ components
on both sides. Although we should impose an extra condition that 
$\phi^*: \Pi_M^{(n)} (PD[{\rm fibr}_M]) \mapsto \Pi_W^{(n)}(PD[{\rm fibr}_W])$,
it is always possible to adjust $\phi^*$ so that this is the case,\footnote{
Although SYZ torus fibration therefore almost disappears from the surface
when Conj. \ref{conj:gen} is applied exclusively to the Calabi--Yau cases
with $B=S^n$, the study in this article reveals that SYZ torus fibration
is one of key concepts in finding characterization on SCFTs that are rational
that cover general Ricci-flat K\"{a}hler manifolds.  
} %
because the level-$n$ Hodge substructure is always simple and of CM-type. 
It should be reminded, however, that existence of a Hodge isomorphism
is equivalent to the isomorphism of the CM fields 
(as in condition (ii) of Conj. \ref{conj:GV-original}) {\it and}
agreement of the decomposition (\ref{eq:dcmp-field-embdngs-4-Hdg-cmp})
on both sides; isomoprhism of the CM fields is not enough.

Finally, there is an alternative to Conjecture \ref{conj:gen}, 
which is to replace condition 3(b) by 
\begin{itemize}
\item [3(b')] {\it the rational Hodge structures on the entire 
$H^k(M;\Q)$ and $H^k(W;\Q)$ are of CM-type.  }
\end{itemize}
The alternative version, Conjecture \ref{conj:gen}' also passes the first test 
(as well as the second one). 
Conditions 3(b) and 3(b') are different, if the answer to 
Question \ref{qstn:weak-vs-srong} "no." 
The Borcea--Voisin orbifold Calabi--Yau threefolds will be good 
test cases both for the math question (Question \ref{qstn:weak-vs-srong}) and 
for finding out which version of the conjecture is right then 
(cf.\ footnote \ref{fn:BV}).

\section{Implications and Discussions}
\label{ssec:distrib}

Whether Conjectures  \ref{conj:forT4} and \ref{conj:gen} are true is 
obviously one of research problems in the future. 
Before closing this article, let us also make a little effort 
to extract some immediate implications when those Conjectures are right. 
Further progress beyond them should still be left for future publications. 

{\bf The Case of a Complex Torus:}
First, consider the family with $M=T^m$. We do have a complete characterization 
\cite{Wendland:2000ye} of the sets of geometry data $(M;G,B)$ where the 
corresponding (S)CFTs are rational (as quoted 
in Prop. \ref{props:Wendland-451}). One may list up 
$(G, B)$ on $M=T^m$ by following the condition (\ref{eq:cond-GnB-rational}).
Alternatively, one may list up a pair $(\Gamma_L, \Gamma_R)$ of even and 
positive definite rank-$m$ lattices where both $\Gamma_L$ and $\Gamma_R[-1]$ 
are primitive sublattices of 
${\rm II}_{m,m} \cong H_1(T^m;\Z) \oplus H^1(T^m;\Z)$ and are orthogonal 
to each other, as in \cite{Hosono:2002yb}.  
Either way, we do not need a Gukov--Vafa like characterization 
(Conj. \ref{conj:GV-original}) on $(G,B)$'s where the (S)CFT is rational.
We used this known characterization---(\ref{eq:cond-GnB-rational}) in 
Prop. \ref{props:Wendland-451} for the case $M=T^4$---in this article
to refine Conj. \ref{conj:GV-original} into Conj. \ref{conj:forT4} 
and test the latter (Thm. \ref{thm:forT4}).  

Conjecture \ref{conj:forT4} along the line of the idea of Gukov and Vafa 
is still useful when we ask, for a given complex manifold $(T^{2n};I)$, which 
choice of a complexified K\"{a}hler form $(B+i\omega)$ yields 
a CFT that is rational. In string theory, we do not have to phrase a question 
in this way, but still this is also a natural way to formulate a question 
when one's favorite subject includes K\"{a}hler geometry, algebraic geometry, 
or arithmetic geometry. Conjecture \ref{conj:forT4} indicates, if it is true, that 
no rational CFT is overlooked by paying attention only to $(T^{2n};I)$ 
that are abelian varieties (i.e., polarizable $I$'s); given the fact 
that all the CM-type abelian varieties have algebraic-geometric 
implementations whose defining equations involve only algebraic numbers, 
we see that $(T^{2n};I)$-target rational CFTs may also have relevance 
to arithmetic geometry (cf the second to last paragraph in 
section \ref{sssec:Intro-background}). 

If Conj. \ref{conj:forT4} is true, as we continue from the previous paragraph, 
then $T^{2n}$-target rational CFTs have the following (mutually 
{\it non}-exclusive) hierarchical classifications. At the crudest level, 
polarizable CM-type rational Hodge structures $(H^1(T^{2n};\Q),I)$ are 
classified by the Galois theoretical properties\footnote{
When $n=2$, the crudest classification is the distinction between the 
cases (A, A', B, C) in this article. There is just one case when $n=1$, 
where the endomorphism field is an imaginary quadratic field.  
} %
of the endomorphism algebra 
${\rm End}(H^1(T^{2n};\Q),I)$, as reviewed in \cite[\S4.1]{Okada:2023udq}. 
There are still infinitely many different choices of the endomorphism 
algebras ${\rm End}(H^1(T^{2n};\Q),I)$ in each case. 
The ring of complex multiplications on $(H^1(T^{2n};\Z), I)$---the 
transpose of ${\rm End}(T^{2n},I)$ 
(cf Notation \ref{notn:cpx-torus-EndRingAlg} and 
Thm. \ref{thm:smCM4CT-via-EndAlgCoh})---is a subring of 
the semi-simple algebra ${\rm End}(H^1(M;\Q), I)$. At the end of 
this three layers of classification,\footnote{
See \cite{Moore:1998pn, Moore:1998zu} or Rmk. \ref{rmk:hierarchical-class-CM-EC} 
for what the latter two layers of classification are in the case $n=1$.
} %
 there are a finite number of 
isomorphism classes of CM-type abelian varieties that share 
the same ring ${\rm End}(T^{2n};I)$. For more review on the distributions 
of CM-type abelian varieties in the moduli space of complex tori, 
readers are referred to the math literatures cited in \cite[app.B]{Kanno:2017nub}. 
This paragraph just provides to string theorists with hints on where to look 
at for more resources; it does not add anything new. 

We are then left with the task of identifying the list of 
the complexified K\"{a}hler forms $(B+i\omega)$ 
on a given CM-type abelian variety $(T^{2n};I)$ so that the (S)CFTs for 
$(T^{2n};I;B+i\omega)$ are rational. Neither the 
condition (\ref{eq:cond-GnB-rational}) in Prop. \ref{props:Wendland-451} 
nor the conditions in Conj. \ref{conj:forT4} provides an immediate 
answer to this question.\footnote{
For complex tori, it will be more involved to exploit the conditions 
in Conj. \ref{conj:forT4} instead of the condition 
(\ref{eq:cond-GnB-rational}) in Prop. \ref{props:Wendland-451}. 
It was not an elementary process for the case $n=2$ already  to 
exploit the compatibility with the torus fibration, as we saw in 
section \ref{ssec:one-more}. There are some works to do. 
} %
The study in the case of $n=2$ in section \ref{sec:metric} of this article, however, 
suggests the following as a natural guess. Suppose that 
\begin{align}
  (T^{2n};I) \sim \prod_{\alpha \in {\cal A}} (A_\alpha )^{n_\alpha}
\end{align}
is the simple component decomposition modulo isogeny (Rmk. \ref{rmk:cmplt-reducibl-abel-var});
a pair $A_\alpha$, $A_\beta$ with distinct $\alpha, \beta \in {\cal A}$ are not mutually isogenous.  
Let $K_\alpha$ be the endomorphism field of the simple CM-type abelian variety $A_\alpha$. 
Then the vector space 
\begin{align}
  {\rm Span}_\Q \left\{ {\cal Q}^{(\alpha,\xi^\alpha)} \; | \; 
    \alpha \in {\cal A}, \; \;
    \xi^\alpha = (\xi^\alpha_{ij}) \in M_{n_\alpha}(K_\alpha), \; \;  
     \overline{\xi^\alpha_{ij}} = -\xi^\alpha_{ji}  \right\}
   \subset H^2(T^{2n};\Q)
\end{align}
of dimension $\sum_{\alpha \in {\cal A}} n_\alpha^2 [K_\alpha:\Q]/2$ over $\Q$, 
where 
\begin{align}
  {\cal Q}^{(\alpha,(\xi^\alpha_{ij}))} & \; := \sum_{i,j=1}^{n_\alpha}
    e^\alpha_{I,i}\wedge e^\alpha_{J,j} 
   {\rm Tr}_{K_\alpha/\Q}\left[ \xi^\alpha_{ij}\eta^\alpha_I \bar{\eta}^\alpha_J
       \right]
  = \sum_{i,j=1}^{n_\alpha} \sum_{a=1}^{[K_\alpha:\Q]/2} 2\tau_a(\xi^\alpha_{ij})
     dz^{\alpha,a}_i \wedge d\bar{z}^{\alpha,\bar{a}}_j, 
\end{align}
is likely to be equal to ${\cal H}^2(T^{2n}_I)$ (we have not proved this for $n>2$ yet); 
no rational CFT is overlooked 
when we scan $B$ only within ${\cal H}^2(T^{2n}_I)$, if Conj. \ref{conj:forT4}
is true.\footnote{
Remember that the proof of Prop. \ref{props:I-polNalgB} is valid only 
for the case $n=2$. 
} %
 Rational metrics can be constructed for a choice of 
\[
 \beta = (\beta^\alpha)_{\alpha \in {\cal A}}, \qquad 
 \beta^\alpha \in M_{n_\alpha}(K_\alpha) \; \; {\rm s.t.} \; \; \overline{\beta^\alpha_{ij}}
 = -\beta^\alpha_{ji}
\]
 (as in \cite[Thm. 2.5]{Chen:2005gm} for the cases $n_\alpha =1$) through 
\begin{align}
  \omega^{(\beta)} & \; :=  \frac{i}{2}
    \sum_{\alpha \in {\cal A}} \sum_{i,j=1}^{n_\alpha} \sum_{a=1}^{[K_\alpha:\Q]/2} 
    \tau_a(\xi_*^\alpha \beta^\alpha_{ij}) 
   ( dz^a_i \otimes d\bar{z}^{\bar{a}}_j - d\bar{z}^{\bar{a}}_j \otimes dz^a_i), 
      \label{eq:basis-KahlerF-forRatG}  \\
 G^{(\beta)} & \; = \sum_{\alpha \in {\cal A}} \sum_{i,j=1}^{n_\alpha} 
   {\rm Tr}_{K_\alpha/\Q}\left[ \xi_*^\alpha \beta^\alpha_{ij}
         \eta^\alpha_I \bar{\eta}^\alpha_J  \right]
     e^\alpha_{I,i}\otimes e^\alpha_{J,j}, 
\end{align}
where $\xi^\alpha_*$ is an element in $K_\alpha$ such that 
$\overline{\xi^\alpha_*}=-\xi^\alpha_*$. The complexified K\"{a}hler forms  
$(B+i\omega^{(\beta)})$ generated in this way satisfy the condition 
(\ref{eq:cond-GnB-rational}) by construction. Our analysis in section 
\ref{sec:metric} proved that all of the K\"{a}hler forms on $(T^{2n};I)$ 
compatible with the condition (\ref{eq:cond-GnB-rational}) are 
generated by $\omega^{(\beta)}$, but the proof is only for $n=2$. 
So, the set of $(B+i\omega)$ on a CM-type $(T^{2n};I)$ described above 
will be a good guess for all that satisfy the 
condition (\ref{eq:cond-GnB-rational}) and $B^{(2,0)}=0$ on a given 
$(T^{2n};I)$, but we do not have a proof yet for $n>2$. 

It is beyond the scope of these notes to explore a possible connection 
between the complex multiplication (such as the ring ${\rm End}(M,I)$) 
and various data of those torus target rational (S)CFTs (such as the coefficients 
of the braiding and modular transformations 
(cf \cite{DeBoer:1990em}, \cite{Coste:1993af}, \cite{Fuchs:1994tc}). 
It has been observed in a very special class of rational SCFTs\footnote{
Ref. \cite{Kondo:2019jpi} only dealt with $T^2$-target 2d rational 
$N=(1,1)$ SCFTs that are diagonal.   
} %
that the multiplicative group of an appropriate quotient of the 
ring of the complex multiplications is regarded as a part of the automorphism of 
the fusion algebra of the diagonal rational SCFTs. It is not known yet 
whether such a statement holds true in a broader class of geometries/SCFTs, 
or what the precise statement is. 

{\bf The case $M=$ K3:} 
Condition 1 is already a non-trivial characterization 
of K3-target SCFTs that are rational, if Conj. \ref{conj:forT4} is true; 
In a general K3-target $N=(1,1)$ SCFT, 
one may choose a complex structure $I$ (an additional $N=(2,2)$ superconformal 
structure) so that either it is polarizable, or 
$B^{(2,0)}=0$, but condition 1 says that a data $(M;G,B)|_{M=K3}$ of a rational SCFT  
will allow $I$ that has both of the properties. 
Condition 3 of Conj. \ref{conj:forT4} implies that the SCFT is rational 
only when $(B+i\omega)$ is not just in $H^{1,1}(M_I;\C)$, but is within 
${\cal H}^2(M_I)\otimes \C$. 
Moreover, the Picard number $\rho$ has to be no less than 10, and 
the K\"ahler form must be within the image of $T_M^v\otimes \R \subset 
A(M_I)\otimes \R$ projected on to the algebraic part  
${\cal H}^2(M_I)\otimes \R$ for any K3 target rational SCFT.
This also means that the CM field $K^r$ on the transcendental lattice 
$T_M$ of $(M;I)$ and its mirror $(W;I^\circ)$ has degree $2n:= [K^r:\Q] \leq 12$, 
if Conj. \ref{conj:forT4} is true.  

For any CM-field $K^r$ with $[K^r:\Q] \leq 20$  (it is enough 
to discuss only those with $[K^r:\Q] \leq 12$ because of the reasoning above, though), 
there are points in the complex structure moduli space of K3 surface $(M;I)=M_I$ where 
$K^r$ is the endomorphism field of the level-2 component $[H^2(M_I;\Q)]_{\ell=2}$ \cite[\S3]{MR3531364}.
We can read out more from \cite[\S3]{MR3531364} in fact; one can construct a rational Hodge 
structure on the $2n$-dimensional vector space $K^r/\Q$, by 
choosing an element $\lambda \in (K^r_0)^\times$ that is 
mapped to $\R_{>0}$ by one embedding $\tau^0_0$ of $K^r_0$ and to $\R_{<0}$ by all the other $(n -1)$ 
embeddings of $K^r_0$; here, $K^r_0$ is the totally real subfield of $K^r$
of degree $n$; a bilinear form $b_\lambda : K^r \times K^r \rightarrow \Q$ 
is determined by using $b(x,y) = {\rm Tr}_{K^r/\Q}[\lambda x\bar{y}]$ for 
$(x,y) \in K^r\times K^r$; the pair of embeddings of $K^r$ whose restriction to $K^r_0$ is $\tau^0_0$ 
are associated with the Hodge (2,0) and (0,2) components on $K\otimes \C$ \cite{piatetski1973arithmetic};
as explained in \cite[\S3]{MR3531364}, there exists 
a quadratic space---a vector space and a bilinear form---over $\Q$ of $22-2n$ dimensions, 
$(V,\varphi)$ so that 
\begin{align}
  (K^r, b_\lambda) \oplus (V, \varphi) \cong H^2(K3;\Q)
\end{align}
as a quadratic space; furthermore, the data of $(K^r, b_\lambda)$ determines $(V,\varphi)$
uniquely (modulo vector-space isomorphisms). 
The Torelli theorem of K3 surface indicates that a projective K3 surface of CM-type is 
specified when one specifies where in the vector space $H^2(K3;\Q)$ the integral part $H^2(K3:\Z)$
lies.\footnote{
It will be more natural for experts of complex K3 surfaces to 
list up transcendental lattices $T_M$ with rank $\leq 12$ first, and 
then to try to list up candidates of CM fields that can be 
endomorphism fields of CM-points in the period domain of $T_M$. 
All that is known to the authors, however, is that there is 
a constraint 
${\rm discr}(T_M) \in (-1)^{[K^r:\Q]/2} D_{K^r/\Q} \times (\Q^\times)^2$ ---(**) 
on such a CM field $K^r$ \cite[Lemma 1.3.2]{MR3330542}. 
The authors have not tried to think whether this constraint 
is enough to guarantee that a CM point with CM by $K^r$ exists 
in the period domain of $T_M$. 
We do not know for a given pair of $K^r$ and $T_M$ satisfying (**) 
whether a basis $\{ \eta_I \}$ of $K^r/\Q$ always exists 
for a basis $\{ e_I \}$ of $T_M$ so that $\Omega := e_I \tau^r_{(20)}(\eta_I)$
satisfies $(\Omega, \Omega^\sigma)_{T_M}=0$ for all $\sigma \in {\rm Gal}((K^r)^{\rm nc}/\Q)$
except for $\sigma = {\rm cc}$. 
} %
The transcendental lattice $T_M$ is $K \cap H^2(K3;\Z)$, the algebraic part ${\cal H}^2(M_I)$ 
is $V$, and $K^r = {\rm End}(T_M\otimes \Q, I)$, by construction.  
All the CM-type projective K3 surfaces are obtained in this way 
(cf \cite[Prop. 1.3.10]{mengchenphd}, \cite[footnote 70]{Kanno:2020kxr}).

For a transcendental lattice $T_M \subset H^2(K3;\Z)$ of a CM-type K3 surface $M$, 
there are infinitely many other CM points in the period domain $D(T_M)$; just like 
in the case of elliptic curves, the group ${\rm Isom}(T_M \otimes \Q)$ acts on $D(T_M)$, 
and the images of a CM point under this group are also CM points. Those CM points 
populate densely within $D(T_M)$. Moreover, different choices of $K^r$ and $\lambda$ 
result in different transcendental lattices $T_M$, in general. 
The two paragraphs so far are just a review. 

We are left with the task of identifying the list\footnote{
Reference \cite{mengchenphd} discusses how to find an appropriate $(B+i\omega)$
for a complex CM-type K3 surface $X$ 
with $\rho(X) \geq 10$ such that $X$ has a mirror that is also of CM-type, 
motivated by Conj. \ref{conj:GV-original}. 
} %
of $(B+i\omega)$ for a given 
K3 surface $(M;I)$ of CM-type, so that the SCFT for $(M;I;B+i\omega)$ 
is rational. One can make a partial progress by allowing ourselves to assume 
that Conj. \ref{conj:forT4} is true; generate $(B+i\omega)$ in ${\cal H}^2(M_I)=(V,\varphi)$ 
with coefficients in the endomorphism field $\tau_{(20)}(K^r)$ 
(condition 3 of Conj. \ref{conj:forT4}), 
generate a subspace $T_M^v \otimes \Q \subset A(M_I)\otimes \Q$ by the 
Galois group action on the coefficients as we have done in 
section \ref{sssec:vert-coh-mapped2-alg-subsp}, 
and impose the conditions that $\{ (e^{(B+i\omega)/2})^\sigma \; | \; \sigma \in 
{\rm Gal}(K^{\rm nc}/\Q)\}$ yields a rational Hodge structure 
under the pairing (\ref{eq:pairing-Mukai}) as we have done in 
Lemmas \ref{lemma:GV-2-RCFT-caseBC-even} and \ref{lemma:GV-2-RCFT-caseAprm-even}. 
Once this analysis is done, then the polarizable rational 
Hodge structure on the level-2 component of the mirror 
is isomorphic to the one on $T_M^v\otimes \Q$ 
(Lemma \ref{lemma:mirrorhHdgH2L2=vHdgOnAlg} works also when $M=K3$);  
it is further Hodge isomorphic to the one on $T_M\otimes \Q$ by construction, 
so a Hodge isomorphism $\phi^*$ in condition 5 of Conj. \ref{conj:forT4}
then exists. 
It will be a subject of a separate research paper to elaborate more on 
these processes and to identify the freedom left in $(B+i\omega)$. 
On top of this analysis, one has to impose the compatibility (\ref{eq:cond-8}) 
of the Hodge isomorphism $\phi^*$ with the SYZ torus fibration. 

Research on K3-target SCFTs is being carried out by some 
of the present authors. One is to use known K3-target 
SCFTs that are rational and check whether the conditions in 
Conj. \ref{conj:forT4} are satisfied.  The other is to develop techniques 
so that we can handle both topology and Hodge structure in an SYZ torus 
fibration $\pi_M: M \rightarrow B$. 

{\bf The case $M$ is a Calabi--Yau $n$-fold with $n>2$:}
here, we think of those with $h^{0,q}(M)=0$ for $q=1,2,\cdots, n-1$. 
This case is completely different in nature from the cases where $M$ is 
either a complex torus or a K3 surface. Here are two major differences.
One is in the fact that the moduli space of complex structure of $M$ 
is much smaller than the space of integral Hodge structures on $H^n(M;\Z)$, 
as has been explained in the literatures. The other is that the strong CM
condition on $M$ may not be the same as the weak CM condition
(Question \ref{qstn:weak-vs-srong}); importance of this Question is the primary 
lesson in this article for applications to the cases of Calabi--Yau $n$-folds 
($n>2$), for now, because that is the difference between conditions 3(b) and 3(b') in Conj. \ref{conj:gen}.
%
%
Practical progress in the study of Calabi--Yau target 
rational SCFTs will be made possible after such progress in math, and
also in practical progress in the case $M$ is K3. 

%

There has been a question of how densely rational SCFTs 
populate the moduli space of Ricci-flat K\"{a}hler 
target SCFTs. 
Reference \cite{Gukov:2002nw} conjectured that 
rational SCFTs might have something to do with CM-type 
rational Hodge structures (Conj. \ref{conj:GV-original}), 
and further combined the observation with Andr\'e--Oort 
conjecture in math \cite{ andre1989g, oort1997canonical, 
andre1997distribution},\footnote{
cf also \cite{tsimerman2015proof} and \cite{moonen2011torelli}. 
} %
 which says that CM points are not dense in the moduli space of such manifolds in general 
(except for the moduli space of abelian varieties and K3 surfaces). 
So, it has been hinted that rational SCFTs do not populate 
densely within the whole moduli space of $N=(1,1)$ SCFTs with 
a Calabi--Yau threefold target space. What is implicit in this 
line of argument is to set $V_M^{(k)} = H^k(M;\Q)$ (in the language 
of Conj. \ref{conj:gen}) and demand that the rational Hodge structure 
is of CM-type on the entire $H^k(M;\Q)$, not just on the level-$n$ 
component $[H^n(M;\Q)]_{\ell=n}$. So, although the inference on scarcity 
of rational SCFTs {\it from} the scarcity of CM-type Calabi--Yau 
manifolds does not have to be questioned at this moment,  
finer understanding on the choice of statements in 
Conj. \ref{conj:gen}, as we discussed in section \ref{sec:towards}, 
and on Question \ref{qstn:weak-vs-srong} / footnote \ref{fn:BV}, 
however, might change this perspective in the future.

The question above may have a consequence beyond mathematical 
physics. Suppose one day that mankind discovers that 
Type IIB flux compactification is theoretically consistent 
only when the SCFT is rational; it is not bad to enjoy 
such a speculation sometimes \cite{Gukov:2002nw}. That may further 
indicate that the vacuum complex structure of the internal Calabi--Yau 
threefold is something captured by a special subvariety 
of Calabi--Yau moduli space interpreted as a Shimura variety, 
if we speculate along the lines of Gukov--Vafa and Andr\`e--Oort.  
When the moduli space has a group action, discrete and/or continuous, 
its isotropy subgroup at the vacuum point may remain in the 
low-energy effective field theory of the moduli fields as 
gauged and/or accidental symmetry. 

\subsection*{Acknowledgments}   

We thank M. Asakura, M. Ashwinkumar Y. Goto, H. Lange, M . Yamazaki and N. Yui for valuable correspondence.  
This work was supported in part by WPI Initiative (AK, MO, TW), 
the Schroedinger Fellowship (AK), ICTP Math section visiting fellowship (AK), FoPM, WINGS Program of the University of Tokyo, JSPS Research Fellowship for Young Scientists, JSPS KAKENHI Grant Number JP23KJ0650 (MO), and 
a Grant-in-Aid for Scientific Research on 
Innovative Areas 6003 (MO and TW), MEXT, Japan. 

AK and TW thank the Abdus Salam ICTP for hospitality during the school and workshop on Number Theory and Physics and the Conference on Arithmetic Geometry (June-July 2024). AK would like to thank the Isaac Newton Institute for Mathematical Sciences, Cambridge, for support and hospitality during the programme Black holes: bridges between number theory and holographic quantum information, where work on this paper was undertaken. This work was supported by EPSRC grant EP/R014604/1.
\appendix

\section{A Guide to the Theory of Complex Multiplication}
\label{sec:guide}

This is a guide to the theory of complex multiplication, 
written for string theorists. It is not a lecture note or textbook. 
It is meant to provide concise presentations of key concepts and objects, 
known facts, and which implies what, beyond what is covered in \cite{Moore:1998pn}, \cite{Moore:1998zu}; 
while proofs are often omitted, large fraction of math that we need in this article are 
collected here for the benefit of readers. If a reader wants to seek for further references 
written by mathematicians, we 
recommend \cite{milnecm}, \cite[\S3]{MR3586372},  \cite{shimura2016abelian} and \cite{birkenhakeabelian}. 

As a research article in string theory, however, it was not an option 
to restrict our attention only to complex projective varieties without a reason 
and ignore general complex analytic manifolds that do not have an embedding into a projective space. 
Most of literature on the theory 
of complex multiplication deals only with complex projective varieties. So, the present 
authors made an effort to extend the widely accepted theory of complex multiplications 
to be able to cover complex analytic manifolds. Reference \cite{birkenhaketori} was useful 
for this purpose. 
Statements in 
Prop. \ref{props:rHstr-ssmpl-cpltRdcbl}, 
Prop. \ref{props:CM-alg-always-ssmpl} and 
Prop. \ref{props:CM-Intuitv-summary4CT}
are not easily found elsewhere for this reason. 

\subsection{Complex Multiplication of Complex Tori: Part I}
\label{ssec:Guide-CM4CT-PartI}

%
\begin{notn}
$M_n(A)$ for an algebra $A$ is the algebra of $A$-valued $n\times n$ matrices.
\end{notn}

\begin{notn}
\label{notn:cpx-torus-EndRingAlg}
Let $X$ be a complex torus of complex $n$-dimensions. It has a 
complex analytic presentation $X \cong \C^n/\Lambda$; here, 
$\C^n$ may be regarded as a group with respect to the sum of 
$n$-component vectors, and $\Lambda$ is a subgroup of $\C^n$ isomorphic 
to $\Z^{2n}$ such that $\Lambda \otimes_\Z \R$ covers the entire 
$\C^n \cong \R^{2n}$. The complex analytic manifold $X$ is also 
regarded as an abelian group under the sum of vectors. 
The set of all the holomorphic maps from $X$ to itself that 
is also a group homomorphism, 
\begin{align}
 {\rm End}(X) \cong \left\{ \varphi \in M_{n}(\C): 
  \C^n \longrightarrow \C^n \; | \; \varphi (\Lambda) \subset \Lambda \right\}, 
  \label{eq:end-ring-torusGeom}
\end{align}
forms a ring. 
The ring ${\rm End}(X)$ may change as complex structure of $X$ changes 
(see Thm. \ref{thm:ellC-criteria-CMorNot} and Ex. \ref{ex:CM-EllC}).

One may also assign another algebraic object 
\begin{align}
 {\rm End}_\Q(X) \cong \left\{ \varphi \in M_{n}(\C):
   \C^n \longrightarrow \C^n \; | \;
      \varphi(\Lambda \otimes \Q) \subset \Lambda \otimes \Q \right\} 
  \label{eq:def-alg-isogn-cpx-tri}
\end{align}
to a complex torus $X$. This ${\rm End}_\Q(X)$ is a finite-dimensional 
algebra over $\Q$.  

In the case of 1-dimensional complex tori ($n=1$), i.e., the case of 
elliptic cures, the maps $\varphi$ in ${\rm End}(X)$ or ${\rm End}_\Q(X)$ 
are to multiply complex numbers $\varphi \in M_1(\C) = \C$ on the 
complex analytic coordinates of $\C^n = \C$. 
\end{notn}
\begin{defn}
\label{def:CM-geom-4CT}
A complex $n$-dimensional torus $X$ ($\dim_\R X=2n$) is said to have 
{\it sufficiently many complex multiplications (CM)}, when the algebra 
${\rm End}_\Q(X)$ contains a subalgebra ${\cal K}$ with 
$\dim_\Q {\cal K} = 2n$ that is isomorphic to the direct sum 
of number fields $\oplus_i F_i$. 
\end{defn}
\begin{thm}
\label{thm:ellC-criteria-CMorNot}
In the case of elliptic curves, i.e., $n=1$, the algebra ${\rm End}_\Q(X)$ 
is known to be either isomorphic to $\Q$, or to a quadratic extension 
field $K$ over $\Q$ that cannot be embedded to $\R$. An elliptic curve 
has sufficiently many complex multiplications if and only if 
${\rm End}_\Q(X)$ is such a quadratic extension field $K$, which 
contains $\varphi \in \C$ that cannot be found within $\R \subset \C$. 
\qed
\end{thm}
\begin{exmpl}
\label{ex:CM-EllC}
Let $X \cong \C/\Lambda$ be an elliptic curve with sufficiently many complex 
multiplication, and let $\Lambda = \Z \oplus \tau \Z$. Then there must be 
a set of mutually prime integers $a, b, c$ such that 
$a\tau^2+b\tau+c=0$ ---(**). A ring 
\begin{align}
  \Z \oplus \Z (a\tau) \cong \left\{ \varphi \in M_{1}(\C), \;
   \varphi : \C \longrightarrow \C   \; | \; \varphi = m + n a\tau, \quad 
    m,n \in \Z \right\}
\end{align}
is in the ring ${\rm End}(X)$, and $K = \Q(\tau)$ is the algebra 
${\rm End}_\Q(X)$; the algebra ${\rm End}_\Q(X)$ is always a field 
in the case $X$ is an elliptic curve. The converse is also true:
where there exists $(a,b,c)$ satisfying (**), then the elliptic curve 
$\C/(\Z \oplus \tau \Z)$ has sufficiently many complex multiplications. 
See \cite{Moore:1998pn}\cite{Moore:1998zu} for more explanation on this. 
\end{exmpl}
\begin{rmk}
\label{rmk:hierarchical-class-CM-EC}
Elliptic curves with sufficiently many complex multiplications 
are therefore classified by their fields $K$. A quadratic extension
field of $\Q$ that cannot be embedded into $\R \subset \C$ is called 
an {\it imaginary quadratic field}. An imaginary quadratic field $K$ 
is always of the form $\Q(\sqrt{p})$ for some negative integer $p$ 
that is free of the square of an integer. So, such elliptic curves are 
classified by the imaginary quadratic fields ${\rm End}_\Q(X)$
(i.e., by the negative and square-free integers $p$). The subring 
${\rm End}(X)$ within the field $\Q(\sqrt{p})$ can be used 
to provide a finer classification; the ring (modulo isomorphism) 
is known to be determined by the value $b^2-4ac$ of $\tau$. 
There can still be a finite number of mutually non-isomorphic elliptic 
curves sharing the same value of $(b^2-4ac)$
({\it discriminant}, {\it conductor}/{\it order} and {\it class number} 
are the relevant jargons; more information is found in \cite{Moore:1998pn}, \cite{Moore:1998zu}, \cite[\S7, \S10.C]{MR4502401}, and various lecture notes available online). We call this {\it hierarchical 
classification} of elliptic curves with complex multiplications. 

The complex structure parameters $\tau$ of elliptic curves with 
sufficiently many complex multiplications populate the 
complex upper half plane ${\cal H}$ densely (or one fundamental 
region of ${\cal H}$ under the action of ${\rm SL}_2\Z$). 
Those $\tau$'s sharing the same ${\rm End}_\Q(X) \cong \Q(\sqrt{p})$
forms an orbit of ${\rm GL}_2\Q$. 
\end{rmk}
\begin{defn}
\label{def:isogeny}
Let $f: X \longrightarrow Y$ be a holomorphic map from one complex torus $X$ to another 
$Y$ that is a homomorphism and is both surjective and of finite kernel. Such a map is called 
an {\it isogeny}.  

For an isogeny $f$, it is known that there always exists an isogeny 
$g: Y \longrightarrow X$ such that $f\circ g = m \; {\rm id}_{Y}$ and 
$g\circ f = m \; {\rm id}_X$ for some positive integer $m$. Such an 
isogeny $g$ is called a {\it dual isogeny of} $f$. 
\end{defn}
\begin{anythng}
A dual isogeny $g$ above is not an inverse of the map $f$, while 
one may think of $m^{-1} g$ as something close to an inverse of $f$. 
The object $m^{-1}g$ is not necessarily a geometric map $Y \rightarrow X$, 
but still it is in ${\rm Hom}_\Q(Y,X) := 
\left\{ M_n(\C) \ni \varphi: \C^n \longrightarrow \C^n \; | \;
   \varphi(\Lambda_Y \otimes \Q) \subset \Lambda_X \otimes \Q \right\}$. 
One may also call an element of ${\rm Hom}_\Q(Y,X)$ as an {\it isogeny}. 
Existence of an isogeny between a pair of complex tori can be used 
to introduce an equivalence relation among complex tori. A pair of 
complex tori for which an isogeny exists are said to be {\it isogenous}. 
\end{anythng}
\begin{rmk}
\label{rmk:CM-vs-isogeny}
When a pair of complex tori $X$ and $X'$ are isogenous, 
$X'$ has sufficiently many complex multiplications if and only if 
$X$ does (because Def. \ref{def:CM-geom-4CT} depends only on 
${\rm End}_\Q(X)$). 
The classification of complex tori with sufficiently many complex multiplications 
by the algebra ${\rm End}_\Q(X)$ agrees with the classification of them 
by isogenies. 
\end{rmk}
\begin{exmpl}
\label{ex:ExE-CM}
Think of $X = E \times E$ where $E = \C/(\Z \oplus \tau \Z)$ is an elliptic 
curve with sufficiently many complex multiplications. Now, 
${\rm End}_\Q(E) \cong \Q(\sqrt{p})$ for some square-free negative integer $p$. 
It is known that 
\begin{align}
 {\rm End}_\Q(X) \cong M_{2}(\Q(\sqrt{p})). 
\end{align}

This algebra ${\rm End}_\Q(X)$ is not commutative. But it contains 
such 4-dimensional subalgebras over $\Q$ as 
\begin{align}
  {\cal K}_1 := \diag \left( \Q(\sqrt{p}), \; \Q(\sqrt{p}) \right),
\end{align}
and 
\begin{align}
  {\cal K}_2 := {\rm Span}_\Q \left\{ {\bf 1}_{2\times 2}, \;
    \sqrt{p}{\bf 1}_{2\times 2}, \; \left( \begin{array}{cc} & d \\ 1 & \end{array} \right), \; \sqrt{p} \left( \begin{array}{cc} & d \\ 1 & \end{array} \right)
   \right\}
\end{align}
for any non-zero integer $d$. For this reason, the complex 2-dimensional 
torus $X = E \times E$ is with sufficiently many complex multiplications
(Def. \ref{def:CM-geom-4CT}).  
\end{exmpl}
%

\subsection{Rational Hodge Structure, Polarization and Algebra}
\label{ssec:rHstr-Pol-alg}

Although the ring ${\rm End}(X)$ of a complex torus $X$ is introduced 
in (\ref{eq:end-ring-torusGeom}) as a set of holomorphic maps of the geometry $X$, 
we may think of it also as a set of linear maps on the homology group $H_1(X;\Z)$. 
The condition $\varphi(\Lambda) \subset \Lambda$ [resp. 
$\varphi(\Lambda \otimes \Q) \subset \Lambda \otimes \Q$] is read 
as $\varphi ( H_1(X;\Z)) \subset H_1(X;\Z)$ [resp. 
$\varphi ( H_1(X;\Q) ) \subset H_1(X;\Q)$]. The condition that 
a map $\varphi: X \longrightarrow X$ is holomorphic means that 
holomorphic tangent vectors of $X$ are mapped by $\varphi$ to 
holomorphic tangent vectors. The pull-back action of the geometric 
maps $\varphi$ map $H^1(X;\Z)$ to $H^1(X;\Z)$ [resp. 
$H^1(X;\Q)$ to $H^1(X;\Q)$], and map (1,0) forms on $X$ to 
(1,0) forms. So, here is a 
\begin{thm}
\label{thm:smCM4CT-via-EndAlgCoh}
A complex torus $X$ of complex $n$-dimensions has sufficiently many complex 
multiplications if and only if the algebra 
\begin{align}
  {\rm End}(H^1(X;\Q))^{\rm Hdg} := \left\{ \varphi^* \in {\rm End}(H^1(X;\Q))
    \; | \; \varphi^*(H^{1,0}(X;\C)) \subset H^{1,0}(X;\C)  \right\}  
 \label{eq:def-End-Hdg-H1-cpx-tri}
\end{align}
contains a subalgebra ${\cal K}^*$ over $\Q$ of the form $\oplus_i F_i$, 
with $\dim_\Q{\cal K}^* = 2n$. \qed
\end{thm}
\begin{rmk}
\label{rmk:announce-CMdef-use-EndAlg-onCoh}
One can use either one of the algebras, 
${\rm End}(H^1(X;\Q))^{\rm Hdg}$ in (\ref{eq:def-End-Hdg-H1-cpx-tri}) 
acting on the cohomology group and 
${\rm End}_\Q(X)$ in (\ref{eq:def-alg-isogn-cpx-tri}) on the geometry, 
to characterize complex tori $X$ 
with sufficiently many complex multiplications. For other 
compact K\"{a}hler manifolds $X$, however, the algebra 
${\rm End}(H^k(X;\Q))^{\rm Hdg}$ makes sense, while ${\rm End}_\Q(X)$ does not. 
One may therefore think of promoting existence of a large enough commutative subalgebra in the 
Hodge-decomposition preserving algebra on the cohomology groups (like 
${\rm End}(H^1(X;\Q))^{\rm Hdg}$) as a definition of complex multiplication. 
We will do so in Def. \ref{def:CM-rat-Hdg-str-byEnd} and 
Rmk. \ref{rmk:geomCM=strCM-CTAV}, after doing necessary preparations. 
\end{rmk}

The crucial ingredients in defining the algebra ${\rm End}(H^1(X;\Q))^{\rm Hdg}$, 
where $X$ is a complex torus, is to keep track of the vector subspace 
of $H^1(X;\C)$ corresponding to the (1,0) forms \underline{relatively} to 
the rational (or integral) subspace $H^1(X;\Q)$ (or $H^1(X;\Z)$) within 
$H^1(X;\C)$; without using $H^1(X;\Q)$ as a reference we would have had just the information 
on the dimension of the subspace, $h^{1,0}=n$. So, the following notion is introduced as a 
generalization.

\vspace{3mm}

\noindent
{\bf \large Rational Hodge Structure:}  
\begin{defn}
\label{def:rHdgStr}
A pair $(V_\Q, \phi)$ is called a {\it rational Hodge structure} of weight-$m$
($m$ is an integer), when $V_\Q$ is a finite dimensional vector space over 
$\Q$, $\phi$ is an isomorphism between vector spaces over $\C$, 
\begin{align}
  \phi: V_\Q \otimes \C \longrightarrow \oplus_{p,q}^{(p+q=m)} V^{p,q}, 
\end{align}
and the complex conjugation operation on the $\C$ tensor factor 
of $V_\Q \otimes \C$ maps $V^{p,q}$ to $V^{q,p}$. The integer 
\[
 {\rm max}(|p-q| \; | \; V^{p,q} \neq \emptyset)
\]
is called the {\it level} of $(V_\Q, \phi)$. 

A {\it Hodge morphism} $f: (V_\Q, \phi_V) \longrightarrow (W_\Q, \phi_W)$
is a homomorphism $f: V_\Q \rightarrow W_\Q$ of vector spaces over $\Q$ 
whose linear extension by $\otimes \C$ maps $V^{p,q}$ to $W^{p,q}$. 
When a pair $f: (V_\Q,\phi_V) \rightarrow (W_\Q, \phi_W)$ and 
$g: (W_\Q, \phi_W) \rightarrow (V_\Q, \phi_V)$ of Hodge morphisms 
satisfy $g \circ f = {\rm id}_{(V_\Q,\phi_V)}$ and 
$f \circ g = {\rm id}_{(W_\Q,\phi_W)}$, then they are called 
{\it Hodge isomorphism}s; we say that the rational Hodge 
structures $(V_\Q, \phi_V)$ and $(W_\Q, \phi_W)$ are {\it (Hodge-)isomorphic}. 
The pullback of an isogeny between a pair of complex tori is 
a Hodge isomorphism of their weight-1 rational Hodge structures. 

The algebra of Hodge morphisms from a rational Hodge structure $(V_\Q, \phi)$
to itself is denoted by ${\rm End}(V_\Q, \phi)$, 
or ${\rm End}(V_\Q)^{\rm Hdg}$ when the choice of $\phi$ is obvious from 
a context, and is called the {\it endomorphism algebra of} $(V_\Q, \phi)$. 
\end{defn}
\begin{rmk}
\label{rmk:rHdgStr-4-cpxKahlerMfd}
Let $(M, J, \omega)$ be a complex K\"{a}hler manifold of 
$\dim_\C M = n$, where $(M, J)$ is a compact complex manifold (with $J$ the 
almost complex structure) and $\omega$ a closed symplectic form. 
Or alternatively, let $M$ be a complex 
projective non-singular variety of $\dim_\C M=n$. Then all the vector spaces 
$H^k(M;\Q)$ with $k=0,1,\cdots, 2n$ are given the unique Hodge decomposition
$\phi_k$, so we have rational Hodge structures $(H^k(M;\Q),\phi_k)$ for 
all $k=0,1,\cdots, 2n$. For a pair of mutually isomorphic complex K\"{a}hler 
manifolds, or of mutually isomorphic complex projective non-singular varieties, 
there exists isomorphisms between their rational Hodge structures. So, 
rational Hodge structures on the cohomology groups are machinery to capture 
some information on the complex structure of those manifolds/varieties
where linear algebra can be applied, even when $X$ is not a complex torus.   
We also use the notation $(H^k(M;\Q), J)$ as well as $(H^k(M;\Q),\phi_k)$ 
for the rational Hodge structure. 
\end{rmk}

\vspace{5mm}

Additional information on a rational Hodge structure called 
{\it polarization} (defined below) is not a natural object in 
the eyes of most of string theorists. It plays an important role 
in algebraic geometry, in the theory of complex multiplication, 
and also in this study. %
\begin{defn}
\label{defn:polrz-HS}
Let $(V_\Q, \phi)$ be a rational Hodge structure of weight-$m$. 
A symmetric [resp. anti-symmetric] bilinear form 
${\cal Q}: V_\Q \times V_\Q \rightarrow \Q$ when $m$ is even [resp. odd] 
is said to be a {\it polarization} of $(V_\Q, \phi)$ when it satisfies 
the following two conditions:
\begin{align}
 \chi \in V^{p,q}, \; \psi \in V^{p',q'} & \quad {\rm then} \; 
     {\cal Q}(\chi, \psi) \neq 0 \quad {\rm only~when~}
     p+p'=q+q'=m, \label{eq:cond-polrztn-1} \\
 \chi \in V^{p,q} \backslash \{ 0\} & \quad {\rm then} \; 
 (-1)^{\frac{m(m+1)}{2}}{\cal Q}(\chi, J^*\bar{\chi}) > 0. 
    \label{eq:cond-polrztn-2}
\end{align}
Here, $J^*$ on $V_\Q \otimes \C$ is the operator multiplying 
$i^{p-q}$ on the $V^{p,q}$ component. The last condition is equivalent 
to the positive definiteness of the symmetric bilinear form 
$H_{\cal Q}(-,-) := (-1)^{\frac{m(m+1)}{2}}{\cal Q}(-,J^*-)$ on $V_\Q \otimes \R$. 

When there exists a polarization ${\cal Q}$ for a rational Hodge structure 
$(V_\Q, \phi)$, we say that $(V_\Q,\phi)$ is {\it polarizable}. A triple 
$(V_\Q, \phi, {\cal Q})$ is called a {\it polarized rational Hodge structure}. 

A symmetric [resp. anti-symmetric] bilinear form ${\cal Q}$ satisfying 
the condition (\ref{eq:cond-polrztn-1}) is called a {\it polarization
of index $k$} of $(V_\Q, \phi)$ when the symmetric bilinear form 
$H_{\cal Q}$ is non-degenerate, but has $k$ negative eigenvalues 
(a polarization is a polarization of index 0). 
\end{defn}
What is relevant in this article is whether a rational Hodge structure 
is polarizable or not, rather than specific choices of a polarization.
Independently of our context, the polarizability is a relevant question 
in algebraic geometry for the following reason. 
\begin{rmk}
\label{rmk:pol-exists-in-algV}
Let $X$ be a complex projective non-singular variety, with an embedding 
to a projective space specified. Then there always exists a unique 
element $D_{P} \in H^2(X;\Z) \cap H^{1,1}(X;\R)$ such that its expression 
as a differential form $D_{P} = i H_{a\bar{b}}(z,\bar{z}) dz^a 
\wedge d\bar{z}^{\bar{b}}$ has a positive definite Hermitian matrix $H_{a\bar{b}}$
everywhere on $X$. Conversely, when such $D_P$ exists for a compact 
complex K\"{a}hler manifold $X$, then there exists an embedding of $X$ into 
a projective space so that we can see $X$ as an algebraic variety. 

For such an $X$, with $D_P$, the rational Hodge structure 
$(H^k(X;\Q), \phi_k)$ (cf Rmk. \ref{rmk:rHdgStr-4-cpxKahlerMfd}) 
is polarizable\footnote{
\label{fn:prim-dcmp}
The vector space $H^k(X;\Q)$ is decomposed into $\oplus_{s \in \Z_{\geq 0}}
[H^k(X;\Q)]^{(s)}$ with ${\rm max}(0,k-n) \leq s \leq k/2$, where the 
$s$-th component is in the image of $D_P^s \wedge: H^{k-2s}(X;\Q) \rightarrow 
H^k(X;\Q)$ and is annihilated by 
$D_P^{n-k+s+1} \wedge: H^k(X;\Q) \rightarrow H^{2n+2s-k+2}(X;\Q)$. A bilinear 
form ${\cal Q}$ on the $s$-th component is given by 
${\cal Q}(\xi, \eta) := (-1)^s \int_X D_P^{n-k} \wedge \xi \wedge \eta$.
} %
 for all of $k=0,1,\cdots, 2\dim_\C X$.
\end{rmk}

\vspace{5mm}

Having introduced the definition of rational Hodge structure and 
polarization, let us know introduce a few more concepts that we need
in order to analyze substructures and possible decomposition of a
rational Hodge structure. 
\begin{defn}
For a rational Hodge structure $(V_\Q, \phi)$, a pair $(W_\Q, \phi|_W)$
of a vector subspace $W_\Q$ of $V_\Q$ and the restriction of $\phi$, 
$\phi|_W: W_\Q \otimes \C \rightarrow \oplus_{p,q} V^{p,q}$
is said to be a {\it rational Hodge substructure} of $(V_\Q, \phi)$ when 
the non-trivial second equality in 
\begin{align}
 \phi(W_\Q \otimes \C) = \phi (W_\Q \otimes \C) \cap (\oplus_{p,q} V^{p,q}) =
    \oplus_{p,q} \left( \phi(W_\Q \otimes \C) \cap V^{p,q} \right) 
\end{align}
holds true. 
\end{defn}
\begin{rmk}
When a rational Hodge structure $(V_\Q, \phi)$ has a rational Hodge 
substructure $(W_\Q, \phi|_W)$, it is always possible to define the 
quotient, $(V_\Q/W_\Q, \phi)$, on the quotient vector space $V_\Q/W_\Q$. 
Although it is always possible to find a vector subspace $W'_\Q$ of 
$V_\Q$ so that $V_\Q \cong W_\Q \oplus W'_\Q$, this does not guarantee 
that there exists a Hodge decomposition $\phi'$ of $W'_\Q$ so that 
$(W'_\Q, \phi')$ is a rational Hodge substructure of $(V_\Q, \phi)$. 
cf Ex. \ref{ex:Shafarevic}.
\end{rmk}
\begin{rmk}
\label{rmk:PrHdgStr-orth-dcmp}
Let $(W_\Q, \phi|_W)$ be a rational Hodge substructure of a rational 
Hodge structure $(V_\Q, \phi)$. Suppose further that $(V_\Q, \phi)$ has 
a polarization ${\cal Q}$ of some index, and its restriction on $W_\Q$ 
is still non-degenerate. Then the pair $(W^\perp_\Q, \phi|_{W^\perp_\Q})$ is 
also a rational Hodge substructure of $(V_\Q, \phi)$ then, where 
$W^\perp_\Q$ is the orthogonal complement of $W_\Q$ in $V_\Q$ with 
respect to the non-degenerate bilinear form ${\cal Q}$. In this situation, 
\begin{align}
 (V_\Q, \phi) \cong (W_\Q, \phi|_W) \oplus (W^\perp_\Q, \phi|_{W^\perp_\Q})
  \label{eq:rHstr-dcmp-to2}
\end{align}
is a Hodge isomorphism. 
\end{rmk}
Due to the subtlety discussed in the last two remarks, it makes sense 
to distinguish the two concepts below. 
\begin{defn}
A rational Hodge structure $(V_\Q, \phi)$ is said to be {\it simple}
when it does not have a rational Hodge substructure $(W_\Q, \phi|_W)$ 
where $W_\Q \neq V_\Q$ and $W_\Q \neq \{0\}$. 

A rational Hodge structure $(V_\Q, \phi)$ is said to be {\it indecomposable}
when there is no pair of rational Hodge substructures $(W', \phi')$ 
and $(W'',\phi'')$ such that $(V_\Q, \phi) \cong (W', \phi')\oplus 
(W'',\phi'')$. See Ex. \ref{ex:Shafarevic} for an indecomposable rational 
Hodge structure that is not simple. 
\end{defn}
\begin{exmpl}
\label{ex:rHdgSubStr}
The weight-1 rational Hodge structure on $H^1(E_1 \times E_2;\Q)$, where 
$E_1$ and $E_2$ are elliptic curves, has non-trivial rational Hodge 
substructures. The vector subspaces $H^1(E_1;\Q)\otimes H^0(E_2;\Q)$ 
and $H^0(E_1;\Q)\otimes H^1(E_2;\Q)$ of $H^1(E_1\times E_2;\Q)$ support 
rational Hodge substructures. 

When $X$ is a complex torus of 2-dimensions that has an isogeny 
$f: X \longrightarrow E_1\times E_2$, then the weight-1 rational 
Hodge structure $H^1(X;\Q)$ also has non-trivial rational Hodge
structures; the pullback $f^*: H^1(E_1\times E_2;\Q) \longrightarrow 
H^1(X;\Q)$ is an isomorphism of rational Hodge structures. 

When $X$ is a complex projective non-singular K3 surface, the algebraic 
part is defined by 
${\cal H}^2(X) := \left\{ x \in H^2(X;\Q) \; | \; (\Omega_X^{2,0},x)=0 \right\}$,
where $\Omega^{(2,0)}_X$ is a nowhere vanishing holomorphic (2,0) form on $X$. 
The transcendental part $T_X \otimes \Q$ is the orthogonal complement 
$T_X \otimes \Q := \left\{ x \in H^2(X;\Q) \; | \; (x, y)= 0 \; {\rm for~} {}^\forall y \in {\cal H}^2(X) \right\}$. The vector subspace $T_X \otimes \Q$ supports 
a level-2 rational Hodge substructure of $(H^2(X;\Q), \phi_2)$. 
The substructure on $T_X\otimes \Q$ is simple. 
$T_X \otimes \Q \cong [H^2(X;\Q)]_{\ell =2}$. 
\end{exmpl}
\begin{props}
\label{props:rHstr-indcmp-dcmp}
\cite[Thm. 7.5]{birkenhaketori}
When a rational Hodge structure $(V_\Q, \phi)$ is given, it is always possible 
to express 
\begin{align}
  (V_\Q, \phi) \cong \oplus_{a \in A} (W_\Q^a, \phi^a)
   \label{eq:rHstr-indcmp-dcmp}
\end{align}
with rational Hodge substructures $(W_\Q^a, \phi^a)$ (labeled by $a \in A$)
that are all indecomposable, by repeating the process of finding 
a substructure and testing whether a complement as 
in (\ref{eq:rHstr-dcmp-to2}) exists
(Rmk. \ref{rmk:PrHdgStr-orth-dcmp} is an example of such process). 
 By grouping the indecomposable substructures 
that are mutually Hodge-isomorphic together, the decomposition can be organized 
into the form 
\begin{align}
 (V_\Q, \phi) \cong \oplus_{\alpha \in {\cal A}} \left(
     \oplus_{\lambda_\alpha=1}^{n_\alpha} (W_\Q^{\alpha, \lambda_\alpha},\phi^\alpha) \right)
  =: \oplus_{\alpha \in {\cal A}} (V_\Q,\phi)^{\alpha},
  \label{eq:rHstr-indcmp-dcmp-by-grpHiso}
\end{align}
where the set ${\cal A}$ labels distinct Hodge-isomorphism classes of indecomposable components. 
Such an expression as (\ref{eq:rHstr-indcmp-dcmp}) [resp. as (\ref{eq:rHstr-indcmp-dcmp-by-grpHiso})] 
is called an {\it indecomposable decomposition} [resp. the {\it isotypic} decomposition]. 
When $(V_\Q, \phi)$ is polarizable, 
an indecomposable decomposition is a simple-component decomposition, 
because we can apply Rmk. \ref{rmk:PrHdgStr-orth-dcmp} for any 
simple rational Hodge substructure. 

There may be multiple indecomposable decompositions for one 
$(V_\Q, \phi)$, but those decompositions are unique modulo Hodge isomorphisms.
\end{props}
\begin{defn}
When an indecomposable decomposition of a rational Hodge structure
$(V_\Q, \phi)$ is also a simple component decomposition, we say 
that $(V_\Q, \phi)$ is {\it completely reducible}. 
\end{defn}

\vspace{3mm}

\noindent
{\bf \large Structure of the Endomorphism Algebra and its Representations:}  
The endomorphism algebra ${\rm End}(V_\Q, \phi)$ of a rational Hodge structure 
$(V_\Q, \Q)$ (a reminder: $\dim_\Q V_\Q  < \infty$) is an algebra over $\Q$ with 
$\dim_\Q {\rm End}(V_\Q, \phi) < \infty$. It is an example 
of Artinian algebras. The vector space $V_\Q$ is also regarded as a 
representation space of the algebra ${\rm End}(V_\Q, \phi)$, which is 
an example of an Artinian module of the algebra ${\rm End}(V_\Q, \phi)$.
The theory of structure of Artinian algebras and their representations is 
therefore very useful in the present context. We do not write down 
the definition of Artinian algebras and modules here; we only need to 
use some results quoted below that are known to be reliable for those objects.  
\begin{defn}
\label{defn:divis-alg}
An algebra $D$ over a field $F$ is a {\it division algebra} 
if any non-zero element $x \in D$ has an element $y \in D$ 
such that $x y = yx = 1$ with respect to the multiplication law of 
the algebra $D$. A division algebra $D$ is regarded as a field if the 
multiplication law of the algebra $D$ is commutative (abelian).
\end{defn}
\begin{lemma}
\label{lemma:div-alg-dim}
(e.g., \cite[Thm. 4.11]{MR998775})
Let $D$ be a division algebra over a field $F$ with $[D:K] < \infty$, where 
$K$ is the center $Z(D)$; $K$ is an extension of $F$. It is then known\footnote{
Here is an example: the quarternions $D= {\rm Span}_\R\{ 1, i, j, k \}$
forms a division algebra, and its center is $K=\R$. 
We know that $D\otimes_\R \C \cong M_{2}(\C)$, and that 
$\dim_\R D = 4=2^2$.  
} %
 that there exists an integer $g$ such that $D\otimes_K \overline{K} \cong 
M_{g}(\overline{K})$. In particular, $\dim_K D = g^2$. 
\end{lemma}
\begin{lemma}
\label{lemma:simpl-Hstr-alg-divs}
When a rational Hodge structure $(V_\Q, \phi)$ is simple,  
the algebra ${\rm End}(V_\Q,\phi)$ is a division algebra 
(see Def. \ref{defn:divis-alg}). 
Furthermore, if there is a Hodge $(p,q)$ component with $h^{p,q}=1$, then 
the division algebra is a field
\cite[Lemma 3.3 + Cor. 3.6]{MR3586372}.

Let $X$ be a complex projective non-singular variety with $\dim_\C X=n$ that 
has trivial canonical bundle (e.g., an abelian variety, a complex algebraic K3 surface and 
a Calabi--Yau threefold). Then $h^{n.0}(X)=1$. So, when we choose 
$(V_\Q,\phi)$ to be the 
smallest rational Hodge substructure of $H^n(X;\Q)$ that contains 
$H^{n,0}(X;\C) \subset H^n(X;\C)$, the algebra ${\rm End}(V_\Q,\phi)$ 
is a field \cite[Cor. 3.6]{MR3586372}. \qed 
\end{lemma}
This Lemma says that the algebra ${\rm End}(V_\Q, \phi)$ has a specific 
property (being a division algebra) when the rational Hodge structure 
$(V_\Q, \phi)$ is simple; to write down corresponding statement 
for the case $(V_\Q, \phi)$ is indecomposable, we need a 
\begin{defn}
A ring $R$ is called a {\it local ring} when the set of all the 
non-invertible elements\footnote{
An element $x \in R$ is said to be {\it invertible} when there exists $y \in R$
such that $xy =1$ and $yx=1$. {\it Non-invertible} elements are those that 
are not invertible. For example, for any nilpotent element $a \in R$, 
$(1-a) \in R$ is always invertible. 
} %
 of $R$ forms an ideal of $R$.
When a local ring $R$ is an algebra over some ring or field, then 
it is called a {\it local algebra}. 
\end{defn}
\begin{rmk}
It is known that the ideal of all the nilpotent elements of a finite 
dimensional local algebra is also the unique maximal left-ideal, and 
also the unique maximal right-ideal. The quotient by this ideal is known 
to be a division algebra. See \cite[Prop. 15.15]{MR1245487} 
and \cite[Thm. 5.7]{MR998775} for more information. 
\end{rmk}
\begin{rmk}
\label{rmk:simple-HgdStr-divs-Alg}
(\cite[Thm. 5.10]{MR998775}, \cite[Prop. 7.3]{birkenhaketori})
A rational Hodge structure $(V_\Q, \phi)$ is indecomposable 
if and only if ${\rm End}(V_\Q, \phi)$ is a local algebra. 
\end{rmk}

We have seen properties of the algebras ${\rm End}(V_\Q, \phi)$
that correspond to whether rational Hodge structures $(V_\Q, \phi)$ 
have substructures. Let us next review the properties of 
${\rm End}(V_\Q, \phi)$ that are related to the complete reducibility of 
$(V_\Q, \phi)$. 
\begin{defn}
Let $\mathfrak{R}$ be an algebra over a field $F$ with 
$\dim_F \mathfrak{R} < \infty$. For such an algebra $\mathfrak{R}$, 
the Jacobson radical $J(\mathfrak{R})$---whose definition we do not write 
down here (if interested, see \cite[Thm. 15.3]{MR1245487}, 
\cite[Thm. 3.3]{MR998775}; this article is readable without knowing 
its definition)---agrees with the union of all the nilpotent 
ideals of $\mathfrak{R}$, and $J(\mathfrak{R})$ itself is also a nilpotent 
ideal (e.g., \cite[Thm. 3.5]{MR998775}). 

An algebra $\mathfrak{R}$ over a field $F$ with $\dim_F \mathfrak{R} < \infty$
is said to be {\it semi-simple} when $J(\mathfrak{R})=0$.
Therefore, $\mathfrak{R}$ being semi-simple is equivalent to 
the absence of non-trivial nilpotent ideals (e.g., \cite[I.\S1]{milnecm}). 
\end{defn}
\begin{exmpl}
The algebra $\mathfrak{R} = \{ a = (a_{ij} ) \in M_n(F) \; | \; 
a_{ij}=0 \; {\rm for} \; i>j \}$ of $F$-valued upper triangular matrices
(where $F$ is a field) is an example of an algebra that is not semi-simple. 
The ideal $I$ of strictly upper triangular 
matrices $I = \{ a=(a_{ij}) \; | \; a_{ij}=0 \; {\rm for} \; i\geq j \}$
is nilpotent. On the other hand, the algebra $\mathfrak{R} = M_n(F)$ 
of $F$-valued matrices is an example of a semi-simple algebra. 
\end{exmpl}
To see that the algebra ${\rm End}(V_\Q, \phi)$ of a polarizable rational Hodge structure 
is semi-simple, the notion Rosati involution is introduced
(cf \cite[\S3]{shimura2016abelian}, \cite[I. \S2]{milnecm},
 \cite[\S2.3]{birkenhaketori}, \cite[eq. (3.3)]{MR3586372}):
\begin{defn}
Let $(V_\Q, \phi)$ be a rational Hodge structure, and ${\cal Q}$ its 
polarization of some index $k$. The {\it Rosati involution} 
$': {\rm End}(V_\Q, \phi) \ni f \longmapsto f' \in {\rm End}(V_\Q, \phi)$ 
is given by 
\begin{align}
  {\cal Q}(f \cdot v, u) = {\cal Q}(v, f' \cdot u), \qquad \qquad 
  {}^\forall v, u \in V_\Q. 
\end{align}
For $f, g \in {\rm End}(V_\Q, \phi)$, $(f g)' = g' f'$. 
When we choose a basis of $V_\Q$ and express $f$ and ${\cal Q}$ 
as matrices, then the matrix for $f'$ is given by 
$f' = {\cal Q}^{-1} \cdot f^T \cdot {\cal Q}$. Note that 
${\cal Q}$ has to be non-degenerate for this definition to be possible. 
\end{defn}
\begin{lemma}
\label{lemma:End-alg-is-CM-wPol-ala-Milne}
Let $(V_\Q, \phi)$ be a rational Hodge structure that is polarizable
(index 0). 
Then the algebra $\mathfrak{R} = {\rm End}(V_\Q, \phi)$ over the field $\Q$ is 
semi-simple. 
\end{lemma}
A sketch of {\it Proof}: 
choose a polarization ${\cal Q}$ for $(V_\Q, \phi)$;
the Rosati involution with respect to ${\cal Q}$ on the algebra 
$\mathfrak{R}$ has the following property: for any non-zero $f \in \mathfrak{R}$
and $e \in V_\Q \otimes \R$, $H_{\cal Q}(e, f'f e) > 0$ unless $f e = 0 \in V\otimes \R$; 
this is because 
$H_{\cal Q}(f e, fe) 
%
%
= H_{\cal Q}(e, f'f e)$ must be positive. 
This then means that 
\begin{align}
  \tr {}_{V_\Q}[f'f] > 0 \qquad \qquad {}^\forall f_{\neq 0} 
   \in \mathfrak{R}; 
\end{align}
to see this, let $\{ e_I \}$ be an orthonormal basis on $V_\Q \otimes \R$ with respect 
to the positive definite symmetric form 
$H_{\cal Q}: V_\Q \otimes \R \times V_\Q \otimes \R \rightarrow \R$.  Then 
${\rm tr}_{V_\Q}[f'f]$ is equal to $\sum_I H_{\cal Q}(f e_I, f e_I)$.

If $J(\mathfrak{R})$ were non-zero (cf, \cite[Prop. 1.36]{milnecm}), there must be 
a non-zero element $a \neq 0$ in $J(\mathfrak{R})$. Then $b:= a'a$ is also in the ideal 
$J(\mathfrak{R})$. All the elements in $J(\mathfrak{R})$, which is 
a nilpotent ideal, are supposed to be nilpotent, on one hand.
On the other, $b \neq 0$ because $\tr_{V_\Q} [b] > 0$; 
$b^2 = b'b \neq 0$ because $\tr_{V_\Q}[b'b]>0$; 
$b^4 = (b^2)'b^2 \neq 0$ because $\tr_{V_\Q}[(b^2)'b^2] > 0$.
In particular, $b^{2^n} \neq 0$ for any $n \in \N$. So, $b\in J(\mathfrak{R})$
is not nilpotent. That is a contradiction. An assumption that 
$J(\mathfrak{R}) \neq \{0\}$ must be wrong. 

Although the Rosati involution can be defined on the algebra 
$\mathfrak{R} = {\rm End}(V_\Q,\phi)$ whenever there exists a polarization 
of index $k$, the proof above works only for a polarization of index $k=0$, 
because the positive-definiteness is essential. 
 \qed 

The following fact---Wedderburn's theorem---is very important and useful, 
because we can use it 
in the case of $\mathfrak{R}={\rm End}(V_\Q,\phi)$ of a polarizable 
rational Hodge structure $(V_\Q, \phi)$.
\begin{lemma}
\label{lemma:Wedderburn}
A semi-simple algebra $\mathfrak{R}$ over $\Q$ with $\dim_\Q \mathfrak{R} 
< \infty$ has a structure 
\begin{align}
 \mathfrak{R} \cong \oplus_{\alpha \in {\cal A}} M_{n_\alpha}(D_\alpha)
   \label{eq:WeddB-alg-str}
\end{align}
for some finite set ${\cal A}$, $n_\alpha \in \N$, and 
a division algebra $D_\alpha$.
\end{lemma}
\begin{cor}
\label{cor:ss-comm-alg-is-sum-ofFields}
Suppose that a semi-simple algebra $\mathfrak{R}$ over $\Q$ with 
$\dim_\Q \mathfrak{R} < \infty$ is commutative. Then $\mathfrak{R}$ is 
of the form of $\oplus_i F_i$ where $F_i$ are number fields 
(with $[F_i:\Q] < \infty$).
\end{cor}
\begin{rmk}
The algebra $\mathfrak{R} = M_n(D)$ with a division algebra $D$ 
is known to be {\it simple}, in that the algebra $\mathfrak{R}$ does not 
have a non-trivial (non 0, non-$\mathfrak{R}$) ideals. 
\end{rmk}
\begin{rmk}
An irreducible representation of a simple algebra $M_n(D)$ on a vector 
space $V$ over $\Q$ is of the form of $V \cong D^{\oplus n}$, where 
we see $D$ as a vector space over $\Q$. 

A faithful representation $V_\Q$ over $\Q$ of a semi-simple algebra 
$\mathfrak{R}$ in (\ref{eq:WeddB-alg-str}) has to contain at least 
\begin{align}
  \oplus_{\alpha \in {\cal A}} (D_\alpha)^{\oplus n_\alpha s_\alpha} 
  \label{eq:modules-simple-Art-alg}
\end{align}
with $s_\alpha \geq 1$ for each $\alpha \in {\cal A}$. So, 
\begin{align}
  \dim_\Q V_\Q \geq \sum_{\alpha \in {\cal A}} n_\alpha [D_\alpha: \Q].
\end{align}
When this lower-bound is combined with Lemma \ref{lemma:div-alg-dim}, then 
\begin{align}
  \dim_\Q V_\Q \geq \sum_{\alpha \in {\cal A}} n_\alpha g_\alpha^2 [K_\alpha:\Q], 
  \label{eq:ineq-dimV-alg}
\end{align}
where $K_\alpha$ is the center of $D_\alpha$. 
\end{rmk}
\begin{rmk}
\label{rmk:EndAlg-str-notby-Wedbb-butby-HdgIsoClssDcmp}
The vector space $V_\Q$ of a polarizable rational Hodge structure 
$(V_\Q, \phi)$ is also a representation space of the semi-simple 
algebra $\mathfrak{R} = {\rm End}(V_\Q, \phi)$; the representation on 
$V_\Q$ is faithful, obviously from how the algebra was defined. 

In fact, Refs.  \cite[\S5]{shimura2016abelian} and 
 \cite[Cor. 5.3.8]{birkenhakeabelian}
derive the structure (\ref{eq:WeddB-alg-str}) of the algebra 
${\rm End}(V_\Q, \phi)$
in a different route without using the semi-simple nature of 
$\mathfrak{R}$ (Lemma \ref{lemma:End-alg-is-CM-wPol-ala-Milne}) and 
Wedderburn's theorem (Lemma \ref{lemma:Wedderburn}). 
The alternative route starts from the fact that $(V_\Q, \phi)$ has 
a simple component decomposition (\ref{eq:rHstr-indcmp-dcmp-by-grpHiso}) 
in Prop. \ref{props:rHstr-indcmp-dcmp}. The zero map is the only possible 
Hodge morphism between the simple Hodge substructures 
$(W^{\alpha, \lambda_\alpha}_\Q, \phi^\alpha)$ and $(W^{\beta,\mu_\beta}_\Q,\phi^\beta)$ 
when $\alpha \neq \beta$ (i.e., they are not Hodge-isomorphic), while 
the set of Hodge morphisms are one-to-one with a division algebra 
${\rm End}(W^{\alpha,\lambda_\alpha}_\Q, \phi^\alpha)$ denoted 
by $D_\alpha$; this is Schur's Lemma.  
That is also how one may derive the structure  (\ref{eq:WeddB-alg-str})
of the algebra $\mathfrak{R}$.  

The division algebra $D_\alpha$ acts on the vector space 
$W^{\alpha, \lambda_\alpha}_\Q$ faithfully. So, $\dim_\Q W^{\alpha, \lambda_\alpha}_\Q$
must be divisible by $\dim_\Q D_\alpha$. The ratio determines 
$s_\alpha$ in (\ref{eq:modules-simple-Art-alg}). 
\end{rmk}
\begin{rmk}
\label{rmk:str-alg-gen-rHstr}
(\cite[Thm. 7.5]{birkenhaketori})
The structure of $\mathfrak{R}= {\rm End}(V_\Q, \phi)$ of a rational Hodge 
structure that is not necessarily polarizable can also be inferred from the 
indecomposable decomposition (\ref{eq:rHstr-indcmp-dcmp-by-grpHiso}).
Given Rmk. \ref{rmk:simple-HgdStr-divs-Alg}, we know that there exists 
a subalgebra of the form 
\begin{align}
  \oplus_{\alpha \in {\cal A}} M_{n_\alpha}(Q_\alpha)  \subset \mathfrak{R}, 
\end{align}
where $Q_\alpha = {\rm End}(W^{\alpha, \lambda_\alpha}_\Q, \phi^\alpha)$ for any 
$\lambda_\alpha = 1,\cdots, n_\alpha$
is a local algebra. When the rational Hodge 
structure $(V_\Q, \phi)$ is not polarizable, 
${\rm Hom}((W^{\alpha,\lambda_\alpha}_\Q, \phi^\alpha), (W^{\beta, \mu_\beta}_\Q, \phi^\beta))$ can be non-zero. All of those off-diagonal Hodge morphisms belong 
to $J(\mathfrak{R})$. $M_{n_\alpha}(J(Q_\alpha))$ also belong to $J(\mathfrak{R})$. 
In the end, 
\begin{align}
  \mathfrak{R}/J(\mathfrak{R}) \cong \oplus_{\alpha \in {\cal A}}
   M_{n_\alpha} (Q_\alpha/J(Q_\alpha)). 
\end{align}
See \cite[Thm. 7.5]{birkenhaketori} for more details. 
\end{rmk}

\begin{rmk}
\label{rmk:Loewy-series-rHstr}
When a rational Hodge structure $(V_\Q, \phi)$ is not necessarily 
polarized, one may exploit an indecomposable decomposition of 
$(V_\Q, \phi)$---Prop. \ref{props:rHstr-indcmp-dcmp}---and its consequence
in the algebra $\mathfrak{R}={\rm End}(V_\Q, \phi)$ on one hand (as we 
have seen in the previous Remark). 
On the other, however, one might also exploit rational Hodge substructures 
that may still be contained in individual indecomposable 
components $(W^a_\Q, \phi^a)$. Here is a way we find useful, applying 
the idea known as {\it Loewy series} in algebra.  
 
Within the rational Hodge structure $(V_\Q, \phi)$ (with $\dim_\Q V_\Q 
< \infty$ and $\dim_\Q \mathfrak{R} < \infty$), a chain of 
rational Hodge substructures is introduced by 
\begin{align}
 \{ 0 \} = (J(\mathfrak{R}))^\ell V_\Q \subset (J(\mathfrak{R}))^{\ell-1}V_\Q
   \subset \cdots 
 \subset (J(\mathfrak{R}))^2V_\Q \subset J(\mathfrak{R}) V_\Q \subset V_\Q,   
\end{align}
where $J(\mathfrak{R})$ is the Jacobson radical of $\mathfrak{R}$;
this chain of substructures ends at a finite length $\ell$. 
At each step in the chain, the quotient 
$(J(\mathfrak{R}))^{i-1}V_\Q/(J(\mathfrak{R}))^{i}V_\Q$ 
is a completely reducible representation of $\mathfrak{R}$, 
and is of non-zero dimensions (\cite[p.346]{MR1245487}, 
Azumaya--Nakayama's lemma in \cite[Thm. 3.6]{MR998775}). 
The algebra $\mathfrak{R}$ acts on $(J(\mathfrak{R}))^{i-1}V_\Q/(J(\mathfrak{R}))^{i}V_\Q$ (the quotient Hodge structure at each step of the chain) 
 through the quotient $\mathfrak{R}/J(\mathfrak{R})$. 
We will use this structure in the appendix \ref{ssec:def-CM-rHstr}.
\end{rmk}
\begin{props}
\label{props:rHstr-ssmpl-cpltRdcbl}
The endomorphism algebra $\mathfrak{R}={\rm End}(V_\Q, \phi)$ 
of a rational Hodge structure $(V_\Q, \phi)$ is semi-simple, 
if and only if $(V_\Q, \phi)$ is completely reducible. 
\end{props}
The claim itself will be obvious from Rmk. \ref{rmk:str-alg-gen-rHstr}.
The statement here sounds similar to the tie between the 
semi-simple nature of an algebra and the complete reducibility 
of its arbitrary modules. The statement here is a little 
different (a variation of the same theme) in a few respects, however. 
First, we have a Hodge structure $(V_\Q, \phi)$ first, and then 
the algebra ${\rm End}(V_\Q, \phi)$ is determined from $(V_\Q, \phi)$,
not in the other way around. Second, we only talk of the complete 
reducibility of the defining representation $(V_\Q, \phi)$, not all of 
the representations of the algebra ${\rm End}(V_\Q, \phi)$. Third, 
a rational Hodge structure may be simple, even when it is not 
an irreducible representation of the algebra ${\rm End}(V_\Q, \phi)$. 
\qed

\subsection{Complex Torus vs Abelian Variety (general ones, not CM)}

\begin{defn}
\label{def:abelianVar-vs-coxTrs}
A complex torus $X$ of $n$-dimensions is said to be an {\it abelian variety}, 
when there exists $D_P \in H^2(X;\Q) \cap H^{1,1}(X;\R)$ whose Hermitian 
form $H_{D_P}(-,-) := D_P(-,J-)$ on $H_1(X;\R)$ is positive definite; 
here, $J$ is the almost complex structure operator of (the tangent space of)
 $X$. 
\end{defn}
\begin{rmk}
When a complex torus $X$ has an embedding into a projective space, 
then Rmk. \ref{rmk:pol-exists-in-algV} implies that it is an abelian variety. 
The converse is true in fact for a complex torus; $D_P$ in the definition 
of an abelian variety can be used to construct an embedding into 
a projective space, and one can regard the image of such an embedding 
as an algebraic variety. So, $D_P$ is often called a {\it polarization of
the abelian variety} $X$.  
\end{rmk}
\begin{rmk}
\label{rmk:refer2GriffHarr}
Abelian varieties therefore form a special class of complex tori. 
A complex torus chosen randomly from the moduli space of complex tori 
is not an abelian variety in general. 
See \cite[II.6]{Griffiths-Harris} for more. 

The fact that abelian varieties can be regarded as algebraic varieties
and that such powerful results as 
Prop. \ref{props:rHstr-indcmp-dcmp} and 
Lemmas \ref{lemma:End-alg-is-CM-wPol-ala-Milne}+\ref{lemma:Wedderburn}  
 can be used will be good enough reasons 
for mathematicians to pay particular attention to this special class 
of objects among complex tori in general. In a study in string theory, 
as in this article however, that is not a good enough reason to 
ignore complex tori that cannot be regarded as algebraic varieties. 

In the study of this article, distinction between this special class 
of objects as opposed to the other general complex tori is still important
because of the observation in Ex. \ref{ex:non-cpt-Hdg-grp}, and also in 
section \ref{ssec:pol}.  
\end{rmk}
\begin{defn}
Let $X \cong \C^n/\Lambda$ be a complex torus of $n$-dimensions. 
Its {\it complex subtorus} $Y$ consists of a vector subspace 
$\C^m \subset \C^n$ where $\C^m \cap \Lambda$ is of rank-$2m$. 
A complex subtorus $Y$ of $X$ is a complex submanifold $X$ as 
well as a subgroup of the addition group $X$. 
The quotient group $X/Y$ is also regarded as a complex torus. 

A complex torus $X$ is said to be {\it simple}, when its complex 
subtorus is either $X$ itself, or the zero-dimensional one $\{0\}$. 

When a complex abelian variety has a complex subtorus, the subtorus 
is an abelian variety, which is called an {\it abelian subvariety}. 
A complex abelian variety $X$ is said to be {\it simple}, when its abelian 
subvariety is either $X$ itself, or $\{0\}$. 

A complex torus $X$ is said to be {\it indecomposable} when there
is no isogeny $X \simeq X_1 \times X_2$ with the product of 
two complex tori (where $X_1, X_2 \neq \{0\}$). Even when $X$ has a 
complex subtorus $X_1$, so that there is an exact sequence 
\begin{align}
  0 \longrightarrow X_1 \longrightarrow X \longrightarrow X_2 
    \longrightarrow 0, 
  \label{eq:cpx-trs-shortExSq}
\end{align}
where $X_2$ is a complex torus, it is not always true  
that there is an isogeny $X \cong X_1 \times X_2$
(not always true that the exact sequence splits).
For more about what goes wrong in a complex torus that is not 
an abelian variety, see Ref. \cite[\S1]{birkenhaketori}. 
\end{defn}
\begin{rmk}
\label{rmk:cmplt-reducibl-abel-var}
When $X$ is a complex abelian variety and $X_1$ its abelian subvariety,
however, 
there exists an abelian variety $X_2$ such that $X$ and $X_1 \times X_2$
are isogenous. 
This means that an indecomposable abelian variety 
is a simple abelian variety, and that any complex abelian variety $X$ 
is isogenous to some product of simple abelian varieties.  
That is not always true when $X$ is a complex torus. 
\end{rmk}
\begin{rmk}
What is stated in the previous remark follows from what has already 
been explained in this appendix, in fact. An isogeny between 
a pair of complex tori $X$ and $Y$ induces a Hodge isomorphism between 
the weight-1 rational Hodge structure on $H^1(X;\Q)$ and $H^1(Y;\Q)$; 
a Hodge isomorphism between them also indicates that there is an isogeny 
between $X$ and $Y$, first of all.  
A complex torus $X$ has a complex subtorus $X_1$, with $X_2$ 
the quotient complex torus, if and only if the rational Hodge 
structure on $H_1(X;\Q)$ has a rational Hodge substructure 
(on the subspace $H_1(X_1;\Q)$), and the weight-1 rational Hodge 
structure on $H^1(X;\Q)$ has a substructure on the pull-backed subspace 
$H^1(X_2;\Q)$. Therefore, a simple complex torus corresponds to 
a simple weight-1 rational Hodge structure, and an indecomposable 
complex torus corresponds to an indecomposable weight-1 rational 
Hodge structure. Complete reducibility of abelian varieties 
(Rmk. \ref{rmk:cmplt-reducibl-abel-var}) corresponds to 
the complete reducibility of their polarizable weight-1 
rational Hodge structures (Prop. \ref{props:rHstr-indcmp-dcmp}). 
\end{rmk}

\begin{exmpl}
\label{ex:Shafarevic}
Given a pair of complex tori $X_1$ and $X_2$, one might be 
interested in the set of complex tori $X$ that has $X_1$ 
as a subtorus and $X_2$ as the quotient, modulo isomorphism 
of complex tori (identification by maps that are both holomorphic 
and group homomorphisms). The classification is given by 
the group ${\rm Ext}^1(X_2, X_1)$, which is the Coker of 
${\rm Hom}(\C^{\dim_\C X_2}, X_1) \longrightarrow {\rm Hom}_\Z(\Lambda_2, X_1)$. 
It is known that the short exact sequence (\ref{eq:cpx-trs-shortExSq})
splits (equivalently, the complex torus $X$ is decomposable, 
isogenous to $X_1 \times X_2$, and the rational Hodge structure on $H^1(X;\Q)$
is decomposable) if and only if the corresponding element in ${\rm Ext}^1(X_2,X_1)$ is a torsion element. See \cite[\S1.5, 1.6]{birkenhaketori}. 
\end{exmpl}

\begin{rmk}
Whereas an indecomposable abelian variety $X$ does not have a subtorus, 
an indecomposable complex torus $X$ may have a non-trivial subtorus. What 
Rmk. \ref{rmk:Loewy-series-rHstr} does in this present context, 
where $(V_\Q, \phi)$ is an indecomposable weight-1 rational Hodge structure 
$(H^1(X;\Q), \phi_1)$, is to introduce a chain of rational Hodge 
substructures of $(H^1(X;\Q), \phi_1)$. There then exists a 
corresponding chain of complex subtori 
\begin{align}
  X \supset S_{\ell-1}X \supset \cdots \supset S_2X \supset S_1X \supset 
  S_0X = \{0\},
\end{align}
such that the vector subspace $(J(\mathfrak{R}))^i V_\Q \subset V_\Q=H^1(X;\Q)$ 
is equal to $H^1(X/S_iX;\Q)$ (pulled back to $H^1(X;\Q)$). 
Each of the complex tori $S_{i}X/S_{i-1}X$ with $i=1,2,\cdots, \ell$ is  
of positive dimensions, and is isogenous to the direct sum of simple complex tori. 
\end{rmk}

\subsection{Useful Lemmas}
\label{ssec:useful}

In this appendix (which for the most part only collects known 
facts from textbooks), we still do not provide an elementary review  
on number fields (finite dimensional algebraic extension 
fields over $\Q$) and Galois theory. Readers with little experience 
dealing with number fields might have a look at textbooks, or 
a quick glance at the appendix A of \cite{Kanno:2017nub}.

The following facts (Lemmas \ref{lemma:very-useful}, \ref{lemma:very-useful-2}) 
are regarded so trivial by mathematicians that we have to 
read that out between the lines in such textbooks. 
The authors are unable to refer to a specific text for this reason. For 
the reader with a background in string theory, it will still be 
better that they are written down explicitly.\footnote{
The appendix B.2 of the preprint version of \cite{Kanno:2017nub}
(main text II.B.3 of the journal version) has a little more 
pedagogical explanation on the first half of Lemma \ref{lemma:very-useful}.
The statement here is slightly polished up from the version there, however. 
} %
\begin{lemma}
\label{lemma:very-useful}
Let $F$ be a number field (with $[F:\Q] < \infty$), and 
$\{\tau_{a=1,\cdots, [F:\Q]} \}$ its embeddings to $\overline{\Q} \subset \C$. 
Let $V_\Q$ be a vector space over $\Q$ with $\dim_\Q V_\Q = [F:\Q]$ that 
has a faithful action of $F$ (where $\Q \subset F$ acts as the scalar 
multiplication on the vector space $V_\Q$). 

The action of $F$ on $V_\Q$ can be diagonalized simultaneously 
because the multiplication in $F$ is commutative. The $\dim_\Q V_\Q$ 
eigenspaces are in one-to-one with the $[F:\Q]$ embeddings 
$F \hookrightarrow \C$; when $v \in V_\Q \otimes \C$ is an eigenvector 
of all $x \in F$ (i.e., $x \cdot v = v \lambda_x$ for some eigenvalue
$\lambda_x \in \C$), the assignment $F \ni x \longmapsto \lambda_x \in \C$
preserves the addition and multiplication laws in $F$, so it is 
an embedding $F \hookrightarrow \C$ (this reasoning is used already 
in Lemma \ref{lemma:simpl-Hstr-alg-divs}).   

There is also a concrete procedure to determine the eigenvectors. 
First, choose any non-zero element $v_* \in V_\Q$, and arbitrary basis 
$\{ \omega_{I=1,\cdots, [F:\Q]} \}$ of the vector space $F/\Q$. 
Then $\{ \omega_I \cdot v_* \}_{I=1,\cdots, [F:\Q]}$ can be used as a basis 
of the vector space $V_\Q$ over $\Q$. Now, we can 
choose the eigenvectors to be\footnote{ 
So, the action of $F$ on $V_\Q$ splits into 1-dimensions on $V_\Q \otimes_\Q k$ 
whenever $k \subset \overline{\Q}$ contains the normal closure of $F$ in 
$\overline{\Q}$. 
}\raisebox{5pt}{,}\footnote{
A sum over an index repeated twice (e.g., $I$ in (\ref{eq:useful-Lmm-egVct}))
is implicit, as in physics literatures. 
} %
\begin{align}
  v_a :=  (\omega_I \cdot v_*) \tau_a(\eta_I), \qquad a=1,\cdots, [F:\Q], 
   \label{eq:useful-Lmm-egVct}
\end{align}
where $\{ \eta_{J=1,\cdots, [F:\Q]} \}$ is the basis of $F/\Q$ dual to 
$\{ \omega_I \}$ with respect to the bilinear form 
$F \times F \ni (x,y) \mapsto {\rm Tr}_{F/\Q}[x y] \in \Q$.  
That is, ${\rm Tr}_{F/\Q}[\omega_I \eta_J] = \delta_{IJ}$.
The matrix $(\tau_a(\eta_I))_{aI}$ is used as the inverse matrix 
of $(\tau_a(\omega_J))_{Ja}$. 
For any element $x \in F$, $v_a \in V_\Q \otimes \overline{\Q}$
is an eigenvector, with the eigenvalue $\tau_a(x)$. 
All those eigenvectors $v_a$ ($a=1,\cdots, [F:\Q]$) are obtained from 
one of them, say, $v_{a*}$, by applying Galois transformations on the 
coefficients $\tau_{a*}(\eta_I)$ of the expansion of $v_{a*}$ with respect 
to the rational basis $\{ (\omega_I\cdot v_*)_{I=1,\cdots, [F:\Q]} \}$ of $V_\Q$, 
because $\tau_a = \sigma_a \cdot \tau_{a*}$ for some 
$\sigma_a \in {\rm Gal}(\overline{\Q}/\Q)$. We may express this
in the form of $v_a = v_{a*}^{\sigma_a}$.

For any basis $\{ \eta'_{I=1,\cdots, [F:\Q]} \}$ of $F/\Q$, there exists 
a basis $\{ v'_I \}$ of $V_\Q$ where the simultaneous eigenvectors 
are in the form of $v_a = v'_I \tau_a(\eta'_I)$. To see this, just find 
the rational coefficient matrix $\eta_I = C_{IJ} \eta'_J$ and set 
$v'_J := (\omega_I \cdot v_*)C_{IJ}$. 
\end{lemma}

\begin{lemma}
\label{lemma:very-useful-2}
Conversely, for any basis $\{ \eta_J \}$ of $F/\Q$ and $\{v_I \}$ of $V_\Q$, 
one may construct a non-trivial action of $F$ on the vector space $V_\Q$ 
over $\Q$ so that $v_a := v_I \tau_a(\eta_I)$ for $a=1,\cdots, [F:\Q]$
are all eigenvectors of the action of $F$. The action of $x \in F$ on $V_\Q$ 
claimed here is given as follows. First, write down the multiplication law 
in $F$ as follows: 
\begin{align}
  (x \cdot ): \quad \omega_I \longmapsto x \cdot \omega_I =
     \omega_K [A(x)]_{KI}, 
\end{align}
where $\{ \omega_I \}$ is the basis of $[F:\Q]$ dual to $\{ \eta_J \}$, 
and $[A(x)]$ is a $\Q$-valued $[F:\Q] \times [F:\Q]$ matrix. 
Using this matrix, the action of $x$ on $V_\Q$ is 
\begin{align}
  x \cdot : \quad v_I \longmapsto v_K [A(x)]_{KI}. 
\end{align}
\qed
\end{lemma}

The facts above in both ways (Lemmas \ref{lemma:very-useful} and \ref{lemma:very-useful-2}) hold for a general number field $F$ not 
necessarily a CM field, or even a totally imaginary field. 

\subsection{CM-type Rational Hodge Structure}
\label{ssec:def-CM-rHstr}

%
\begin{defn}
\label{def:CM-rat-Hdg-str-byEnd}
A rational Hodge structure $(V_\Q, \phi)$ is said to have {\it sufficiently 
many complex multiplications} when the endomorphism algebra ${\rm End}(V_\Q, \phi)$ 
contains a commutative semi-simple subalgebra ${\cal K}$ such that 
$\dim_\Q {\cal K} = \dim_\Q V_\Q$. When $(V_\Q, \phi)$ is also polarizable
and has sufficiently many complex multiplications, we say that it is 
{\it of CM-type}. 
\end{defn}
As announced in Rmk. \ref{rmk:announce-CMdef-use-EndAlg-onCoh}
(cf also Cor. \ref{cor:ss-comm-alg-is-sum-ofFields}),
this is a way to generalize the notion of complex multiplications 
from complex tori to a broader class of compact K\"{a}hler manifolds 
/ projective non-singular varieties. For a given manifold/variety $X$, 
there are multiple components of rational Hodge structures; 
each of the cohomology groups $H^k(X;\Q)$, with $k=0,1,\cdots, 2\dim_\C X$, 
has a rational Hodge structure, and each of them may also have substructures. 
Not all of those rational Hodge substructures are independent (when we choose 
$X$ from a deformation family of manifolds in the same topological class);
complex multiplication in one rational Hodge substructure may (or may not) 
imply complex multiplication in other substructures. This appendix 
does not dig into the question; interested readers might have a look at 
Ref. \cite{Okada:2023udq} and references therein for more information. 
\begin{lemma}
\label{lemma:cm-fields-mostly-totImgry}
Let ${\cal K} \cong \oplus_i F_i$ be a commutative semi-simple subalgebra 
of the algebra ${\rm End}(V_\Q, \phi)$, required in the definition 
of a weight-$m$ rational Hodge structure $(V_\Q,\phi)$ with sufficiently 
many complex multiplications; each of $F_i$ is a number field. 
Then each $F_i$ acts on a rational Hodge substructure $(W^i_\Q, \phi|_{W^i})$ 
of $(V_\Q, \phi)$ faithfully, with $\dim_\Q W^i_\Q = [F_i:\Q]$, and the 
number of real embeddings of $F_i$ into $\C$ is not more than 
$h^{m/2,m/2}(W^i)$; when the weight $m$ is odd, in particular, 
all the $F_i$'s are totally imaginary. 
\end{lemma}
\proof The subalgebra ${\cal K}$ must be represented faithfully 
on the vector space $V_\Q$ (by def of the algebra ${\rm End}(V_\Q, \phi)$), 
on one hand, and $\dim_\Q{\cal K} = \dim_\Q V_\Q$ on the other. This 
is possible if and only if there is a decomposition $V_\Q \cong \oplus_i W^i_\Q$ 
of vector spaces over $\Q$ such that each $F_i$ acts faithfully 
on the corresponding $W^i_\Q$ and trivially on other $W^j_\Q$'s ($j\neq i$), 
and $\dim_\Q F_i = \dim_\Q W^i_\Q$. It is not hard to see that 
the rational Hodge structure $(V_\Q, \phi)$ also decomposes into 
$\oplus_i (W^i_\Q, \phi|_{W^i})$. So, we can deal with each $F_i$ acting 
on $(W^i_\Q,\phi|_{W^i})$ separately. 

Now, one may apply Lemma \ref{lemma:very-useful} to $F_i$ and $W^i_\Q$. 
The field $F_i$ acts linearly on each of the Hodge components 
$[W^i_\Q\otimes \C]^{p,q}$, so simultaneous eigenspaces of the commutative 
algebra $F_i$ are found within the individual Hodge components 
with distinct $(p,q)$'s. Each of the eigenvectors $v_a$ in one-to-one with 
the embeddings $\tau_a: F_i \hookrightarrow \overline{\Q} \subset \C$
is assigned to one of the Hodge components $[W^i_\Q \otimes \C]^{p,q}$. 

When one embedding $\tau_a$ of $F_i$ has its eigenstate $v_a$ in the 
Hodge $(p,q)$ component, the embedding ${\rm cc} \circ \tau_a$ of $F_i$ 
corresponds to the eigenstate $\bar{v}_a \in W^i_\Q \otimes \C$ obtained 
by the complex conjugation on the $\otimes \C$ factor. Thus, 
$\bar{v}_a$ is in the Hodge $(q,p)$ component. The embedding $\tau_a$ 
is real (i.e., ${\rm cc} \circ \tau_a = \tau_a$) only when $(p,q)=(q,p)$. 
\qed

\begin{rmk}
\label{rmk:Phi-p-q}
Let $F$ be a number field, $(W_\Q,\phi)$ a weight-$m$ rational Hodge 
structure such that $F \subset {\rm End}(W_\Q, \phi)$ and 
$[F:\Q] = \dim_\Q W_\Q$. Then each of the simultaneous eigenvectors $v_a$ of the 
action of $F$, in one to one correspondence with the embeddings $\tau_a$ 
of the field $F$, belongs to one of the Hodge $(p, q)$ component of 
$(W_\Q, \phi)$, as we have already argued within the proof of the 
previous Lemma. This observation introduces a decomposition 
\begin{align}
  {\rm Hom}_{\rm field}(F,\overline{\Q}) = \amalg_{p=0}^m \Phi^{(p,m-p)},
  \label{eq:dcmp-field-embdngs-4-Hdg-cmp}
\end{align}
where $\Phi^{(p,q)}$ consists of embeddings whose corresponding eigenvectors 
are in the Hodge $(p,q)$ component. The complex conjugate embeddings of 
those in $\Phi^{(p,q)}$ are those in $\Phi^{(q,p)}$. 
\end{rmk}

\begin{props}
\label{props:CM-alg-always-ssmpl}
Think of a rational Hodge structure $(V_\Q, \phi)$ with sufficiently 
many complex multiplications. Let ${\cal K} = \oplus_i F_i$ be 
a semi-simple commutative algebra in $\mathfrak{R}= {\rm End}(V_\Q, \phi)$
with $\sum_i [F_i:\Q] = \dim_\Q V_\Q$. Then the algebra $\mathfrak{R}$ 
is semi-simple, and each indecomposable component $(W^a_\Q, \phi^a)$ of 
$(V_\Q, \phi)$ is in fact a simple rational Hodge substructure. 
That is, $(V_\Q, \phi)$ is completely reducible. 
\end{props}
\proof
Note first that the subalgebra ${\cal K}\cong \oplus_i F_i$ in 
$\mathfrak{R}$ is mapped injectively into the semi-simple algebra 
$\mathfrak{R}/J(\mathfrak{R})$; that is verified by only noting 
that none of the non-zero elements of ${\cal K}$ are nilpotent. 
Secondly, each of $(J(\mathfrak{R}))^{i-1}V_\Q/(J(\mathfrak{R}))^iV_\Q$ 
with $i=1,2,\cdots, \ell$ supports a faithful representation of the 
division algebra $\mathfrak{R}/J(\mathfrak{R})$, which contains ${\cal K}$. 
So, 
\begin{align}
 \dim_\Q \left[ (J(\mathfrak{R}))^{i-1}V_\Q/(J(\mathfrak{R}))^iV_\Q \right]
   & \; \geq \sum_j [F_j:\Q], \qquad i=1,2,\cdots, \ell, \\
 \dim_\Q V_\Q  & \; \geq \ell \dim_\Q {\cal K}. 
\end{align}
Therefore, the condition that $\dim_\Q V_\Q = \dim_\Q {\cal K}$ implies that 
$\ell =1$, and also $J(\mathfrak{R}) V_\Q=0$.  

Noting that $V_\Q$ is the defining representation of the algebra 
$\mathfrak{R}$, not a general representation of $\mathfrak{R}$, this 
$J(\mathfrak{R})V_\Q=0$ implies that $J(\mathfrak{R})=0$. That is, 
the algebra $\mathfrak{R}$ is semi-simple. The local algebra 
$Q_\alpha = {\rm End}(W^{\alpha_\Q,\lambda_\alpha},\phi^\alpha)$ common
to any $\lambda_\alpha \in \{1,\cdots, n_\alpha\}$ 
of an indecomposable decomposition is a division algebra 
for any Hodge-isomorphism class $\alpha \in {\cal A}$, in fact. 
See also Prop. \ref{props:rHstr-ssmpl-cpltRdcbl}.
\qed

For this reason, we restrict our attention to the cases 
the algebra $\mathfrak{R}$ is semi-simple in the rest of this 
appendix \ref{sec:guide}. As is well-known in the case 
of a polarizable rational Hodge structure $(V_\Q, \phi)$, 
we may use the following 
\begin{rmk}
\label{rmk:max-subfield-in-divAlg-dim}
It is known that a maximal subfield $F$ in a division algebra 
$D$ is of degree $[F:Z(D)]=g$, where $Z(D)$ is the center of $D$, 
when $\dim_{Z(D)}D = g^2$ (cf Lemma \ref{lemma:div-alg-dim}), 
\end{rmk}
\noindent to extract more information about the endomorphism algebra 
of a (not necessarily polarized) rational Hodge structure with 
sufficiently many complex multiplications. 
\begin{props}
\label{props:str-endAlg-whenCM+smCM}
The semi-simple algebra $\mathfrak{R} = {\rm End}(V_\Q, \phi)$ of 
a (not necessarily polarizable) rational Hodge structure with 
sufficiently many complex multiplications is of the form 
of 
\[
  \mathfrak{R} \cong \oplus_{\alpha \in {\cal A}} M_{n_\alpha}(K_\alpha), 
\]
where $K_\alpha$ is a number field with $[K_\alpha:\Q] =
 \dim_\Q W^{\alpha,\lambda_\alpha}_\Q$. The set ${\cal A}$ labels 
the Hodge-isomorphism classes of the simple rational Hodge substructures 
in $(V_\Q, \phi)$. 
\end{props}
Given Prop. \ref{props:CM-alg-always-ssmpl}, the proof well-known\footnote{
e.g., \cite[Props. II.4 and II.6 (\S5)]{shimura2016abelian},
 \cite[Prop. 1.2 (\S I.1) and Prop. 3.1+3.3(\S I.3)]{milnecm}
} %
 in the case of a polarizable $(V_\Q, \phi)$  
works as it is also for $(V_\Q, \phi)$ without a polarization.  
A commutative subalgebra ${\cal K}$ available within the semi-simple 
algebra $\mathfrak{R}$ has a dimension bounded from above by 
\begin{align}
 \dim_\Q {\cal K} & \; \leq \sum_{\alpha \in {\cal A}} n_\alpha g_\alpha [Z(D_\alpha):\Q]
  \leq \sum_{\alpha \in {\cal A}} n_\alpha g_\alpha^2 [Z(D_\alpha): \Q]
  = \sum_{\alpha \in {\cal A}} n_\alpha [D_\alpha:\Q] \nonumber \\
 & \;  \leq \sum_{\alpha \in {\cal A}} n_\alpha s_\alpha [D_\alpha:\Q]
  =  \dim_\Q V_\Q;
\end{align}
the first inequality and the equality in the middle are from 
Lemma \ref{lemma:div-alg-dim} and Rmk. \ref{rmk:max-subfield-in-divAlg-dim}, 
and $s_\alpha$'s in the last line are the ones in (\ref{eq:modules-simple-Art-alg}) 
for $V_\Q$.
The condition $\dim_\Q {\cal K} = \dim_\Q V_\Q$ for ${\cal K}$ of a 
rational Hodge structure $(V_\Q, \phi)$ is therefore equivalent to 
$g_\alpha =1$ and $s_\alpha =1$ for all $\alpha \in {\cal A}$. 
In other words, the endomorphism algebra $D_\alpha = 
{\rm End}(W^{\alpha, \lambda_\alpha}_\Q, \phi^\alpha)$ of each simple component 
is a field, $D_\alpha = Z(D_\alpha)$, of dimension 
$[Z(D_\alpha):\Q] = \dim_\Q W^{\alpha, \lambda_\alpha}_\Q$.  \qed

Lemma \ref{lemma:cm-fields-mostly-totImgry} is all that we can say 
about the field $K_\alpha$ for a general rational Hodge structure 
$(V_\Q, \phi)$ with sufficiently many complex multiplications. 
When $(V_\Q, \phi)$ is polarizable, the following is well-known. 
\begin{rmk}
Suppose that there is a number field $F \hookrightarrow \mathfrak{R} 
= {\rm End}(V_\Q, \phi)$ of a polarizable rational Hodge 
structure $(V_\Q, \phi)$. When the Rosati involution of a 
polarization maps $F \subset \mathfrak{R}$ to $F$ (not necessarily 
individual elements in $F$, but $F$ as a whole), then one can 
prove that $F$ is either totally real, or a CM-field. 
See \cite[Lemma II.2 + Prop. II.5 (\S5)]{shimura2016abelian},
 \cite[Prop. 1.39]{milnecm} and \cite[Thm. 3.7]{MR3586372} for a proof. 
The positivity of a polarization (involution) is vital in the proof, 
so a polarization of a positive index is not enough here. 

This applies to $F = Z(D_\alpha) {\bf 1}_{n_\alpha \times n_\alpha}$, 
in particular; this is because the center $Z(\mathfrak{R}) \cong \oplus_\alpha 
Z(D_\alpha) {\bf 1}_{n_\alpha \times n_\alpha}$ of the algebra $\mathfrak{R}$
is mapped by an involution to the center itself. 
\end{rmk}
\begin{rmk}
When a polarizable simple rational Hodge structure $(V_\Q, \phi)$ of 
level $\ell > 0$ is of CM-type, the number field 
$K = Z(D) = D= {\rm End}(V_\Q, \phi)$ is a CM-field rather than a totally real field. 
See \cite[Prop. 3.6 (\S I.3)]{milnecm} and \cite[Rmk. 3.14]{MR3586372}
for the proof. 
\end{rmk}

\vspace{3mm}
Compared with the endomorphism field $K_\alpha=Z(D_\alpha)$ of the individual 
simple Hodge substructures, there is less importance in 
commutative subalgebras ${\cal K} \subset \mathfrak{R}$ in the 
definition of sufficiently many complex multiplications. 
Even when a Hodge structure $(V_\Q, \phi)$ with sufficiently many 
complex multiplications is given, there can be considerable 
freedom/arbitrariness
%
 in the choice of ${\cal K}$ with $\dim_\Q {\cal K}=\dim_\Q V_\Q$. 
See Ex. \ref{ex:ExE-CM}. 
When $(V_\Q, \phi)$ is polarizable, and a choice of polarization ${\cal Q}$
is fixed, then there is a smaller class of a choice of 
${\cal K} \cong \oplus_i F_i$ that is better motivated: a subalgebra 
${\cal K}$ that is mapped to itself (not necessarily identically, though) 
by the Rosati involution with respect to ${\cal Q}$. 
\begin{lemma}
\label{lemma:numbr-field-assign2-isotypicCmp}
In a commutative semi-simple subalgebra ${\cal K} \cong \oplus_{i \in I} F_i$ 
required in the Definition of a rational Hodge structure with sufficiently 
many complex multiplications, each number field $F_i$ is within one 
of the simple factors $M_{n_\alpha}(K_\alpha)$; there is a map 
$I \ni i \mapsto \alpha \in {\cal A}$. 
\end{lemma}
\proof Suppose that $F_i$ is embedded in $M_{m_{\alpha}}(K_{\alpha}) \oplus 
M_{m_{\beta}}(K_{\beta})$, where $\alpha, \beta \in {\cal A}$, 
 with $m_{\alpha}\leq n_{\alpha}$ and $m_{\beta}\leq n_{\beta}$,  
and $[F_i:\Q] = [K_{\alpha}:\Q]m_{\alpha} + [K_{\beta}:\Q]m_{\beta}$.
The only possible such a commutative algebra is of the form of 
$F^{\alpha} \oplus F^{\beta}$ where $F^{\alpha}/K_{\alpha}$ and 
$F^{\beta}/K_{\beta}$ are extensions of degree $m_{\alpha}$ and 
$m_{\beta}$, respectively.  Such an algebra is not a field, 
when both $m_{\alpha}$ and $m_{\beta}$ are positive (non-zero). 
\qed

In the case of a polarizable $(V_\Q, \phi)$ that is of CM-type,  
the fields $F_i$ in ${\cal K}$ required in the definition of 
complex multiplications are subject only to 
Lemma \ref{lemma:cm-fields-mostly-totImgry}, although we know that 
$K_{\alpha}$'s are all CM fields. It is possible, however, 
to choose $F_i$'s in ${\cal K} \cong \oplus_{i\in I} F_i$ so that 
they are all CM-fields when $(V_\Q, \phi)$ is of CM-type (see \cite[Prop. 3.6(b) (\S I.3)]{milnecm}
for a construction). Moreover,
\begin{rmk}
\label{rmk:build-pol-from-CMfield}
Let $(W^i_\Q, \phi^i)$ be a rational Hodge structure with 
sufficiently many complex multiplications, with just one 
Hodge isomorphism class of simple substructures. Let 
$F_i$ be a CM-field within the simple algebra 
${\rm End}(W^i_\Q, \phi^i)$  (already 
Lemmas \ref{lemma:cm-fields-mostly-totImgry} 
and \ref{lemma:numbr-field-assign2-isotypicCmp} are implicit), 
with $[F_i:\Q] = \dim_\Q W^i_\Q$.  
Then a polarization ${\cal Q}_i: W^i_\Q \times W^i_\Q \longrightarrow \Q$
is constructed,  by fixing one isomorphism $F_i \cong W^i_\Q$ as 1-dimensional 
vector space over $F_i$ and setting it by 
\begin{align}
 {\cal Q}: W^i_\Q \times W^i_\Q \cong F_i \times F_i \ni (x, y) \longmapsto
   \tr {}_{F_i/\Q}[ \lambda x\bar{y}]    \in \Q; 
\end{align}
$\lambda \in F_i^\times$ has to be invariant [resp. odd] under the 
complex conjugation in the CM-field for ${\cal Q}$ to be symmetric 
(even weight) [resp. anti-symmetric (odd weight)]; inequalities 
have to be imposed\footnote{
The inequalities to be imposed depend on the type $\{ h^{p,q}(W^i) \}$ 
of the Hodge structure $(W^i_\Q,\phi^i)$, but in an obvious way. 
} %
 further on the embedding images of $\lambda$ for the positive definiteness of the 
Hermitian form $H_{\cal Q}$. Otherwise, any $\lambda$ is fine. 
Furthermore, the subfield $F_i$ in ${\rm End}(W^i_\Q, \phi^i)$ 
is mapped to itself (non-trivially) under the Rosati involution 
with respect to those ${\cal Q}$'s.

When $F$ is not a CM-field, there is no algorithm to find a polarization 
${\cal Q}$ on $(V_\Q, \phi)$ whose Rosati involution is the identity 
on $F_0$, even when $(V_\Q, \phi)$ is polarizable.  
\end{rmk}
%

\subsection{Complex Multiplication of Complex Tori: Part II}
\label{ssec:cpx-trs-II}

The notion of CM-type has been generalized from the original 
one with geometric intuition to the property of rational Hodge 
structures, as reviewed in the appendices \ref{ssec:rHstr-Pol-alg} 
\ref{ssec:useful} and \ref{ssec:def-CM-rHstr}. 
While Def. \ref{def:CM-rat-Hdg-str-byEnd} is applicable to any 
rational Hodge (sub)structures on the cohomology groups of any 
compact K\"{a}hler manifolds / non-singular projective varieties $M$,
that is not (yet) a definition of a property of $M$. There are 
at least a few different versions in trying to define a ``CM'' property 
of such varieties, as reviewed around  
Defs. \ref{def:strongCM} and \ref{def:weakCM-level}. 

\begin{rmk}
\label{rmk:geomCM=strCM-CTAV}
When $X$ is a complex torus (or even an abelian variety), their 
complex multiplication property in Def. \ref{def:CM-geom-4CT}
is equivalent to that in Def. \ref{def:CM-rat-Hdg-str-byEnd}  
applied to the weight-1 rational Hodge structure on $H^1(X;\Q)$.  
When the rational Hodge structure on $H^1(X;\Q)$ is with sufficiently 
many complex multiplications / of CM-type, then so are the 
rational Hodge structures on $H^k(X;\Q)$ for $k=2,3,\cdots, 2n$;
this follows\footnote{
The proof in \cite{borcea1998calabi} is written only for polarizable 
rational Hodge structures, but one can confirm that the polarizability 
is not essential in the proof. See also \cite[A.3]{Okada:2023udq} for the 
Hodge group for rational Hodge structures that are not necessarily polarized. 
} %
 from Prop. 1.21 of \cite{borcea1998calabi}.  So, an abelian variety 
$X$ of CM-type in Def. \ref{def:CM-geom-4CT} is of strong CM-type 
in Def. \ref{def:strongCM} and vice versa. 
\end{rmk}

Here, in the appendix \ref{ssec:cpx-trs-II}, we focus on complex tori and 
abelian varieties denoted by $X$ that are with sufficiently many CMs 
/ (strong) CM-type. Consequences of the materials in the appendices 
\ref{ssec:rHstr-Pol-alg}--\ref{ssec:def-CM-rHstr} on the Hodge structure 
of $H^1(X;\Q)$ are now summarized and written down here. 
Consequences on the Hodge structure on $H^k(X;\Q)$ with $k\geq 2$ can be 
worked out from that on the $H^1(X;\Q)$, as we do in 
section \ref{ssec:CM-abel-surface}. 

We begin with the case of a complex abelian variety $X$, 
instead of a complex torus that does not necessarily have 
a polarization. Definition \ref{def:CM-geom-4CT} for $X$ to be of CM-type 
is equivalent to Def. \ref{def:CM-rat-Hdg-str-byEnd} for its 
rational Hodge structure $(H^1(X;\Q), \phi_1)$ to be of CM-type; 
the latter is equivalent to the following: 
\begin{rmk}
\label{rmk:CM-Intuitv-summary4AV}
The weight-1 rational Hodge structure $(H^1(X;\Q), \phi_1)$ 
of a complex abelian variety is of CM-type (by 
Def. \ref{def:CM-rat-Hdg-str-byEnd}), 
if and only if its all the simple components $(W^a_\Q, \phi^a)$ 
(Prop. \ref{props:rHstr-indcmp-dcmp}) are of CM-type. The condition that 
the simple component $(W^a_\Q, \phi^a)$ is of CM-type is equivalent 
to the condition that ${\rm End}(W^a_\Q, \phi^a)$ is a CM-field $K_\alpha$ 
with $[K_\alpha:\Q] = \dim_\Q W^a_\Q$.
Lemma \ref{lemma:very-useful} provides more detailed information 
on the rational Hodge structure on $(W^a_\Q,\phi^a)$.

In the language of geometry, in a modulo-isogeny decomposition 
$X \sim \prod_{\alpha \in {\cal A}} (X_\alpha)^{n_\alpha}$ into simple abelian 
varieties (Rmk. \ref{rmk:cmplt-reducibl-abel-var}), $X$ is of CM-type 
(by Def. \ref{def:CM-geom-4CT}) if and only if ${\rm End}_\Q(X_\alpha)$ 
is a CM field $K_\alpha$ with $[K_\alpha:\Q] = 2\dim_\C X_\alpha$. 
%
%
\qed
\end{rmk}
\begin{props}
\label{props:CM-Intuitv-summary4CT}
Let $X$ be a complex torus. 
It has sufficiently many complex multiplications, 
or equivalently, its weight-1 rational Hodge structure 
$(H^1(X;\Q), \phi_1)$ has sufficiently many complex multiplications, 
if and only if $X$ [resp. $(H^1(X;\Q), \phi_1)$] is completely reducible, 
and each simple factor $X_\alpha$ [resp. $(W^a, \phi^a)$] has sufficiently 
many complex multiplications; this condition on each simple factor 
is also equivalent to the condition that the division algebra 
${\rm End}(W^a,\phi^a)$ is a totally imaginary field $K_\alpha$ 
with $[K_\alpha:\Q] = 2\dim_\C X_\alpha$ [resp. $[K_\alpha:\Q] = \dim_\Q W^a$]. 
The algebra $\mathfrak{R}={\rm End}(V_\Q, \phi)$ as a whole is 
semi-simple, $\oplus_{\alpha \in {\cal A}} M_{n_\alpha}(K_\alpha)$. 
All of those also follow from Prop. \ref{props:CM-alg-always-ssmpl}, 
Prop. \ref{props:str-endAlg-whenCM+smCM} and 
Lemma \ref{lemma:cm-fields-mostly-totImgry}. 
\end{props}
%

\subsection{Hodge Group of a Non-polarizable Rational Hodge Structure}
\label{ssec:Hdg-grp-ex}

Whether a rational Hodge structure [resp. a polarizable one] has 
sufficiently many complex multiplications [resp. CM-type] or not 
can be characterized either by the endomorphism algebra (Def. \ref{def:CM-rat-Hdg-str-byEnd}), 
or alternatively, by the Hodge group (or Mumford--Tate group); 
cf p. \pageref{pg:CM-def-1n2}. In this article, we use the 
endomorphism algebras for analysis and derivations most of the time;
the review in this appendix \ref{sec:guide} has not referred to 
the Hodge group so far for this reason.   

There is a definition for the Hodge group and Mumford--Tate group of 
a rational Hodge structure widely accepted in the literatures. 
In this article, we used Hodge group just once in section \ref{ssec:pol}; 
the Hodge group is used there not to build an argument in this article 
but to read out the (limitation of the) consequences of the argument 
made elsewhere; little intuition on a Hodge group is required in following 
the arguments there.  So, the authors decided not to repeat a formal 
definition of the Hodge group in this article; string theorists 
in need of an introduction on this subject written by string theorists
might still have a look at \cite[A.3]{Okada:2023udq}. 

There is not an essential difference between abelian varieties of CM-type
and generic complex tori with sufficiently many complex multiplications, 
so far as the characterization conditions (i) and (ii) in 
p. \pageref{pg:CM-def-1n2} are concerned. When it comes to the 
characterization condition (iii) in p. \pageref{pg:CM-def-1n2}, however, 
there is a crucial difference between rational Hodge structures that 
are polarizable, and those that are not. This difference was primarily why the 
authors are led to focus on polarizable complex structures in this article. 

Given the importance of this difference in the property (iii), it will be 
a popular attitude among string theorists to try to appreciate what is 
happening not just in abstract definition and logic, but also with 
concrete examples. So, in this appendix \ref{ssec:Hdg-grp-ex}, 
an example is presented for illustration purpose. We assume that 
readers of this appendix \ref{ssec:Hdg-grp-ex} are familiar with 
the definition of the Hodge group of a rational Hodge structure 
(that is not necessarily polarizable). We will use the same 
notations as in \cite[A.3]{Okada:2023udq} here.

\begin{exmpl}
\label{ex:non-cpt-Hdg-grp}
Examples of totally imaginary fields that are not 
CM fields are found\footnote{
Math StackExchange entry ``totally imaginary number field of degree 4'' \\
\url{https://math.stackexchange.com/questions/4372232/}
} %
 in the database LMFDB (\url{www.lmfdb.org}). 
For example, $F = \Q[x]/(x^4-2x^2+2)$. For any totally imaginary 
field $F$ and a decomposition of the $[F:\Q]$ embeddings 
$\Phi^{(1,0)} \amalg \Phi^{(0,1)}$ such that the complex conjugation of 
$\Phi^{(1,0)}$ is $\Phi^{(0,1)}$, a rational Hodge structure 
$(V_\Q, h_{\Phi^{(1,0)}})$ is given on the $[F:\Q]$-dimensional vector space 
$V_\Q =F$ so that the field $F$ is contained in the endomorphism 
algebra ${\rm End}(F, h_{\Phi^{(1,0)}})$ (as we have reviewed in 
Lemma \ref{lemma:very-useful-2}). So, we can mass-produce examples of 
a complex torus with sufficiently many complex multiplications.  

Let us take the totally imaginary field $F \cong \Q[x]/(x^4-2x^2+2)$ 
as an example here. The four embeddings $F \hookrightarrow \C$ are 
given by 
\begin{align}
 \tau_{\epsilon \epsilon'}: x \longmapsto 
  \epsilon' 2^{\frac{1}{4}} e^{\epsilon \pi i/8}, \qquad 
   \epsilon \in \{ \pm 1\}, \quad \epsilon' \in \{ \pm 1\}.
\end{align}
There are two rational Hodge structures that can be introduced on 
the $[F:\Q]$-dimensional vector space $V_\Q = F$; one is for 
$\Phi^{(1,0)} = \{ \tau_{++}, \tau_{--} \}$ and the other for 
$\Phi^{(1,0)} = \{ \tau_{++}, \tau_{+-} \}$. To describe the Hodge 
group of the two rational Hodge structures, we use 
$\{ \eta_{I=1,2,3,4} \} = \{ 1,x,x^2, x^3\}$ as a basis of the extension field 
$F/\Q$; its dual basis $\{ \omega_{I=1,2,3,4} \}$ of 
$F/\Q$ is used as a basis for the vector space $V_\Q = F$ (as in 
Lemmas \ref{lemma:very-useful} and \ref{lemma:very-useful-2}). Endomorphisms  
$x, x^2, x^3 \in F^\times \subset [{\rm End}(V_\Q, h_{\Phi^{(1,0)}})]^\times
 \subset {\rm GL}(V_\Q)$ act on $V_\Q$ 
as 
\begin{align}
 x = \left( \begin{array}{cc|cc}
    & 1 & & \\ & & 1 & \\ \hline & & & 1 \\ -2 & & 2 & \end{array} \right), 
 \quad 
 x^2 = \left( \begin{array}{cc|cc}
   & & 1 & \\ & & & 1 \\ \hline -2 & & 2 & \\ & -2 & & 2 \end{array} \right), 
 \quad 
 x^3 = \left( \begin{array}{cc|cc}
   & & & 1 \\ -2 & & 2 & \\ \hline & -2 & & 2 \\ -4 & & 2 & \end{array} \right)
\end{align}
in the matrix representation when the basis $\{ \omega_I \}$ is used for $V_\Q$. Using this convention, we can compute the Hodge group of each of the 
rational Hodge structures by following the definition honestly: 
\begin{align}
 {\rm Hdg}(V_\Q, h_{\{ \tau_{++}, \tau_{--}\}}) & \;
   = \left\{ A+Bx+Dx^3 \; | \; B^2+4BD+2D^2=0, \; {\rm Nm}_{F/\Q}=1  \right\}, 
   \label{eq:Hdg-grp-ex-A} \\
 {\rm Hdg}(V_\Q, h_{\{\tau_{++}, \tau_{+-} \}}) & \;
   = \left\{ A + C x^2  \; | \; A^2+2AC + 2C^2 =1  \right\} , 
   \label{eq:Hdg-grp-ex-B} 
\end{align}
where ${\rm Nm}_{F/\Q} = A^4 -2A^2 (B^2-2D^2) + 8(BD)^2$. 
Note that all the coefficients of the defining equations in 
(\ref{eq:Hdg-grp-ex-A}, \ref{eq:Hdg-grp-ex-B}) are in the field $\Q$, 
so both are algebraic subgroups of ${\rm GL}(V_\Q)$ defined over $\Q$. 
Both are subgroups of the group ${\rm Res}_{F/\Q}(\mathbb{G}_m) \subset 
{\rm GL}(V_\Q)$ (roughly speaking, it is $F^\times$; 
see \cite[A.3]{Okada:2023udq} for more pedagogical explanations), so 
both are commutative, as in the property/condition (ii) in 
p. \pageref{pg:CM-def-1n2}. 

For the rational Hodge structure $(V_\Q, h_{\{ \tau_{++}, \tau_{+-} \}})$, 
the set of real points of the Hodge group (\ref{eq:Hdg-grp-ex-B})
is compact; it is an ellipse $\{ (A, C) \in \R^2 \; | \; A^2+2AC+2C^2=1\}$. 
This Hodge group has the property (iii) in p. \pageref{pg:CM-def-1n2}. 
For the rational Hodge structure $(V_\Q, h_{\{ \tau_{++}, \tau_{--} \}})$, on 
the other hand, the set of real points of the Hodge group 
(\ref{eq:Hdg-grp-ex-A}) is non-compact in the limit 
\begin{align}
 \frac{B}{D} = (-2-\sqrt{2}), \qquad 
 \frac{A^2}{D^2} \simeq 4(\sqrt{2}+1), \qquad 
  A, B, D \in \R, \quad |A|, |B|, |D| \rightarrow +\infty.
\end{align}
The property (iii) is not satisfied. As we argued in section \ref{ssec:pol}, 
the properties (i) and (ii) are guaranteed for rational Hodge structures 
with sufficiently many complex multiplications, but the property (iii) is 
not, in general. 

It is not an accident that the set of real points 
${\rm Hdg}(\R)$ for $(V_\Q, h_{\{ \tau_{++}, \tau_{+-} \}})$ is compact, in fact. 
The degree-4 field $F = \Q[x]/(x^4-2x^2+2)$ contains a degree-2 subfield 
$K := \Q[\theta]/(\theta^2-2\theta +2) \cong \Q(i)$, embedded through 
$K \ni \theta \mapsto x^2 \in F$. The pair $(F, \{ \tau_{++}, \tau_{+-}\})$
is not primitive in the sense of \cite[\S8]{shimura2016abelian}, because 
both of the embeddings $\tau_{++}$ and $\tau_{+-}$ of $F$ become identical 
when restricted to the subfield $K \subset F$. This reveals that the 
weight-1 rational Hodge structure $(V_\Q, \{ \tau_{++}, \tau_{+-} \})$ 
is Hodge isomorphic to that of $H^1(E\times E;\Q)$ of an elliptic curve 
$E$ with CM by $\Q(i)$. The degree-4 number field $F = \Q[x]/(x^4-2x^2+2)$ 
is yet another way to choose a subalgebra ${\cal K} \subset M_2(\Q(i))$
in Ex. \ref{ex:ExE-CM}; cf also the comments just before 
Rmk. \ref{rmk:build-pol-from-CMfield}.
So, the rational Hodge structure 
$(V_\Q, \{ \tau_{++}, \tau_{+-} \})$ is polarizable in fact; the property 
(iii) is therefore guaranteed in this case as discussed in 
section \ref{ssec:pol}.  
\end{exmpl}
%

\section{Details of the Analysis}
\label{sec:details}

This appendix collects technical computations whose results are used 
in the main text. Lemmas \ref{lemma:GV-2-RCFT-caseBC-even} 
and \ref{lemma:GV-2-RCFT-caseAprm-even} are quoted in section \ref{sssec:5classes-B+iomega}, 
while Lemmas \ref{lemma:list-SYZ-mirror-caseBC}, \ref{lemma:list-SYZ-mirror-caseAprm} 
and \ref{lemma:list-SYZ-mirror-caseA} are used in section \ref{ssec:list-SYZ-mirror}.
Section \ref{ssec:one-more} uses the result of Lemma \ref{lemma:mirror-H1-HdgStr-caseBC}.

\subsection{Case (B, C)}

\begin{lemma}
\label{lemma:GV-2-RCFT-caseBC-even}
Let $(T^4;I)$ be a CM-type abelian surface in the case (B, C), 
and $(B+i\omega)$ be in ${\cal H}^2(T^4_I)\otimes
\tau^r_{(20)}(K^r)$; the reflex field $K^r$ and its embeddings 
are described in Discussion \ref{statmnt:rflx-field-levelN}.
Suppose that $(B+i\omega)$ introduces a polarized rational Hodge structure 
on $T_M^v\otimes \Q \subset A(T^4_I)\otimes \Q$ and the pairing (\ref{eq:pairing-Mukai}), 
in the way described in Lemma \ref{lemma:mirrorhHdgH2L2=vHdgOnAlg}/section \ref{ssec:1st-step-4-converse}. 
Then $(B+i\omega)$ must be of the form 
\begin{align*}
 (B+i\omega)/2 & \; = Z_1 e_1 + Z_2^\pm e_2, 
   \qquad \qquad \qquad \qquad \qquad \quad (\ref{eq:BC-B+iomega})     
\end{align*}
where 
\begin{align}
 Z_1 & \; := \tau^r_{++}\left( A+\frac{C'}{2}\xi^r +\frac{D'}{2d}\frac{2qd}{\xi^r}\right), \qquad 
 Z_2^{\pm} := \tau^r_{++} \left( \widetilde{A} \pm \frac{D'}{2}\xi^r
   \pm \frac{C'}{2}\frac{2qd}{\xi^r} \right), \\
 & \qquad \left( A, \widetilde{A}, C', D' \in \Q, \quad 
   (C', D') \neq (0,0) \right). \nonumber  
\end{align}
\end{lemma}
\proof
%
%
The fact that $\mho := e^{(B+i\omega)/2}$ is the only generator 
of the Hodge $(2,0)$ component of the CM-type Hodge structure 
on $T^v_M\otimes \Q$, with the CM field $K^r$ and embedding 
$\tau^r_{(20)}=\tau^r_{++}$, implies that there must be a basis 
$\{ 1, \eta_1, \eta_2, \eta_4\}$ of $K^r/\Q$, so that 
\begin{align}
 \mho & \; = \tau^r_{++} \left[ 1 + e_1 \eta_1 + e_2 \eta_2
     + (\hat{\alpha}^1\hat{\beta}_1\hat{\alpha}^2\hat{\beta}_2) \eta_4
    \right], 
\end{align}
where a rational basis $\{ e_1, e_2 \}$ of ${\cal H}^2(M_I)$
is the one introduced in (\ref{eq:def-ratBasis-H2-caseBC});  
we must set $\eta_4 = (d\eta_1^2-\eta_2^2) \in K^r$ so that 
$(\mho,\mho)=0$ in the pairing (\ref{eq:pairing-Mukai}). 
So, there are eight rational parameters for $\eta_1, \eta_2 \in K^r$, 
for the moment. The Hodge (0,2) component should be given by 
\begin{align}
  \overline{\mho} = \tau^r_{+-}\left[ 1 + e_1 \eta_1 + e_2 \eta_2
  + (\hat{\alpha}^1\hat{\beta}_1\hat{\alpha}^2\hat{\beta}_2) \eta_4 \right], 
\end{align}
and the (1,1) components by the two vectors
\begin{align}
 \Sigma & \; =  \tau^r_{-+}\left[ 1 + e_1 \eta_1 + e_2 \eta_2
  + (\hat{\alpha}^1\hat{\beta}_1\hat{\alpha}^2\hat{\beta}_2) \eta_4 \right], \\
 \overline{\Sigma} & \; = 
    \tau^r_{--}\left[ 1 + e_1 \eta_1 + e_2 \eta_2
  + (\hat{\alpha}^1\hat{\beta}_1\hat{\alpha}^2\hat{\beta}_2) \eta_4 \right].  
\end{align}

This Hodge decomposition must be polarized with respect to 
(\ref{eq:pairing-Mukai}). The condition $(\mho, \mho)=0$
is built in by construction, $\mho = e^{2^{-1}(B+i\omega)}$. The 
remaining non-trivial information from the polarization is that 
$(\mho, \Sigma)=0$ and $(\mho, \overline{\Sigma})=0$. The two conditions 
are equivalent to 
\begin{align}
  -2^{-1} \left( \tau^r_{++}(X) - \tau^r_{-\pm}(X),
    \tau^r_{++}(X) - \tau^r_{-\pm}(X) \right)_{{\cal H}^2} = 0
\end{align}
for $X=e_1\eta_1 + e_2 \eta_2$, using just the pairing in ${\cal H}^2(M_I)$; 
those conditions are further rewritten as 
\begin{align}
  d \left( \tau^r_{++}(\eta_1)-\tau^r_{-\pm}(\eta_1) \right)^2
 - \left( \tau^r_{++}(\eta_2)-\tau^r_{-\pm}(\eta_2) \right)^2 = 0
   \label{eq:cond-vert-Pol-caseBC}
\end{align}
in the normal closure of the number field $K^r$. 

The eight rational parameters for $\eta_{1,2} \in K^r$, that is, 
$A,B,C,D,\widetilde{A}, \widetilde{B}, \widetilde{C}, \widetilde{D} \in \Q$ in 
\begin{align}
 \eta_1 =: A + B y' + C\xi^r + D \xi^r y', \qquad 
 \eta_2 =: \widetilde{A} + \widetilde{B} y'
     + \widetilde{C} \xi^r + \widetilde{D} \xi^ry', 
\end{align}
should satisfy the conditions (\ref{eq:cond-vert-Pol-caseBC}). 
Straightforward computation translates the conditions to 
\begin{align}
 d BC = \widetilde{B}\widetilde{C}, \quad 
 d BD = \widetilde{B}\widetilde{D}, \quad 
 d(D^2d'-C^2) = (\widetilde{D}^2d' -\widetilde{C}^2), 
\end{align}
along with 
\begin{align}
 d\left[ d'(B^2-2CD)+p(C^2+d'D^2)\right] = 
 \left[d'(\tilde{B}^2-2\widetilde{C}\widetilde{D})
        + p(\widetilde{C}^2+d'\widetilde{D}^2)\right].
\end{align}
There are four conditions on the eight parameters. 

First, one can immediately see that the rational parameters $A$ 
and $\widetilde{A}$ dropped out. So any $A,\widetilde{A} \in \Q$
has no conflict with the condition (\ref{eq:cond-vert-Pol-caseBC})
for the consistency of the Hodge structure (of $B+i\omega$) with 
the polarization (\ref{eq:pairing-Mukai}). 

Second, we prove that $\widetilde{B}=0$ by contradiction. 
If $\widetilde{B}\neq 0$, then $\widetilde{C}$ and $\widetilde{D}$ 
can be solved in terms of $C$, $D$ and $B/\widetilde{B}$. Then
\begin{align}
   (D^2d'-C^2)\left( \frac{B^2}{\widetilde{B}^2}d-1\right)=0.
\end{align}
This is a contradiction\footnote{
If $D=C=\widetilde{D}=\widetilde{C}=0$, then $[T^v_M\otimes \C]^{(2,0)}
=[T^v_M \otimes \C]^{(0,2)} = \C \subset T^v_M\otimes \C$. 
This is not appropriate as a Hodge decomposition. In physics terminology,
this corresponds to $\omega = 0$, and ${\rm volume}(T^4)=0$.  
} %
because neither $d$ nor $d'$ is a square of a rational number. 

Thirdly, $\widetilde{B}=0$ implies that either $B=0$ or $C=D=0$ holds
true. The latter is not possible, however, 
because $\widetilde{D}^2d'-\widetilde{C}^2=0$ would follow, although $d'$ 
is not a square of a rational number. So, $B=0$. We have now proved 
that the $B$-field is 
\begin{align}
  \tau^r_{++}\left[ e_1 (A + By')
    + e_2 (\widetilde{A}+\widetilde{B}y')\right] = A e_1 + \widetilde{A}e_2
\end{align}
for free $A,\widetilde{A} \in \Q$. This is the same as saying that the 
$B$-field is in ${\cal H}^2(M_I)$.
The rationality condition of the $B$-field (\ref{eq:cond-GnB-rational}) 
follows from conditions 1--4 and a part of 5 in Thm. \ref{thm:forT4}. 

Next, change the parametrization as follows.
\begin{align}
  C = \frac{1}{2}\left( C' + \frac{p D'}{qd}\right), \quad 
  D = \frac{D'}{2qd}, \quad 
 \widetilde{C} = \frac{1}{2}\left( \widetilde{C}'
      + \frac{p}{q} \widetilde{D}'\right), \quad \widetilde{D}= \frac{\widetilde{D}'}{2q}, 
\end{align}
or equivalently, 
\begin{align}
  \xi^r(C+Dy') = D' \frac{q}{\xi^r} + \frac{C'}{2}\xi^r, \qquad 
 \xi^r(\widetilde{C} + \widetilde{D} y') =
       \frac{\widetilde{C}'}{2}\xi^r+\widetilde{D}'\frac{qd}{\xi^r} .
\end{align}
Then the remaining two conditions on $C, D, \widetilde{C}, \widetilde{D}$ 
are rewritten as 
\begin{align}
  d(C')^2 + \frac{2p}{q}(C'D') + (D')^2 & \; = 
        d (\widetilde{D}')^2+\frac{2p}{q}\widetilde{C}'\widetilde{D}'
           +(\widetilde{C}')^2 , \\
 (C')^2 pd + (D'C')2qd + (D')^2 p & \; = 
   (\widetilde{D}')^2 dp + (\widetilde{D}'\widetilde{C}') 2qd
      + (\widetilde{C}')^2 p . 
\end{align}
So, this is equivalent to 
\begin{align}
  D' C' = \widetilde{C}' \widetilde{D}', \qquad 
  d (C')^2 + (D')^2 = d (\widetilde{D}')^2 + (\widetilde{C}')^2. 
\end{align}
This coupled quadratic equations seem to allow two possibilities, 
\begin{align}
 \frac{\widetilde{C'}}{\widetilde{D}'} = d \frac{C'}{D'}, \qquad 
 \frac{\widetilde{C}'}{\widetilde{D}'} = \frac{D'}{C'}, 
\end{align}
including $D'=\widetilde{D}'=0$ and 
$\widetilde{D}'=C'=0$, respectively. 
The first case is impossible, because 
$(\widetilde{C}')^2 = d (C')^2$ is a 
contradiction for the parameters $C', \widetilde{C}' \in \Q$ for 
$d$ that is not a square. The only option is 
\begin{align}
  (C', D') = (\widetilde{D}', \widetilde{C}'), 
  \label{eq:cond-polHstr-CM-to-kahler-OK}
\end{align}
and 
\begin{align}
     (C', D') = - (\widetilde{D}', \widetilde{C}').
   \label{eq:cond-polHstr-CM-to-kahler-opp}
\end{align}
There are two kinds of solutions, (\ref{eq:cond-polHstr-CM-to-kahler-OK}) 
and (\ref{eq:cond-polHstr-CM-to-kahler-opp}), for the Hodge structure 
on $[T^v_M\otimes \Q]$ to be compatible with the polarization 
(\ref{eq:pairing-Mukai}); for solutions of both kinds, 
there are two free rational parameters $C', D' \in \Q$ for $\omega$
(besides the two free parameters $A,\widetilde{A} \in \Q$ for the $B$-field). 
\qed 

\begin{lemma}
\label{lemma:list-SYZ-mirror-caseBC}
Let $(T^4;I)$ be a CM abelian surface in the case (B, C), and 
the complexified K\"{a}hler parameter $(B+i\omega)$ is in either 
(\ref{eq:BC-B+iomega};$+$) or (\ref{eq:BC-B+iomega};$-$). 
A vector subspace $\Gamma_{f\Q} \subset H_1(T^4;\Q)$ satisfies 
the condition (\ref{eq:cond-Enckevort}) if and only if it is of the form\footnote{
apologies for our poor choice of notations: $d \in \Gamma_{f\Q}$
here and a positive integer $d$ that generates $\Q(\sqrt{d})$
} %
\begin{align*}
\Gamma_{f\Q}={\rm Span}_\Q\{c,d\}, \quad
   & c:=c_1\alpha_1+c_2\beta^1+c_3\alpha_2+c_4\beta^2,
\nonumber\\
&d:=dc_3\alpha_1+dc_4\beta^1+c_1\alpha_2+c_2\beta^2,
   \qquad \qquad \qquad (\ref{eq:BC-GammafQ}) \\
&c_1,\ldots,c_4\in\Q, c_1^2+c_2^2+c_3^2+c_4^2\neq0. 
    \qquad \qquad \qquad (\ref{eq:BC-GammafQ-prm})
\end{align*}
\end{lemma}
\proof Let $\Gamma_f={\rm Span}_\Q\{c',d'\}$ be such an $n$-dimensional  
subspace of $H_1(T^{2n};\Q)$. The condition (\ref{eq:cond-Enckevort}) 
is then equivalent to
\begin{align}
e_1(c',d')=e_2(c',d')=0,
\label{eq:BC-SYZcond-e1e2}
\end{align}
in the case of both (\ref{eq:BC-B+iomega};$+$) and (\ref{eq:BC-B+iomega};$-$);
we have used the fact that the volume of $T^4$ must be positive 
($\omega\neq0$, i.e.\ $(C',D')\neq(0,0)$), and that the integer $d$ is 
square-free.

Let us rewrite the condition (\ref{eq:BC-SYZcond-e1e2}) in a way useful for later analysis, by 
parametrizing the generators $c',d'$ of $\Gamma_{f\Q}$ as
\begin{align}
c':=\ &c_1\alpha_1+c_2\beta^1+c_3\alpha_2+c_4\beta^2,\\
d':=\ &d_1\alpha_1+d_2\beta^1+d_3\alpha_2+d_4\beta^2,\\
&c_1,\ldots,c_4,d_1,\ldots,d_4\in\Q\,.
\end{align}
To proceed, note that there are linear combinations of $e_1$ and $e_2$ that are decomposable:
\begin{align}
e_1+\sqrt{d}e_2=(\hat{\alpha}^1+\sqrt{d}\hat{\alpha}^2)(\hat{\beta}_1+\sqrt{d}\hat{\beta}_2)=\hat{\gamma}^1\hat{\delta}_1,\\
e_1-\sqrt{d}e_2=(\hat{\alpha}^1-\sqrt{d}\hat{\alpha}^2)(\hat{\beta}_1-\sqrt{d}\hat{\beta}_2)=\hat{\gamma}^2\hat{\delta}_2,
\end{align}
where 
\begin{align}
\hat{\gamma}^1:=\hat{\alpha}^1+\sqrt{d}\hat{\alpha}^2,\quad & \hat{\delta}_1:=\hat{\beta}_1+\sqrt{d}\hat{\beta}_2,\\
\hat{\gamma}^2:=\hat{\alpha}^1-\sqrt{d}\hat{\alpha}^2,\quad & \hat{\delta}_2:=\hat{\beta}_1-\sqrt{d}\hat{\beta}_2.
\end{align}
The condition (\ref{eq:BC-SYZcond-e1e2}) is therefore equivalent to 
$\hat{\gamma}^1\hat{\delta}_1(c',d')=\hat{\gamma}^2\hat{\delta}_2(c',d')=0$.
This translates to 
\begin{align}
&\left\{\begin{array}{l}
(c_1+\sqrt{d}c_3):(c_2+\sqrt{d}c_4)=(d_1+\sqrt{d}d_3):(d_2+\sqrt{d}d_4)\\
(c_1-\sqrt{d}c_3):(c_2-\sqrt{d}c_4)=(d_1-\sqrt{d}d_3):(d_2-\sqrt{d}d_4)
\end{array}\right..
\label{eq:BC-SYZcond-cd}
\end{align}

The generator $d'$ for a given $c'$ is therefore constrained by the relation (\ref{eq:BC-SYZcond-cd}); let us see how, by translating the relation (\ref{eq:BC-SYZcond-cd}) further. 
An immediate consequence of the relation (\ref{eq:BC-SYZcond-cd}) is that there exist $k_1,k_2\in\C$ 
such that $d'$ can be rewritten as
\begin{align}
d'=\,&\frac{1}{2}\Bigl((c_1+\sqrt{d}c_3)k_1+(c_1-\sqrt{d}c_3)k_2\Bigr)\alpha_1+\frac{1}{2\sqrt{d}}\Bigl((c_1+\sqrt{d}c_3)k_1-(c_1-\sqrt{d}c_3)k_2\Bigr)\alpha_2
\nonumber\\
&+\frac{1}{2}\Bigl((c_2+\sqrt{d}c_4)k_1+(c_2-\sqrt{d}c_4)k_2\Bigr)\beta^1+\frac{1}{2\sqrt{d}}\Bigl((c_2+\sqrt{d}c_4)k_1-(c_2-\sqrt{d}c_4)k_2\Bigr)\beta^2.
\label{eq:BC-d'-rational}
\end{align}
Since the generator $d'$ is an element of $H_1(T^4;\Q)$, the coefficients in (\ref{eq:BC-d'-rational}) must be 
rational numbers.
It follows from this fact and some easy calculation that $k_1,k_2\in\Q(\sqrt{d})$, and moreover
\begin{align}
&k_1=a+b\sqrt{d},\ k_2=a-b\sqrt{d}\\
&(a,b\in\Q,(a,b)\neq(0,0)).
\end{align}
So, the generator $d'$ has to be of the form 
\begin{align}
d'=(ac_1+bdc_3)\alpha_1+(bc_1+ac_3)\alpha_2+(ac_2+bdc_4)\beta^1+(bc_2+ac_4)\beta^2
\label{eq:BC-d'-result}
\end{align}
for the conditions (\ref{eq:BC-SYZcond-e1e2}, \ref{eq:BC-SYZcond-cd}) to be satisfied. 
This (\ref{eq:BC-d'-result}) is also sufficient. 
In the statement of this Lemma, $c=c'$ and $d = d' -a c'$.

%
%
\qed

\begin{lemma}
\label{lemma:mirror-H1-HdgStr-caseBC}
Let $(T^4;I)$ be a CM-type abelian surface in the case (B, C),  
$(B+i\omega)$ a complexified K\"{a}hler form on $(T^4;I)$ in one 
of the two classes (\ref{eq:BC-B+iomega};$+$) and (\ref{eq:BC-B+iomega};$-$). 
Choose one $\Gamma_{f\Q} \subset H_1(T^4;\Q)$ from (\ref{eq:BC-GammafQ}), fix 
one $\Gamma_b \subset H_1(T^4;\Q)$, and think of the T-duality 
along $\Gamma_f$ while the directions $\Gamma_b$ are fixed; 
let $(T^4_\circ;G^\circ, B^\circ;I^\circ)$ be the geometric data of the 
corresponding SYZ-mirror SCFT. 
In the rational Hodge structure $(H^1(T^4_\circ;\Q), I^\circ)$, its Hodge 
(1,0) component is generated by the two vectors 
in (\ref{eq:BC-mirror-hol-basis-vs-rat-basis}). 
\end{lemma}
\proof Note, first, that 
\begin{align}
 W^3 := g^*(H^3(T^4_\circ;\Q)) & \; = 
 {\rm Span}_\Q \left\{ \hat{e}, \hat{f},
      \hat{c}\hat{e}\hat{f}, \hat{d}\hat{e}\hat{f} \right\} = 
{\rm Span}_\Q \{ \hat{e}, \hat{f}, e_1\hat{f}, e_2\hat{f} \};  
\end{align}
The vector space $g^*(H^1(T^4_\circ;\Q)) \subset H^{\rm odd}(T^4;\Q)$ 
has a basis $\{ \hat{c}, \hat{d}, \hat{c}\hat{d}\hat{e}, 
\hat{c}\hat{d}\hat{f}\}$; there is also an advantage in dealing with 
the vector space $W^1/W^3$ instead, where 
\[
W^1 := g^*(H^1(T^4_\circ;\Q)\oplus H^3(T^4_\circ;\Q)) \subset H^{\rm odd}(T^4;\Q), 
\]
because $W^1/W^3$ depends only 
on the choice of the directions $\Gamma_{f\Q}$ in which T-duality 
is taken, not on the directions $\Gamma_{b\Q}$ that are fixed. 
The vector space $W^1/W^3$ has a basis 
%
\begin{align}
 \left\{ [\hat{c}'], \;  [\hat{d}'], \; [\hat{c}'\hat{d}'\hat{e}],  
   [\hat{c}'\hat{d}'\hat{f}] \right\}
  = \left\{ \hat{c}'+W^3, \;  \hat{d}'+W^3, 
  \; \hat{c}'\hat{d}'\hat{e} + W^3,  
   \hat{c}'\hat{d}'\hat{f} + W^3  \right\}. 
\end{align}
%
The isomorphism $H^1(T^4_\circ;\Q) \hookrightarrow W^1 \rightarrow W^1/W^3$, 
where the first one is $g^*$ and the second one the projection, is 
proportional to 
\begin{align}
  \left( c, \; d, \; \hat{e}, \; \hat{f} \right) \longmapsto
 \left( \frac{[\hat{d}']}{c_1^2-dc_3^2}, \; \frac{-[\hat{c}']}{c_1^2-dc_3^2}, 
       \; \frac{[\hat{c}'\hat{d}'\hat{e}]}{(c_1^2-dc_3^2)^2},
       \; \frac{[\hat{c}'\hat{d}'\hat{f}]}{(c_1^2-dc_3^2)^2} \right). 
\end{align}

The Hodge (1,0) component of $(H^1(T^4_\circ;\Q), I^\circ)$ is such that 
is identified under the isomorphism above with the subspace of 
$(W^1/W^3)\otimes \C$ generated by the two vectors
\begin{align}
  \mho \hat{c}' + W^3
      = \hat{c}' + Z_1 e_1\hat{c}' + Z_2^{\pm} e_2 \hat{c}' + W^3, 
  \qquad 
  \mho \hat{d}' + W^3
      = \hat{d}' + Z_1 e_1 \hat{d}' + Z_2^{\pm} e_2 \hat{d}' + W^3. 
\end{align}
One may reorganize these basis elements of the Hodge component over $\C$
to 
\begin{align}
& 
\begin{pmatrix}
[\mho\hat{c}'] & [\mho\hat{d}']
\end{pmatrix}
\begin{pmatrix}
1 & 1\\
\mp\sqrt{d} & \pm\sqrt{d}
\end{pmatrix}\\
&=\begin{pmatrix}
[\hat{c}'] & [\hat{d}'] & -[e_1\hat{d}'] & -[e_2\hat{d}']
\end{pmatrix}
\begin{pmatrix}
1 & 1\\
\tau_{++}(\mp y) & \tau_{-+}(\mp y) \\
\tau_{++}(\Xi_\pm) & \tau_{-+}(\Xi_\pm)\\
\tau_{++}(\pm\Xi_\pm y) & \tau_{-+}(\pm\Xi_\pm y)
\end{pmatrix}
\label{eq:BC-W1-CM}\\
&=
\begin{pmatrix}
[\hat{c}'] & [\hat{d}'] & \frac{[\hat{c}'\hat{d}'\hat{e}]}{(c_1^2-dc_3^2)^2} & \frac{[\hat{c}'\hat{d}'\hat{f}]}{(c_1^2-dc_3^2)^2}
\end{pmatrix}
\begin{pmatrix}
1 & & & \\
& 1 & & \\
& & c_1 & c_3\\
& & dc_3 & c_1
\end{pmatrix}
\begin{pmatrix}
1 & 1\\
\tau_{++}(\mp y) & \tau_{-+}(\mp y) \\
\tau_{++}(\Xi_\pm) & \tau_{-+}(\Xi_\pm)\\
\tau_{++}(\pm\Xi_\pm y) & \tau_{-+}(\pm\Xi_\pm y)
\end{pmatrix},
\label{eq:BC-mhoc'mhod'-c'd'c'd'ec'd'f-CM}
\end{align}
where 
\begin{align}
\Xi_\pm:=\tilde{A}\pm Ay\pm D'x\pm C'xy\in K;
\label{eq:Xi}
\end{align}
%
%
The sign choice below ($-$ in $\pm$ and $+$ in $\mp$) is for the 
case (\ref{eq:BC-B+iomega};$-$) and the sign choice above for the 
case (\ref{eq:BC-B+iomega}; $+$).

Therefore, the Hodge (1,0) component of $(H^1(T^4_\circ;\Q), I^\circ)$ has 
a basis obtained by mapping a basis above by the isomorphism 
$H^1(T^4_\circ;\Q) \leftrightarrow W^1/W_3$ above. 
Namely, we may use 
%
\begin{align}
\begin{pmatrix}
dz^{1'}_\circ & dz^{2'}_\circ
\end{pmatrix}
=&
\begin{pmatrix}
-s_0d & s_0c & \hat{e} & \hat{f}
\end{pmatrix}
\begin{pmatrix}
1 & & & \\
& 1 & & \\
& & c_1 & c_3\\
& & dc_3 & c_1
\end{pmatrix}
\begin{pmatrix}
1 & 1\\
\tau_{++}(\mp y) & \tau_{-+}(\mp y) \\
\tau_{++}(\Xi_\pm) & \tau_{-+}(\Xi_\pm)\\
\tau_{++}(\pm\Xi_\pm y) & \tau_{-+}(\pm\Xi_\pm y)
\end{pmatrix}   \nonumber \\
=&
\begin{pmatrix}
c & d & \hat{e} & \hat{f}
\end{pmatrix}
\begin{pmatrix}
\tau_{++}(\mp s_0y) & \tau_{-+}(\mp s_0y) \\
- s_0 & - s_0 \\
\tau_{++}((c_1\pm c_3y)\Xi_\pm) & \tau_{-+}((c_1\pm c_3y)\Xi_\pm)\\
\tau_{++}((dc_3\pm c_1y)\Xi_\pm) & \tau_{-+}((dc_3\pm c_1y)\Xi_\pm)
\end{pmatrix},
\qquad \qquad (\ref{eq:BC-mirror-hol-basis-vs-rat-basis}) \nonumber 
\end{align}
where $s_0 := (c_1^2-dc_3^2)$. 

It is obvious from this expression (cf.\ Lemma \ref{lemma:very-useful-2})
that the endomorphism algebra of the rational Hodge structure 
$(H^1(T^4_\circ;\Q),I^\circ)$ is the degree-4 CM field $K$, and 
both $dz^{1'}_\circ$ and $dz^{2'}_\circ$ are eigenvectors of the action 
of $K$. That this rational Hodge structure is of CM-type had been guaranteed 
already in the argument ``Immediate Consequences'' in 
section \ref{ssec:list-SYZ-mirror}, but we also have an alternative proof 
by a brute force (down to earth) computations here. \qed 

\subsection{Case (A')}

\begin{lemma}
\label{lemma:GV-2-RCFT-caseAprm-even}
Let $(T^4;I)$ be a CM-type abelian surface in the case (A'), 
and $(B+i\omega)$ be in ${\cal H}^2(T^4_I)\otimes
\tau^r_{(20)}(K^r)$; the reflex field $K^r$ and its embeddings 
are described in Discussion \ref{statmnt:rflx-field-levelN}.
Suppose that the polarized rational Hodge structure 
introduced on $T_M^v \otimes \Q \subset A(T^4_I)\otimes \Q$ by 
$\mho =e^{(B+i\omega)/2}$, as discussed in section \ref{ssec:1st-step-4-converse}, 
is of CM-type in the way described in 
Lemma \ref{lemma:mirrorhHdgH2L2=vHdgOnAlg}. Then $(B+i\omega)$ must 
be of the form 
\begin{align*}
(B+i\omega)/2 = \left(A + C \sqrt{p_1} \right) \hat{\alpha}^1\hat{\beta}_1
    + \left(\widetilde{A} + \widetilde{C} \sqrt{p_2} \right)
       \hat{\alpha}^2\hat{\beta}_2,       \qquad 
    A, \widetilde{A} \in \Q, \;  C, \widetilde{C} \in \Q_{\neq 0},
   \qquad (\ref{eq:Aprime-B+iomega-RCFT})
\end{align*}
or 
\begin{align*}
(B+i\omega)/2 = \left(A + C \sqrt{p_2} \right) \hat{\alpha}^1\hat{\beta}_1
    + \left(\widetilde{A} + \widetilde{C} \sqrt{p_1} \right)
       \hat{\alpha}^2\hat{\beta}_2,       \qquad 
    A, \widetilde{A} \in \Q, \;  C, \widetilde{C} \in \Q_{\neq 0} .
    \qquad (\ref{eq:Aprime-B+iomega-nonRCFT})
\end{align*}
\end{lemma}
\proof
%
%
Because the Hodge structure on $T_M^v\otimes \Q$ is of CM-type, 
and $\mho := e^{2^{-1}(B+i\omega)}$ is the generator 
of the 1-dimensional Hodge (2, 0) component, $\mho$ must be 
in the form of 
\begin{align}
 \mho = \tau^r_{++} \left[ 1 + (\hat{\alpha}^1\hat{\beta}_1) \eta_1
      + (\hat{\alpha}^2\hat{\beta}_2) \eta_2
      + (\hat{\alpha}^1\hat{\beta}_1\hat{\alpha}^2\hat{\beta}_2)  \eta_4 \right]
\end{align}
for some basis $\{ 1,\eta_1, \eta_2, \eta_4\}$ of $K^r/\Q$. The property 
$(\mho, \mho) = 0$ implies that $\eta_4 = \eta_1 \eta_2$, so 
there are eight rational parameters for $\eta_1$ and $\eta_2$ at this 
moment. 

Let us parametrize the freedom by $A,B,C,D,\widetilde{A},
 \widetilde{B},\widetilde{C}, \widetilde{D} \in \Q$, where 
\begin{align}
 \eta_1 = A+By'+C\xi^r+D\xi^ry', \qquad 
 \eta_2 = \widetilde{A} +\widetilde{B}y'
     +\widetilde{C}\xi^r+\widetilde{D}\xi^ry'.
\end{align}
For the Hodge decomposition to be compatible with its polarization 
(\ref{eq:pairing-Mukai}), we impose $(\mho, \Sigma)=0$ and 
$(\mho, \overline{\Sigma})=0$. As a result, we obtain 
\begin{align}
  B\widetilde{B} + p_1 D \widetilde{D} = 0, \quad
  B\widetilde{D} + \widetilde{B}D=0, \\
  B\widetilde{B} p_2 + C\widetilde{C}=0, \quad 
  B\widetilde{C} + \widetilde{B}C = 0. 
\end{align}
Now, we have four conditions on the eight rational parameters. 

One can prove that $B=\widetilde{B}=0$ (or otherwise we should accept 
an unphysical zero-volume situation (such as $C=D=0$)); the proof is
similar to the case (B, C), so we omit the detail. The four 
conditions above are reduced to $D\widetilde{D}=0$ and $C\widetilde{C}=0$. 
So, there are two classes of solutions (apart from the zero-volume situations):
\begin{align}
  D=0, \quad \widetilde{C}=0, & \; {\rm so} \; 
   2^{-1}(B+i\omega) & \; = \hat{\alpha}^1\hat{\beta}_1 (A+C\sqrt{p_1})
    + \hat{\alpha}^2\hat{\beta}_2
               (\widetilde{A} + \widetilde{D}\sqrt{p_1}\sqrt{p_1p_2}), 
    \label{eq:cond-vHst-B+iw-OK}  \\
  C=0, \quad \widetilde{D}=0, & \; {\rm so} \; 
   2^{-1}(B+i\omega) & \; =
       \hat{\alpha}^1\hat{\beta}_1 (A+D\sqrt{p_1}\sqrt{p_1p_2})
     + \hat{\alpha}^2\hat{\beta}_2 (\widetilde{A} + \widetilde{C}\sqrt{p_1}).
    \label{eq:cond-vHst-B+iw-opp}  
\end{align}
Set $\widetilde{C}:= \widetilde{D}|p_1|$ in (\ref{eq:cond-vHst-B+iw-OK})
and $C := D |p_1|$ in (\ref{eq:cond-vHst-B+iw-opp}) to obtain 
(\ref{eq:Aprime-B+iomega-RCFT}) and (\ref{eq:Aprime-B+iomega-nonRCFT}), 
respectively.  \qed 

\begin{lemma}
\label{lemma:list-SYZ-mirror-caseAprm}
Let $(T^4;I)$ be a CM abelian surface in the case (B, C), and 
the complexified K\"{a}hler parameter $(B+i\omega)$ is in either 
(\ref{eq:Aprime-B+iomega-RCFT}) or (\ref{eq:Aprime-B+iomega-nonRCFT}). 
A vector subspace $\Gamma_{f\Q} \subset H_1(T^4;\Q)$ satisfies 
the condition (\ref{eq:cond-Enckevort}) if and only if it is of the form 
\begin{align*}
\Gamma_{f\Q}={\rm Span}_\Q\{c,d\},\quad &c:=c_1\alpha_1+c_2\beta^1,\quad d:=c_3\alpha_2+c_4\beta^2,
 \qquad \qquad \qquad (\ref{eq:Aprime-GammafQ})\\
&c_1,\ldots,c_4\in\Q, (c_1,c_2),(c_3,c_4)\neq(0,0).
 \qquad \qquad \qquad (\ref{eq:Aprime-GammafQ-prm})
\end{align*}
\end{lemma}
\proof 
To obtain the list of such $\Gamma_f$'s, we first list up all the rank-$n$ 
subspaces $\Gamma_{f\Q}\subset H_1(T^{2n};\Q)$ satisfying 
$\omega|_{\Gamma_{f\Q}\otimes\R}=0$ and $B|_{\Gamma_{f\Q}\otimes\R}=0$.
Then $\Gamma_{f\Q}\cap H_1(T^{2n};\Z)$ for these $\Gamma_{f\Q}$ constitute 
the list of all $\Gamma_f$ satisfying (\ref{eq:cond-Enckevort}).

Let $\Gamma_f={\rm Span}_\Q\{c',d'\}$ be such a rank-$(n=2)$ subspace of 
$H_1(T^{2n=4};\Q)$. We parameterize the generators by
\begin{align}
c':=\ &c_1\alpha_1+c_2\beta^1+c_3\alpha_2+c_4\beta^2,\\
d':=\ &d_1\alpha_1+d_2\beta^1+d_3\alpha_2+d_4\beta^2,\\
&c_1,\ldots,c_4,d_1,\ldots,d_4\in\Q\,.
\end{align}
The above condition $\omega|_{\Gamma_{f\Q}\otimes\R}=B|_{\Gamma_{f\Q}\otimes\R}=0$ 
for $B+i\omega$ in either (A')--(\ref{eq:Aprime-B+iomega-RCFT}) or 
(A')--(\ref{eq:Aprime-B+iomega-nonRCFT}) is then equivalent to
\begin{align}
&c_1:c_2=d_1:d_2 \quad \text{and} \quad c_3:c_4=d_3:d_4,
\end{align}
where we used the positive volume condition $C, \widetilde{C} \neq 0$. 
Therefore, we can always reorganize the basis $\{c', d'\}$ of $\Gamma_{f\Q}$ 
to $\{ c,d\}$ in (\ref{eq:Aprime-GammafQ}). 
\qed

\subsection{Case (A)}

\begin{lemma}
\label{lemma:list-SYZ-mirror-caseA}
Let $(T^4;I)$ be a CM abelian surface in the case (A), and 
the complexified K\"{a}hler parameter $(B+i\omega)$ is in  
(\ref{eq:A-B+iomega}). Then one can always find a rank-$2$ primitive 
subgroup $\Gamma_f$ of $H_1(T^4;\Z)$ that 
satisfies the condition (\ref{eq:cond-Enckevort}); i.e., there always 
exists a geometric SYZ-mirror. 
\end{lemma}
\proof  we have seen in (\ref{eq:parametr-Kahler-caseA}) 
how the K\"{a}hler form is parametrized. A rational $B=B^{\rm alg}$ 
in ${\cal H}(T^4_I)\otimes \Q$ should also have 4 parameters in $\Q$ 
because $\dim_\Q {\cal H}^2(T^4_I)=4$.
Details in Discussion \ref{statmnt:Tm} reveal that 
\begin{align}
 B^{\rm alg} & \; = \sqrt{p} (dz^1, dz^2) \wedge \left( \begin{array}{cc}
     h_1^B & c_1^B-c_2^B \sqrt{p} \\ c_1^B + c_2^B \sqrt{p} & h_2^B 
   \end{array} \right)
   \left( \begin{array}{c} d\bar{z}^{\bar{1}} \\ d\bar{z}^{\bar{2}}
   \end{array} \right), \qquad 
   h_{1,2}^B, c_{1,2}^B \in \Q.
\end{align}

With a straightforward computation, one can show that 
$\Gamma_{f\Q} = {\rm Span}_\Q \{ \alpha'_1, \alpha'_2\}$ with 
\begin{align}
  \alpha'_1 = \alpha_1, \quad 
  \alpha'_2 = \alpha_2 - \left\{ \frac{c_2}{h_1}- \frac{c_1}{h_1}
     \frac{(h_1^Bc_2-h_1c_2^B)}{(h_1^Bc_1-h_1c_1^B)} \right\} \beta^1
   - \frac{h_1^Bc_2-h_1c_2^B}{h_1^Bc_1-h_1c_1^B} \beta^2 
  \label{eq:SYZ-directn-caseA-gen}
\end{align}
satisfies the condition $\omega|_{\Gamma_{f\Q}}=0$ and $B|_{\Gamma_{f\Q}}=0$. 
In the case $h_1^Bc_1-h_1c_1^B=0$, the same conditions are satisfied 
when 
\begin{align}
  \alpha'_1 = \alpha_1 + \frac{1}{pv} \beta^1, \qquad 
  \alpha'_2 = \alpha_2 - \frac{c_1}{h_1}\left( v - \frac{1}{pv} \right) \beta^1
   + v \beta^2, \qquad v \in \Q, \; v \neq 0. 
   \label{eq:SYZ-directn-caseA-spc}
\end{align}
The T-duals in the directions (\ref{eq:SYZ-directn-caseA-gen},
 \ref{eq:SYZ-directn-caseA-spc}) are geometric SYZ-mirrors. 

The choices of $\Gamma_f$ above still come with a variety of choices 
of $\Gamma_b$, and moreover, there will be more choices of $\Gamma_f$ 
other than the one above; they are only meant to be examples. \qed






%
%
%

\bibliographystyle{alpha}             
\bibliography{complexMultiplication}  

\end{document}